\documentclass[10pt,preprint]{sigplanconf}
\pdfoutput=1

\usepackage{amsmath}
\usepackage{amssymb}
\usepackage{amsthm}
\usepackage{fancyhdr}
\usepackage{mathpartir}
\usepackage{xspace}
\usepackage{graphicx}
\usepackage{placeins} %
\usepackage{color,soul} %
\usepackage{verbatim}
\usepackage{enumerate}
\usepackage{booktabs}
\usepackage{subfigure}
\usepackage{tikz}
\usetikzlibrary{calc,fit,shapes,arrows,shadows,trees,backgrounds}
\usetikzlibrary{decorations.pathreplacing}
\usepackage[pdftex,breaklinks=true]{hyperref} %
\usepackage[compact]{titlesec}

\newcommand{\ttt}[1]{\texttt{#1}}

\newcommand{\out}[1] {}

\newcounter{codeLineCntr}

\newcommand{\codeLineL}[1]
 {\refstepcounter{codeLineCntr}\label{#1}{\thecodeLineCntr}}
\newcommand{\codeLineNN}{}    %
\newcommand{\codeLineLN}[1]{} %

\newenvironment{codeListing}
 {\setcounter{codeLineCntr}{0}
 \fontsize{9}{11}
  \fontsize{8}{8}
  \vspace{-.1in}
  \ttfamily\begin{tabbing}}
  {\end{tabbing}
   \vspace{-.1in}}

\newenvironment{codeListing9}
 {\setcounter{codeLineCntr}{0}
  \fontsize{9}{11}
  \vspace{-.075in}
  \ttfamily
  \begin{tabbing}}
 {\end{tabbing}
 \vspace{-.075in}
}

\newenvironment{codeListing10}
 {\setcounter{codeLineCntr}{0}
  \fontsize{10}{12}
  \ttfamily
  \begin{tabbing}}
 {\end{tabbing}
}

\newcommand{\fixedCodeFrame}[1]
{
\begin{center}
\fbox{
\parbox[t]{0.9\columnwidth}{
#1
}
}\end{center}
}

\reversemarginpar
\newif\ifnotes
\notestrue

\newcommand{\punt}[1]{}

\newcommand{\secref}[1]{Section~\ref{sec:#1}}

\newcommand{\secreftwo}[2]{Sections \ref{sec:#1} and~\ref{sec:#2}}

\newcommand{\appref}[1]{Appendix~\ref{app:#1}}
\newcommand{\figref}[1]{Figure~\ref{fig:#1}}
\newcommand{\figreftwo}[2]{Figures \ref{fig:#1} and~\ref{fig:#2}}

\newcommand{\tabref}[1]{Table~\ref{tab:#1}}

\newcommand{\lineref}[1]{line~\ref{line:#1}}
\newcommand{\linereftwo}[2]{lines \ref{line:#1} and~\ref{line:#2}}

\newcommand{\thmref}[1]{Theorem~\ref{thm:#1}}

\newcommand{\lemref}[1]{Lemma~\ref{lem:#1}}

\renewcommand{\eqref}[1]{Equation~(\ref{eq:#1})}

\newcommand{\proc}[1]{\ifmmode\mbox{\textsc{#1}}\else\textsc{#1}\fi}

  \newcommand{\func}[1]{\ifmmode\mathrm{#1}\else\textrm{#1}fi} %
\newcounter{remark}[section]

\newcommand{\myremark}[3]{
\refstepcounter{remark}
\[
\left\{
\sf 
\parbox{\columnwidth}{
{\bf {#1}'s remark~\theremark:} 
{#3}
}
\right.
\]
\marginpar{\bf {#2}.~\theremark}
}

\newcommand{\uremark}[1]{\myremark{Umut}{U}{#1}}

\newcommand{\mremark}[1]{\myremark{Matt}{M}{#1}}

\definecolor{lightgreen}{rgb}{.85,.95,.85}
\definecolor{lightblue}{rgb}{.85,.90,1}
\definecolor{lightred}{rgb}{.95,.85,.85}
\definecolor{lightgrey}{rgb}{.95,.95,.95}
\sethlcolor{lightgrey}

\newcommand{\hlm}[1]{\text{\hl{$#1$}}}
\newcommand{\hldiff}{\sethlcolor{lightgreen}\hlm}

\newcommand{\etnode}[1]{\textbf{#1}}

\newcommand{\TODO}[1]{\sethlcolor{lightred}\hl{{\bf TODO}}~#1}
\newcommand{\im}[1]{\ensuremath{#1}}

\newcommand{\stepsarrow}{\longrightarrow}

\renewcommand{\MathparLineskip}{\lineskiplimit=0.5em\lineskip=0.4em
  plus 0.1em}

\newcommand{\mycaptionrule}{}

\newcommand{\bnfdef}{\mathrel{\colon\!\!\!\colon\!\!\mathord{=}}}
\newcommand{\bnfalt}{\mathrel{\mid}}

\newcommand{\kw}[1]{\text{\bf #1}}

\newcommand{\seqof}[1]{\overline{#1}}
\newcommand{\empseq}{\left<\right>}

\renewcommand{\vec}[1]{\overline{#1}}

\newcommand{\descr}[1]{\text{#1}}

\newcommand{\cons}{\mathop{::}}
\newcommand{\append}{\mathop{@}}
\newcommand{\fromto}[3]{{#1}_{#2}^{#3}}

\theoremstyle{plain}
\newtheorem{thm}{Theorem}[section]
\newtheorem{lem}[thm]{Lemma}

\newtheorem*{cor}{Corollary}

\theoremstyle{definition}
\newtheorem{defn}{Definition}[section]

\theoremstyle{remark}

\newcommand{\DITTO}{\textsf{DITTO}\xspace}

\newcommand{\CEAL}{\textsf{CEAL}\xspace}
\newcommand{\IL}{\textsf{IL}\xspace}
\newcommand{\il}{\textsf{IL}\xspace}
\newcommand{\CEALIL}{\textsf{CEAL-IL}\xspace}
\newcommand{\C}{\text{C}\xspace}

\newcommand{\SRCC}{\textsf{C$_\textsf{src}$}\xspace}
\newcommand{\srcc}{\SRCC}
\newcommand{\DeltaML}{\textsf{DeltaML}\xspace}
\newcommand{\kone}{\texttt{eval\_right}\xspace}
\newcommand{\ktwo}{\texttt{eval\_op}\xspace}

\newcommand{\ilparen}[1]{\texttt{(}{#1}\texttt{)}}

\newcommand{\iluexp}{e^{\sf u}}
\newcommand{\iltexp}{e^{\sf t}}

\newcommand{\ilalloc}[1]{\kw{alloc}\ilparen{#1}}
\newcommand{\ilapp}[2]{{#1}\,\ilparen{#2}}

\newcommand{\ilfun}[3]{\kw{fun}~#1\kw{(}#2\kw{)}.#3}
\newcommand{\ilif}[3]{\kw{if}~{#1}~\kw{then}~{#2}~\kw{else}~{#3}}
\newcommand{\illet}[2]{\kw{let}~{#1}~\kw{in}~{#2}}
\newcommand{\illetnb}[1]{\kw{let}~{#1}~\kw{in}} %
\newcommand{\ilmemo}[1]{\kw{memo}~{#1}}
\newcommand{\ilpop}[1]{\kw{pop}~#1}
\newcommand{\ilprimop}[1]{\kw{$\oplus$}\ilparen{#1}}
\newcommand{\ilprimapp}[1]{\kw{primapp}\ilparen{\kw{$\oplus$},{#1}}}
\newcommand{\ilpush}[2]{\kw{push}~#1~\kw{do}~#2}

\newcommand{\iltuple}[2]{\ilparen{#1,\!#2}}

\newcommand{\ilread}[2]{\kw{read}\ilparen{#1[#2]}}
\newcommand{\ilvar}[2]{{#1}~\kw{=}~{#2}}
\newcommand{\ilupdate}[1]{\kw{update}~{#1}}
\newcommand{\ilwake}[1]{\ilupdate{#1}}
\newcommand{\ilwrite}[3]{\kw{write}\iltuple{{#1}[{#2}]}{#3}}
\newcommand{\ilprop}{\kw{prop}}

\newcommand{\emptyenv}{\varepsilon}
\newcommand{\storegarb}{\diamond}

\newcommand{\traction}[1]{\mbox{\textsf{#1}}}
\newcommand{\tralloc}[2]{\traction{A}_{#1,#2}}
\newcommand{\trread}[3]{\traction{R}_{#2[#3]}^{#1}}
\newcommand{\trwrite}[3]{\traction{W}_{#2[#3]}^{#1}}
\newcommand{\trmemo}[2]{\traction{M}_{{#1},{#2}}}
\newcommand{\trupdate}[2]{\traction{U}_{{#1},{#2}}}
\newcommand{\trwake}[2]{\trupdate{#1}{#2}}
\newcommand{\trpop}[1]{#1}
\newcommand{\trpush}[1]{({#1})}

\newcommand{\tr}{\im{T}}
\newcommand{\trcons}[2]{{#1}\!\cdot\!{#2}}

\newcommand{\pushmark}{\square}
\newcommand{\propmark}{\boxplus}
\newcommand{\undomark}{\boxminus}

\newcommand{\trx}{\Pi}
\newcommand{\trxcons}[2]{{#1}\!\cdot\!{#2}}
\newcommand{\trxpushmark}{\pushmark}
\newcommand{\trxpropmark}[1]{\propmark_{#1}}
\newcommand{\trxundomark}[1]{\undomark_{#1}}

\newcommand{\trz}[2]{\left<\smash{{#1}\statesep{#2}}\right>}

\newcommand{\stcons}[2]{{#1}\!\cdot\!{#2}}
\newcommand{\mkstframe}[2]{\lfloor {#1},{#2}\rfloor}

\newcommand{\emptystack}{\emptyenv} %

\newcommand{\statesep}{,}
\newcommand{\ilustate}[4]{{#1}\statesep{#2}\statesep{#3}\statesep{#4}}

\newcommand{\staterel}[3]{{#1} \sim^{#2} {#3}}

\newcommand{\ilusteprel}{\stackrel{\textsf{r}}{\stepsarrow}}

\newcommand{\ilustep}[6]{
  {#1}\statesep{#2}\statesep{#3} \ilusteprel
  {#4}\statesep{#5}\statesep{#6}}

\newcommand{\storestin}[3]{{#1}\statesep{#2}\statesep{#3}}
\newcommand{\storestout}[2]{{#1}\statesep{#2}}
\newcommand{\storestep}{\stackrel{s}{\stepsarrow}}

\newcommand{\ilustepm}[6]{
  {#1}\statesep{#2}\statesep{#3} \ilusteprel^\ast
  {#4}\statesep{#5}\statesep{#6}}

\newcommand{\ilustepmr}[3]{
  \ilusteprel^\ast {#1}\statesep{#2}\statesep{#3}}

\newcommand{\domof}[1]{\text{\sf dom}(#1)}

\newcommand{\trzip}[2]{\left<\smash{#1}, \smash{#2}\right>}

\newcommand{\trsteppop}[2]{(\sf Pop)}

\newcommand{\trstart}{\varepsilon}
\newcommand{\trend}{\varepsilon}

\newcommand{\kwprop}{\kw{prop}}

\newcommand{\refwrap}[1]{\textbf{#1}\xspace}

\newcommand{\refrefall}{\refwrap{R.1}--\refwrap{11}}
\newcommand{\refreffun}{\refwrap{R.1}}
\newcommand{\refrefvar}{\refwrap{R.2}}
\newcommand{\refrefift}{\refwrap{R.3}}
\newcommand{\refrefiff}{\refwrap{R.4}}
\newcommand{\refrefapp}{\refwrap{R.5}}
\newcommand{\refrefstore}{\refwrap{R.6}}
\newcommand{\refrefmemo}{\refwrap{R.7}}
\newcommand{\refrefwake}{\refwrap{R.8}}
\newcommand{\refrefpush}{\refwrap{R.9}}
\newcommand{\refrefpop}{\refwrap{R.10}}
\newcommand{\refrefv}{\refwrap{R.11}}

\newcommand{\refstorealloc}{\refwrap{S.1}}
\newcommand{\refstoreread}{\refwrap{S.2}}
\newcommand{\refstorewrite}{\refwrap{S.3}}

\newcommand{\refevalall}{\refwrap{E.0}--\refwrap{8}}
\newcommand{\refevaltoeval}{\refwrap{E.0}--\refwrap{6}}
\newcommand{\refevalu}{\refwrap{E.0}}
\newcommand{\refevalalloc}{\refwrap{E.1}}
\newcommand{\refevalread}{\refwrap{E.2}}
\newcommand{\refevalwrite}{\refwrap{E.3}}
\newcommand{\refevalmemo}{\refwrap{E.4}}
\newcommand{\refevalwake}{\refwrap{E.5}}
\newcommand{\refevalpush}{\refwrap{E.6}}
\newcommand{\refevalpop}{\refwrap{E.7}}
\newcommand{\refevalv}{\refwrap{E.8}}

\newcommand{\refpropeval}{\refwrap{P.E}}
\newcommand{\refevalprop}{\refwrap{E.P}}

\newcommand{\refpropall}{\refwrap{P.1}--\refwrap{8}}
\newcommand{\refproptoprop}{\refwrap{P.1}--\refwrap{6}}
\newcommand{\refpropalloc}{\refwrap{P.1}}
\newcommand{\refpropread}{\refwrap{P.2}}
\newcommand{\refpropwrite}{\refwrap{P.3}}
\newcommand{\refpropmemo}{\refwrap{P.4}}
\newcommand{\refpropwake}{\refwrap{P.5}}
\newcommand{\refproppush}{\refwrap{P.6}}
\newcommand{\refproppop}{\refwrap{P.7}}
\newcommand{\refpropv}{\refwrap{P.8}}

\newcommand{\refundoall}{\refwrap{U.1}--\refwrap{4}}
\newcommand{\refundoalloc}{\refwrap{U.1}}
\newcommand{\refundostep}{\refwrap{U.2}}
\newcommand{\refundopush}{\refwrap{U.3}}
\newcommand{\refundomark}{\refwrap{U.4}}

\newcommand{\reftracingall}{\refevalall,\refpropeval,\refevalprop,\refpropall,\refundoall}

\newcommand{\iltstate}[5]{{#2}\statesep{#1}\statesep{#3}\statesep{#4}\statesep{#5}}

\newcommand{\iltevalstep}{\stackrel{\textsf{t}}{\stepsarrow}}

\newcommand{\iltpropstep}{\iltevalstep}
\newcommand{\iltundostep}{\iltevalstep}
\newcommand{\iltevalpropstep}{\iltevalstep}
\newcommand{\iltpropevalstep}{\iltevalstep}

\newcommand{\iltsteprel}{\stackrel{\textsf{t}}{\stepsarrow}}

\newcommand{\iltstepm}[8]{
  {#1}\statesep{#2}\statesep{#3}\statesep{#4} \iltevalstep^\ast
  {#5}\statesep{#6}\statesep{#7}\statesep{#8}}

\newcommand{\iltbigstep}[6]{
  {#1}\statesep{#2}\statesep{#3} \Downarrow {#4}\statesep{#5}\statesep{#6}
}

\newcommand{\iltbigprop}[5]{
  {#1}\statesep{#2} \curvearrowright {#3}\statesep{#4}\statesep{#5}
}

\newcommand{\trrewind}{\circlearrowleft}
\newcommand{\rewindstep}[4] {
  {#1} ; {#2} \trrewind {#3} ; {#4}
}
\newcommand{\rewindstepm}[4] {
  {#1} ; {#2} \trrewind^\ast {#3} ; {#4}
}

\newcommand{\rcmd}{{\alpha_{\sf r}}}

\newcommand{\tcmd}{{\alpha_{\sf t}}}

\newcommand{\FreeLabels}[1]{\textsf{FF}(#1)\xspace}

\newcommand{\isdpsenv}[2]{#2 \propto #1}

\newcommand{\arityof}[1]{\mathsf{Arity}(#1)}

\newcommand{\DPSexp}[2]{[\![#1]\!]_{#2}\xspace}
\newcommand{\DPSenv}[1]{[\![#1]\!]}
\newcommand{\DPSfun}[2]{\mathsf{D}_{#1}^{#2}}

\newcommand{\storeagnostic}{store agnostic\xspace}
\newcommand{\compositionallystoreagnostic}{compositionally store agnostic\xspace}

\newcommand{\okay}[1]{#1~\textsf{ok}\xspace}
\newcommand{\fsc}[1]{#1~\textsf{fsc}\xspace}

\newcommand{\nongarbof}[1]{#1|_{\text{\sf gc}}}

\newcommand{\SA}{SA\xspace}
\newcommand{\CSA}{CSA\xspace}
\newcommand{\CSAexp}[1]{\textsf{isCSA}(#1)\xspace}
\newcommand{\LiveVars}[1]{\textsf{LV}(#1)\xspace}
\newcommand{\sa}[3]{\mathsf{SA}(#3, #1, #2)}
\newcommand{\csa}[3]{\mathsf{CSA}(#3, #1, #2)}
\newcommand{\noreuse}[1]{\mathsf{noreuse}(#1)}

\newcommand{\numof}[2]{\#_{#1}({#2})}

\newcommand{\dropum}[1]{\mathsf{drop}_{\undomark}({#1})}
\newcommand{\trlast}[1]{\mathsf{last}({#1})}
\newcommand{\prefixes}[1]{\mathsf{Prefixes}({#1})}

\newcommand{\tuple}[1]{\left<#1\right>}
\newcommand{\zero}{{\bf 0}}
\newcommand{\usup}[1]{{#1}\ensuremath{^{\sf u}}}
\newcommand{\tsup}[1]{{#1}\ensuremath{^{\sf t}}}

\tikzstyle{binnode}=[draw, fill=gray!10, 
    rounded corners, thick,
    inner sep=2pt,
    text centered, drop shadow]

\tikzstyle{modbinnode}=[draw, fill=red!50, 
    rounded corners, thick,
    inner sep=2pt,
    text centered, drop shadow]

\tikzstyle{leafnode}=[draw, fill=gray!10,
    circle, thick,
    inner sep=2pt,
    text centered, drop shadow]

\tikzstyle{modleafnode}=[draw, fill=red!50,
    circle, thick,
    inner sep=2pt,
    text centered, drop shadow]

\tikzstyle{level 1}=[level distance=0.75cm, sibling distance=1.5cm]
\tikzstyle{level 2}=[level distance=0.75cm, sibling distance=0.75cm]
\tikzstyle{level 3}=[level distance=0.75cm, sibling distance=0.75cm]

\tikzstyle{trnode}=[
  rounded corners,
  draw=black,
  thick, 
  inner sep=2pt, 
  text centered, 
  minimum width=0.5cm,
  minimum height=0.4cm,
]

\tikzstyle{myarrow}=[->]
\tikzstyle{control}=[myarrow,draw=black,thick]
\tikzstyle{push}=[myarrow,draw=black,thick,dotted]
\tikzstyle{pop}=[myarrow,draw=black,thick]
\tikzstyle{new}=[fill=red!20]
\tikzstyle{affected}=[fill=red!20]

\newcommand{\tracerowsep}{2mm}
\newcommand{\tracecolsep}{2.5mm}

\newcommand{\trnodel}[2]{{\footnotesize ${\bf #2}_{#1}$}}

\newcommand{\maketracea}{{
  \begin{tikzpicture}
  \matrix[row sep=\tracerowsep, column sep=\tracecolsep,ampersand replacement=\&] {
    \node [trnode] (A_a) {\trnodel{1}{a}};
    \& \& \& \& \& 
    \node [trnode] (B_a) {\trnodel{2}{a}};
    \& \& \& 
    \node [trnode] (C_a) {\trnodel{3}{a}};
    \\
    \node [trnode] (A_b) {\trnodel{1}{b}};
    \& \& \&
    \node [trnode] (B_b) {\trnodel{2}{b}};
    \&
    \node [trnode] (C_b) {\trnodel{3}{b}};
    \&
    \node [trnode] (A_g) {\trnodel{1}{g}};
    \&
    \node [trnode] (B_g) {\trnodel{2}{g}};
    \&
    \node [trnode] (C_g) {\trnodel{3}{g}};
    \\
    \node [trnode] (A_c) {\trnodel{1}{c}};
    \&
    \node [trnode] (B_c) {\trnodel{2}{c}};
    \&
    \node [trnode] (C_c) {\trnodel{3}{c}};
    \&
    \node [trnode] (A_f) {\trnodel{1}{f}};
    \& \&
    \node [trnode] (A_h) {\trnodel{1}{h}};
    \&
    \node [trnode] (A_i) {\trnodel{1}{i}};
    \\
    \node [trnode] (A_d) {\trnodel{1}{d}};
    \&
    \node [trnode] (A_e) {\trnodel{1}{e}};
    \\
  };

  \path[control]
    (A_a) edge       (A_b)
    (A_b) edge       (A_c)
    (A_c) edge       (A_d)
    (A_d) edge [pop] (B_c)
    (B_c) edge       (A_e) 
    (A_e) edge [pop] (C_c)
    (C_c) edge [pop] (B_b) 
    (B_b) edge       (A_f)
    (A_f) edge [pop] (C_b)
    (C_b) edge [pop] (B_a)
    (B_a) edge       (A_g)
    (A_g) edge       (A_h)
    (A_h) edge [pop] (B_g)
    (B_g) edge       (A_i)
    (A_i) edge [pop] (C_g)
    (C_g) edge [pop] (C_a);

  \path[push] (A_a) edge (B_a) (B_a) edge (C_a);
  \path[push] (A_b) edge (B_b) (B_b) edge (C_b);
  \path[push] (A_c) edge (B_c) (B_c) edge (C_c);
  \path[push] (A_g) edge (B_g) (B_g) edge (C_g);

 \end{tikzpicture}
}}

\newcommand{\maketraceb}{{
  \begin{tikzpicture}[remember picture]
  \matrix[row sep=\tracerowsep, column sep=\tracecolsep,ampersand replacement=\&] {
    \node [trnode] (A_a) {\trnodel{1}{a}};
    \& \& \& \& \& 
    \node [trnode,affected] (B_a) {\trnodel{2}{a}};
    \& \& \& \& \&
    \node [trnode,affected] (C_a) {\trnodel{3}{a}};
    \\
    \node [trnode] (A_b) {\trnodel{1}{b}};
    \& \& \&
    \node [trnode] (B_b) {\trnodel{2}{b}};
    \&
    \node [trnode] (C_b) {\trnodel{3}{b}};
    \&
    \node [trnode,new] (A_j) {\trnodel{1}{j}};
    \& \& \&
    \node [trnode,new] (B_j) {\trnodel{2}{j}};
    \& 
    \node [trnode,new] (C_j) {\trnodel{3}{j}};
    \\
    \node [trnode] (A_c) {\trnodel{1}{c}};
    \&
    \node [trnode] (B_c) {\trnodel{2}{c}};
    \&
    \node [trnode] (C_c) {\trnodel{3}{c}};
    \&
    \node [trnode] (A_f) {\trnodel{1}{f}};
    \& \&
    \node [trnode] (A_g) {\trnodel{1}{g}};
    \&
    \node [trnode] (B_g) {\trnodel{2}{g}};
    \&
    \node [trnode] (C_g) {\trnodel{3}{g}};
    \&
    \node [trnode,new] (A_k) {\trnodel{1}{k}};
    \\
    \node [trnode] (A_d) {\trnodel{1}{d}};
    \&
    \node [trnode] (A_e) {\trnodel{1}{e}};
    \& \& \& \&
    \node [trnode] (A_h) {\trnodel{1}{h}};
    \&
    \node [trnode] (A_i) {\trnodel{1}{i}};
    \\
  };

  \path[control]
    (A_a) edge       (A_b)
    (A_b) edge       (A_c)
    (A_c) edge       (A_d)
    (A_d) edge [pop] (B_c)
    (B_c) edge       (A_e) 
    (A_e) edge [pop] (C_c)
    (C_c) edge [pop] (B_b) 
    (B_b) edge       (A_f)
    (A_f) edge [pop] (C_b)
    (C_b) edge [pop] (B_a)
    
    (B_a) edge       (A_j)
    (A_j) edge       (A_g)

    (A_g) edge       (A_h)
    (A_h) edge [pop] (B_g)
    (B_g) edge       (A_i)
    (A_i) edge [pop] (C_g)
    
    (C_g) edge [pop] (B_j)
    (B_j) edge       (A_k)
    (A_k) edge [pop] (C_j)
    (C_j) edge [pop] (C_a)
    ;

  \path[push] (A_a) edge (B_a) (B_a) edge (C_a);
  \path[push] (A_b) edge (B_b) (B_b) edge (C_b);
  \path[push] (A_c) edge (B_c) (B_c) edge (C_c);
  \path[push] (A_g) edge (B_g) (B_g) edge (C_g);
  \path[push] (A_j) edge (B_j) (B_j) edge (C_j);

 \end{tikzpicture}
}}

\begin{document}

\title{Low-Level Self-Adjusting Computation}
\title{Compiling Implicit, Low-Level Self-Adjusting Computation}
\title{An Intermediate Language for Self-Adjusting Computation}
\title{Compiling Low-Level Self-Adjusting Computation}
\title{Self-Adjusting Computation with Stacks}
\title{A Self-Adjusting Intermediate Language}
\title{An Intermediate Language for Self-Adjusting Computation}
\title{An Intermediate Language for Sound Self-Adjusting Computation}
\title{A Sound Semantics for Self-Adjusting Stack Machines}
\title{A Sound Semantics for Self-Adjusting Store'n'Stack Machines}
\title{A Consistent Semantics for Self-Adjusting Stack Machines}
\title{A Sound Semantics for Self-Adjusting Store-Agnostic Small-Step Stack Machines}
\title{A Consistent Semantics for Self-Adjusting Stack Machines}

\title{Self-Adjusting Stack Machines}

\authorinfo{
  Matthew A. Hammer
  \and
  Georg Neis
  \and
  Yan Chen
  \and
  Umut A. Acar
  }
  {Max Planck Institute for Software Systems}  
  {\{hammer,neis,chenyan,umut\}@mpi-sws.org}

\maketitle

\newcommand{\techreportonly}[1]{#1}

\newcommand{\coloronly}[1]{#1}

\begin{abstract}
Self-adjusting computation offers a language-based approach to writing
programs that automatically respond to dynamically changing data.
Recent work made significant progress in developing sound semantics
and associated implementations of self-adjusting computation for
high-level, functional languages.
These techniques, however, do not address issues that arise for
low-level languages, i.e., stack-based imperative languages that lack
strong type systems and automatic memory management.

In this paper, we describe techniques for self-adjusting computation
which are suitable for low-level languages.
Necessarily, we take a different approach than previous work: instead
of starting with a high-level language with additional primitives to
support self-adjusting computation, we start with a low-level
intermediate language, whose semantics is given by a stack-based
abstract machine.
We prove that this semantics is sound: it always updates computations
in a way that is consistent with full reevaluation.
We give a compiler and runtime system for the intermediate language
used by our abstract machine.
We present an empirical evaluation that shows that our approach is
efficient in practice, and performs favorably compared to prior
proposals.
\end{abstract}

\section{Introduction}
\label{sec:introduction}
\label{sec:intro}

Many applications operate on data that changes incrementally, i.e., by
a small amount, over time.  Such incremental changes often require
only incremental updates to the output, making it possible to respond
to dynamically changing data more efficiently than recomputing the
output from scratch.  These improvements are often asymptotically
significant, providing as much as a linear factor of speedup.  To
exploit this potential, one can develop ``dynamic'' or ``kinetic''
algorithms that are designed to deal with particular forms of changing
input by taking advantage of the particular structure of the problem
at hand~\cite{ChiangTa92,EppsteinGaIt99,motion-survey02}.  This manual
approach often yields updates that are asymptotically faster than full
reevaluation, but carries inherent complexity and non-compositionality
that makes the algorithms difficult to design, analyze, and use.

As an alternative to manual design of dynamic and kinetic algorithms,
the programming languages community has developed techniques that
either automate or mostly automate the process of translating an
implementation of an algorithm for fixed input into a version for
changing input.  This is a challenging problem because the compiler is
expected to improve the asymptotic complexity of the program.  Many
different approaches have been considered; for more detail on previous
work we refer the reader to Ramalingam and Reps'
survey~\cite{RamalingamRe93} and to \secref{related-work}.  Recent
advances on self-adjusting computation~\cite{Acar09} made substantial
progress on this problem by proposing techniques that allow both
purely functional and imperative programs to automatically respond to
changes in their data.  The approach has been shown to be effective in
a reasonably broad range of areas including computational geometry,
invariant checking, motion simulation, and machine learning
(e.g.,~\cite{ShankarBo07,AcarBlTaTu08,AcarIhMeSu07}) and has even
helped solve challenging open problems~\cite{AcarCoHuTu10}.

Self-adjusting computation typically relies on programmer help to
identify the data that can change over time, called {\em changeable
  data}, and the dependencies between this data and program code.
This changeable data is typically stored in special memory cells
referred to as \emph{modifiable references} (\emph{modifiables} for
short), so called because they can undergo incremental modification.
The read and write dependencies of modifiables are recorded in a
dynamic \emph{execution trace} (or \emph{trace}, for short), which
effectively summarizes the self-adjusting computation.
When modifiables change, the trace is automatically edited through a
\emph{change propagation} algorithm: some portions of the trace are
reevaluated (when the corresponding subcomputations are affected by a
changed value), some portions are discarded (e.g., when reevaluation
changes control paths) and some portions are reused (when a
subcomputation remains unaffected, i.e., when it remains consistent
with the values of modifiables).
We typically say that a semantics for self-adjusting computation is
\emph{sound} (alternatively, \emph{consistent}), if the change
propagation mechanism always yields a result consistent with full
reevaluation.

The initial approaches for self-adjusting computation offer
programming interfaces within existing functional languages, namely,
SML and Haskell, either via a library~\cite{Carlsson02,AcarBlBlHaTa09}
or with special compiler support~\cite{Ley-WildFlAc08}.
However, in all these systems, self-adjusting programs have a
purely-functional flavor, as modifiables must be written exactly once.
Later, Acar et al. lifted this write-once restriction by giving a
higher-order imperative semantics for self-adjusting
computation~\cite{AcarAhBl08}.

Unfortunately, this imperative semantics is not well-suited for
modeling low-level languages---by \emph{low-level} we mean (here and
throughout) stack-based imperative languages that lack strong type
systems and automatic memory management.
First, the imperative semantics assumes that only modifiables are
mutable: all other data is implicitly assumed to be immutable.
While a strong type system can enforce this policy, in a low-level
setting, all data is mutable by default, and there is no strong type
system to enforce other policies.
Next, the imperative semantics implicitly assumes that all garbage is
collected automatically.  This includes garbage from the
self-adjusting program itself, as well as from updating its trace
via change propagation.
Such automatic collection cannot be assumed for low-level languages.
Finally, and perhaps most importantly, the imperative semantics
provides no account of how execution traces should be incrementally
edited by the system for reuse.
Instead, the semantics effectively relies on an oracle to generate
reusable traces, and leaves the internal behavior of this oracle
unspecified.
Consequently, the oracle hides many of the pratical issues that would
otherwise arise, such as how memory allocation and collection interact
with trace reuse.

Based on their imperative semantics, Acar et al. describe a
library-based implementation for SML~\cite{AcarAhBl08}.  Following
this library interface, \CEAL~\cite{HammerAcCh09} provides compiler
support to write self-adjusting computations in C.
However, because of the issues raised above, the soundness property
proven for the semantics generally does not hold for \CEAL programs
unless they adhere to various correct-usage restrictions.
In particular, \CEAL programs must only mutate modifiables and local
variables---global variables, return values\footnote{The imperative
  semantics restricts return types to unit (i.e., \texttt{void}).},
and user-defined data structures must be immutable (and hence,
non-modifiable).
Furthermore, since even immutable data must first be initialized in a
low-level setting, and since this initialization is itself a case of
mutation, \CEAL programs are required to treat such initialization
code in a special way.
Namely, they must separate it into designated ``initialization
functions'', as introduced in previous work on automatic memory
management for self-adjusting computation~\cite{HammerAc08}.

Failing to follow the correct-usage restrictions given above, a \CEAL
program could crash, or alternatively, fail to provide correct
updates.
As a simple example, consider a trivial program that calls two
functions: the first copies some input from modifiable $m_\text{in}$ to
a global variable~$g$; the second copies the value of $g$ into another
modifiable $m_\text{out}$ as output.
The computational dependencies of modifiable references~$m_\text{in}$
and $m_\text{out}$ are traced, but those of global variable~$g$ are
not.  Consequently, when $m_\text{in}$ changes, $m_\text{out}$ will not
be updated, since doing so requires knowledge of its dependency on
$g$.
An analogous scenario can be constructed using any non-modifiable
memory in place of global~$g$ (e.g., a user-defined data type).

At present, we are aware of no generally sound implementation of
self-adjusting computation for low-level languages, nor a semantics
that suggests one.

\paragraph{Self-adjusting stack machines.}

In this paper, we describe techniques for sound self-adjusting
computation which are suitable for low-level languages.  
To achieve soundness without losing generality, we take a
fundamentally different approach than previous work: instead of
starting with a high-level language with additional primitives to
support self-adjusting computation, we start with a low-level
intermediate language called \il.

We give two semantics to \IL by defining two abstract machines: the
\emph{reference machine} models conventional evaluation semantics,
while the \emph{tracing machine} models self-adjusting semantics.
Each machine is defined by a \emph{transition relation} between
\emph{machine configurations}.
Our low-level setting is reflected by the reference machine's
configurations: each consists of a store, a stack, an environment and
a program.
The tracing machine extends these configurations with an execution
trace.
We define \emph{traced evaluation} and \emph{change propagation}
within the tracing machine by including transitions that incrementally
edit the trace (i.e., transitions that either insert, remove or replay
traced execution steps).
We show that automatic memory management is a natural aspect of
automatic change propagation by defining a notion of garbage
collection.

\paragraph{Contributions.} Our contributions are as follows:
\begin{enumerate}
\item We provide an abstract machine semantics for self-adjusting
  computation.
  This includes accounts of how change propagation interacts with a
  control stack, with return values and with memory management.
  We prove that this semantics is sound.

\item We describe and implement a compiler and runtime system for \IL,
  the intermediate language used by our abstract machines.
  Additionally, we give two automatic optimizations to reduce the
  overhead of the approach.

\item We describe and implement a front-end that translates a large
  subset of C into \IL, and perform an empirical evaluation of our
  implementation.

\end{enumerate}

\newcommand{\trofnd}[2]{$\textbf{#1}_{#2}$}

\section{Overview}
\label{sec:overview}

We introduce the challenges for giving self-adjusting computation
support to programs written in low-level languages.
In particular, we consider two example programs and consider
strategies for incrementally updating their computations.
We introduce our approach, in which we restructure these programs
in \IL, our intermediate language for self-adjusting computation.
We informally describe a change propagation semantics for \IL
programs that addresses the challenges from the examples.

\subsection{Example 1: Reducing Trees}
\label{sec:example1}

\begin{figure}
\centering

\begin{minipage}{\columnwidth}
\fixedCodeFrame{\parbox{0.8\columnwidth}{
\begin{codeListing}
\=XX\=XX\=XX\=XX\=XX\=\kill
\codeLineNN\>
\kw{typedef} \kw{struct} node\_s* node\_t;
\\
\codeLineNN\>
\kw{struct} node\_s \{
\\
\codeLineNN\>\>
\kw{enum} \{ LEAF, BINOP \} tag;
\\
\codeLineNN\>\>
\kw{union} \{ int leaf\_val;
\\
\codeLineNN\>\>
~~~~~~~~\kw{struct} \{ \kw{enum} \{ PLUS, MINUS \} op;
\\
\codeLineNN\>\>
~~~~~~~~~~~~~~~~node\_t left, right;~\} binop;
\\
\codeLineNN\>\>\>
~~~~~\} u; \};
\end{codeListing}}}
\nocaptionrule
\vspace*{-2mm}
\caption{Type declarations for expression trees in C.}
\label{fig:exptrees-c-types}
\end{minipage}
\\[2mm]

\begin{minipage}{\columnwidth}
\fixedCodeFrame{\parbox{0.8\columnwidth}{
\begin{codeListing}
XX\=XX\=XX\=XX\=\kill
\codeLineL{evalc::line:evalc::def}\>
\kw{int} eval (node\_t root) \{
\\
\codeLineL{line:evalc::ifleaf}\>\>
\kw{if} (root->tag == LEAF) 
\\
\codeLineL{line:evalc::thenleaf}\>\>\>
\kw{return} root->u.leaf\_val;
\\
\codeLineL{line:evalc::ifleafelse}\>\>
\kw{else} \{
\\
\codeLineL{line:evalc::firsteval}\>\>\>
\kw{int} l = eval (root->u.binop.left);
\\
\codeLineL{line:evalc::secondeval}\>\>\>
\kw{int} r = eval (root->u.binop.right);
\\
\codeLineL{line:evalc::ifplus}\>\>\>
\kw{if} (root->u.binop.op == PLUS) \kw{return} (l + r); 
\\
\codeLineL{line:evalc::elseminus}\>\>\>
\kw{else} \kw{return} (l - r);
\\
\codeLineL{line:evalc::ifleafelseend}\>
\}~\}
\end{codeListing}
}}
\nocaptionrule
\vspace*{-2mm}
\caption{The \ttt{eval} function in C.}
\label{fig:exptrees-c-eval}
\end{minipage}
\\[2mm]

{
\newcommand{\offsettag}{TAG}
\newcommand{\offsetval}{LEAF\_VAL}
\newcommand{\offsetop}{OP}
\newcommand{\offsetleft}{LEFT}
\newcommand{\offsetright}{RIGHT}

\newcommand{\hlbgl}[1]{#1}
\newcommand{\hlbgd}[1]{#1}

\newcommand{\hlboxl}[1]{
  \colorbox[gray]{0.95}{#1}
}
\newcommand{\hlboxd}[1]{
  \colorbox[gray]{0.88}{#1}
}

\begin{minipage}{\columnwidth}
\centering
\includegraphics[height=1.5in]{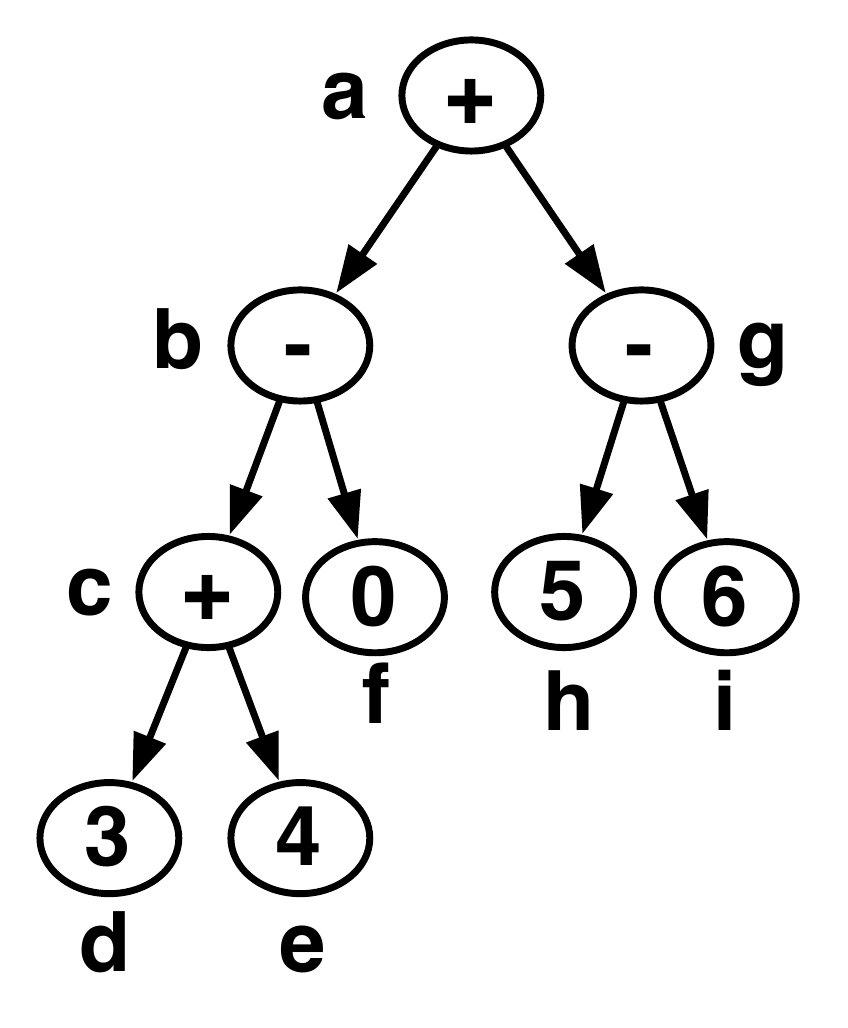}
\quad\quad
\includegraphics[height=1.5in]{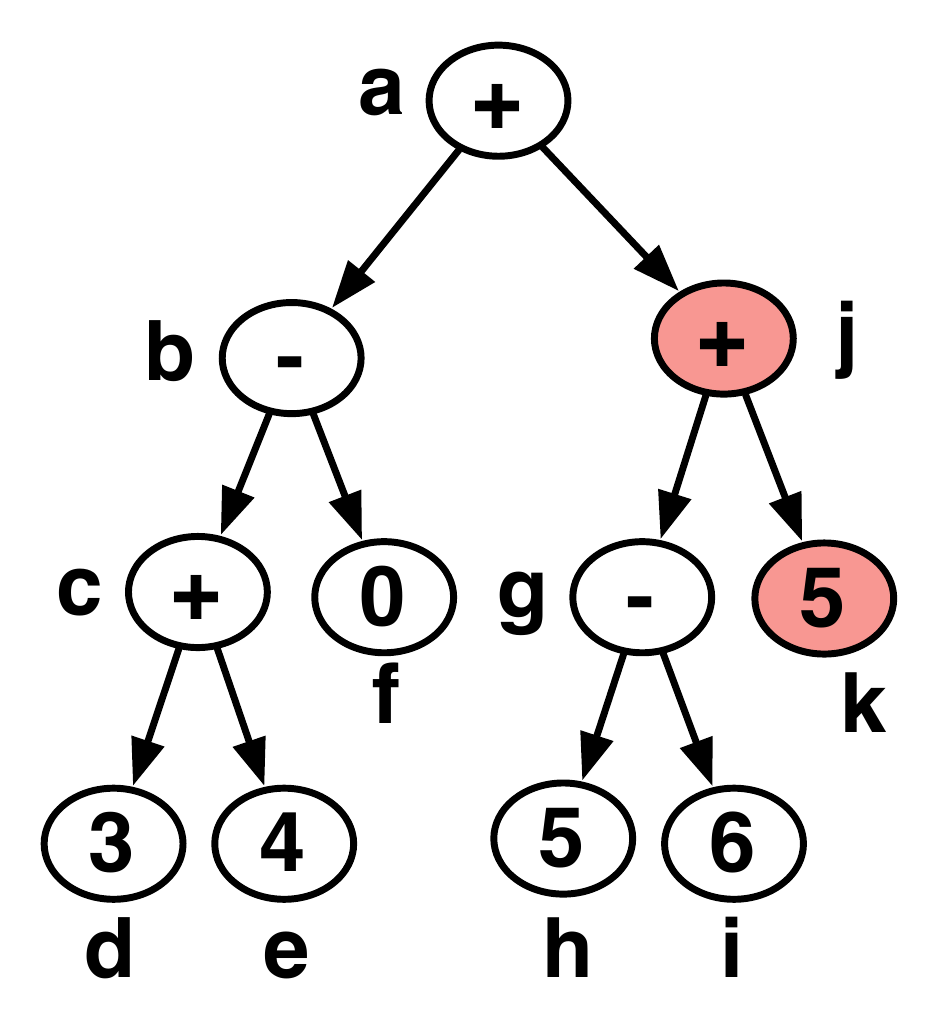}
\nocaptionrule
\vspace*{-2mm}
\caption{Example expression trees.}
\label{fig:exptrees-example}
\end{minipage}
\\[2mm]

\begin{minipage}{\columnwidth}
\maketracea
\maketraceb
\nocaptionrule
\vspace*{-5mm}
\caption{Example execution traces of \texttt{eval}.}
\label{fig:exptrees-traces}
\end{minipage}
\\[2mm]

\begin{minipage}{\columnwidth}
\fixedCodeFrame{\parbox{0.8\columnwidth}{
\setlength{\fboxsep}{0pt}%
\begin{codeListing}
XXXX\=XX\=XX\=XX\=XX\=XX\=XX\=XX\=XX\=XX\=XX\=
\kill
\codeLineL{line:fun}\>\kw{let} eval (root) = \kw{memo}
\\  
\codeLineL{line:eval::ktwo::def}\>\>
\kw{let} \kone(l) =
\\
\codeLineL{line:kone::let}\>\>\>
\hlbgl{\kw{let}} \ktwo(r) = \kw{update}
\\
\codeLineL{line:eval::ktwo::readop}\>\>\>\>
\kw{let}~op = \kw{read} (root[\offsetop])~\kw{in}
\\
\codeLineL{line:eval::ktwo::ifop}\>\>\>\>
\kw{if} (op == PLUS) \kw{then} \kw{pop} (l+r)
\\
\codeLineL{line:eval::ktwo::else}\>\>\>\>
\kw{else} \kw{pop} (l-r)
\\
\codeLineL{line:kone::in}\>\>\>
\hlbgl{\kw{in}}
\\
\codeLineL{line:kone::in::push}\>\>\>
\hlbgl{\kw{push} \ktwo \kw{do}} \kw{update}
\\
\codeLineL{line:kone::read}\>\>\>\>
\hlbgl{\kw{let}~right $=$ \kw{read} (root[\offsetright])~\kw{in}}
\\
\codeLineL{line:kone::eval}\>\>\>\>
\hlbgl{eval (right)}
\\
\codeLineL{line:eval::in}\>\>
\kw{in}
\\
\codeLineL{line:eval::update}\>\>
\kw{update}
\\
\codeLineL{line:eval::readtag}\>\>\>
\kw{let}~tag $=$ \kw{read} (root[\offsettag])~\kw{in}
\\
\codeLineL{line:eval::iftag}\>\>\>
\kw{if} (tag == LEAF) 
\\
\codeLineL{line:eval::iftagleaf}\>\>\>\>
\kw{let}~leaf\_val $=$ \kw{read} (root[\offsetval])~\kw{in} 
\\
\codeLineL{line:eval::iftagleaf}\>\>\>\>
\kw{pop} (leaf\_val)
\\
\codeLineL{line:eval::else}\>\>\>
\kw{else} 
\\
\codeLineL{line:eval::else::push}\>\>\>\>
\kw{push} \kone~\kw{do} \kw{update}
\\
\codeLineL{line:eval::else::push::do::read}\>\>\>\>\>
\kw{let}~left $=$ \kw{read} (root[\offsetleft])~\kw{in}
\\
\codeLineL{line:eval::else::push::do::end}\>\>\>\>\>
eval (left)
\end{codeListing}}}
\nocaptionrule
\vspace*{-2mm}
\caption{The \ttt{eval} function in \IL.}
\label{fig:exptrees-il-code}
\end{minipage}
}

\end{figure}

For our first example, we consider a simple evaluator for expression
trees, as expressed with user-defined C data structures.
These expression trees consist of integer-valued leaves and internal
nodes that represent the binary operations of addition and subtraction.
\figref{exptrees-c-types} shows their representation in \C.  The
\ttt{tag} field (either \ttt{LEAF} or \ttt{BINOP}) distinguishes
between the \ttt{leaf\_val} and \ttt{binop} fields of the
\kw{union}~\ttt{u}.
\figref{exptrees-c-eval} gives a simple \C function that evaluates
these trees.

Suppose we first run \ttt{eval} with an expression tree as shown on
the left in \figref{exptrees-example}; evaluating $((3+4)-0) + (5-6)$,
the execution will return the value $6$.
Suppose we then change the expression tree to $((3+4)-0) +
  ((5-6){\bf+5})$ as shown in \figref{exptrees-example} on the right.
How shall change propagation efficiently update the output?

\paragraph{Strategy for change propagation.}
We first consider the computation's structure, of which
\figref{exptrees-traces} gives a summary: the upper and lower versions
summarize the computation before and after the change, respectively.
Their structure reflects the stack behavior of \texttt{eval}, which
divides each invocation into (up to) three fragments: Fragment one
checks the tag of the node, returning the leaf value, if present, or
else recurring on the left subtree (lines 2--5); fragment two recurs
on the right subtree (line 6); and fragment three combines and returns
the results (lines 7--8).

In \figref{exptrees-traces}, each fragment is labeled with a tree
node, e.g., \trofnd{b}{2} represents fragment two's execution on
node~\textbf{b}.
The dotted horizontal arrows indicate pushing a code fragment on the
stack for later.  Solid arrows represent the flow of control from one
fragment to the next; when diagonal, they indicate popping the stack
to continue evaluation.

Based on these two computations' structure, we informally sketch a
strategy for change propagation.
First, since the left half of the tree is unaffected, the left half of
the computation (\trofnd{a}{1}--\trofnd{b}{3}) is also unaffected, and
as such, change propagation should reuse it.
Next, since the right child for \textbf{a} has changed, the
computation that reads this value, fragment~\trofnd{a}{2}, should be
reevaluated.
This reevaluation recurs to node~\textbf{g}, whose subtree has not
changed.  Hence, change propagation should reuse the corresponding
computation (\trofnd{g}{1}--\trofnd{g}{3}), including its return
value,~$-1$.
Comparing \trofnd{j}{1}--\trofnd{j}{3} against
\trofnd{g}{1}--\trofnd{g}{3}, we see that \textbf{a}'s right subtree
evaluates to~$4$ rather than~$-1$.
Hence, change propagation should reevaluate \trofnd{a}{3}, to yield
the new output of the program,~$11$.

\paragraph{Challenges.} 
For change propagation to use the strategy sketched above, it must
identify dependencies among data and the three-part structure of this
code, including its call\-/\-return dependencies.
In particular, it must identify where previous computations should be
reused, reevaluated or discarded\footnote{ To see an example where
  computation is discarded, imagine the change in reverse; that is,
  changing the lower computation into the upper one.  }.
In \secref{introtoil}, we discuss how the \IL code of
\figref{exptrees-il-code}, which represents \figref{exptrees-c-eval},
informs the change propagation strategy described above.

\subsection{Example 2: Reducing Arrays}
\label{sec:example2}

\begin{figure}
\centering
\begin{minipage}{\columnwidth}
\fixedCodeFrame{\parbox{0.9\columnwidth}{
\begin{codeListing}
XX\=XX\=XX\=XX\=XX\=XX\=\kill

\codeLineNN\>
\kw{int} MAX;
\\[2mm]
\codeLineNN\>
\kw{void} array\_max(\kw{int}* arr, \kw{int} len) \{
\\
\codeLineNN\>\>
\kw{while}(len > 1) \{
\\
\codeLineNN\>\>\>
\kw{for}(\kw{int} i = 0; i < len - 1; i += 2) \{
\\
\codeLineNN\>\>\>\>
\kw{int} m;
\\
\codeLineNN\>\>\>\>
max(arr[i], arr[i + 1], \&m);
\\
\codeLineNN\>\>\>\>
arr[i / 2] = m;
\\
\codeLineNN\>\>\> 
\}
\\
\codeLineNN\>\>\>
len = len / 2;
\\
\codeLineNN\>\>
\}
\\
\codeLineNN\>\>
MAX = arr[0];
\\
\codeLineNN\>
\}
\end{codeListing}
}}
\end{minipage}
\nocaptionrule
\caption{Iteratively compute the maximum of an array.}
\label{fig:arrayred-c-code}
\end{figure}

\tikzstyle{ac}=[draw=black,minimum height=1.3em,text centered]

\tikzstyle{acw}=[ac,fill=grey!20]

\tikzstyle{acs}=[ac,draw=white,fill=black!10]

\tikzstyle{acc}=[ac,fill=red!30]

\begin{figure}
\begin{tabular}{cc}
\begin{tikzpicture}
\matrix[row sep=0.1cm, column sep=0cm] {
  \node [ac] {2}; 
  & 
  \node [ac] {9}; 
  & 
  \node [ac] {3}; 
  & 
  \node [ac] {5};
  & 
  \node [ac] {4};
  & 
  \node [ac] {7};
  & 
  \node [ac] {1};
  &
  \node [ac] {6};
  
  \\[2mm]
  
  \node [ac] {9};
  & 
  \node [ac] {5};
  & 
  \node [ac] {7};
  & 
  \node [ac] {6};
  &
  \node [acs] {4};
  & 
  \node [acs] {7};
  & 
  \node [acs] {1};
  &
  \node [acs] {6};
  \\

  \node [ac] {9};
  & 
  \node [ac] {7};
  &
  \node [acs] {7};
  & 
  \node [acs] {6};
  &
  \node [acs] {4};
  & 
  \node [acs] {7};
  & 
  \node [acs] {1};
  &
  \node [acs] {6};
  \\

  \node [ac] {9};
  &    
  \node [acs] {7};
  &
  \node [acs] {7};
  & 
  \node [acs] {6};
  &
  \node [acs] {4};
  & 
  \node [acs] {7};
  & 
  \node [acs] {1};
  &
  \node [acs] {6};
  \\
};
\end{tikzpicture}
&
\begin{tikzpicture}
\matrix[row sep=0.1cm, column sep=0cm] {
  \node [ac] {2}; 
  & 
  \node [acc] {0}; 
  & 
  \node [ac] {3}; 
  & 
  \node [ac] {5};
  & 
  \node [ac] {4};
  & 
  \node [ac] {7};
  & 
  \node [ac] {1};
  &
  \node [ac] {6};

  \\[2mm]
  
  \node [acc] {2};
  & 
  \node [ac] {5};
  & 
  \node [ac] {7};
  & 
  \node [ac] {6};
  &
  \node [acs] {4};
  & 
  \node [acs] {7};
  & 
  \node [acs] {1};
  &
  \node [acs] {6};
  \\

  \node [acc] {5};
  & 
  \node [ac] {7};
  &
  \node [acs] {7};
  & 
  \node [acs] {6};
  &
  \node [acs] {4};
  & 
  \node [acs] {7};
  & 
  \node [acs] {1};
  &
  \node [acs] {6};
  \\

  \node [acc] {7};
  &    
  \node [acs] {7};
  &
  \node [acs] {7};
  & 
  \node [acs] {6};
  &
  \node [acs] {4};
  & 
  \node [acs] {7};
  & 
  \node [acs] {1};
  &
  \node [acs] {6};
  \\
};

\end{tikzpicture}
\end{tabular}
\nocaptionrule
\caption{Snapshots of the array from \figref{arrayred-c-code}.}
\label{fig:arrayred-data}
\end{figure}

As a second example, \figref{arrayred-c-code} gives C code for
(destructively) computing the maximum element of an array.
Rather than perform a single linear scan, it finds this maximum
iteratively by performing a logarithmic number of \emph{rounds}, in
the style of a (sequentialized) data-parallel algorithm.
For simplicity, we assume that the length of arrays is always a power
of two.
Each round combines pairs of adjacent elements in the array,
producing a sub-sequence with half the length of the original.
The remaining half of the array contains inactive elements no longer
accessed by the function.

Rather than return values directly, we illustrate commonly used
imperative features of C by returning them indirectly:
function~\texttt{max} returns its result by writing to a provided
pointer, and \texttt{array\_max} returns its result by assigning it to
a special global variable~\texttt{MAX}.

\figref{arrayred-data} illustrates the computation for two
(closely-related) example inputs.  Below each input, each computation
consists of three snapshots of the array, one per round.
For readability, the inactive elements of the array are still shown
but are greyed, and the differences between the right and left
computation are highlighted on the right.

\paragraph{Strategy for change propagation.} 
We use \figref{arrayred-data} to develop a strategy for change
propagation.  Recall that each array snapshot summarizes one round of
the outer \texttt{while} loop.  Within each snapshot, each (active)
cell summarizes one iteration of the inner \texttt{for} loop.
That \texttt{ar\-ray\-\_max} uses an iterative style affects the
structure of the computation, which consequently admits an efficient
strategy for change propagation: reevaluate each affected iteration of
the inner \ttt{for} loop, that is, those summarized by the highlighted
cells in \figref{arrayred-data}.

It is simple to (manually) check that each active cell depends on
precisely two cells in the previous round, affects at most one cell in
the next round, and is computed independently of other cells in the
same round.
Hence, for a single input change, at most one such iteration is
affected per round.
Since the number of rounds is logarithmic in the length of the input
array, this change propagation strategy is efficient.

\paragraph{Challenges.} To efficiently update the computation, change
propagation should reevaluate each affected iteration, being careful
not to reevaluate any of the unaffected iterations.

\subsection{Introduction to \IL}
\label{sec:introtoil}

\begin{figure*}
\centering
\subfigure[]{
\label{fig:arrayred-il-code-a}
\begin{minipage}{\columnwidth}
\begin{codeListing}
XX\=XX\=XX\=XX\=XX\=XX\=\kill
\codeLineNN\>
\kw{let} for\_loop (i) = 
\\
\codeLineNN\>\>
\kw{let} m\_ptr = \kw{alloc}(1) \kw{in}
\\
\codeLineNN\>\>
\kw{let} after\_max() = \kw{update}
\\
\codeLineNN\>\>\>
\kw{let} m\_val = \kw{read}(m\_ptr[0]) \kw{in}
\\
\codeLineNN\>\>\>
\kw{let} \_ = \kw{write}(arr[i/2], m\_val) \kw{in} 
\\
\codeLineNN\>\>\>
\kw{if} (i < len - 1) 
\\
\codeLineNN\>\>\>\>
\kw{then} for\_loop(i + 2) 
\\
\codeLineNN\>\>\>\>
\kw{else} \ldots
\\
\codeLineNN\>\>
\kw{in} 
\\
\codeLineNN\>\>
\kw{push} after\_max \kw{do} \kw{update}
\\
\codeLineNN\>\>\>
\kw{let} a = \kw{read}(arr[i]) \kw{in}
\\
\codeLineNN\>\>\>
\kw{let} b = \kw{read}(arr[i + 1]) \kw{in}
\\
\codeLineNN\>\>\>
max(a, b, m\_ptr)
\\
\codeLineNN\>
\kw{in} for\_loop(0)
\\
\end{codeListing}
\end{minipage}
}
\subfigure[]{
\label{fig:arrayred-il-code-b}
\begin{minipage}{\columnwidth}
\begin{codeListing}
XX\=XX\=XX\=XX\=XX\=XX\=\kill
\codeLineNN\>
\kw{let} for\_loop (i) = 
\\
\codeLineNN\>\>
\kw{let} m\_ptr = \kw{alloc}(1) \kw{in}
\\
\codeLineNN\>\>
\kw{let} after\_max() = \kw{update}
\\
\codeLineNN\>\>\>
\kw{let} m\_val = \kw{read}(m\_ptr[0]) \kw{in}
\\
\codeLineNN\>\>\>
\kw{let} \_ = \kw{write}(arr[i/2], m\_val) \kw{in} 
\\
\codeLineNN\>\>\>
\hldiff{\kw{memo}}
\\
\codeLineNN\>\>\>\>
\kw{if} (i < len - 1) 
\\
\codeLineNN\>\>\>\>\>
\kw{then} for\_loop(i + 2) 
\\
\codeLineNN\>\>\>\>\>
\kw{else} \ldots
\\
\codeLineNN\>\>
\kw{in} 
\\
\codeLineNN\>\>
\kw{push} after\_max \kw{do} \kw{update}
\\
\codeLineNN\>\>\>
\kw{let} a = \kw{read}(arr[i]) \kw{in}
\\
\codeLineNN\>\>\>
\kw{let} b = \kw{read}(arr[i + 1]) \kw{in}
\\
\codeLineNN\>\>\>
max(a, b, m\_ptr)
\\
\codeLineNN\>
\kw{in} for\_loop (0)
\\
\end{codeListing}
\end{minipage}
}
\subfigure[]{
\label{fig:arrayred-il-code-c}
\begin{minipage}{\columnwidth}
\begin{codeListing}
XX\=XX\=XX\=XX\=XX\=XX\=\kill

\codeLineNN\>
\kw{let} for\_loop (i) = 
\\
\codeLineNN\>\>
\hldiff{\kw{let}~\texttt{for\_next () =}}
\\
\codeLineNN\>\>\>
\kw{if} (i < len - 1) \kw{then} for\_loop(i + 2) 
\\
\codeLineNN\>\>\>\>
\kw{else} \ldots
\\
\codeLineNN\>\>
\hldiff{\kw{in}}
\\
\codeLineNN\>\>
\hldiff{\kw{push}~\texttt{for\_next~\kw{do}}}
\\
\codeLineNN\>\>\>
\kw{let} m\_ptr = \kw{alloc}(1) \kw{in}
\\
\codeLineNN\>\>\>
\kw{let} after\_max() = \kw{update}
\\
\codeLineNN\>\>\>\>
\kw{let} m\_val = \kw{read}(m\_ptr[0]) \kw{in}
\\
\codeLineNN\>\>\>\>
\kw{let} \_ = \kw{write}(arr[i/2], m\_val) \kw{in} 
\\
\codeLineNN\>\>\>\>
\hldiff{\texttt{\kw{pop} ()}}
\\
\codeLineNN\>\>\>
\kw{in} 
\\
\codeLineNN\>\>\>
\kw{push} after\_max \kw{do} \kw{update}
\\
\codeLineNN\>\>\>\>
\kw{let} a = \kw{read}(arr[i]) \kw{in}
\\
\codeLineNN\>\>\>\>
\kw{let} b = \kw{read}(arr[i + 1]) \kw{in}
\\
\codeLineNN\>\>\>\>
max(a, b, m\_ptr)
\\
\codeLineNN\>
\kw{in} for\_loop(0)
\\
\end{codeListing}
\end{minipage}
}
\caption{Three versions of \IL code for the \texttt{for} loop in
  \figref{arrayred-c-code}; highlighting indicates their
  slight differences. }
\label{fig:arrayred-il-code}
\end{figure*}

The primary role of \IL is to make precise the computational
dependencies and possible change propagation behaviors of a
low-level self-adjusting program.
In particular, it is easy to answer the following questions for a
program when expressed in \IL:

\begin{itemize}
\item Which data dependencies are \emph{local} versus \emph{non-local}?
\item Which code fragments are saved on the \emph{control stack}?
\item Which computation fragments are saved in the computation's
  \emph{trace}, for later reevaluation or reuse?
\end{itemize}

We informally introduce the syntax and semantics of \IL by addressing
each of these questions for the examples in
\secreftwo{example1}{example2}.  In \secref{il}, we make the syntax
and semantics precise.

\paragraph{Static Single Assignment.} 
To clearly separate local and non-local dependencies, \IL employs a
(functional variant of) static single assignment form
(SSA)~\citep{Appel98}.
Within this representation, the control-flow constructs of C are
represented by locally-defined functions, \emph{local state} is
captured by let-bound variables and function parameters, and all
\emph{non-local state} (memory content) is explicitly allocated within
the store and accessed via \kw{read}s and \kw{write}s.

For example, we express the \texttt{for} loop from
\figref{arrayred-c-code} as the recursive function \texttt{for\_loop}
in \figref{arrayred-il-code-a}.
This function takes an argument for each variable whose definition is
dependent on the \ttt{for} loop's control flow\footnote{Where
  traditional SSA employs $\phi$-operators to express
  control-dependent variable definitions, functional SSA uses ordinary
  function abstraction.}, in this case, just the iteration variable
\texttt{i}.
Within the body of the loop, the local variable \texttt{m} is encoded
by an explicit store allocation bound to a temporary
variable~\texttt{m\_ptr}.
Although not shown, global variable~\texttt{MAX} is handled
analogously.
This kind of indirection is necessary whenever assignments can occur
non-locally (as with global variables like \ttt{MAX}) or via pointer
indirection (as with local variable \ttt{m}).
By contrast, local variables \texttt{arr}, \texttt{i} and \texttt{len}
are only assigned directly and locally, and consequently, each is a
proper SSA variable in \figref{arrayred-il-code-a}.
Similarly, in \figref{exptrees-c-eval} the assignments to~\texttt{l}
and~\texttt{r} are direct, and hence, we express each as a proper SSA
variable in \figref{exptrees-il-code}.
We explain the other \IL syntax from
\figreftwo{exptrees-il-code}{arrayred-il-code-a} below (\kw{push},
\kw{pop}, \kw{update}, \kw{memo}).

\paragraph{Stack operations.}
As our first example illustrates (\secref{example1}), the control
stack necessarily breaks a computation into multiple fragments.
In particular, before control flow follows a function call, it first
pushes on the stack a code fragment (a local continuation) which later
takes control when the call completes.

The stack operations of \IL make this code fragmentation explicit: the
expression $\kw{push}~f~\kw{do}~e$ saves function~$f$ (a code fragment
expecting zero or more arguments) on the stack and continues by
evaluating $e$; when this subcomputation \kw{pop}s the stack, the
saved function~$f$ is applied to the (zero or more) arguments of the
\kw{pop}.

In \figref{exptrees-il-code}, the two recursive calls to \texttt{eval}
are preceded by \kw{push}es that save functions \ttt{eval\_right} and
\ttt{eval\_op}, corresponding to code fragments for evaluating the
right subtree (fragment two) and applying the binary operator
(fragment three), respectively.
Similarly, in \figref{arrayred-il-code-a}, the call to \texttt{max} is
preceded by a \kw{push} that saves function~\texttt{after\_max},
corresponding to the code fragment following the call.
We note that since \texttt{max} returns no values, \texttt{after\_max}
takes no arguments.

\paragraph{Reevaluation and reuse.}
To clearly mark which computations are saved in the trace---which in
turn defines which computations can be reevaluated and reused---\IL
uses the special forms \kw{update} and \kw{memo}, respectively.

The \IL expression $\kw{update}~e$, which we call an \emph{update
  point}, has the same meaning as $e$, except that during change
propagation, the computation of $e$ can be recovered from the
program's original computation and reevaluated.  This reevaluation is
necessary exactly when the original computation of $e$ contains
\kw{read}s from the store that are no longer consistent within the
context of new computation.

Dually, the \IL expression $\kw{memo}~e$, which we call a \emph{memo
  point}, has the same meaning as $e$, except that during
reevaluation, a previous computation of $e$ can be reused in place the
present one, provided that they \emph{match}.  Two computations of the
same expression~$e$ match if they begin in locally-equivalent states
(same local state, but possibly different non-local state).
This notion of memoization is similar to function
caching~\citep{PughTe89} in that it reuses past computation to avoid
reevaluation, but it is also significantly different in that impure
code is supported, and non-local state need not match (a matching
computation may contain inconsistent \kw{read}s).
We correct inconsistencies by reevaluating each inconsistent \kw{read}
within the reused computation.

We can insert \kw{update} and \kw{memo} points freely within an
existing \IL program without changing its meaning (up to reevaluation
and reuse behavior).  Since they allow more fine-grained reevaluation
and reuse, one might want to insert them before and after every
instruction in the program.  Unfortunately, each such insertion incurs
some tracing overhead, as \kw{memo} and \kw{update} points each
necessitate saving a snapshot of local state.

Fortunately, we can automatically insert a smaller yet equally
effective set of \kw{update} points by focusing only on \kw{read}s.
\figreftwo{exptrees-il-code}{arrayred-il-code-a} show examples of
this: since each \kw{read} appears within the body of an \kw{update}
point, we can reevaluate these \kw{read}s, including the code that
depends on them, should they become inconsistent with memory.
We say that each such \kw{read} is \emph{guarded} by an \kw{update}
point.

For memo points, however, it is less clear how to automatically strike
the right balance between too many (too much overhead) and not enough
(not enough reuse).  Instead, we expose surface syntax to the C
programmer, who can insert them as statements (\texttt{memo;}) as well
as expressions (e.g., \texttt{memo(f(x))}).
In \secref{introtoprop}, we discuss where to place memo points within
our running examples.

\subsection{Change Propagation Strategies Revisited}
\label{sec:introtoprop}

In \secreftwo{example1}{example2}, we sketched strategies for updating
computations using change propagation. 
Based on the \IL representations described in \secref{introtoil}, we
informally describe our semantics for change propagation in greater detail.
The remainder of the paper makes this semantics precise and describes
our current implementation.

\paragraph{Computations as traces.} We represent computations using an
execution trace, which records the \kw{memo} and \kw{update} points,
store operations (\kw{alloc}s, \kw{read}s and \kw{write}s), and stack
operations (\kw{push} and \kw{pop}).

To a first approximation, change propagation of these traces has two
aspects: reevaluating inconsistent subtraces, and reusing consistent
ones.
Operationally, these aspects mean that we need to decide not only
which computations in the trace to reevaluate, but also where this
reevaluation should cease.

\paragraph{Beginning a reevaluation.} 
In order to repair inconsistencies in the trace, we begin
reevaluations at \kw{update} points that guard inconsistent
\kw{read}s.
We identify \kw{read}s as inconsistent when the memory location they
depend on is affected by \kw{write}s being inserted into or removed
from the trace.  That is, a \kw{read} is identified as \emph{affected}
in one of two ways: when inserting a newly traced \kw{write} (of a
different value) that becomes the newly read value, or when removing a
previously traced \kw{write} that had been the previously read value.
In either case, the \kw{read} in question becomes inconsistent and
cannot be reused in the trace without first being reevaluated.
To begin such a reevalaution, we restore the local state from the
trace and reevaluate within the context of the current memory and
control stack, which generally both differ from those of the original
computation.

\paragraph{Ending a reevaluation.}
We end a reevaluation in one of two ways.
First, recall that we begin reevaluation with a different control
stack than that used by the original computation.  Hence, we will
eventually encounter a \kw{pop} that we cannot correctly reevaluate,
as doing so requires knowing the contents of the original
computation's stack.
Instead, we cease reevaluation at such \kw{pop}s.  We justify this
behavior below and describe how it still leads to a sound approach.

Second, as described in \secref{introtoil}, when we encounter a
\kw{memo} point, we may find a matching computation to reuse.
If so, we cease the current reevaluation and begin reevaluations that 
repair inconsistencies within the reused computation, if any.

\paragraph{Example 1 revisited.}
The strategy from \secref{example1} requires that the previous
computation be reevaluated in some places, and reused in others.
First, as \figref{exptrees-il-code} shows, we note that however an
input tree is modified, \kw{update} points guard the computation's
affected \kw{read}s.  We reevaluate these update points.
For instance, in the given change (of the right subtree
of~\textbf{a}), line 9 has the first affected \kw{read}, which is
guarded by an \kw{update} point on line 8; this point corresponds to
\trofnd{a}{2}, which we reevaluate first.
Second, our strategy reuses computation \trofnd{g}{1}--\trofnd{g}{3}.
To this end, we can insert a \ttt{memo} statement at the beginning of
function~\ttt{eval} in \figref{exptrees-c-eval} (not shown), resulting
in the \kw{memo} point shown on line 1 in \figref{exptrees-il-code}.
Since it precedes each invocation, this \kw{memo} point allows
for the desired reuse of unaffected subcomputations.

\paragraph{Example 2 revisited.}
Recall that our strategy for \secref{example2} consists of
reevaluating iterations of the inner \ttt{for}~loop that are affected,
and reusing those that are not.
To begin each reevaluation within this
loop~(\figref{arrayred-il-code-a}), we reevaluate their \kw{update}
points.  

Now we consider where to cease reevaluation.  Note that the
\kw{update} point in \ttt{after\_max} guards a \kw{read}, as well as
the recursive use of \ttt{for\_loop}, which evaluates the remaining
(possibly unaffected) iterations of the loop.
However, recall that we do not want reevaluation to continue with the
remaining iterations---we want to reuse them.  

We describe two ways to cease reevaluation and enable reuse.
First, we can insert a \ttt{memo} statement at the end of the inner
\ttt{for} loop in \figref{arrayred-c-code}, resulting in the \kw{memo}
point shown in \figref{arrayred-il-code-b}.  Second, we can wrap the
\ttt{for} loop's body with a \emph{cut block}, written
\ttt{cut\{\ldots\}}, resulting in the additional \kw{push}-\kw{pop}
pair in \figref{arrayred-il-code-c}.
Cut blocks are optional but convenient syntactic sugar: their use is
equivalent to moving a code block into a separate function (hence the
\kw{push}-\kw{pop} pair in \figref{arrayred-il-code-c}).
Regardless of which we choose, the new \kw{memo} and \kw{pop} both
allow us to cease reevaluation immediately after an iteration is
reevaluated within \figreftwo{arrayred-il-code-b}{arrayred-il-code-c},
respectively.

\paragraph{Call/return dependencies.} 
Recall from \secref{example1} that we must be mindful of call/return
dependencies among the recursive invocations.  In particular, after
reevaluating a subcomputation whose return value changes, the consumer
of this return value (another subcomputation) is affected and should
be reevaluated (\trofnd{a}{3} in the example).

Our general approach for call/return dependencies has three
parts. First, when proving consistency (\secref{consistency}), we
restrict our attention to programs whose subcomputations' return
values do not change, a crucial property of programs that we make
precise in \secref{consistency}.  Second, in \secref{dps}, we provide
an automatic transformation of arbitrary programs into ones that have
this property.
Third, in \secref{opt}, we introduce one simple way to refine this
transformation to reduce the overhead that it adds to the transformed
programs. With more aggressive analysis, we expect that further
efficiency improvements are possible.

Contrasted with proving consistency for a semantics where a fixed
approach for call/return dependencies is ``baked in'', our consistency
proof is more general.
It stipulates a property that can be guarenteed by either of the two
transformations that we describe (\secreftwo{dps}{opt}).  Furthermore,
it leaves the possibility open for future work to improve the
currently proposed transformations, e.g., by employing more
sophosticated static analysis to further reduce the overhead that they
introduce.

\subsection{Guide for the Paper}

\secref{il} presents the abstract machine semantics for \IL, including
our change propagation semantics.
\secref{consistency} presents our consistency theorem.
\secref{dps} presents a \emph{destination-passing style}
transformation whose target programs meet our side condition for
consistency.
\secref{compiling} gives compilation and runtime techniques for
our semantics.
\secref{implementation} describes our implementation.
\secref{evaluation} gives an empirical evaluation.
\secreftwo{relatedwork}{conclusion} give related work and conclude.

\section{A Self-Adjusting Intermediate Language}
\label{sec:il}
\label{sec:cealil}

We present \IL, a self-adjusting intermediate language, as well as two
abstract machines that evaluate \IL syntax.
We call these the \emph{reference machine} and the \emph{tracing machine},
respectively.
As its name suggests, we use the first machine as a reference when
defining and reasoning about the tracing machine.
Each machine is defined by its own transition relation
over similar machine components.  The tracing machine mirrors the
reference machine, but includes additional machine state components
and transition rules that work together to generate and edit execution
traces.
This tracing behavior formalizes the notion of \IL as a self-adjusting
language.

\subsection{Abstract Syntax of \IL}
\label{sec:il-syntax}
\begin{figure}
\[
\hspace{-18pt}
\begin{array}{rrcll}
& e 
& \bnfdef 
& \iluexp
& \descr{Untraced expression}
\\
&
& \bnfalt
& \iltexp
& \descr{Traced expression}
\\[1mm]

& \iluexp 
& \bnfdef 
& \illet{\ilfun{f}{\vec x}{e_1}}{e_2}
& \descr{Function definition}
\\
& 
& \bnfalt 
& \illet{\ilvar{x}{\ilprimop{\vec v}}}{e}
& \descr{Primitive operation}
\\ 
& 
& \bnfalt 
& \ilif{x}{e_1}{e_2}
& \descr{Conditional}
\\ 
& 
& \bnfalt 
& \ilapp{f}{\vec{x}}
& \descr{Function application}
\\[1mm]
& \iltexp 
& \bnfdef 
& \illet{\ilvar{x}{\iota}}{e}
& \descr{Store instruction}
\\ 
& 
& \bnfalt 
& \ilmemo{e}
& \descr{Memo point}
\\ 
& 
& \bnfalt 
& \ilwake{e}
& \descr{Update point}
\\ 
& 
& \bnfalt 
& \ilpush{f}{e}
& \descr{Stack push}
\\ 
& 
& \bnfalt
& \ilpop{\vec x}
& \descr{Stack pop}
\\[1mm]
& \iota 
& \bnfdef 
& \ilalloc{x}
& \descr{Allocate an array of size~$x$}
\\ 
& 
& \bnfalt 
& \ilread{x}{y}
& \descr{Read $y$th entry at $x$}
\\ 
& 
& \bnfalt 
& \ilwrite{x}{y}{z}
& \descr{Write $z$ as $y$th entry at $x$}

\\[1mm]
& v
& \bnfdef 
& n \bnfalt x
& \descr{Natural numbers, variables}
\end{array}
\]
\mycaptionrule
\caption{\IL syntax.}
\label{fig:il-syntax}
\end{figure}

\figref{il-syntax} shows the abstract syntax for~\IL.
Programs in \IL are expressions, which we partition into
traced~$\iltexp$ and untraced~$\iluexp$.  This distinction does not
constrain the language; it merely streamlines the technical
presentation.  Expressions in \IL follow an administrative normal form
(ANF)~\cite{FlanaganSaDuFe93} where (nearly) all values are variables.

Expressions consist of function definitions, primitive operations,
conditionals, function calls, store instructions~($\iota$), \kw{memo}
points, \kw{update} points, and operations for pushing (\kw{push}) and
popping (\kw{pop}) the stack.
Store instructions~($\iota$) consist of operations for allocating
(\kw{alloc}), reading (\kw{read}) and writing (\kw{write}) memory.
Values~$v$ include natural numbers and variables (but not function
names).
Each expression ends syntactically with either a function call or a
stack \kw{pop} operation.
Since the form for function calls is syntactically in tail position,
the \IL program must explicitly push the stack to perform non-tail
calls.
Expressions terminate when they \kw{pop} on an empty stack---they
yield the values of this final pop.

Notice that \IL programs are first-order: although functions can nest
syntactically, they are not values; moreover, function names~$f,g,h$
are syntactically distinct from variables~$x,y,z$.
Supporting either first-class functions (functions as values) or
function pointers is beyond the scope of the current work, though we
believe our semantics could be adapted for these settings\footnote{
  For example, to model function pointers, one could adapt this
  semantics to allow a function~$f$ to be treated as a value if $f$ is
  closed by its arguments; this restriction models the way that
  functions in C admit function pointers, a kind of ``function as a
  value'', even though C does not include features typically
  associated with first-class functions (e.g. implicitly-created
  closures, partial application).
}.

In the remainder, we restrict our attention to programs (environments
$\rho$ and expressions $e$) that are well-formed in the following
sense:
\begin{enumerate}
\item They have a unique arity (the length of the value sequence they
  potentially return) that can be determined syntactically.
\item All variable and function names therein are distinct.  (This can
  easily be implemented in a compiler targeting \IL.)  Consequently we
  don't have to worry about the fact that \IL is actually dynamically
  scoped.
\end{enumerate}

\subsection{Machine Configurations and Transitions}
\label{sec:machineconfig}

In addition to sharing a common expression
language~(viz. \IL,~\secref{il-syntax}), the reference and tracing
machines share common \emph{machine components}; they also have
related \emph{transition relations}, which specify how these machines
change their components as they run \IL programs.

\paragraph{Machine configurations.} 
Each machine configuration consists of a handful of components.
\figref{il-state} defines the common components of two machines: a
store~($\sigma$), a stack~($\kappa$), an environment~($\rho$) and a
command~($\rcmd$ for the reference machine, and $\tcmd$ for the
tracing machine).
The tracing machine has an additional component---its trace---which we
describe in \secreftwo{traces}{contextszippers}.

A store~$\sigma$ maps each store \emph{entry}~($\ell[n]$) to either
\emph{uninitialized} contents~(written $\bot$) or a machine
value~$\nu$.
Each entry~$\ell[n]$ consists of a store location~$\ell$ and a
(natural number) offset~$n$.
In addition, a store may mark a location as garbage, denoted as $\ell
\mapsto \storegarb$, in which case all store entries for $\ell$ are
undefined.  These garbage locations are not used in the reference
semantics; in the tracing machine, they help to define a notion of
garbage collection.
A stack~$\kappa$ is a (possibly empty) sequence of \emph{frames},
where each frame~$\mkstframe \rho f$ saves an evaluation context that
consists of an environment~$\rho$ and a function~$f$~(defined
in~$\rho$).
An {\em environment} $\rho$ maps variables to machine values and
function names to their definitions.

In the case of the reference machine, a \emph{(reference)
  command}~$\rcmd$ is either an \IL expression~$e$ or a sequence of
machine values~$\vec \nu$; for the tracing machine, a \emph{(tracing)
  command}~$\tcmd$ is either $e$, $\vec \nu$, or an additional
command~\kwprop, which indicates that the machine is performing
\emph{change propagation} (i.e., replay of an existing trace).

Each \emph{machine value}~$\nu$ consists of a natural number~$n$ or a
store location~$\ell$.
Intuitively, we think of machine values as corresponding to machine
words, and we think of the store as mapping location-offset pairs
(each of which is itself a machine word) to other machine words.

For convenience, when we do not care about individual components of a
machine configuration (or some other syntactic object), we often use
underscores (\_) to avoid giving them names.  The quantification
should always be clear from context.

\paragraph{Transition relations.}

In the reference machine, each machine configuration, written
$\ilustate \sigma \kappa \rho \rcmd$, consists of four components: a
store, a stack, an environment and a command, as described above.
In \secref{il-rmachine}, we formalize the following stepping relation
for the reference machine:
$$
\ilustate \sigma \kappa \rho \rcmd
\ilusteprel
\ilustate {\sigma'} {\kappa'} {\rho'} {\rcmd'}
$$
Intuitively, the command~$\rcmd$ tells the reference machine what to
do next.  In the case of an expression~$e$, the machine proceeds by
evaluating~$e$, and in the case of machine values~$\vec \nu$, the
machine proceeds by popping a stack frame~$\mkstframe \rho f$ and
using it as the new evaluation context.  
If the stack is empty, the machine terminates and the command $\vec
\nu$ can be viewed as giving the machine's results.
Since these results may consist of store locations, the complete
extensional result of the machine must include the store (or at least,
the portion reachable from $\vec \nu$).

The tracing machine has similar machine configurations, though it also
includes a pair~$\trzip \trx \tr$ that represents the current trace,
which may be in the midst of adjustment; we describe this component
separately in \secreftwo{traces}{contextszippers}.
In \secref{il-tmachine}, we formalize the following stepping relation
for the tracing machine:
$$
\iltstate \sigma {\trz{\trx}{\tr}} \kappa \rho 
\tcmd
\iltsteprel
\iltstate  {\sigma'} {\trz{\trx'}{\tr'}} {\kappa'} {\rho'}
{\tcmd'}
$$
At a high level, this transition relation accomplishes several things:
(1) it ``mirrors'' the semantics of the reference machine when
evaluating \IL expressions; (2) it traces this evaluation, storing the
generated trace within its trace component; and (3) it allows
previously-generated traces to be either reused (during change
propagation), or discarded (when they cannot be reused).
To accomplish these goals, the tracing machine distinguishes machine
transitions for change propagation from those of normal execution by
giving change propagation the distinguished command~\kw{prop}.

\subsection{Reference Machine Transitions}
\label{sec:il-rmachine}
\begin{figure}
\[
\begin{array}{rrcl}  

\textit{Store} & \sigma & \bnfdef & \emptyenv
\bnfalt 
\sigma[\ell[n] \mapsto \bot]
\\
&&\bnfalt&
\sigma[\ell[n] \mapsto \nu]
\bnfalt
\sigma[\ell \mapsto \storegarb]
\\[1mm]

\textit{Stack} & \kappa & \bnfdef & \emptystack
\bnfalt 
\stcons{\kappa}{\mkstframe{\rho}{f}}
\\[1mm]

\textit{Environment} & \rho & \bnfdef & \emptyenv 
\bnfalt  \rho[ x \mapsto \nu ]
\\
&& \bnfalt & \rho[ f \mapsto \ilfun{f}{\vec x}{e} ]
\\[1mm]

\textit{Reference command} & \rcmd & 
\bnfdef & e
\bnfalt   \vec \nu
\\[1mm]

\textit{Tracing command} & \tcmd & 
\bnfdef & \rcmd
\bnfalt   \kwprop
\\[1mm]

\textit{Machine value} & \nu
&  \bnfdef & n \bnfalt \ell
\end{array}
\]

\mycaptionrule
\caption{Common machine components.}
\label{fig:il-state}
\end{figure}

\begin{figure}[t]

\begin{mathpar}
\noindent
\inferrule*[right=\refreffun]
{
  \rho' = \rho[f \mapsto \ilfun{f}{\vec x}{e_1}]
}
{
  \ilustate
  {\sigma}
  {\kappa}
  {\rho}
  {\illet{\ilfun{f}{\vec x}{e_1}}{e_2}}
  \ilusteprel 
  \ilustate
  {\sigma}
  {\kappa}
  {\rho'}
  {e_2}
}

\inferrule*[right=\refrefvar]
{
  \rho\fromto{(v_i)}{i=1}{|\vec v|} = \vec \nu
  \\
  \rho' = \rho[x\mapsto \ilprimapp{\vec \nu}]
}
{
  \ilustate
  {\sigma}
  {\kappa}
  {\rho}
  {\illet{\ilvar{x}{\ilprimop{\vec{v}}}}{e}}
  \ilusteprel
  \ilustate
  {\sigma}
  {\kappa}
  {\rho'}
  {e}
}

\inferrule*[right=\refrefift]
{\rho(x) \ne 0}
{
  \ilustate
  {\sigma}
  {\kappa}
  {\rho}
  {\ilif{x}{e_1}{e_2}}
  \ilusteprel
  \ilustate
  {\sigma}
  {\kappa}
  {\rho}
  {e_1}
}

\inferrule*[right=\refrefiff]
{\rho(x) = 0}
{
  \ilustate
  {\sigma}
  {\kappa}
  {\rho}
  {\ilif{x}{e_1}{e_2}}
  \ilusteprel
  \ilustate
  {\sigma}
  {\kappa}
  {\rho}
  {e_2}
}

\inferrule*[right=\refrefapp]
{
  \rho(f) = \ilfun{f}{\vec x}{e}
  \\
  \rho' = \rho\fromto{[x_i\mapsto\rho(x_i)]}{i=1}{|\vec x|}
}
{
  \ilustate
  {\sigma}
  {\kappa}
  {\rho}
  {\ilapp{f}{\vec x}}
  \ilusteprel 
  \ilustate
  {\sigma}
  {\kappa}
  {\rho'}
  {e}
}

\inferrule*[right=\refrefstore]
{
  \storestin
  {\sigma}
  {\rho} 
  {\iota}
  \storestep 
  \storestout
  {\sigma'}{\nu}
}
{
  \ilustate
  {\sigma}
  {\kappa}
  {\rho}
  {\illet{\ilvar{x}{\iota}}{e}}
  \ilusteprel 
  \ilustate
  {\sigma'}
  {\kappa}
  {\rho[x \mapsto \nu]}
  {e}
}

\inferrule*[right=\refrefmemo]
{ }
{
  \ilustate
  {\sigma}
  {\kappa}
  {\rho}
  {\ilmemo{e}}
  \ilusteprel 
  \ilustate
  {\sigma}
  {\kappa}
  {\rho}
  {e}
}

\inferrule*[right=\refrefwake]
{ }
{
  \ilustate
  {\sigma}
  {\kappa}
  {\rho}
  {\ilwake{e}}
  \ilusteprel 
  \ilustate
  {\sigma}
  {\kappa}
  {\rho}
  {e}
}

\inferrule*[right=\refrefpush]
{ }
{
  \ilustate
  {\sigma}
  {\kappa}
  {\rho}
  {\ilpush{f}{e}}
  \ilusteprel 
  \ilustate
  {\sigma}
  {\stcons{\kappa}{\mkstframe{\rho}{f}}}
  {\rho}
  {e}
}

\inferrule*[right=\refrefpop]
{
  \vec \nu = \rho\fromto{(x_i)}{i=1}{|\vec x|}
}
{
  \ilustate
  {\sigma}
  {\kappa}
  {\rho}
  {\ilpop{\vec x}}
  \ilusteprel
  \ilustate
  {\sigma}
  {\kappa}
  {\emptyenv}
  {\vec \nu}
}

\inferrule*[right=\refrefv]
{
  \rho(f) = \ilfun{f}{\vec x}{e}
  \\
  \rho' = \rho\fromto{[x_i \mapsto \nu_i]}{i=1}{|\vec x|}
}
{
  \ilustate
  {\sigma}
  {\stcons{\kappa}{\mkstframe{\rho}{f}}}
  {\emptyenv}
  {\vec \nu}
  \ilusteprel 
  \ilustate
  {\sigma}
  {\kappa}
  {\rho'}
  {e}
}
\end{mathpar}

\caption{Stepping relation for reference machine ($\ilusteprel$).}
\label{fig:il-eval}
\end{figure}

\begin{figure}
  \begin{mathpar}
    \inferrule*[right=\refstorealloc]
    {
      \ell \not\in \domof{\sigma} 
      \\
      \sigma' = \sigma\fromto{[\ell[i] \mapsto \bot]}{i=1}{\rho(x)}
    }
    {
      \storestin
      {\sigma}
      {\rho}
      {\ilalloc{x}}
      \storestep
      \storestout
      {\sigma'}
      {\ell}
    }

    \inferrule*[right=\refstoreread]
    {\sigma(\rho(x)[\rho(y)]) = \nu}
    {
      \storestin
      {\sigma}
      {\rho}
      {\ilread{x}{y}}
      \storestep
      \storestout
      {\sigma}
      {\nu}
    }

    \inferrule*[right=\refstorewrite]
    {
      \sigma' = \sigma[\rho(x)[\rho(y)] \mapsto \rho(z)]
    }
    {
      \storestin
      {\sigma}
      {\rho}
      {\ilwrite{x}{y}{z}}
      \storestep
      \storestout
      {\sigma'}
      {0}
    }
  \end{mathpar}
  \caption{Stepping relation for store instructions ($\storestep$).}
  \label{fig:il-eval-store}
\end{figure}

\figref{il-eval} specifies the transition relation for the reference
machine, as introduced in \secref{machineconfig}.
A function definition updates the environment, binding the function name
to its definition.
A primitive operation first converts each value argument~$v_i$ into a
machine value $\nu_i$ using the environment.  Here we abuse notation
and write $\rho(v)$ to mean $\rho(x)$ when $v = x$ and $n$ when $v =
n$.  The machine binds the result of the primitive operation (as
defined by the abstract \kw{primapp}~function) to the given variable
in the current environment.
A conditional steps to the branch specified by the scrutinee.
A function application steps to the body of the specified function
after updating the environment with the given arguments.
A store instruction~$\iota$ steps using an auxiliary
judgement~(\figref{il-eval-store}) that allocates in, reads from and
writes to the current store.
An \kw{alloc} instruction allocates a fresh location $\ell$ for which
each offset (from 1 to the specified size) is marked as uninitialized.
A \kw{read} (resp. \kw{write}) instruction reads (resp. writes) the
store at a particular location and offset.
A \kw{push} expression saves a return context in the form of a stack
frame~$\mkstframe \rho f$ and steps to the body of the \kw{push}.
A \kw{pop} expression steps to a machine value sequence~$\vec \nu$, as specified
by a sequence of variables.
If the stack is non-empty, the machine passes control to function~$f$,
as specified by the topmost stack frame~$\mkstframe \rho f$, by
applying $f$ to $\vec \nu$; it recovers the environment $\rho$ before
discarding this frame.
Otherwise, if the stack is empty, the value sequence~$\vec \nu$
signals the termination of the machine with results~$\vec \nu$.

\begin{figure}

\[
\begin{array}{@{}rr@{\;\;}c@{\;\;}l@{}}

\textit{Trace} & \tr & 
\bnfdef & 
\trcons t \tr 
\bnfalt \trend
\\[1mm]

\textit{Tr. Action} & t & 
\bnfdef &
\tralloc{\ell}{n} 
\bnfalt  \trread{\nu}{\ell}{n}
\bnfalt  \trwrite{\nu}{\ell}{n}
\bnfalt  \trmemo{\rho}{e} 
\bnfalt  \trwake{\rho}{e} 
\bnfalt  \trpush{\tr} 
\bnfalt  \trpop{\vec\nu}
\\[1mm]

\textit{Tr. Context} & \trx & 
\bnfdef &
\trstart
\bnfalt \trxcons{\trx}{t}
\bnfalt \trxcons{\trx}{\trxpushmark}
\bnfalt \trxcons{\trx}{\trxpropmark{\tr}}
\bnfalt \trxcons{\trx}{\trxundomark{\tr}}
\end{array}
\]
\mycaptionrule
\caption{Traces, trace actions and trace contexts.}
\label{fig:il-trace}
\end{figure}

\subsection{The Structure of the Trace}
\label{sec:traces}

The structure of traces used by the tracing machine is specified by
\figref{il-trace}.
They each consist of a (possibly empty) sequence of zero or more
\emph{trace actions}~$t$.
Each action records a transition for a corresponding traced
expression~$\iltexp$.

In the case of store instructions, the corresponding action indicates
both the instruction and each machine value involved in its evaluation.
For \kw{alloc}s, the action~$\tralloc{\ell}{n}$ records the allocated
location as well as its size~(i.e., the range of offsets it defines).
For \kw{read}s ($\trread{\nu}{\ell}{n}$) and \kw{write}s
($\trwrite{\nu}{\ell}{n}$) the action stores the location and offset
being accessed, as well as the machine value being read or written,
respectively.
For \kw{memo} expressions, the trace action~$\trmemo \rho e$ records
the body of the memo point, as well as the current environment at this
point; \kw{update} expressions are traced analogously.
For \kw{push} expressions, the action~$\trpush T$ records the trace of
evaluating the \kw{push} body; it is significant that in this case,
the trace action is not atomic: it consists of the arbitrarily large
subtrace~$\tr$.
For \kw{pop} expressions, the action~$\trpop{\vec{\nu}}$ records the
machine values being returned via the stack.

There is a close relationship between the syntax of traced expressions
in \IL and the structure of their traces.
For instance, in nearly all traced expressions, there is exactly one
subexpression, and hence their traces~$\trcons{t}{\tr}$ contain
exactly one subtrace,~$\tr$.
The exception to this is \kw{push}, which can be thought
of as specifying two subexpressions:
the first subexpression is given by the body of the \kw{push}, and
recorded within the push action as $\trpush T$; the second
subexpression is the body of the function being pushed, which is
evaluated when the function is later popped.
Hence, push expressions generate traces of the
form~$\trcons{\trpush{T}}{T'}$, where $\tr'$ is the trace generated by
evaluating the pushed/popped function.

\subsection{Trace Contexts and the Trace Zipper}
\label{sec:contextszippers}
\begin{figure}[t]
\centering
\begin{minipage}{\columnwidth}
\includegraphics[width=\columnwidth,page=4]{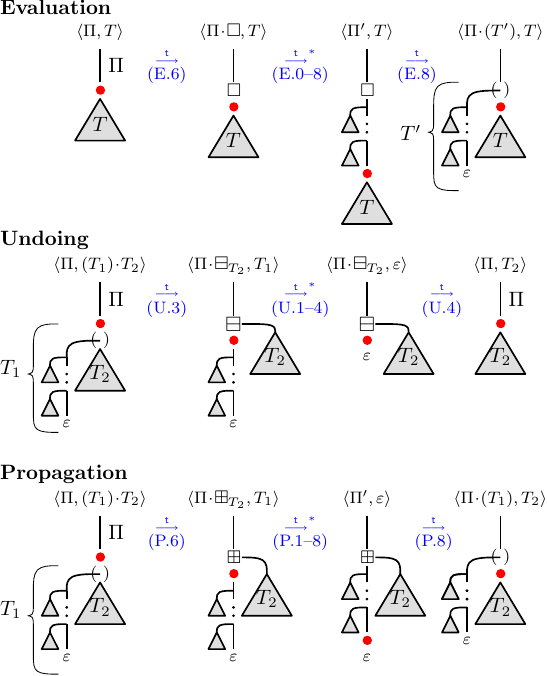}
\caption{Tracing transition modes, across push actions.}
\label{fig:il-tracing-machine-subtraces}  
\end{minipage}
\\[5mm]

\begin{minipage}{\columnwidth}
\centering
\includegraphics[width=0.75\columnwidth,page=2]{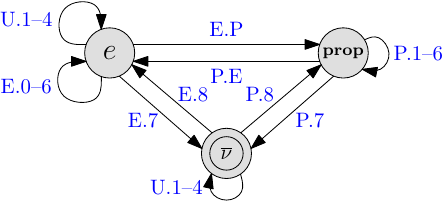}
\caption{Tracing machine: commands and transitions.}
\label{fig:il-tracing-machine-commands}  
\end{minipage}

\end{figure}

As described above, our traces are not strictly sequential structures:
they also consist of nested subtraces created by \kw{push}.
This fact poses a technical challenge for transition semantics (and by
extension, an implementation).
For instance, while generating such a subtrace, 
how should we maintain the context of the trace that
will eventually enclose it?

To address this, the machine augments the trace with a
\emph{context}~(\figref{il-trace}), maintaining in each configuration
both a \emph{reuse trace}~$\tr$, which we say is \emph{in focus}, as
well as an unfocused trace context~$\trx$.
The trace context effectively records a path from the focus
back to the start of the trace.
To move the focus in a consistent manner, the machine places
additional markings $\pushmark$, $\undomark$,
$\propmark$ into the context; two of these markings (viz. $\undomark$, $\propmark$)
also carry a subtrace.  We describe these markings and their subtraces
in more detail below.

This pair of components~$\trz{\trx}{\tr}$ forms a kind of \emph{trace
  zipper}.  More generally, a zipper augments a data structure with a
\emph{focus} (for zipper $\trz{\trx}{\tr}$, we say that $\tr$ is
\emph{in focus}), the ability to perform local edits at the
focus and the ability to move this focus throughout the
structure~\citep{Huet97,AbbottThMcGh04}.
A particularly attractive feature of zippers is that the ``edits'' can
be performed in a non-destructive, incremental
fashion.

To characterize focus movement using trace zippers, we define the
\emph{transition modes} of the tracing machine:

\begin{itemize}

\item \textbf{Evaluation} mirrors the transitions of the reference machine
  and generates new trace actions, placing them behind the focus,
  i.e., $\trz \trx \tr$ becomes $\trz {\trxcons \trx t} \tr$.

\item \textbf{Undoing} removes actions from the reuse trace, just
  ahead of the focus, i.e., $\trz \trx {\trcons t \tr}$ becomes $\trz
  \trx \tr$.

\item \textbf{Propagation} replays the actions of the reuse trace; it
  moves the focus through it action by action, i.e., $\trz \trx
  {\trcons t \tr}$ becomes $\trz {\trxcons \trx t} \tr$.
\end{itemize}
If we ignore \kw{push} actions and their nested subtraces~$\trpush{T}$,
the tracing machine moves the focus in the manner just described,
either generating, undoing or propagating at most one trace action
for each machine transition.
However, since \kw{push} actions consist of an entire subtrace~$\tr$,
the machine cannot generate, undo or propagate them in a single step.
Rather, the machine must make a series of transitions, possibly
interleaving transition modes.
When this process completes and the machine moves its focus out of the
subtrace, it is crucial that it does so in a manner consistent with
its mode upon entering the subtrace.
To this end, the machine may extend the context~$\trx$ with one of three
possible markings, each corresponding to a mode.

For each transition mode, \figref{il-tracing-machine-subtraces} gives
both syntactic and pictorial representations of the focused traces and
illustrates how the machine moves its focus.
The transitions are labeled with corresponding
\coloronly{(\textcolor{blue}{blue})~}transition rules from the tracing
machine, but at this time the reader can ignore them.
For each configuration, the (initial) trace context is illustrated
with a vertical line, the focus is represented by a
\coloronly{(\textcolor{red}{red})} filled circle and the (initial)
reuse trace is represented by a tree-shaped structure that hangs below
the focus.

\paragraph{Evaluation.}
To generate a new subtrace in evaluation mode (via a \kw{push}), the
machine extends the context $\trx$ to $\trxcons \trx \trxpushmark$;
this effectively marks the beginning of the new subtrace.
The machine then performs evaluation transitions that extend the
context, perhaps recursively generating nested subtraces in the
process (drawn as smaller, unlabeled triangles hanging to the left).
After evaluating the \kw{pop} matching the initial \kw{push}, the
machine \emph{rewinds} the current context~$\trx'$, moving the focus
back to the mark~$\trxpushmark$, gathering actions and building a
completed subtrace~$\tr'$; it replaces the mark with a push
action~$\trpush{\tr'}$ (consisting of the completed subtrace), and it
keeps reuse trace~$\tr$ in focus.
We specify how this rewinding works in \secref{il-tmachine};
intuitively, it simply moves the focus backwards, towards the start of
the trace.

\paragraph{Undoing.}
To undo a subtrace~$\tr_1$ of the reuse trace~$\trcons
{\trpush{\tr_1}} {\tr_2}$, the machine extends the context $\trx$ to
$\trxcons{\trx}{\trxundomark{\tr_2}}$; this effectively saves the
remaining reuse trace~$\tr_2$ for either further undo transitions or
for eventual reuse.
Assuming that the machine undoes all of ${\tr_1}$, it will eventually
focus on an empty trace~$\trend$.  In this case, the machine can
move the saved subtrace~$\tr_2$ into focus~(again, for either further
undo transitions or for reuse).

\paragraph{Propagation.}
Finally, to propagate a subtrace~$\tr_1$, the machine uses an approach
similar to undoing: it saves the remaining trace~$\tr_2$ in the
context using a distinguished mark~$\trxpropmark{\tr_2}$, moves the
focus to the end of $\tr_1$ and eventually places~$\tr_2$ into focus.
In contrast to the undo transitions, however, propagation transitions
do not discard the reuse trace, but only move the focus by moving
trace actions from the reuse trace into the trace context.
Just as in evaluation mode, in propagation mode we rewind these
actions from the context and move the focus back to the propagation
mark~($\propmark$).

We note that while our semantics characterizes change propagation
using a step-by-step replay of the trace, this does not yield an
efficient algorithm.  In \secref{runtime}, we give an efficient
implementation that is faithful to this replay semantics, but in which
the change propagation transitions have zero cost.

\subsection{Tracing Machine Transitions}
\label{sec:il-tmachine}
\newcommand{\transitionrule}[5]{
  #1 & #3 & #4 & #5 & #2
  \\
}
\newcommand{\whenclause}[1]{
  \text{when}&#1
}
\newcommand{\andclause}[1]{
  \\
  &&&&\text{and}&#1
}
\newcommand{\ruleheading}[1]{
  \\
  \multicolumn{6}{l}{\textbf{#1}}
  \\
}
\begin{figure*}
\small
\centering
\begin{minipage}{\textwidth}
\[
\begin{array}{lrclrl}
\noindent
\ruleheading{Evaluation}
\transitionrule{\refevalu}
{
\whenclause{
\ilustate
{\sigma}
{\kappa}
{\rho}
{\iluexp}
\ilusteprel
\ilustate
{\sigma}
{\kappa}
{\rho'}
{e}
}
}
{
\iltstate
{\sigma}
{\trz{\trx}{T}}
{\kappa}
{\rho}
{\iluexp}
}
{\iltevalstep}
{
\iltstate
{\sigma}
{\trz{\trx}{T}}
{\kappa}
{\rho'}
{e}
}

\transitionrule{\refevalalloc}
{
\whenclause{
\storestin{\sigma}{\rho}{\ilalloc{y}} 
\storestep 
\storestout{\sigma'}{\ell}
}
}
{
\iltstate
{\sigma}
{\trz{\trx}{T}}
{\kappa}
{\rho}
{\illet{\ilvar{x}{\ilalloc{y}}}{e}}
}
{\iltevalstep}
{
\iltstate
{\sigma'}
{\trz{\trcons{\trx}{\tralloc{\ell}{\rho(y)}}}{T}}
{\kappa}
{\rho[x\mapsto \ell]}
{e}
}

\transitionrule{\refevalread}
{
\whenclause{
\storestin{\sigma}{\rho}{\ilread{y}{z}} \storestep
\storestout{\sigma}{\nu}
}
}
{
\iltstate
{\sigma}
{\trz{\trx}{T}}
{\kappa}
{\rho}
{\illet{\ilvar{x}{\ilread{y}{z}}}{e}}
}
{\iltevalstep}
{
\iltstate
{\sigma}
{\trz{\trcons{\trx}{\trread{\nu}{\rho(y)}{\rho(z)}}}{T}}
{\kappa}
{\rho[x \mapsto \nu]}
{e}
}

\transitionrule{\refevalwrite}
{
\whenclause{
\storestin{\sigma}{\rho}{\ilwrite{x}{y}{z}} \storestep
\storestout{\sigma'}{0}
}
}
{
\iltstate
{\sigma}
{\trz{\trx}{T}}
{\kappa}
{\rho}
{\illet{\ilvar{\_}{\ilwrite{x}{y}{z}}}{e}}
}
{\iltevalstep}
{
\iltstate
{\sigma'}
{\trz{\trcons{\trx}{\trwrite{\rho(z)}{\rho(x)}{\rho(y)}}}{T}}
{\kappa}
{\rho}
{e}
}

\transitionrule{\refevalmemo}
{
\strut
}
{
\iltstate
{\sigma}
{\trz{\trx}{T}}
{\kappa}
{\rho}
{\ilmemo{e}}
}
{\iltevalstep}
{
\iltstate
{\sigma}
{\trz{\trcons{\trx}{\trmemo{\rho}{e}}}{T}}
{\kappa}
{\rho}
{e}
}

\transitionrule{\refevalwake}
{
\strut
}
{
\iltstate
{\sigma}
{\trz{\trx}{T}}
{\kappa}
{\rho}
{\ilwake{e}}
}
{\iltevalstep}
{
\iltstate
{\sigma}
{\trz{\trcons{\trx}{\trwake{\rho}{e}}}{T}}
{\kappa}
{\rho}
{e}
}

\transitionrule{\refevalpush}
{
}
{
\iltstate
{\sigma}
{\trz{\trx}{T}}
{\kappa}
{\rho}
{\ilpush{f}{e}}
}
{\iltevalstep}
{
\iltstate
{\sigma}
{\trz{\trcons{\trx}{\trxpushmark}}{T}}
{\stcons{\kappa}{\mkstframe{\rho}{f}}}
{\rho}
{e}
}

\transitionrule{\refevalpop}
{
\whenclause{
  \vec \nu = \rho\fromto{(x_i)}{i=1}{|\vec x|}  
}
}
{
\iltstate
{\sigma}
{\trz{\trx}{T}}
{\kappa}
{\rho}
{\ilpop {\vec x}}
}
{\iltevalstep}
{
\iltstate
{\sigma}
{\trz{\trcons{\trx}{\trpop{\vec \nu}}}{T}}
{\kappa}
{\emptyenv}
{\vec \nu}
}

\transitionrule{\refevalv}
{
\whenclause{
\trz{\trx}{T_2} ; \trend
  \trrewind^\ast
  \trz{\trcons{\trx'}{\trxpushmark}}{T'_2}; T_1 
}
\andclause{
  \rho(f) = \ilfun{f}{\vec x}{e}
}
\andclause{
  \rho' = \rho\fromto{[x_i \mapsto \nu_i]}{i=1}{|\vec x|}
}
}
{
\iltstate
{\sigma}
{\trz{\trx}{T_2}}
{\stcons{\kappa}{\mkstframe{\rho}{f}}}
{\emptyenv}
{\vec \nu}
}
{\iltevalstep}
{
\iltstate
{\sigma}
{\trz{\trcons{\trx'}{\trpush{T_1}}}{T'_2}}
{\kappa}
{\rho'}
{e}
}

\vspace{-5mm}

\ruleheading{Reevaluation and reuse}

\transitionrule{\refpropeval}
{ }
{
\iltstate
{\sigma}
{\trz{\Pi}{\trcons{\trwake{\rho}{e}}{T}}}
{\kappa}
{\emptyenv}
{\ilprop}
}
{\iltpropevalstep}
{
\iltstate
{\sigma}
{\trz{\trxcons{\Pi}{\trwake{\rho}{e}}}{T}}
{\kappa}
{\rho}
{e}
}

\transitionrule{\refevalprop}
{ }
{
\iltstate
{\sigma}
{\trz{\trx}{\trcons{\trmemo{\rho}{e}}{T}}}
{\kappa}
{\rho}
{\ilmemo{e}}
}
{\iltevalpropstep}
{
\iltstate
{\sigma}
{\trz{\trxcons{\trx}{\trmemo{\rho}{e}}}{T}}
{\kappa}
{\emptyenv}
{\ilprop}
}

\ruleheading{Propagation}

\transitionrule{\refpropalloc}
{
\whenclause{
\storestin{\sigma}{\emptyenv}{\ilalloc{n}} \storestep
\storestout{\sigma'}{\ell}
}
}
{
\iltstate
{\sigma}
{\trz{\trx}{\trcons{\tralloc{\ell}{n}}{T}}}
{\kappa}
{\emptyenv}
{\ilprop}
}
{\iltpropstep}
{
\iltstate
{\sigma'}
{\trz{\trcons{\trx}{\tralloc{\ell}{n}}}{T}}
{\kappa}
{\emptyenv}
{\ilprop}
}

\transitionrule{\refpropread}
{
\whenclause{
\storestin{\sigma}{\emptyenv}{\ilread{\ell}{n}} \storestep
\storestout{\sigma}{\nu}
}
}
{
\iltstate
{\sigma}
{\trz{\trx}{\trcons{\trread{\nu}{\ell}{n}}{T}}}
{\kappa}
{\emptyenv}
{\ilprop}
}
{\iltpropstep}
{
\iltstate
{\sigma}
{\trz{\trcons{\trx}{\trread{\nu}{\ell}{n}}}{T}}
{\kappa}
{\emptyenv}
{\ilprop}
}

\transitionrule{\refpropwrite}
{
\whenclause{
\storestin{\sigma}{\emptyenv}{\ilwrite{\ell}{n}{\nu}} \storestep
\storestout{\sigma'}{0}
}
}
{
\iltstate
{\sigma}
{\trz{\trx}{\trcons{\trwrite{\nu}{\ell}{n}}{T}}}
{\kappa}
{\emptyenv}
{\ilprop}
}
{\iltpropstep}
{
\iltstate
{\sigma'}
{\trz{\trxcons{\trx}{\trwrite{\nu}{\ell}{n}}}{T}}
{\kappa}
{\emptyenv}
{\ilprop}
}

\transitionrule{\refpropmemo}
{\strut}
{
\iltstate
{\sigma}
{\trz{\trx}{\trcons{\trmemo{\rho}{e}}{T}}}
{\kappa}
{\emptyenv}
{\ilprop}
}
{\iltpropstep}
{
\iltstate
{\sigma}
{\trz{\trcons{\trx}{\trmemo{\rho}{e}}}{T}}
{\kappa}
{\emptyenv}
{\ilprop}
}

\transitionrule{\refpropwake}
{\strut}
{
\iltstate
{\sigma}
{\trz{\trx}{\trcons{\trwake{\rho}{e}}{T}}}
{\kappa}
{\emptyenv}
{\ilprop}
}
{\iltpropstep}
{
\iltstate
{\sigma}
{\trz{\trcons{\trx}{\trwake{\rho}{e}}}{T}}
{\kappa}
{\emptyenv}
{\ilprop}
}

\transitionrule{\refproppush}
{\strut}
{
\iltstate
{\sigma}
{\trz{\trx}{\trcons{\trpush{T_1}}{T_2}}}
{\kappa}
{\emptyenv}
{\ilprop}
}
{\iltpropstep}
{
\iltstate
{\sigma}
{\trz{\trcons{\trx}{\trxpropmark{T_2}}}{T_1}}
{\kappa}
{\emptyenv}
{\ilprop}
}

\transitionrule{\refproppop}
{\strut}
{
\iltstate
{\sigma}
{\trz{\trx}{\trpop{\vec \nu}}}
{\kappa}
{\emptyenv}
{\ilprop}
}
{\iltpropstep}
{
\iltstate
{\sigma}
{\trz{\trcons{\trx}{\trpop{\vec \nu}}}{\trend}}
{\kappa}
{\emptyenv}
{\vec \nu}
}

\transitionrule{\refpropv}
{
\whenclause{
\trz{\trx}{\trend}; \trend
\trrewind^\ast
\trz{\trcons{\trx'}{\trxpropmark{T_2}}}{\trend}; T_1
}
}
{
\iltstate
{\sigma}
{\trz{\trx}{\trend}}
{\kappa}
{\emptyenv}
{\vec \nu}
}
{\iltpropstep}
{
\iltstate
{\sigma}
{\trz{\trcons{\trx'}{\trpush{T_1}}}{T_2}}
{\kappa}
{\emptyenv}
{\ilprop}
}

\ruleheading{Undoing}

\transitionrule{\refundoalloc}
{\strut}
{
\iltstate
{\sigma}
{\trz{\trx}{\trcons{\tralloc{\ell}{n}}{\tr}}}
{\kappa}
{\rho}
{\rcmd}
}
{\iltundostep}
{
\iltstate
{\sigma[\ell \mapsto \storegarb]}
{\trz{\trx}{\tr}}
{\kappa}
{\rho}
{\rcmd}
} 

\transitionrule{\refundostep}
{
\whenclause{
(t=\trread{\_}{\_}{\_}~
  |~\trwrite{\_}{\_}{\_}~
  |~\trmemo{\_}{\_}~
  |~\trwake{\_}{\_}~
  |~\trpop{\vec \nu}
)
}}
{
\iltstate
{\sigma}
{\trz{\trx}{\trcons{t}{\tr}}}
{\kappa}
{\rho}
{\rcmd}
}
{\iltundostep}
{
\iltstate
{\sigma}
{\trz{\trx}{\tr}}
{\kappa}
{\rho}
{\rcmd}
}

\transitionrule{\refundopush}
{\strut}
{
\iltstate
{\sigma}
{\trz{\trx}{\trcons{\trpush{\tr_1}}{\tr_2}}}
{\kappa}
{\rho}
{\rcmd}
}
{\iltundostep}
{
\iltstate
{\sigma}
{\trz{\trcons{\trx}{\trxundomark{\tr_2}}}{\tr_1}}
{\kappa}
{\rho}
{\rcmd}
}

\transitionrule{\refundomark}
{\strut}
{
\iltstate
{\sigma}
{\trz{\trcons{\trx}{\trxundomark{\tr}}}{\trend}}
{\kappa}
{\rho}
{\rcmd}
}
{\iltundostep}
{
\iltstate
{\sigma}
{\trz{\trx}{\tr}}
{\kappa}
{\rho}
{\rcmd}
}
\end{array}
\]
\caption{Stepping relation for tracing machine ($\iltsteprel$).}
\label{fig:il-tracing-steps}
\end{minipage}
\end{figure*}

We use the components and transitions of the reference
machine~(\secreftwo{machineconfig}{il-rmachine}, respectively) as a
basis for defining the transitions of the tracing machine.
You may recall from \secref{machineconfig} that the tracing machine
extends the reference machine in two important ways.

First, the machine configurations of the tracing machine extend the
reference configurations with an extra component~$\trzip{\trx}{\tr}$,
the trace zipper~(\secref{contextszippers}), which augments the trace
structure~$\tr$~(\secref{traces}) with a trace context and a movable
focus.

Second, a tracing command $\tcmd$ consists of either a reference command
$\rcmd$ or the additional propagation command~\kw{prop}, which
indicates that the machine is doing change propagation.
Using these two extensions of the reference machine, the tracing
machine generates traces of execution (during evaluation transitions),
discards parts of previously-generated traces (during undoing
transitions), and reuses previously-generated traces (during
propagation transitions).

These three transition modes (evaluation, undoing and propagation) can
interact in ways that are not straightforward.
\figref{il-tracing-machine-commands} helps illustrate their
interrelationships, giving us a guide for the transition rules of the
tracing machine.
The arcs indicate the machine command before and after the machine
applies the indicated transition rule\coloronly{~(written in
  \textcolor{blue}{blue})}.
\figref{il-tracing-steps} gives the complete transition relation for
the tracing machine.  Recall that each transition is of the form:
$$
\iltstate \sigma {\trz{\trx}{\tr}} \kappa \rho 
\tcmd
\iltsteprel
\iltstate  {\sigma'} {\trz{\trx'}{\tr'}} {\kappa'} {\rho'}
{\tcmd'}
$$
We explain \figref{il-tracing-steps} using
\figref{il-tracing-machine-commands} as a guide.
Under an expression command~$e$, the machine can take both
evaluation~(\refevaltoeval) and undo~(\refundoall) transitions while
remaining in evaluation mode, as well as transitions \refevalprop
and \refevalpop, which each change to another command.
Under the propagation command~$\kw{prop}$, the machine can take
propagation transitions~(\refproptoprop) while remaining in
propagation mode, as well as transitions \refpropeval and \refproppop,
which each change to another command.

Propagation can transition into evaluation (\refpropeval) when it's
focused on an \kw{update} action that it (non-deterministically)
chooses to activate; it may also (non-deterministically) choose to
ignore this opportunity and continue propagation.
Dually, evaluation can transition directly into
propagation~(\refevalprop) when its command is a \kw{memo} point that
matches a \kw{memo} point currently focused in the reuse trace (and in
particular, the environment~$\rho$ must also match); it may also
(non-deterministically) choose to ignore this opportunity and continue
evaluation.
We describe a deterministic algorithms for change propagation and
memoization in \secref{runtime}.

Evaluation (respectively, propagation) transitions into a value
sequence~$\vec{\nu}$ after evaluating~(respectively, propagating) a
\kw{pop} operation under \refevalpop~(respectively, \refproppop).
Under the value sequence command, the machine can continue to undo the
reuse trace (\refundoall).
To change commands, it rewinds its trace context and either resumes
evaluation~(\refevalv) upon finding the mark~$\pushmark$, or resumes
propagation~(\refpropv) upon finding the mark~$\propmark$.
The machine rewinds the trace using the following \emph{trace
  rewinding} relation:
\[
\begin{array}{lcl}
{\trz{\trcons{\Pi}{t}}{T}} ; {T'}
& \trrewind &
{\trz{\Pi}{T}} ; \trcons{t}{T'}
\\
{\trz{\trcons{\Pi}{\trxundomark{T_2}}}{\trend}} ; T'
& \trrewind &
{\trz{\Pi}{T_2}} ; T'
\\
{\trz{\trcons{\Pi}{\trxundomark{T_2}}}{\trcons{t}{T_1}}} ; T'
& \trrewind &
{\trz{\Pi}{\trcons{\trpush{\trcons{t}{T_1}}}{T_2}}} ; T'
\end{array}
\]
This relation simultaneously performs two functions.  First, it moves
the focus backwards across actions (towards the start of the trace)
while moving these actions into a new subtrace~$\tr'$; the first case
captures this behavior.  Second, when moving past a leftover undo
mark~$\trxundomark{T_2}$, it moves the subtrace~$T_2$ back into the
reuse trace; the second and third cases capture this behavior.
Note that unlike~$\undomark$, there is no way to rewind beyond either
$\pushmark$ or $\propmark$ marks.  This is intentional: rewinding is
meant to stop when it encounters either of these marks.

\section{Consistency}
\label{sec:consistency}
\label{sec:il-consistency}

In this section we formalize a notion of consistency between the
reference machine and tracing machine.
As a first step, we show that when run from scratch (without a
reuse trace), the results of the tracing machine are consistent with
the reference machine, i.e., the final machine values and stores
coincide.
To extend this property beyond from-scratch runs, it is necessary to
make an additional assumption: we require each \IL program run in the
tracing machine to be \emph{compositionally store agnostic} (CSA, see
below).
We then show that, for CSA \IL programs, the tracing machine reuses
computations in a consistent way: its final trace,
store, and machine values are consistent with a from-scratch run of
the tracing machine, and hence, they are consistent with a run of the
reference machine.

\techreportonly{
  Finally, we discuss some interesting invariants of the tracing
  machine~(\secref{invariants}) that play a crucial role in the
  consistency proof.
}

\subsection{Compositional Store Agnosticism (CSA)}
\label{sec:csa}

The property of compositional store agnosticism characterizes the
programs for which our tracing machine runs consistently.
We build this property from a less general property that we call
\emph{store agnosticism}.
Intuitively, an \IL program is store agnostic iff, whenever an
\kw{update} instruction is performed during its execution, then the
value sequence that will eventually be popped is already determined at
this point and, moreover, independent of the current store.

\begin{defn}
  Formally, we define $\sa{\rho}{e}{\sigma}$ to mean: \\ If $
  \ilustepm {\sigma}{\epsilon}{\rho \statesep e} {\_}{\_}{\rho'
    \statesep \ilupdate{e'}}$, then there exists $\vec{\nu}$ such that
  $\seqof w = \vec{\nu}$ whenever $\ilustepm {\_}{\epsilon}{\rho'
    \statesep {e'}} {\_}{\epsilon}{\epsilon \statesep \seqof{w}}$.
\end{defn}

To see why this property is significant, recall how the tracing
machine deals with intermediate results.  In stepping
rule~\refevalv, the tracing machine mirrors the reference machine:
it passes the results to the function on the top of the control stack.
However, in stepping rule~\refpropv, the tracing machine does
\emph{not} mirror the reference machine: it essentially discards the
intermediate results and continues to process the remaining reuse
trace.
This behavior is not generally consistent with the reference machine:
If \refpropv is executed after switching to evaluation mode
(\refpropeval) and performing some computation in order to adjust to a
modified store, then the corresponding intermediate result may be
different.  However, if the subprogram that generated the reuse trace
was store agnostic, then this new result will be the same as the
original one; consequently, it is then safe to continue processing the
remaining reuse trace.

Compositional store agnosticism is a generalization of store
agnosticism that is preserved by execution.
\begin{defn}
  We define $\csa{\rho}{e}{\sigma}$ to mean: \\ If
  $\ilustepm{\sigma}{\emptystack}{\rho \statesep
    e}{\sigma'}{\kappa}{\rho' \statesep e'}$, then
  $\sa{\rho'}{e'}{\sigma'}$.
\end{defn}
\begin{lem}
  \label{lem:paper-csa-preservation-untraced}
  If $\ilustate{\sigma}{\emptystack}{\rho}{e} \ilusteprel^\ast
  \ilustate{\sigma'}{\kappa'}{\rho'}{e'}$ and $\csa{\rho}{e}{\sigma}$,
  then $\csa{\rho'}{e'}{\sigma'}$.
\end{lem}

\subsection{Consistency of the Tracing Machine}

The first correctness property says that, when run from scratch
(i.e.\ without a reuse trace), the tracing machine mirrors the
reference machine.
\begin{thm}[Consistency of from-scratch runs]~\\
\label{thm:consistency-fromscratch}
If $\iltstepm
{\trz \trstart \trend}{\sigma}{\emptystack}{\rho\statesep\rcmd}
{\trz \_ \_}{\sigma'}{\emptystack}{\emptyenv\statesep\vec \nu}$
\\
then
$\ilustepm
{\sigma}{\emptystack}{\rho\statesep\rcmd}
{\sigma'}{\emptystack}{\emptyenv\statesep\vec \nu}$.
\end{thm}

In the general case, the tracing machine does not run from scratch,
but with a reuse trace generated by a from-scratch run.
To aid readability for such executions we introduce some notation. We
call a machine reduction \emph{balanced} if the initial and final
stacks are each empty, and the initial and final trace contexts are
related by the trace rewinding relation.
If know that the stack and trace context components of a machine
reduction meet this criteria, we can specify this (balanced) reduction
more concisely.
\begin{defn}[Balanced reductions]
  \begin{mathpar}
    \inferrule*{
      \iltstepm
      {\trz{\trstart}{\tr}}{\sigma}{\epsilon}{\rho\statesep\rcmd}
      {\trz{\trx}{\trend}}{\sigma'}{\epsilon}{\epsilon\statesep\vec\nu}
      \\\\
      \rewindstepm{\trz{\trx}{\trend}}{\trend}{\trz{\trstart}{\trend}}{\tr'}
    }
    {
      \iltbigstep{\tr}{\sigma}{\rho\statesep\rcmd}{\tr'}{\sigma'}{\vec\nu}
    }
    \and
    \inferrule*{
      \iltstepm
      {\trz{\trstart}{\tr}}{\sigma}{\epsilon}{\epsilon\statesep\ilprop}
      {\trz{\trx}{\trend}}{\sigma'}{\epsilon}{\epsilon\statesep\vec\nu}
      \\\\
      \rewindstepm{\trz{\trx}{\trend}}{\trend}{\trz{\trstart}{\trend}}{\tr'}
    }
    {
      \iltbigprop{\tr}{\sigma}{\tr'}{\sigma'}{\vec\nu}
    }
  \end{mathpar}
\end{defn}

We now state our second correctness result.  It uses an auxiliary
function that collects garbage: $\nongarbof{\sigma}(\ell) =
\sigma(\ell)$ for $\ell \in \domof{\nongarbof{\sigma}} =
\{\ell~|~\ell \in \domof{\sigma}~\text{and}~\sigma(\ell) \ne \storegarb\}$.
\begin{thm}[Consistency]~\\
  \label{thm:general-consistency}
  Suppose~$\iltbigstep{\trend}{\sigma_1}{\rho_1\statesep\rcmd_1}{\tr_1}{\sigma_1'}{\vec\nu_1}$
  and $\csa{\rho_1}{\rcmd_1}{\sigma_1}$.
  \begin{enumerate}
  
  \item
    If~$\iltbigstep{\tr_1}{\sigma_2}{\rho_2\statesep\rcmd_2}{\tr_1'}{\sigma_2'}{\vec\nu_2}$
    \\
    then~$\iltbigstep{\trend}{\nongarbof{\sigma_2}}{\rho_2\statesep\rcmd_2}{\tr_1'}{\nongarbof{\sigma_2'}}{\vec\nu_2}$
    \vspace{2mm}

  \item
    If~$\iltbigprop{\tr_1}{\sigma_2}{\tr_1'}{\sigma_2'}{\vec\nu_2}$
    \\
    then~$\iltbigstep{\trend}{\nongarbof{\sigma_2}}{\rho_1\statesep\rcmd_1}{\tr_1'}{\nongarbof{\sigma_2'}}{\vec\nu_2}$
  \end{enumerate}
\end{thm}

The first statement says that, when run with an arbitrary from-scratch
generated trace~$\tr_1$, the tracing machine produces a final trace,
store and return value sequence that are consistent with a
from-scratch run of the same program.
The second statement is analogous, except that it concerns change
propagation: when run over an arbitrary from-scratch generated
trace~$\tr_1$, the machine produces a result consistent with a
from-scratch run of the program that generated~$\tr_1$.  Note that in
each case the initial store may be totally different from the one used
to generate~$\tr_1$.

Finally, observe how each part of
Theorem~\ref{thm:general-consistency} can be composed with
Theorem~\ref{thm:consistency-fromscratch} to obtain a corresponding
run of the reference machine.

\paragraph{Collecting the garbage.} The tracing machine may undo portions of the
reuse trace in order to adjust it to a new store.  Whenever it undoes
an allocation (rule \refundoalloc), it marks the corresponding
location as garbage ($\ell \mapsto \storegarb$).

In order for this to make sense we better be sure that these locations
are not live in the final result, i.e., they neither appear in
$\tr_1'$ nor $\vec{\nu}_2$ nor are referenced from the live portion
of $\sigma_2'$.  In fact, this is a consequence of the consistency
theorem: the from-scratch run in the conclusion produces the same
$\tr_1'$ and $\vec{\nu}_2$.  Moreover, since its final store is
$\nongarbof{\sigma_2'}$, it is clear that these components and
$\nongarbof{\sigma_2'}$ itself cannot refer to garbage.

\techreportonly{
  \subsection{Invariants}
\label{sec:invariants}

The proof of Theorem~\ref{thm:general-consistency} is by induction on
the length of the given from-scratch run producing $\tr_1$.  It
requires numerous lemmas and, moreover, the theorem statement
needs to be strengthened in several ways.  In the remainder of this
section, we explain the main generalizations as they expose invariants
of the tracing machine that are crucial for its correct
functioning\footnote{To our knowledge, this is the first work that
  characterizes the entire trace (both in and out of focus), in the
  midst of adjustment.  Such characterizations may be useful, for
  example, to verify efficient implementations of the tracing
  machine.}.  Full details of this and all other proofs mentioned
later on can be found in the accompanying technical appendix.

\paragraph{Non-empty trace context and stack.}  Neither the trace
context nor the stack will stay empty during execution, so we need to
account for that.  In part 2 of the generalized version of the theorem
we therefore assume the following about the given from-scratch run
(see below for part 1):
\begin{enumerate}[a)]
\item $\csa{\rho_1}{\rcmd_1}{\sigma_1}$
\item $\iltstate {\sigma_1} {\trz{\trx_1}{\trend}} {\kappa_1} {\rho_1}
  {\rcmd_1} \iltsteprel^\ast \iltstate {\sigma_1'}
  {\trz{\trx_1'}{\trend}} {\kappa_1} {\emptyenv} {\vec{\nu_1}}$
\item $\rewindstepm {\trz{\trx_1'}{\trend}} {\trend}
  {\trz{\trx_1}{\trend}} {\tr_1}$
\item $\trx_1$ contains neither undo ($\undomark$) nor propagation ($\propmark$) marks
\end{enumerate}
When these conditions are all met, we say $\fsc{\tr_1}$ (``from-scratch
consistent'').  
Condition (a) is the same as in the theorem statement.  Conditions (b)
and (c) are similar to the assumptions stated in the theorem, except
more general: they allow a non-empty trace context and a non-empty stack.
The new condition (d), ensures that the trace context mentioned in (b)
and (c) only describes past evaluation steps, and not past or
pending undoing or propagating steps.
Apart from the assumption, we also must generalize the rest of part 2
accordingly but we omit the details here.

\paragraph{Reuse trace invariants.}  While it is intuitively clear
that propagation (part 2) must run with a from-scratch generated trace
in order to generate one, this is not strictly necessary for
evaluation (part 1).  In fact, here the property $\fsc{\tr_1}$ is not
always preserved: Recall that in evaluation mode the machine may undo
steps in $\tr_1$.  Doing so may lead to a reuse trace that is no
longer from-scratch generated!  In particular, if $\tr_1 =
\trcons{\trpush{\trcons{t}{\tr_2}}}{\tr_3}$, then, using steps
\refundopush, \refundostep and eventually \refevalv, the machine
may essentially transform this into $\trcons{\trpush{\tr_2}}{\tr_3}$,
which in general may not have a corresponding from-scratch run.

In order for the induction to go through, we therefore introduce a
weaker property, $\okay{\tr_1}$, for part 1.  It is defined as
follows:
\begin{mathpar}
\inferrule*
{
  \strut
}
{
  \okay{\trend}
}
\and
\inferrule*
{
  \fsc{\tr}
}
{
  \okay{\tr}
}
\and
\inferrule*
{
  \okay{\tr}
  \\
  \okay{\tr'}
}
{
  \okay{\trcons{\trpush{\tr}}{\tr'}}
}
\end{mathpar}
Note that if $\tr_1 = \trcons{\trmemo{\rho}{e}}{\tr_2}$ and
$\okay{\tr_1}$, then $\fsc{\tr_1}$ (and thus $\fsc{\tr_2}$) follows by
inversion.  This comes up in the proof precisely when in part 1
evaluation switches to propagation (step \refevalprop) and we
therefore want to apply the inductive hypothesis of part 2, where we
need to know that the new reuse trace is $\fsc$ (not ``just''
$\okay$).

\paragraph{Trace context invariant.}  In order for $\okay{\tr_1}$ and
$\fsc{\tr_1}$ to be preserved by steps \refundomark and \refpropv,
respectively, we also require $\okay{\trx_1}$, defined as follows:
\begin{mathpar}
\inferrule*
{
  \strut
}
{
  \okay{\trstart}
}
\and
\inferrule*
{
  \okay{\trx}
}
{
  \okay{\trxcons{\trx}{t}}
}
\and
\inferrule*
{
  \okay{\trx}
}
{
  \okay{\trxcons{\trx}{\trxpushmark}}
}
\\
\inferrule*
{
  \okay{\trx} 
  \\
  \fsc{\tr}
}
{
  \okay{\trxcons{\trx}{\trxpropmark{\tr}}}
}
\and
\inferrule*
{
  \okay{\trx} 
  \\
  \okay{\tr}
}
{
  \okay{\trxcons{\trx}{\trxundomark{\tr}}}
}
\end{mathpar}
Note the different assumptions about $\tr$ in the last two rules.
This corresponds exactly to the different assumptions about $\tr_1$ in
part 1 and part 2.

}

\section{Destination-Passing Style}
\label{sec:dps}
\begin{figure}[t]
\[
\begin{array}{@{}l@{\;\;\,}r@{\;\;\,}l@{}}

\DPSexp{
\illet{\ilfun{f}{\vec x}{e_1}}{e_2}} y
& = &
\illet{\ilfun{f}{\vec x \append z}{\DPSexp{e_1}{z}}} {\DPSexp{e_2} y}
\\

\DPSexp{\illet{\ilvar{x}{\ilprimop{\vec y}}}{e}} y
& = &
\illet{\ilvar{x}{\ilprimop{\vec y}}}{\DPSexp{e} y}
\\

\DPSexp{
  \kw{if}~x
  ~\kw{then}~e_1
  ~\kw{else}~e_2
} y
& = &
\kw{if}~x
~\kw{then}~\DPSexp{e_1} y
~\kw{else}~\DPSexp{e_2} y
\\

\DPSexp{\ilapp f {\vec x}} y
& = &
\ilapp f {\vec x \append y}
\\

\DPSexp{
  \illet{\ilvar{x}{\iota}}{e}
} y
& = &
\illet{\ilvar{x}{\iota}}
{\DPSexp{e} y}
\\

\DPSexp{
  \ilmemo{e}
} y
& = &
\ilmemo{\DPSexp{e} y}
\\

\DPSexp{
  \ilupdate{e}
} y
& = &
\ilwake{\DPSexp{e} y}

\\[2mm]

\!
\begin{aligned}[t]
&\DPSexp{
  \ilpush f e
} y
\\
&~~\text{when}~\arityof{f} = n
\end{aligned}

& = &

\begin{aligned}[t]
&\kw{let~fun}~f'(z).~\ilupdate{}
\\[-1mm]
&~~~~\illet{\ilvar{x_1}{\ilread{z}{1}}} {~\cdots}
\\[-1mm]
&~~~~\illet{\ilvar{x_n}{\ilread{z}{n}}} {}
\\[-1mm]
&~~~~\ilapp{f}{x_1, \ldots, x_n, y}
\\[-1mm]
&\kw{in}
\\[-1mm]
&\ilpush {f'} {\ilmemo{}}
\\[-1mm]
&~~~~\illet{\ilvar{z}{\ilalloc{n}}} {\DPSexp{e}{z}}
\end{aligned}

\\~\\

\begin{aligned}[@{}t]
&\DPSexp { \ilpop {\vec x} } y
\\
&~~\text{when}~|\vec x| = n
\end{aligned}

& = &

\!
\begin{aligned}[t]
&\illet{\ilvar{\_}{\ilwrite{y}{1}{x_1}}}{\cdots}
\\[-1mm]
&\illet{\ilvar{\_}{\ilwrite{y}{n}{x_n}}}{}
\\[-1mm]
&\ilpop \left<y\right>
\end{aligned}

\\[8mm]

  \DPSenv{\emptyenv} & = & \emptyenv \\
  \DPSenv{\rho[x \mapsto \nu]} & = & \DPSenv{\rho}[x \mapsto \nu] \\
  \DPSenv{\rho[f \mapsto \ilfun{f}{\vec x}{e}]} & = & \DPSenv{\rho}[f \mapsto \ilfun{f}{\vec x \append y}{\DPSexp{e}{y}}] \\
\end{array}
\]

\mycaptionrule
\caption{
 Destination-passing-style (DPS) conversion.
}
\label{fig:il-dps}
\end{figure}

In \secref{csa}, we defined the CSA property that the tracing machine
requires of all programs for consistency.
In this section, we describe a \emph{destination-passing-style}
transformation and show that it transforms arbitrary \IL programs into
CSA \IL programs, while preserving their semantics.  The idea is as
follows: A DPS-converted program takes an additional parameter~$x$
that acts as its destination.  Rather than return its results
directly, the program then instead writes them to the memory
specified by~$x$.

\figref{il-dps} defines the DPS transformation for an expression $e$
and a destination variable~$x$, written~$\DPSexp e x$.
Naturally, to DPS-convert an expression closed by an
environment~$\rho$, we must DPS-convert the environment as well,
written $\DPSenv \rho$.  In order to comply with our assumption that
all function and variable names are distinct, the conversion actually
has to thread through a set of already-used names.  For the sake of
readability we do not include this here.

Most cases of the conversion are straightforward.  The interesting
ones include function definition, function application, \kw{push}, and
\kw{pop}.
For function definitions, the conversion extends the function
arguments with an additional destination parameter~$z$~(we write $\vec
x \append z$ to mean $\vec x$ appended with $z$).
Correspondingly, for application of a function~$f$, the conversion
additionally passes the current destination to~$f$.
For \kw{push}es, we allocate a fresh destination~$z$ for the \kw{push}
body; we memoize this allocation with a \kw{memo} point.  When the
\kw{push}~body terminates, instead of directly passing control to $f$,
the program calls a wrapper function $f'$ that reads the destination
and finally passes the values to the actual function $f$.  Since these
\kw{read}s may become inconsistent in subsequent runs, we prepend them
with an \kw{update} point.
For \kw{pop}s, instead of directly returning its result, the converted
program writes it to its destination and then returns the latter.

As desired, the transformation yields \CSA programs (here and later on
we assume that $n$ is the arity of the program being transformed):
\begin{thm}[DPS programs are CSA]~\\
  $\csa{\DPSenv{\rho}}{\illet{\ilvar{x}{\ilalloc{n}}}{\DPSexp{e}{x}}}{\sigma}$
\end{thm}

Moreover, the transformation preserves the extensional semantics of
the original program:
\begin{thm}[DPS preserves extensional semantics]~\\
  \label{thm:paper-dps-extensional}
  If $\ilustepm{\sigma_1}{\emptyenv}{\rho \statesep
    e}{\sigma_1'}{\emptyenv}{\emptyenv \statesep \vec \nu}$ \\ then
  $\ilustepm{\sigma_1}{\emptyenv}{\DPSenv{\rho} \statesep
    \illet{\ilvar{x}{\ilalloc{n}}}{\DPSexp{e}{x}}}{\sigma_1' \uplus
    \sigma_2'}{\emptyenv}{\emptyenv \statesep \ell}$ \\ with
  $\sigma_2'(\ell, i) = \nu_i$ for all $i$.
\end{thm}

Because it introduces destinations, the transformed program allocates
additional store locations~$\sigma_2'$.
These locations are disjoint from the original store $\sigma_1'$,
whose contents are preserved in the transformed program.
If we follow one step of indirection, from the returned location to
the values it contains, we recover the original results~$\vec \nu$.

\subsection{An Example}

As a simple illustrative example, consider the source-level
expression~{\ttt{$f$(max(*$p$,*$q$))}}, which applies function~$f$ to
the maximum of two dereferenced pointers \ttt{*$p$} and \ttt{*$q$}.
Our front end translates this expression into the following:

\begin{codeListing9}
XX\=XX\=XX\=XX\=XX\=XX\=\kill
  \ilpush{$f$}{}
  \\\>
  \ilupdate{}
  \\\>\>
  \illet{\ilvar{$x$}{\ilread{$p$}{0}}}{}
  \\\>\>
  \illet{\ilvar{$y$}{\ilread{$q$}{0}}}{}
  \\\>\>
  \ilif{$x > y$}
  {\ilpop{$x$}}
  {\ilpop{$y$}}
\end{codeListing9}

Notice that the body of this \kw{push} is not store agnostic---when
the memory contents of either pointer is changed, the \kw{update} body
can evaluate to a different return value, namely the new maximum of
$x$ and $y$.
To address this, the DPS transformation converts this fragment into
the following:
\techreportonly{\\[2mm]}
\begin{codeListing9}
XX\=XX\=XX\=XX\=XX\=XX\=\kill
\hldiff{\kw{let}~\ilfun{f'}{m}{~\ilupdate{}}}
\\
\>
\hldiff{\illet{\ilvar{m'}{\ilread{m}{0}}}{f\ilparen{m',z}}}
\\
\hldiff{\kw{in}}
\\
\ilpush{\hldiff{f'}}
\hldiff{\kw{memo}}
\\\>
\hldiff{\illet{\ilvar{m}{\ilalloc{\texttt{1}}}}{}}
\\\>
\ilupdate{}
\\\>\>
\illetnb{\ilvar{$x$}{\ilread{$p$}{0}}}{}
\\\>\>
\illetnb{\ilvar{$y$}{\ilread{$q$}{0}}}{}
\\\>\>
\kw{if}~{$x > y$}~\kw{then}
\\\>\>\>
\hldiff{\illetnb{\ilvar{\_}{\ilwrite{m}{\texttt{0}}{x}}}}
{\ilpop{\hldiff{m}}}
\\\>\>
\kw{else}
\\\>\>\>
\hldiff{\illetnb{\ilvar{\_}{\ilwrite{m}{\texttt{0}}{y}}}}
{\ilpop{\hldiff{m}}}
\end{codeListing9}

Notice that instead of returning the value of either $x$ or $y$ as
before, the body of the \kw{push} now returns the value of $m$, a
pointer to the maximum of $x$ and $y$.
In this case, the push body is indeed store agnostic---though $x$ and
$y$ may change, the pointer value of~$m$ remains fixed, since it is
defined outside of the \kw{update} body.

The astute reader may wonder why we place the allocation of $m$ within
the bodies of the \kw{push} and \kw{memo} point, rather than ``lift it''
outside the definition of function~$f'$.
After all, by lifting it, we would not need to return~$m$ to $f'$ via
the stack \kw{pop}---the scope of variable~$m$ would include that of
function~$f'$.
We place the allocation of~$m$ where we do to promote reuse of
nondeterminism: by inserting this \kw{memo} point, the DPS
transformation effectively associates local input state~(the
values of~$p$ and~$q$) with the local output state~(the value
of~$m$).
Without this \kw{memo} point, every \kw{push} body will generate a
fresh destination each time it is reevaluated, and in general, this
nondeterministic choice will prevent reuse of any subcomputation,
since this subcomputation's local state includes a distinct,
previously chosen destination.
To avoid this behavior and to allow these subcomputations to instead
be reused during change propagation, the DPS conversion inserts
\kw{memo} points that enclose each (non-deterministic) allocation of a
destination.

\techreportonly{

\section{Cost Models}
\label{sec:costmodel}

We define a generic framework for modeling various dynamic costs of
our \IL abstract machines (both reference and tracing).  By
instantiating the framework with different concrete cost models, we
show several cost equivalences between the \IL reference machine and
the \IL tracing machine~(\secref{cealil}), show that our DPS
conversion~(\secref{dps}) respects the intensional reference semantics
of \IL up to certain constant factors, and give a cost model for our
implementation~(\secreftwo{compiling}{implementation}).

\paragraph{Cost model framework.}
We define machine steps and step sequences generically for both the
reference and tracing machines.
Let $S$ be a (finite) set of steps, where each step~$s \in S$
corresponds to precisely one stepping rule available to the machine in
question.
For the reference machine, these steps consist of
\refrefall~(\figref{il-eval}), though sometimes we distinguish between
the sub-cases of \refrefstore~(\figref{il-eval-store}).  For the
tracing machine, the steps consist of
\reftracingall~(\figref{il-tracing-steps}).
Given an initial machine state, we define a \emph{step sequence}~$\vec
s$ as the zero or more steps~$s_i \in S$ taken by some execution of
the machine until it terminates (with an empty stack).  No step
sequence is defined when the machine fails to terminate with an empty
stack (i.e., when it either diverges or becomes stuck).  
Note that when the machine permits non-deterministic steps, the
initial machine state does not fix a unique step sequence.

A \emph{cost model} is a triple~$M = \tuple{C, \zero, \gamma}$ where:
type~$C$ is the type of costs;
the zero cost~$\zero \in C$ is the cost of an empty step sequence;
and
the cost function $\gamma : S \rightarrow C \rightarrow C$ assigns to
each step $s \in S$ a function that maps the cost before the step~$s$
is taken to the cost after~$s$ is taken.
Given an execution sequence $\vec s = \tuple{s_1, \ldots, s_n}$, we
define the cost function of $\vec s$ under $M$ as the following
composition of cost functions:
$\gamma~{\vec s} = (\gamma~s_n) \circ \cdots \circ (\gamma~s_1)$.
By assuming zero initial cost, we can evaluate this composition of
cost functions to a yield \emph{total cost} for $\vec s$ as
$\gamma~{\vec{s}}~\zero = c \in C$.

\paragraph{Models for steps, stacks and stores.} 
We define several basic cost models for measuring machine steps, store
usage and stack usage.
Cost model~$M_s = \tuple{C_s, \zero_s, \gamma_s}$ counts machine steps:
$C_s = \mathcal{N}$,
$\zero_s = 0$
and
$\gamma_s~s~n = n + 1$.
Cost model~$M_\sigma = \tuple{C_\sigma, \zero_\sigma,
  \gamma_\sigma}$ measures store usage as the number of
allocations~($a$), reads~($r$), and writes~($w$), respectively.
We represent these in a triple:
$C_\sigma = \mathcal{N}^3$,
$\zero_\sigma = \tuple{0, 0, 0}$,
and $\gamma_\sigma$ is:
\[
\begin{aligned}
\gamma_\sigma~s_{\text{alloc}}~\tuple{a, r, w} 
&= \tuple{a+1, r, w}
\\[-2mm]
\gamma_\sigma~s_{\text{read}}~\tuple{a, r, w} 
&= \tuple{a, r+1, w}
\\[-2mm]
\gamma_\sigma~s_{\text{write}}~\tuple{a, r, w} 
&= \tuple{a, r, w+1}
\\[-2mm]
\gamma_\sigma~s_{\text{nostore}}~\tuple{a, r, w} 
&= \tuple{a, r, w}
\end{aligned}
\]
To instantiate the model for the reference machine we set
$s_{\text{alloc}} = \refrefstore/\refstorealloc$, 
$s_{\text{read}} = \refrefstore/\refstoreread$ and 
$s_{\text{write}} = \refrefstore/\refstorewrite$;
similarly, for the tracing machine we set
$s_{\text{alloc}} = \refevalalloc$, 
$s_{\text{read}} = \refevalread$ and 
$s_{\text{write}} = \refevalwrite$.
For both machines, we instantiate the case of
$\gamma_\sigma~s_{\text{nostore}}$ for each of the remaining steps.
Cost model~$M_\kappa = \tuple{C_{\kappa}, \zero_{\kappa},
  \gamma_{\kappa}}$ measures the stack usage as the number of times
the stack is pushed~($u$), the number of times it is popped~($d$), the
current stack height~($h$), and the maximum stack height~($m$).
We represent these as a 4-tuple so that
$C_{\kappa} = \mathcal{N}^4$
and 
$\zero_{\kappa} = \tuple{0, 0, 0, 0}$.
We define $\gamma_{\kappa}$ as:
\[ 
\begin{aligned}
  \gamma_{\kappa}~s_{\text{push}}~\tuple{u,d,h,m}
  &= \tuple{u + 1, d, h + 1,\textrm{max}(m, h + 1)}
  \\[-2mm]
  \gamma_{\kappa}~s_{\text{pop}}~\tuple{u,d,h,m} 
  &= \tuple{u, d + 1, h - 1, m}
  \\[-2mm]
  \gamma_{\kappa}~{s_{\text{nostack}}}~\tuple{u,d,h,m}
  &= \tuple{u, d, h, m}
\end{aligned}
\]
To instantiate the model for the reference machine we set
$s_{\text{push}} = \refrefpush$ and 
$s_{\text{pop}} = \refrefv$;
similarly, for the tracing machine we set
$s_{\text{push}} = \refevalpush$ and 
$s_{\text{pop}} = \refevalv$.
Note that the stack is actually popped by $\refrefv$ rather than
$\refrefpop$, (resp. $\refevalv$ versus $\refevalpop$).  The
latter steps---which each evaluate a \kw{pop} expression to a sequence
of machine values---always precede the actual stack pop by one step.

In from-scratch runs, the costs of the tracing machine are equivalent
to that of the reference machine.
\begin{thm}
  \label{thm:costsofrefequivtotracing}
  Fix an initial machine state $\sigma, \epsilon, \rho, e$.  Run under
  the reference machine to yield step sequence $\usup{\vec{s}}$.  Run
  under the tracing machine with an empty reuse trace to yield step
  sequence $\tsup{\vec{s}}$.  The following hold for $\usup{\vec{s}}$
  and $\tsup{\vec{s}}$:
  (1) the step counts under $M_s$ are equal;
  (2) the stack usage under $M_\kappa$ is equal;
  and (3) the store usage under $M_\sigma$ is equal.
\end{thm}

\paragraph{DPS costs.}
Recall that before \IL programs can adjust in a consistent way (in the
tracing machine), we have to ensure that they are compositionally
store agnostic, e.g., by DPS-converting them~(\secref{dps}).
Below we bound the overhead introduced by this transformation in terms
of the reference machine.
By appealing to \thmref{costsofrefequivtotracing}, this bound
equivalently applies to the tracing machine as well.

\begin{thm}[DPS preserves intensional semantics]
\label{thm:dps-intensional}~
\\
Consider the evaluations of expression~$e$ and $\DPSexp e x$ as
given in \thmref{paper-dps-extensional}.
The following hold for their respective step sequences, $\vec s$ and
${\vec s}'$:
(1) the stack usage under $M_\kappa$ is equal. Let $u$ be the number
of $\kw{push}$es performed in each;
(2) the number of allocations under $M_\sigma$ differs by exactly
$u$~(ignoring the initial allocation), the number of reads and writes
under $M_\sigma$ each differs by at most $a \cdot u$ and $a \cdot (u +
1)$, respectively, where $a$ is the maximum arity of any \kw{pop}
taken in $\vec s$;
(3) the number of steps taken under $M_s$ differs by at most $(2 \cdot
a + 5) \cdot u + a$.
\end{thm}

\paragraph{Realized costs.}
Realized costs closely resemble those of a real implementation.
We model them with~$M_t = \tuple{C_t, \zero_t, \gamma_t}$, which
partitions step counts of the tracing machine into evaluation~($e$),
undo~($u$) and propagation~($p$) step counts.
As in previous work, our implementation does \emph{not} incur any
cost for any propagation steps taken---these steps are
effectively skipped.
Therefore, we define the \emph{realized} cost of $\tuple{e,p,u} \in
C_t$ as $(e + u) \in \mathcal{N}$.
These realized costs are proportional to the actual work performed by
\IL programs compiled by our
implementation~(\secreftwo{compiling}{implementation})%
\footnote{ The implementation cost may involve an additional
  logarithmic factor, e.g., to maintain a persistent view of the store
  for every point in the trace.%
}.
Each cost is a triple:
\vspace{-1em}
\[
\begin{aligned}
  C_t &= \mathcal{N}^3 \\
  \zero_t &= \tuple{0,0,0} \\
  \gamma_{t}~s_{\text{eval}}~\tuple{e,p,u} &= \tuple{e + 1, p, u} \\
  \gamma_{t}~s_{\text{prop}}~\tuple{e,p,u} &= \tuple{e, p + 1, u} \\
  \gamma_{t}~{s_{\text{undo}}}~\tuple{e,p,u} &= \tuple{e, p, u + 1} \\
\end{aligned}
\]
(Here $s_{\text{eval}}$ matches steps \refevalall, $s_{\text{prop}}$
matches steps \refpropall, and $s_{\text{undo}}$ matches steps
\refundoall.)

}
\section{Compiling \IL}
\label{sec:compiling}
\label{sec:realizationtechniques}

While the semantics of \IL are given by an abstract
machine~(\secref{il}), in actuality we want to run \IL programs with a
more conventional machine---e.g., a machine that does not support
tracing or change propagation directly.
As such, our compilation process can be thought of as building a
specialized tracing machine for a given \IL program.
At a high level, realizing this machine requires realizing each of its
components, i.e., realizing its store, stack, environment, trace and
stepping rules.

\subsection{Runtime data structures and algorithms}
\label{sec:runtime}

The primary role of the runtime system is to provide realized versions
of the abstract machine's trace and store, an efficient search for
matching \kw{memo} points, and an efficient algorithm for change
propagation.
To give an efficient change propagation algorithm, it is crucial that
the runtime trace and store be ``entangled'', i.e., mutually
referential: the runtime store references certain runtime trace
actions, and the runtime representation of \kw{read} and \kw{write}
trace actions each reference the runtime store.

\paragraph{The runtime trace.}
At a high level, the trace provides an ordering to trace actions.
For efficiency, we use a (total) order maintenance data
structure~\cite{DietzSl87} which bestows each trace action~$t$ an
associated time stamp~$s(t)$; these timestamps admit an efficient
predicate for checking if $t_1 \le t_2$ by checking if $s({t_1}) \le
s({t_2})$.
Concretely, a \emph{trace node} is a record consisting of a time
stamp~$s$ and (at least one) trace action~$t$.  
As a refinement to this approach, below we also consider when and how
several trace actions can share a single trace node.
Most of the trace actions are straightforward to represent during
runtime, though extra care is needed for \kw{read} and \kw{write}
actions, which we describe below.

\paragraph{The runtime store.}
While the store of the abstract machine only retains the current value
of each location-offset entry~$\ell[i]$ (hereafter, just an
\emph{entry}), this generally requires traversing and replaying the
entire trace during change propagation, which is prohibitively
expensive.
As such, the runtime store takes a different tack: for each entry, it
persistently maintains all the corresponding read and written values,
across the entire trace, including the corresponding trace action.
Given a particular point in the trace, we quickly access the current
value of any entry based on when it was last read or written, in terms
of the time stamps described above.
For this purpose, the runtime uses a self-balancing search tree for
each changeable store entry; each node of the tree corresponds to a
\kw{read} or \kw{write} trace action.

\paragraph{The runtime memo table.} 
In the abstract machine, memoization permits trace reuse by matching
\kw{memo} points in the reuse trace.
In the runtime, a hash table indexes each such \kw{memo} point in the
trace.
While evaluating a new \kw{memo} point, the runtime system attempts to
locate matches using this hash table.
Once matched, the change propagation algorithm begins working on the
reused trace.

\paragraph{Change propagation as an algorithm.}
In the abstract machine, change propagation has two purposes: to
replay store effects (viz.~\kw{alloc}s, \kw{write}s) and to ensure
that every reused \kw{read} action is consistent with the current
store.
However, the machine specifies change propagation as a complete
traversal of the trace, while in practice this is not efficient.
As a result, the algorithm performs change propagation somewhat
differently, while still accomplishing its two high-level goals:
replaying store effects and ensuring that \kw{read}s are consistent.

First, to replay store effects, the algorithm relies on the runtime
store being \emph{retained} from one run to the next.
That is, the final store of one run becomes the initial store of the
next run.
This retention is modulo changes in some store entries, and the
reclamation of locations marked as garbage.
Consequently, since the runtime store keeps every traced effect---not
just the most recent ones, as in the abstract machine---it is not
necessary to replay these effects for the benefit of updating the
store.

Second, to find and reevaluate inconsistent \kw{read} actions, the
runtime maintains a priority queue~$Q$ of them, ordered by their
appearance in the trace.
Rather than find them one-by-one via trace traversal, when the runtime
store is changed, it uses the runtime store representation described
above to identify any inconsistent \kw{read}s and enqueue the smallest
enclosing \kw{update} point into $Q$.  To find this \kw{update} point
quickly, each \kw{read} action maintains a reference to this
(unique) enclosing \kw{update} action.
The change propagation algorithm consists of a loop that reevaluates
the \kw{update} points in $Q$, in trace order.

\subsection{Compilation}
\label{sec:compilation}

We compile \IL programs in several phases.  
First, we convert them into destination-passing style; this ensures
that they will replay correctly during change propagation.
Next, we implement each traced expression with a corresponding call
into the runtime, described above.
For most traced forms, this is very straightforward;
however, handling the \kw{update} and \kw{memo} forms requires more
care, which we discuss below.
Finally, we translate the resulting \IL program into our target
language,~C, and compile this code with \texttt{gcc}.

\paragraph{Compiling \kw{update} and \kw{memo}.}
In contrast to the other traced forms, \kw{memo} and \kw{update} each
save and restore the local state of an \IL program---an
environment~$\rho$ and an \IL expression~$e$.
To compile these forms, the following questions arise: How much of the
environment~$\rho$ should be recorded in the runtime trace and/or memo
table?  Once an \IL expression~$e$ is translated into a target
language, how do we reevaluate it during change propagation?

First, we address how we save the environment.  At each of these
points we use a standard analysis~(e.g.,~\cite{Muchnick97}) to
approximate the live variables~$\LiveVars{e}$ at each such $e$, and
then save not $\rho$, but rather $\rho |_{\LiveVars{e}}$, i.e., $\rho$
limited to $\LiveVars{e}$.
This has two important consequences: we save space by not storing dead
variables in the trace, and we (monotonically) increase the potential
for \kw{memo}~matches, as non-matching values of dead variables do not
cause a potential match to fail.

Second, we address the issue of fine-grained reevaluation.  This poses
a problem since languages such as C do not allow programs to jump to
arbitrary control points, e.g., into the middle of a procedure.
To address this limitation, we adapt the ``lambda-lifting''
technique used in earlier work~\cite{HammerAcCh09}.
Originally this technique transformed the control flow graphs of \C
code;
we modify it for \IL such that after being applied, all \kw{update}
points have the form~$\ilupdate~{f(\vec{x})}$ where $f$ is a top-level
function and where variables~$x_i \in \vec x$ close the body of~$f$.
In this form, we implement each \kw{update} point as an
explicitly-constructed function closure, i.e., a record consisting of
a function pointer and values for its arguments.

\subsection{Optimizations}
\label{sec:opt}

We refine the basic approach above with two optimizations.

\paragraph{Trace node sharing~(\textsf{share}).}
The basic runtime system (\secref{runtime}) assigns each trace
action~$t$ to a distinct trace node, with a distinct time stamp~$s$.
Since each trace node brings some overhead, it is desirable if
sequences of consecutive trace actions $\vec t = t_1, \ldots t_n$ can
share a single trace node with a single time stamp.
However, this optimization is complicated by a few issues.

First, how do we realize the comparison~$t_i < t_j$ when $t_i$ and $t_j$
share a single time stamp?
We can accomplish this by following the order of $\vec t$ when placing
the actions into the trace node; this allows us to efficiently compare
$t_i$ with $t_j$ by comparing their addresses.

Second, how do we avoid breaking the sequence when it uses a single
trace node?  This can happen in one of two ways: by either
\kw{memo}-matching some action in the middle of~$\vec t$, thereby
discarding its prefix; or by reevaluating an \kw{update} point in the
middle of~$\vec t$ when this reevaluation takes a new control path.
We avoid these scenarios by packing sequence~$\vec t$ into a single
trace node only when the following criteria are met: if $\vec t$
contains a \kw{memo} point, then it appears first; if $\vec t$
contains an \kw{update} point, then the remaining suffix of $\vec t$
is generated by straight-line code.

\paragraph{Selective destination-passing style~(\textsf{seldps}).}
The DPS conversion~(\figref{il-dps}) introduces extra \IL code for
\kw{push} and \kw{pop} expressions: an extra \kw{alloc}, \kw{update},
\kw{memo}, and some \kw{write}s and \kw{read}s.
Since each of these expressions are traced, this can introduce
considerable overhead for subcomputations that do not interact with
changing data.
In fact, without an \kw{update} point, propagation over the trace
of~$e$ will always yield the same return
values\techreportonly{~(\lemref{prop-vals})}.
Moreover, it is clear from the definition of store
agnosticism~(\secref{csa}) that any computation without an \kw{update}
point is trivially CSA, hence, there is no need to DPS-convert it.
By doing a conservative static analysis, our compiler estimates
whether each expression~$e$ appearing in the form $\ilpush f e$ can
reach an \kw{update} point during evaluation.  If not, we do not apply
the DPS conversion to $\ilpush f e$.
We refer to this refined transformation as \emph{selective} DPS
conversion.

\section{Implementation and a \C Front End}
\label{sec:implementation}
\label{sec:srcc}

Our current implementation consists of a compiler and an associated
runtime system, as outlined in \secref{compiling}.
Additionally, we also implement the optimizations from \secref{opt}.
After compiling and optimizing \IL, our implementation translates it
to C, which we compile using \texttt{gcc}.
In all, our compiler consists of a 10k line extension to CIL and our
runtime system consists of about 6k lines of C code.
We plan to publicly release the system in summer 2011; in the
meantime, we happily offer it to reviewers upon request.

As a front-end to \IL, we support a C-like source language, \srcc.
We use CIL~\cite{NeculaMcRaWe02} to parse \SRCC source into a
control-flow graph representation.
To bridge the gap between this representation and \IL, we utilize a
known relationship between static single assignment~(SSA) form and
lexically-scoped, functional programming~\cite{Appel98}.

Before this translation, we move \SRCC variables to the heap if either
they are globally-scoped, aliased by a pointer (via \SRCC's address-of
operator, {\tt \&}), or are larger than a single machine word.
When such variables come into scope, we allocate space for them in the
heap (via \kw{alloc}); for global variables, this allocation only
happens once, at the start of execution.

As apart of the translation to \IL, we automatically place \kw{update}
points before each \kw{read} (or consecutive sequence of \kw{read}s).
Though in principle we can automatically place \kw{memo} points
anywhere, we currently leave their placement to the programmer by
providing a \kw{memo} keyword in \SRCC; this keyword can be used as a
\SRCC statement, as well as a wrapper around arbitrary
\SRCC~expressions.

\subsection{Current Limitations}

Our source language~\SRCC is more restricted than \C in a few ways,
though most of these restrictions are merely for technical reasons and
could be solved with further compiler engineering.
First, while \SRCC programs may use variadic functions provided by
external libraries (e.g., \texttt{printf}), \SRCC does not currently
support the definition of new variadic functions.
Furthermore, function argument and return types must be scalar
(pointer or base types) and not composite types (\ttt{struct} and
\ttt{union} types).
Removing these restrictions may pose engineering challenges, but
should not require a fundamental change to our approach.

Second, our \SRCC front-end assumes that the program's memory accesses
are word aligned.  This assumption greatly simplifies the translation
of pointer dereferencing and assignment in \SRCC into the \kw{read}
and \kw{write} instructions in \IL, respectively.
To lift this restriction, we could dynamically check the alignment of
each pointer before doing the access, and decompose those accesses
that are not word-aligned into one (or two) that are.

Third, as a more fundamental challenge, \SRCC does not currently
support features of \C that change the stack discipline of the
language, such as \texttt{setjmp}/\texttt{longjmp}.
In \C, these functions are often used to mimic the control operators
and/or exception handling found in higher-level languages.
Supporting these features is beyond the scope of this paper, but
remains of interest for future work.

Finally, to improve efficiency, programs written in \SRCC can be mixed
with foreign C code (e.g., from a standard C library).
Since foreign C code is not traced, it allows those parts of the
program to run faster, as they do not incur the tracing overhead that
would otherwise be incurred within \SRCC.
However, mixing of \SRCC and foreign C code results in a
programming setting that is not generally sound, and contains
potential pitfalls.
In particular, in this setting programs must adhere to the following
correct usage restriction to ensure the consistency of change
propagation: each memory location is either accessed exclusively by
foreign C code (not by \SRCC code) or exclusively by \SRCC code (not
by foreign C code).
While a skilled programmer can observe this restriction (we mix
foreign C code with \SRCC code for some of our benchmarks), we
currently provide no static or dynamic check that this restriction is
met.
Such checks pose interesting challenges for future work.

\newcommand{\appname}[1]{\textsf{#1}\xspace}

\section{Evaluation}
\label{sec:experiments}
\label{sec:evaluation}

We empirically evaluate our approach by considering a number of
benchmarks written in \SRCC~(\secref{srcc}), compiled with our
compiler~(\secref{compiling}).  
Our experiments are very encouraging, showing that our approach can
yield asymptotic speedups, resulting in orders of magnitude speedups
in practice; it does this while incurring only moderate overheads for
pre-processing or initial executions.
We evaluate our compiler and runtime optimizations~(\secref{opt}),
showing that they improve performance of both from-scratch evaluation
as well as of change propagation.
Comparisons with previous work using the unsound \CEAL library and the
\DeltaML language shows that our approach performs competitively.

\subsection{Benchmarks and Measurements}

Our benchmarks consist of expression tree evaluation (i.e., the
example from \secref{overview}), some list primitives, two sorting
algorithms and several computational geometry algorithms.  For our
timings, we used a Linux box running on a 1.8 GHz Intel Xeon (4-core)
processor with 512GB memory.  All our benchmarks are sequential and
are compiled with \texttt{gcc -O3} after translation to \C.

For each benchmark, we measure the \emph{from-scratch time}, the time
to run the benchmark from-scratch on a particular input, and the
average \emph{update time}, the average time required by change
propagation to update the output after inserting or deleting an
element from its input.
We compute this average by iterating over the initial input, deleting
each input element, updating the output by change propagation,
inserting the element again and updating the output by change
propagation.

\paragraph{List primitives.}
These benchmarks include \appname{filter}, \appname{map} (performs
integer additions per element), \appname{reverse}, \appname{minimum}
(integer comparison), and \appname{sum} (integer addition), and the
sorting algorithms \appname{quicksort} (string comparison) and
\appname{mergesort} (string comparison).
We generate lists of $n$ (uniformly) random integers as input for the
list primitives.  For sorting algorithms, we generate lists of $n$
(uniformly) random, 32-character strings.
We implement each list benchmark mentioned above by using an external \C
library for lists, which our compiler links against the self-adjusting
code after compilation.

\paragraph{Computational geometry.}
These benchmarks include \appname{quickhull}, \appname{diameter}, and
\appname{distance}; \appname{quickhull} computes the convex hull of a
point set using the standard quickhull algorithm; \appname{diameter}
computes the diameter, i.e., the maximum distance between any two
points of a point set; \appname{distance} computes the minimum
distance between two sets of points.  Our implementations of
\appname{diameter} and \appname{distance} use \appname{quickhull} to
compute first the convex hull and then compute the diameter and the
distance of the points on the hull (the furthest away points lie on
the convex hull).
For \appname{quickhull} and \appname{distance}, input points are
selected from a uniform distribution over the unit square in
$\mathbb{R}^2$. For \appname{distance}, we select equal numbers of
points from two non-overlapping unit squares in $\mathbb{R}^2$. We
represent real numbers with double-precision floating-point numbers.
As with the list benchmarks, each computational geometry benchmark
uses an external \C library; in this case, the external library
provides geometric primitives for creating points and lines, and
computing simple properties about them (e.g., line-point distance).

\paragraph{Benchmark targets.}
In order to study the effectiveness of the compiler and runtime
optimizations~(\secref{opt}), for each benchmark we generate several
\emph{targets}.  Each target is the result of choosing to use some
subset of our optimizations.
\tabref{targets} lists and describes each target that we consider.
Before measuring the performance of these targets, we use regression
tests to verify that their self-adjusting semantics are consistent
with conventional (non-self-adjusting) versions.
These tests empirically verify our consistency
theorem (\thmref{general-consistency}).

\begin{table}[h]
\begin{center}
\begin{tabular}{lp{0.75\columnwidth}}
  \toprule
  \textbf{\textsf{Target}}
  & 
  \textbf{\textsf{Optimizations used}}
  \\
  \midrule
  \textsf{no-opt} & No optimization is used.
  \\
  \textsf{share}
  &
  Same as \textsf{no-opt} except that certain trace actions can share
  trace nodes.
  \\
  \textsf{seldps}
  &  
  Same as \textsf{no-opt} except that the DPS transformation 
  is selective---only certain functions are transformed.
  \\
  \textsf{opt} & Both \textsf{seldps} and \textsf{share} are used.
  \\
  \bottomrule
\end{tabular}
\end{center}
\nocaptionrule
\caption{Targets and their optimizations~(\secref{opt}).}
\label{tab:targets}
\end{table}

\begin{figure}
\includegraphics[width=\columnwidth]{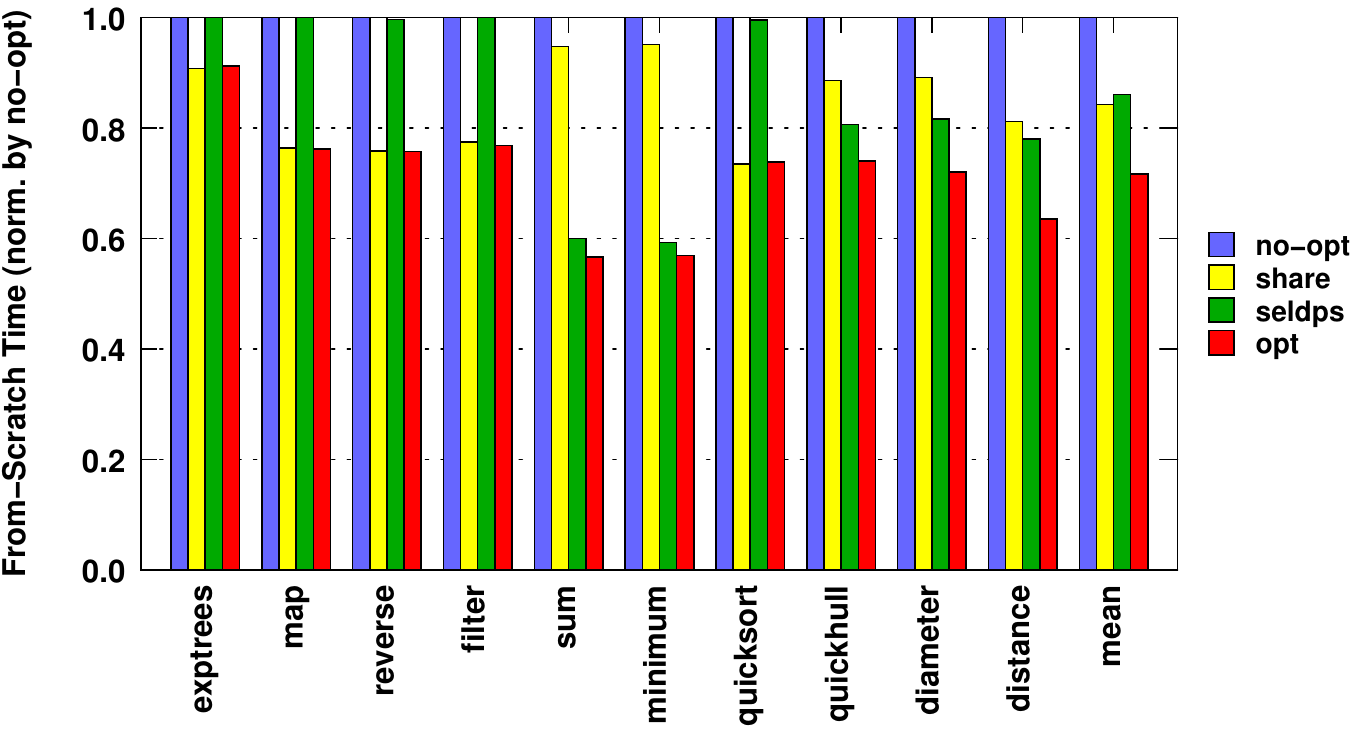}
\includegraphics[width=\columnwidth]{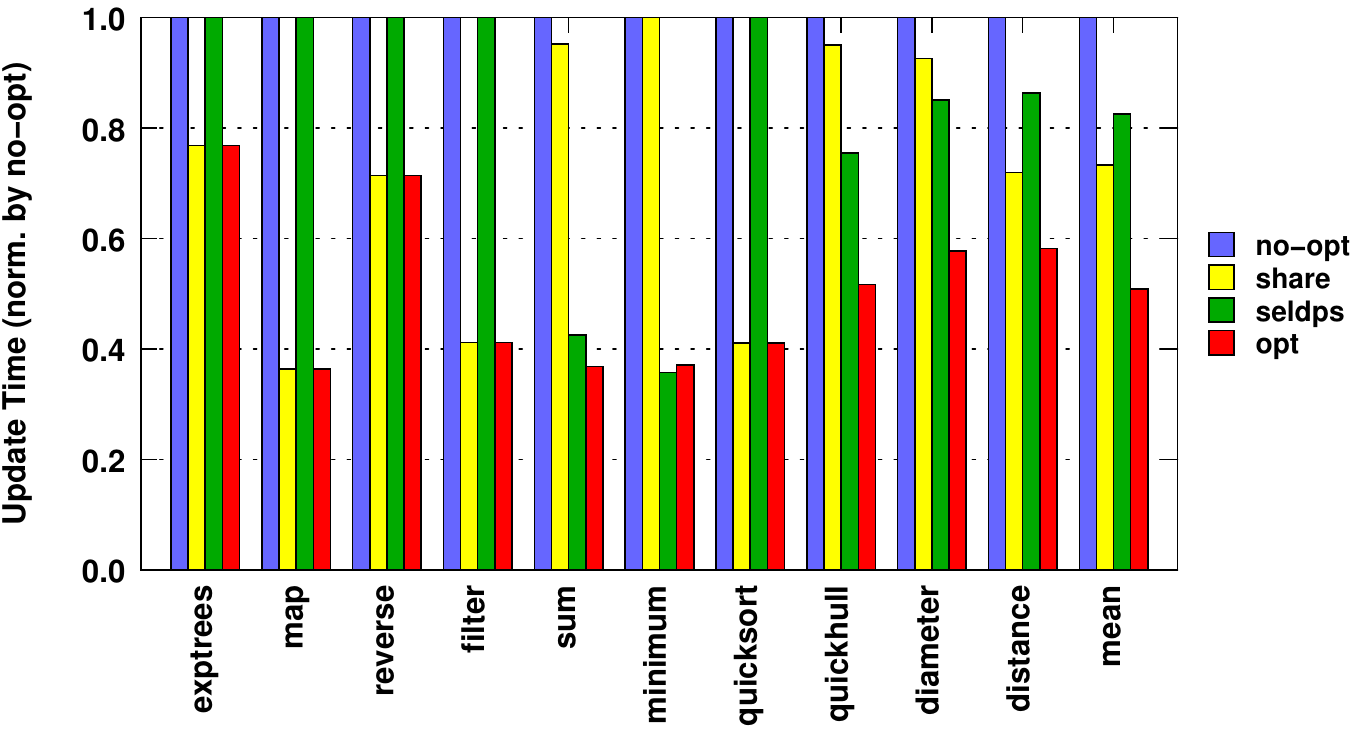}
\caption{Comparison of benchmark targets.}
\label{fig:bargraphs}
\end{figure}

\begin{table*}
\begin{center}
\begin{tabular}{lcrrrrr}
\toprule
{\textsf{\textbf{Benchmark}}}
&
{\textsf{\textbf{N}}}
&
{\textsf{\textbf{Conv}}}
&
{\textsf{\textbf{FS}}}
&
{\textsf{\textbf{Overhead}}}
&
{\textsf{\textbf{Ave.~Update}}}
&
{\textsf{\textbf{Speed-up}}}
\\
& & 
({sec})
& 
({sec})
& 
({\footnotesize \textsf{\textbf{FS} / \textbf{Conv}}})
&
({sec})
&
({\footnotesize \textsf{{\textbf{Conv} / \textbf{AU}}}})
\\
\midrule
\textsf{exptrees} 
&
$10^6$
&
0.18
&
1.53
&
8.5
&
$1.3 \times 10^{-5}$
&
$1.4 \times 10^{4}$
\\
\textsf{map} 
&
$10^6$
&
0.10
&
1.87
&
18.4
&
$3.4 \times 10^{-6}$
&
$3.0 \times 10^{4}$
\\
\textsf{reverse} 
&
$10^6$
&
0.10
&
1.81
&
18.4
&
$2.6 \times 10^{-6}$
&
$3.8 \times 10^{4}$
\\
\textsf{filter} 
&
$10^6$
&
0.13
&
1.42
&
10.7
&
$2.7 \times 10^{-6}$
&
$4.9 \times 10^{4}$
\\
\textsf{sum} 
&
$10^6$
&
0.14
&
1.35
&
9.6
&
$9.3 \times 10^{-5}$
&
$1.5 \times 10^{3}$
\\
\textsf{minimum} 
&
$10^6$
&
0.18
&
1.36
&
7.7
&
$1.3 \times 10^{-5}$
&
$1.4 \times 10^{4}$
\\
\textsf{quicksort} 
&
$10^5$
&
0.40
&
3.30
&
8.2
&
$5.8 \times 10^{-4}$
&
$6.9 \times 10^{2}$
\\
\textsf{mergesort} 
&
$10^5$
&
0.74
&
5.31
&
7.2
&
$9.5 \times 10^{-4}$
&
$7.8 \times 10^{2}$
\\
\textsf{quickhull} 
&
$10^5$
&
0.26
&
0.97
&
3.7
&
$1.2 \times 10^{-4}$
&
$2.2 \times 10^{3}$
\\
\textsf{diameter} 
&
$10^5$
&
0.26
&
0.90
&
3.4
&
$1.5 \times 10^{-4}$
&
$1.8 \times 10^{3}$
\\
\textsf{distance} 
&
$10^5$
&
0.24
&
0.81
&
3.4
&
$3.0 \times 10^{-4}$
&
$7.9 \times 10^{2}$
\\
\bottomrule
\end{tabular}

\end{center}
\nocaptionrule
\vspace{-2mm}
\label{tab:results-summary}
\caption{Summary of benchmark results (using {\sf opt} target of each
  benchmark).} 
\end{table*}

\begin{figure*}
  \begin{center}
    \begin{tabular}{ccc}
    \includegraphics[width=0.65\columnwidth]{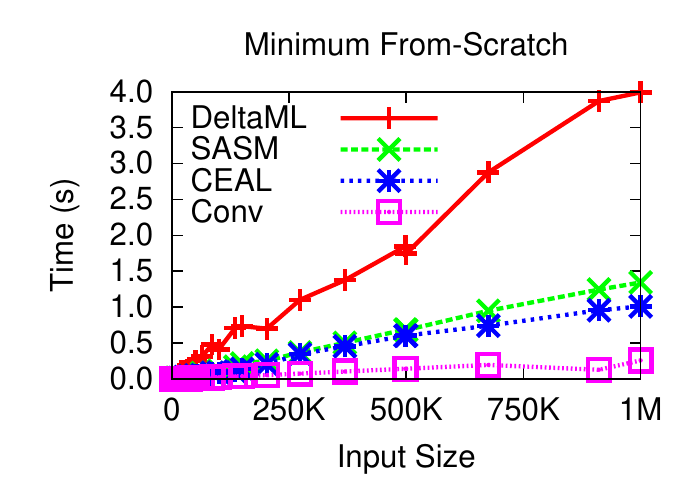}
    &
    \includegraphics[width=0.65\columnwidth]{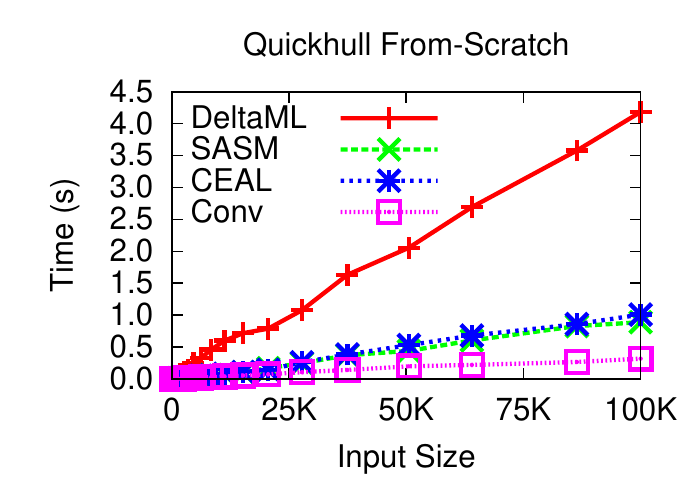}
    &
    \includegraphics[width=0.65\columnwidth]{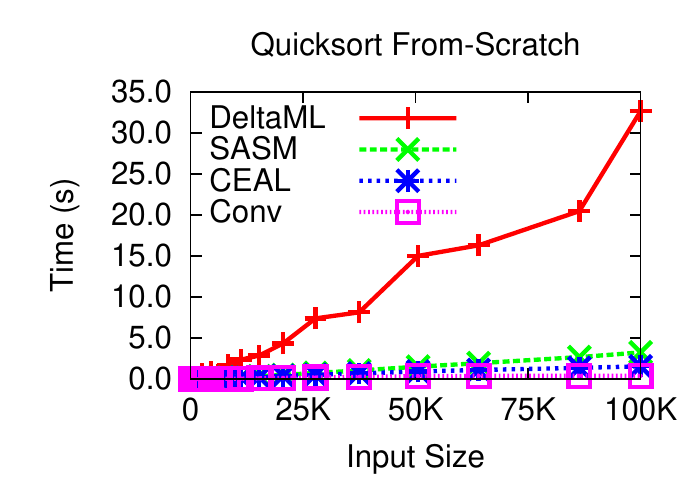}
    \\
    \includegraphics[width=0.65\columnwidth]{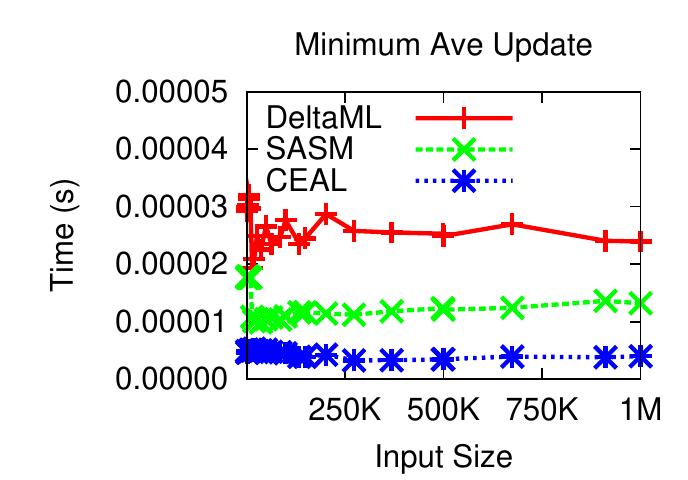}
    &
    \includegraphics[width=0.65\columnwidth]{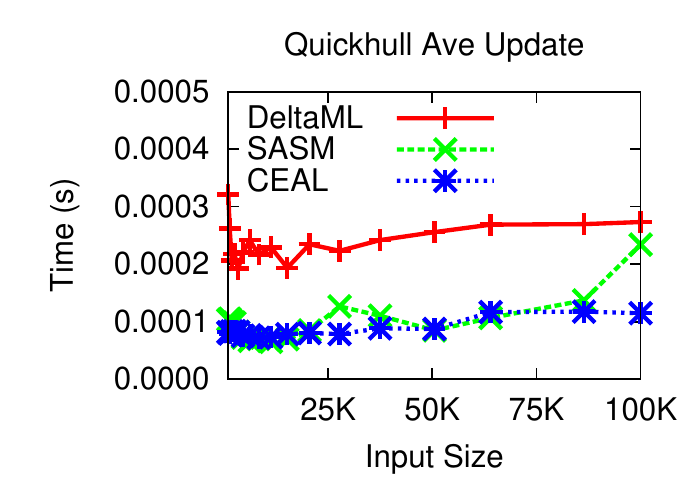}
    &
    \includegraphics[width=0.65\columnwidth]{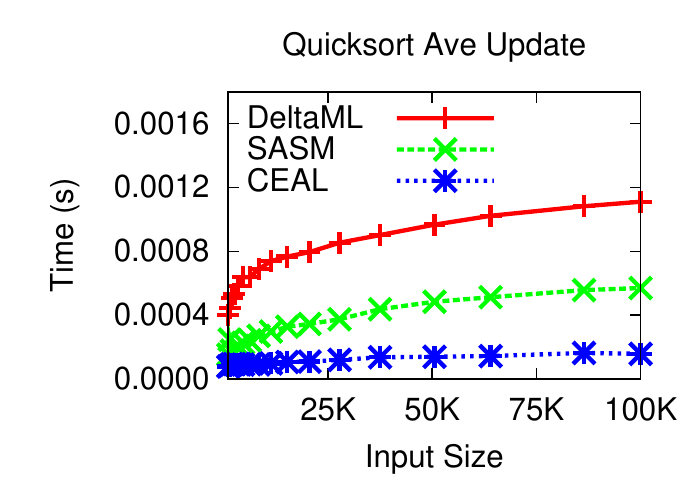}
  \end{tabular}
\end{center}
\caption{\appname{minimum}, \appname{quickhull} and
  \appname{quicksort} performance in \DeltaML, \CEAL and our own
  \textsf{opt} versions~(labeled~\textsf{SASM}).}
\label{fig:comparison-to-past}
\end{figure*}

\subsection{Optimizations}

\figref{bargraphs} compares our targets' from-scratch running time and
average update time.  Each bar is normalized to the \textsf{no-opt}
target.
The rightmost column in each bar graph shows the mean over all
benchmarks.  To estimate the efficacy of an optimization~$X$, we can
compare target~\textsf{no-opt} with the target where~$X$ is turned on.

In the mean, the fully optimized targets~(\textsf{opt}) are nearly
30\% faster from-scratch, and nearly 50\% faster during automatic
updates (via change propagation), when compared to the unoptimized
versions~(\textsf{no-opt}).
These results demonstrate that our optimizations, while conceptually
straightforward, are also practically effective: they significantly
improve the performance of the self-adjusting targets, especially
during change propagation.

\subsection{Summary of Experimental Results}
\tabref{results-summary} summarizes the self-adjusting performance of
the benchmarks by comparing them to conventional, non-self-adjusting
\C~code.  From left to right, the columns show the benchmark name, the
input size we considered (\textsf{N}), the time to run the
conventional (non-self-adjusting) version (\textsf{Conv}), the
from-scratch time of the self-adjusting version (\textsf{FS}), the
\emph{preprocessing overhead} associated with the self-adjusting
version (\textsf{Overhead} is the ratio $\textsf{FS} /
\textsf{Conv}$), the average update time for the self-adjusting
version (\textsf{Ave.~Update}) and the \emph{speed-up} gained by using
change propagation to update the output versus rerunning the
conventional version (\textsf{Speed-up} is the ratio $\textsf{Conv} /
\textsf{Ave.~Update}$).  All reported times are in seconds.  For the
self-adjusting versions, we use the optimized (\textsf{opt}) target of
each benchmark.

The preprocessing overheads of most benchmarks are less than a factor
of ten; for simpler list primitives benchmarks, this overhead is about
18 or less.
However, even at these only moderate input sizes~(viz. $10^5$ and
$10^6$), the self-adjusting versions deliver speed-ups of two, three
or four orders of magnitude.
Moreover, as we illustrate below~(\secref{compare-to-past}), these
speedups increase with input size.

\subsection{Comparison to Past Work}
\label{sec:compare-to-past}

To illustrate how our implementation compares with past systems,
\figref{comparison-to-past} gives representative examples.  It
compares the from-scratch and average update times for three
self-adjusting benchmarks across three different implementations: one
in \DeltaML~\cite{Ley-WildFlAc08}, one in \CEAL~\cite{HammerAcCh09}
and the \textsf{opt} target of our
implementation~(labeled~\textsf{SASM}, for
\emph{S}elf-\emph{A}djusting \emph{S}tack \emph{M}achines).
In the from-scratch graphs, we also compare with the conventional
(non-self-adjusting) \C implementations of each
benchmark~(labeled~\textsf{Conv}).

The three benchmarks shown~(viz.~\textsf{minimum}, \textsf{quickhull}
and \textsf{quicksort}) illustrate a general trend.
First, in from-scratch runs, the \textsf{SASM} implementations are
only slightly slower than that of \CEAL, while the \DeltaML
implementations are considerably slower than both.
For instance, in the case of \textsf{quicksort}, the \DeltaML
implementation is a factor of ten slower than our own.
While updating the computation via change propagation, the performance
of the \textsf{SASM} implementations lies somewhere between that of
\DeltaML and \CEAL, with \CEAL consistently being either faster than
the others, or comparable to \textsf{SASM}.
Although not reported here, we obtain similar results with other
benchmarks.

\section{Related Work}
\label{sec:relatedwork}
\label{sec:related-work}

We discuss most closely related work in the previous sections of the
paper, especially \secref{intro}.
Here, we briefly characterize earlier work on incremental computation and
more recent work on generalizing self-adjusting-computation techniques
to support parallel computation.

Of the many techniques proposed to support incremental computation
(see the survey~\cite{RamalingamRe93}), the most effective ones are
dependence graphs, memoization, and partial evaluation.  Dependence
graphs record the dependencies between data in a computation and rely
on a change-propagation algorithm to update the computation when the
input is modified (e.g.,~\cite{DemersReTe81,Hoover87}).  Dependence
graphs are effective in some applications, e.g. syntax-directed
computations, but are not general-purpose because change propagation
does not update the dependencies.  For example, the INC
language~\cite{YellinSt91}, which uses dependence graphs, does not
permit recursion.  Memoization (also called function caching) (e.g.,
~\cite{PughTe89, AbadiLaLe96,HeydonLeYu00}) applies to any purely
functional program and therefore is more broadly applicable than
dependence graphs.  This classic idea dating back to the late
1950's~\cite{Bellman57,McCarthy63,Michie68} can improve efficiency
when executions of a program with similar inputs perform similar
function calls.  It turns out, however, that even a small input
modification can prevent reuse via memoization, e.g., when they affect
computations deep in the call tree~\cite{AcarBlBlHaTa09}.  Partial
evaluation approaches~\cite{SundareshHu91,FieldTe90} require the user
to fix a part of the input and specialize the program to speedup
modifications to the remaining unfixed part.  The main limitation of
this approach is that it allows input modifications only within a
predetermined partition.

In addition to the early systems discussed above, a more recent
system, DITTO~\cite{ShankarBo07}, offers support for incremental
invariants-checking in Java.
It requires no programmer annotations but only supports a
purely-functional subset of Java.
DITTO also places further restrictions on the programs; while these
restrictions are reasonable for expressing invariant checks, they also
narrow the scope of the approach.

More recent work generalized self-adjusting computation techniques to
support parallel computations.  A paper presents an algorithm for
parallel change propagation~\cite{HammerAcRaGh07}; other papers
consider apply parallel self-adjusting computation to individual
problems~\cite{AcarCoHuTu11pd,SumerAcIhMe11}, as well the map-reduce
framework~\cite{BhatotiaWiRoAcPa11}, a more general setting.

\section{Conclusion}
\label{sec:conclusion}

We described a sound abstract machine semantics for self-adjusting
computation based on a low-level intermediate language.
We implemented this language by presenting compilation and
optimization techniques, including a C-like front end.
Our experiments confirm that the self-adjusting programs produced with
our approach often perform asymptotically faster than full
reevaluation, resulting in orders of magnitude speedups in practice.
We also confirmed that our approach is competitive with past approaches,
which are either unsound or unsuited to low-level settings.

\bibliographystyle{abbrvnat}
\bibliography{../../../../bibliography/main}
\techreportonly{
  \newpage
  \clearpage
  \appendix
  \onecolumn
  \section{Proofs for Consistency}

\begin{defn}[\SA]
  We define $\sa{\rho}{e}{\sigma}$ to mean the following:\\
  If $ \ilustepm {\sigma}{\epsilon}{\rho \statesep e} {\_}{\_}{\rho'
    \statesep \ilupdate{e'}}$, then there exists $\vec{\nu}$ such that
  $\seqof w = \vec{\nu}$ whenever $\ilustepm {\_}{\epsilon}{\rho'
    \statesep {e'}} {\_}{\epsilon}{\epsilon \statesep \seqof{w}}$.
\end{defn}

\begin{defn}[\CSA]
  We define $\csa{\rho}{e}{\sigma}$ to mean the following:\\
  If $\ilustepm{\sigma}{\emptystack}{\rho \statesep
    e}{\sigma'}{\kappa}{\rho' \statesep e'}$, then
  $\sa{\rho'}{e'}{\sigma'}$.
\end{defn}

\begin{defn}
  \label{def:no-reusable-trace}
  We write \emph{$\noreuse{\trz \trx \tr}$} if and only if
  \begin{enumerate}
  \item $\tr = \trend$ 
  \item $\trxundomark{\_} \notin \trx$
  \item $\trxpropmark{\_} \notin \trx$
  \end{enumerate}
\end{defn}

\begin{defn}[From-scratch consistent traces]
  A trace~$\tr$ is from-scratch consistent, written~$\fsc{\tr}$, if
  and only if there exists a closed
  command~$\left<\rho,\rcmd\right>$, store~$\sigma$, and trace
  context~$\trx$ such that      %
\begin{enumerate}
\item $\csa{\rho}{\rcmd}{\sigma}$
\item $\noreuse{\trz{\trx}{\trend}}$
\item $\iltstate {\sigma} {\trz{\trx}{\trend}} {\kappa} {\rho}
  {\rcmd} \iltsteprel^\ast \iltstate {\sigma'}
  {\trz{\trx'}{\trend}} {\kappa} {\emptyenv} {\vec{\nu}}$
\item  $\rewindstepm {\trz{\trx'}{\trend}} {\trend}
  {\trz{\trx}{\trend}} {\tr}$
\end{enumerate}
\end{defn}

\begin{lem}[From traced to untraced]
  \label{lem:traced-to-untraced}
  If 
  $\iltstate{\sigma}{\trz{\trx}{\tr}}{\kappa}{\rho}{\rcmd}
  \iltsteprel^{n}
  \iltstate{\sigma'}{\trz{\trx'}{\tr'}}{\kappa'}{\rho'}{\rcmd'}$
  using only {\bf E.*} and {\bf U.*}, then we also have
  $\ilustate{\sigma}{\kappa}{\rho}{\rcmd} \ilusteprel^{\ast}
  \ilustate{\sigma'}{\kappa'}{\rho'}{\rcmd'}$.
\end{lem}
\begin{proof}
  By inducton on $n$.  When $n = 0$, the claim is obvious.  For $n >
  0$, we inspect the first step taken.  In each possible case, it is
  easy to verify that the claim follows by induction.
\end{proof}

\begin{defn}[Okay traces]
\begin{mathpar}
\inferrule*
{
  \strut
}
{
  \okay{\trend}
}
\and
\inferrule*
{
  \okay{\tr_1}
  \\
  \okay{\tr_2}
}
{
  \okay{\trcons{\trpush{\tr_1}}{\tr_2}}
}
\and
\inferrule*
{
  \fsc{\tr}
}
{
  \okay{\tr}
}
\end{mathpar}
\end{defn}

\begin{defn}[Okay trace contexts]
\begin{mathpar}
\inferrule*
{
  \strut
}
{
  \okay{\trstart}
}
\and
\inferrule*
{
  \okay{\trx}
}
{
  \okay{\trxcons{\trx}{t}}
}
\and
\inferrule*
{
  \okay{\trx}
}
{
  \okay{\trxcons{\trx}{\trxpushmark}}
}
\and
\inferrule*
{
  \okay{\trx} \\
  \fsc{\tr}
}
{
  \okay{\trxcons{\trx}{\trxpropmark{\tr}}}
}
\and
\inferrule*
{
  \okay{\trx} \\
  \okay{\tr}
}
{
  \okay{\trxcons{\trx}{\trxundomark{\tr}}}
}
\end{mathpar}
\end{defn}

\begin{defn}[Okay trace zippers]
\begin{mathpar}
\inferrule*
{
  \okay{\trx} \\
  \okay{\tr}
}
{
  \okay{\trz{\trx}{\tr}}
}
\end{mathpar}
\end{defn}

\begin{lem}[Rewinding is okay]
\label{lem:rewind-okay}
  If $\rewindstepm {\trz{\trx}{\tr}} \_ {\trz{\trx'}{\tr'}} \_$ and
  $\okay{\trz{\trx}{\tr}}$ then $\okay{\trz{\trx'}{\tr'}}$
\end{lem}
\begin{proof} Trivial induction around the following case analysis of
  a rewind step.
  \begin{itemize}
  \item Case
    ${\trz{\trxcons{\trx_1}{t}}{\tr}} ; \_
    \trrewind
    {\trz{\trx_1}{\tr}} ; \_$
    \begin{itemize}
    \item By assumption, $\okay{\trxcons{\trx_1}{t}}$ and $\okay{\tr}$
    \item By inversion, $\okay{\trx_1}$
    \item Hence, $\okay{\trz{\trx_1}{\tr}}$.  
    \end{itemize}

  \item Case
    ${\trz{\trxcons{\trx_1}{\trxundomark{\tr_1}}}{\trend}} ; \_
    \trrewind
    {\trz{\trx_1}{\tr_1}} ; \_$
    \begin{itemize}
    \item By assumption $\okay{\trxcons{\trx_1}{\trxundomark{\tr_1}}}$
    \item By inversion, $\okay{\trx_1}$ and $\okay{\tr_1}$
    \item Hence, $\okay{\trz{\trx_1}{\tr_1}}$.  
    \end{itemize}
    
  \item Case
    ${\trz{\trxcons{\trx_1}{\trxundomark{\tr_2}}}{\trcons{t}{\tr_1}}} ; \_
    \trrewind
    {\trz{\trx_1}{\trcons{\trpush{\trcons{t}{\tr_1}}}{\tr_2}}} ; \_$
    \begin{itemize}
    \item By assumption $\trxcons{\trx_1}{\trxundomark{\tr_2}}$ 
      and $\okay{\trcons{t}{\tr_1}}$
    \item By inversion, $\okay{\trx_1}$ and $\okay{\tr_2}$
    \item Hence, $\okay{\trcons{\trpush{\trcons{t}{\tr_1}}}{\tr_2}}$
    \item And $\okay{\trz{\trx_1}{\trcons{\trpush{\trcons{t}{\tr_1}}}{\tr_2}}}$.
    \end{itemize}
  \end{itemize}
\end{proof}

\begin{lem}[Purity]
  \label{lem:purity}
  If $\iltstepm{\trz{\trx_1}{\tr_1}}{\sigma_1}{\kappa_1}{\rho
    \statesep \rcmd}{\trz{\trx_1}{\tr_1}}{\sigma_1}{\kappa_1}{\rho'
    \statesep \rcmd'}$ using {\bf E.0} only, then for any $\trx_2$,
  $\tr_2$, $\sigma_2$, $\kappa_2$ we have
  $\iltstepm{\trz{\trx_2}{\tr_2}}{\sigma_2}{\kappa_2}{\rho \statesep
    \rcmd}{\trz{\trx_2}{\tr_2}}{\sigma_2}{\kappa_2}{\rho' \statesep
    \rcmd'}$.
\end{lem}
\begin{proof}
  Trivial induction.
\end{proof}

\begin{lem}[Rewinding]\label{lem:rewind-trace-context}
  If $\rewindstepm {\trz{\trx'}{\_}}{\_}{\trz{\trx}{\_}}{\_}$, then
  \begin{enumerate}
  \item $\trx \in \prefixes{\trx'}$,
  \item $\numof{\pushmark}{\trx'} = \numof{\pushmark}{\trx}$, and
  \item $\numof{\propmark}{\trx'} = \numof{\propmark}{\trx}$.
  \end{enumerate}
\end{lem}
\begin{proof}~
  By induction on the number $n$ of rewinding steps.  If $n = 0$, then
  $\trx = \trx'$ and the claim holds trivially.  Suppose $n = 1 + n'$.
  Case analysis on the first step:
  \begin{itemize}
  \item Case
    $\rewindstep{\trz{\trx'}{\tr_1'}}{\tr_2'}{\trz{\trx''}{\tr_1'}}{\trcons{t}{\tr_2'}}
    \trrewind^{n'} \trz{\trx}{\tr_1};{\tr_2}$ with $\trx' =
    \trcons{\trx''}{t}$:
    \begin{itemize}
    \item By induction we get $\trx \in \prefixes{\trx''} \land
      \numof{\pushmark}{\trx''} = \numof{\pushmark}{\trx} \land
      \numof{\propmark}{\trx''} = \numof{\propmark}{\trx}$.
    \item This implies the claims.
    \end{itemize}
  \item Case
    $\rewindstep{\trz{\trx'}{\tr_1'}}{\tr_2'}{\trz{\trx''}{\tr}}{\tr_2'}
    \trrewind^{n'} \trz{\trx}{\tr_1};{\tr_2}$ with $\tr_1' = \trend$
    and $\trx' = \trcons{\trx''}{\trxundomark{\tr}}$:
    \begin{itemize}
    \item By induction we get $\trx \in \prefixes{\trx''} \land
      \numof{\pushmark}{\trx''} = \numof{\pushmark}{\trx} \land
      \numof{\propmark}{\trx''} = \numof{\propmark}{\trx}$.
    \item This implies the claims.
    \end{itemize}
  \item Case
    $\rewindstep{\trz{\trx'}{\tr_1'}}{\tr_2'}{\trz{\trx''}{\trcons{\trpush{\tr_1'}}{\tr}}}{\tr_2'}
    \trrewind^{n'} \trz{\trx}{\tr_1};{\tr_2}$ with $\tr_1' =
    \trcons{t}{\tr_1''}$ and $\trx' =
    \trcons{\trx''}{\trxundomark{\tr}}$:
    \begin{itemize}
    \item By induction we get $\trx \in \prefixes{\trx''} \land
      \numof{\pushmark}{\trx''} = \numof{\pushmark}{\trx} \land
      \numof{\propmark}{\trx''} = \numof{\propmark}{\trx}$.
    \item This implies the claims.
    \end{itemize}
  \end{itemize}
\end{proof}

\begin{lem}
  \label{lem:prefix-dropum}
  If $\trx \in \prefixes{\trx'}$, then $\dropum{\trx} \in
  \prefixes{\dropum{\trx'}}$.
\end{lem}

\begin{lem}
  \label{lem:rewind-dropum}
  If $\rewindstepm{\trz{\trx}{\tr_1}}{\tr}{\trz{\trx'}{\_}}{\tr'}$, then
  $\rewindstepm{\trz{\dropum{\trx}}{\tr_1}}{\tr}{\trz{\dropum{\trx'}}{\_}}{\tr'}$.
\end{lem}
\begin{proof}
  By induction on the number $n$ of rewinding steps.  If $n = 0$, then
  $\trx = \trx'$ and $\tr = \tr'$, so the claim holds obviously.  Now
  suppose $n > 0$.  We inspect the last step:
  \begin{itemize}
  \item Case
    $\rewindstepm{\trz{\trx}{\tr_1}}{\tr}{\trz{\trxcons{\trx'}{t}}{\_}}{\tr_2}
    \trrewind \trz{\trx'}{\_};{\trcons{t}{\tr_2}}$ with $\tr' =
    \trcons{t}{\tr_2}$:
    \begin{itemize}
    \item By induction,
      $\rewindstepm{\trz{\dropum{\trx}}{\tr_1}}{\tr}{\trz{\dropum{\trxcons{\trx'}{t}}}{\_}}{\tr_2}
      \trrewind \trz{\dropum{\trx'}}{\_};{\trcons{t}{\tr_2}}$.
    \end{itemize}
  \item Case
    $\rewindstepm{\trz{\trx}{\tr_1}}{\tr}{\trz{\trxcons{\trx'}{\trxundomark{\_}}}{\_}}{\tr'}
    \trrewind \trz{\trx'}{\_};{\tr'}$:
    \begin{itemize}
    \item By induction,
      $\rewindstepm{\trz{\dropum{\trx}}{\tr_1}}{\tr}{\trz{\dropum{\trxcons{\trx'}{\trxundomark{\_}}}}{\_}}{\tr'}
      = \trz{\dropum{\trx'}}{\_};{\tr'}$.
    \end{itemize}
  \end{itemize}
\end{proof}

\begin{lem}[Trace actions stick around (prefix version)]
\label{lem:prefix}
\label{lem:prefix-single-action}
If
\begin{enumerate}
\item
  $\iltstate{\_}{\trz{\trcons{\trx_1}{\trcons{\tr_2}{\trx_2}}}{\tr_1}}{\_}{\_}{\_}
  \iltsteprel^\ast \iltstate{\_}{\trz{\trx_3}{\_}}{\_}{\_}{\_}$
\item $\dropum{\trx_1} \in \prefixes{\dropum{\trx_3}}$
\end{enumerate}
then $\dropum{\trcons{\trx_1}{\tr_2}} \in \prefixes{\dropum{\trx_3}}$.
\end{lem}
\begin{proof}
  By induction on the length $n$ of the reduction chain.  If $n = 0$,
  then $\trx_3 = \trcons{\trx_1}{\trcons{\tr_2}{\trx_2}}$ and thus the
  claim is obvious.  Now consider $n = 1 + n'$.  We inspect the first
  step:
  \begin{itemize}
  \item Case {\bf E.0}, {\bf U.1-2}:
    \begin{itemize}
    \item Then
      $\iltstate{\_}{\trz{\trcons{\trx_1}{\trcons{\tr_2}{\trx_2}}}{\tr_1}}{\_}{\_}{\_}
      \iltsteprel^{n'} \iltstate{\_}{\trz{\trx_3}{\_}}{\_}{\_}{\_}$.
    \item The claim then follows by induction.
    \end{itemize}
  \item Case {\bf E.1--5,7}, {\bf E.P}, {\bf P.E,1--5,7}:
    \begin{itemize}
    \item Then
      $\iltstate{\_}{\trz{\trcons{\trx_1}{\trcons{\tr_2}{\trcons{\trx_2}{t}}}}{\tr_1}}{\_}{\_}{\_}
      \iltsteprel^{n'} \iltstate{\_}{\trz{\trx_3}{\_}}{\_}{\_}{\_}$,
      for some $t$.
    \item The claim then follows by induction.
    \end{itemize}
  \item Case {\bf E.6}:
    \begin{itemize}
    \item Then
      $\iltstate{\_}{\trz{\trcons{\trx_1}{\trcons{\tr_2}{\trcons{\trx_2}{\trxpushmark}}}}{\tr_1}}{\_}{\_}{\_}
      \iltsteprel^{n'} \iltstate{\_}{\trz{\trx_3}{\_}}{\_}{\_}{\_}$.
    \item The claim then follows by induction.
    \end{itemize}
  \item Case {\bf P.6}:
    \begin{itemize}
    \item Then
      $\iltstate{\_}{\trz{\trcons{\trx_1}{\trcons{\tr_2}{\trcons{\trx_2}{\trxpropmark{\tr}}}}}{\tr_1}}{\_}{\_}{\_}
      \iltsteprel^{n'} \iltstate{\_}{\trz{\trx_3}{\_}}{\_}{\_}{\_}$.
    \item The claim then follows by induction.
    \end{itemize}
  \item Case {\bf U.3}:
    \begin{itemize}
    \item Then
      $\iltstate{\_}{\trz{\trcons{\trx_1}{\trcons{\tr_2}{\trcons{\trx_2}{\trxundomark{\tr}}}}}{\tr_1}}{\_}{\_}{\_}
      \iltsteprel^{n'} \iltstate{\_}{\trz{\trx_3}{\_}}{\_}{\_}{\_}$.
    \item The claim then follows by induction.
    \end{itemize}
  \item Case {\bf E.8}:
    \begin{itemize}
    \item Subcase
      $\trz{\trcons{\trx_1}{\trcons{\tr_2}{\trx_2}}}{\tr_1};\trend
      \trrewind^\ast \trz{\trcons{\trx_1}{\trcons{\tr_2}{\trcons{\trx_2'}{\trxpushmark}}}}{\tr_1'}; \tr_3$:
      \begin{itemize}
      \item Then
        $\iltstate{\_}{\trz{\trcons{\trx_1}{\trcons{\tr_2}{\trcons{\trx_2'}{\trpush{\tr_3}}}}}{\tr_1'}}{\_}{\_}{\_}
        \iltsteprel^{n'} \iltstate{\_}{\trz{\trx_3}{\_}}{\_}{\_}{\_}$.
      \item The claim then follows by induction.
      \end{itemize}
    \item Subcase
      $\trz{\trcons{\trx_1}{\trcons{\tr_2}{\trx_2}}}{\tr_1};\trend
      \trrewind^\ast \trz{\trx_1}{\_}; \_
      \trrewind^\ast \trz{\trcons{\trx_1'}{\trxpushmark}}{\tr_1'}; \tr_3$:
      \begin{itemize}
      \item Then
        $\iltstate{\_}{\trz{\trcons{\trx_1'}{\trpush{\tr_3}}}{\tr_1'}}{\_}{\_}{\_}
        \iltsteprel^{n'} \iltstate{\_}{\trz{\trx_3}{\_}}{\_}{\_}{\_}$.
      \item By Lemma~\ref{lem:rewind-trace-context} we have
        $\trxcons{\trx_1'}{\trxpushmark} \in \prefixes{\trx_1}$.
      \item Hence, using Lemma~\ref{lem:prefix-dropum},
        $\dropum{\trx_1'}, \dropum{\trxcons{\trx_1'}{\trxpushmark}}
        \in \prefixes{\dropum{\trx_1}} \subseteq
        \prefixes{\dropum{\trx_3}}$.
      \item Hence $\dropum{\trxcons{\trx_1'}{\trpush{\tr_3}}} \in
        \prefixes{\dropum{\trx_3}}$ by induction, contradicting
        $\dropum{\trcons{\trx_1'}{\trxpushmark}} \in
        \prefixes{\dropum{\trx_3}}$.
      \end{itemize}
    \end{itemize}
  \item Case {\bf P.8}:
    \begin{itemize}
    \item Subcase
      $\trz{\trcons{\trx_1}{\trcons{\tr_2}{\trx_2}}}{\tr_1};\trend
      \trrewind^\ast
      \trz{\trcons{\trx_1}{\trcons{\tr_2}{\trcons{\trx_2'}{\trxpropmark{\tr}}}}}{\trend};
      \tr_3$ where $\tr_1 = \trend$:
      \begin{itemize}
      \item Then
        $\iltstate{\_}{\trz{\trcons{\trx_1}{\trcons{\tr_2}{\trcons{\trx_2'}{\trpush{\tr_3}}}}}{\tr}}{\_}{\_}{\_}
        \iltsteprel^{n'} \iltstate{\_}{\trz{\trx_3}{\_}}{\_}{\_}{\_}$.
      \item The claim then follows by induction.
      \end{itemize}
    \item Subcase
      $\trz{\trcons{\trx_1}{\trcons{\tr_2}{\trx_2}}}{\tr_1};\trend
      \trrewind^\ast \trz{\trx_1}{\_}; \_ \trrewind^\ast
      \trz{\trcons{\trx_1'}{\trxpropmark{\tr}}}{\trend}; \tr_3$ where
      $\tr_1 = \trend$:
      \begin{itemize}
      \item Then
        $\iltstate{\_}{\trz{\trcons{\trx_1'}{\trpush{\tr_3}}}{\tr}}{\_}{\_}{\_}
        \iltsteprel^{n'} \iltstate{\_}{\trz{\trx_3}{\_}}{\_}{\_}{\_}$.
      \item By Lemma~\ref{lem:rewind-trace-context} we have
        $\trxcons{\trx_1'}{\trxpropmark{\tr}} \in \prefixes{\trx_1}$.
      \item Hence, using Lemma~\ref{lem:prefix-dropum},
        $\dropum{\trx_1'},
        \dropum{\trxcons{\trx_1'}{\trxpropmark{\tr}}} \in
        \prefixes{\dropum{\trx_1}} \subseteq
        \prefixes{\dropum{\trx_3}}$.
      \item Hence $\dropum{\trxcons{\trx_1'}{\trpush{\tr_3}}} \in
        \prefixes{\dropum{\trx_3}}$ by induction, contradicting
        $\dropum{\trcons{\trx_1'}{\trxpropmark{\tr}}} \in
        \prefixes{\dropum{\trx_3}}$.
      \end{itemize}
    \end{itemize}
  \item Case {\bf U.4}:
    \begin{itemize}
    \item Subcase
      $\trz{\trcons{\trx_1}{\trcons{\tr_2}{\trx_2}}}{\tr_1} =
      \trz{\trcons{\trx_1}{\trcons{\tr_2}{\trcons{\trx_2'}{\trxundomark{\tr}}}}}{\trend}$:
      \begin{itemize}
      \item Then
        $\iltstate{\_}{\trz{\trcons{\trx_1}{\trcons{\tr_2}{\trx_2'}}}{\tr}}{\_}{\_}{\_}
        \iltsteprel^{n'} \iltstate{\_}{\trz{\trx_3}{\_}}{\_}{\_}{\_}$.
      \item The claim then follows by induction.
      \end{itemize}
    \item Subcase
      $\trz{\trcons{\trx_1}{\trcons{\tr_2}{\trx_2}}}{\tr_1} =
      \trz{\trcons{\trx_1'}{\trxundomark{\tr}}}{\trend}$ where $\trcons{\tr_2}{\trx_2} = \trstart$:
      \begin{itemize}
      \item Then the claim is (2).
      \end{itemize}
    \end{itemize}
  \end{itemize}
\end{proof}

\begin{lem}[Trace actions stick around (rewinding version)]
\label{lem:rewind-single-action}
  If
  \begin{itemize}
  \item
    $\iltstate{\_}{\trz{\trxcons{\trx}{t}}{\tr}}{\_}{\_}{\_}
    \iltsteprel^\ast
    \iltstate{\_}{\trz{\trx'}{\tr'}}{\_}{\_}{\_}$
  \item
    $\rewindstepm{\trz{\dropum{\trx'}}{\tr_1}}{\trend}{\trz{\dropum{\trx}}{\tr_2}}{\tr_3}$
  \end{itemize}
  then $\rewindstepm
  {\trz{\dropum{\trx'}}{\tr_1}}{\trend}{\trz{\dropum{\trxcons{\trx}{t}}}{\tr_2}}{\tr_3'}
  \trrewind \trz{\dropum{\trx}}{\tr_2};{\tr_3}$.
\end{lem}
\begin{proof}
  Note that the rewinding takes at least one step, otherwise
  Lemmas~\ref{lem:rewind-trace-context} and
  \ref{lem:prefix-single-action} would yield $\trxcons{\trx}{t} \in
  \prefixes{\trx}$, a contradiction.  We inspect the last step:
  \begin{itemize}
  \item Case
    $\rewindstepm{\trz{\dropum{\trx'}}{\tr_1}}{\trend}{\trz{\trxcons{\dropum{\trx}}{t'}}{\tr_2}}{\tr_3'}
    \trrewind \trz{\dropum{\trx}}{\tr_2};{\tr_3}$ with $\tr_3 =
    \trcons{t'}{\tr_3'}$:
    \begin{itemize}
    \item Lemmas~\ref{lem:rewind-trace-context}
      and~\ref{lem:prefix-single-action} yield
      $\dropum{\trcons{\trx}{t}} \in \prefixes{\dropum{\trx'}}$.
    \item Lemma~\ref{lem:rewind-trace-context} also yields
      $\dropum{\trcons{\trx}{t'}} \in \prefixes{\dropum{\trx'}}$.
    \item Hence $t = t'$ and we are done.
    \end{itemize}
  \item Case $\rewindstepm{\trz{\dropum{\trx'}}{\tr_1}}{\trend}{\trz{\trxcons{\dropum{\trx}}{\trxundomark{\widehat{\tr}}}}{\tr_2'}}{\tr_3}
    \trrewind \trz{\dropum{\trx}}{\tr_2};{\tr_3}$:
    \begin{itemize}
    \item Lemma~\ref{lem:rewind-trace-context} yields
      $\trxcons{\dropum{\trx}}{\trxundomark{\widehat{\tr}}} \in
      \prefixes{\dropum{\trx'}}$, which is a contradiction.
    \end{itemize}
  \end{itemize}
\end{proof}

\begin{lem}
  \label{lem:noreuse-preservation}
  If $\iltstate{\sigma}{\trz{\trx}{\tr}}{\kappa}{\rho}{\rcmd}
  \iltsteprel^\ast
  \iltstate{\sigma'}{\trz{\trx'}{\tr'}}{\kappa'}{\rho'}{\rcmd'}$ and
  $\noreuse{\trz{\trx}{\tr}}$, then
  $\noreuse{\trz{\trx'}{\tr'}}$.
\end{lem}
\begin{proof}
  Easy induction on the length of the reduction.
\end{proof}

\begin{lem}
  \label{lem:extended-stack}
  If $\ilustepm{\sigma}{\kappa}{\rho \statesep
    \rcmd}{\sigma'}{\kappa'}{\rho' \statesep \rcmd'}$, then
  $\ilustepm{\sigma}{\kappa_0 @ \kappa}{\rho \statesep
    \rcmd}{\sigma'}{\kappa_0 @ \kappa'}{\rho' \statesep \rcmd'}$ for
  any $\kappa_0$.
\end{lem}
\begin{proof}
  Easy induction on the length of the reduction.
\end{proof}

\begin{lem}[CSA preservation (untraced)]
  \label{lem:csa-preservation-untraced}
  If $\ilustate{\sigma}{\emptystack}{\rho}{e} \ilusteprel^\ast
  \ilustate{\sigma'}{\kappa'}{\rho'}{e'}$ and $\csa{\rho}{e}{\sigma}$,
  then $\csa{\rho'}{e'}{\sigma'}$.
\end{lem}
\begin{proof}~
  \begin{itemize}
  \item Suppose $\ilustate{\sigma'}{\emptystack}{\rho'}{e'}
    \ilusteprel^\ast \ilustate{\sigma''}{\kappa''}{\rho''}{e''}$.
  \item We must show $\sa{\rho''}{e''}{\sigma''}$.
  \item By Lemma~\ref{lem:extended-stack} we get
    $\ilustate{\sigma'}{\kappa'}{\rho'}{e'} \ilusteprel^\ast
    \ilustate{\sigma''}{\stcons{\kappa'}{\kappa''}}{\rho''}{e''}$.
  \item Hence $\ilustate{\sigma}{\emptystack}{\rho}{e}
    \ilusteprel^\ast
    \ilustate{\sigma''}{\stcons{\kappa'}{\kappa''}}{\rho''}{e''}$.
  \item The goal then follows from $\csa{\rho}{e}{\sigma}$.
  \end{itemize}
\end{proof}

\begin{lem}
  \label{lem:pushmark-decomposition}
  Suppose $\dropum{\trx} \in \prefixes{\dropum{\trx'}}$,
  $\dropum{\trxcons{\trx}{\trxpushmark}} \notin
  \prefixes{\dropum{\trx'}}$ and $|\widetilde{\kappa}| =
  \numof{\pushmark}{\widetilde{\trx}}$.
  \begin{enumerate}
  \item If
    $\iltstate{\sigma}{\trz{\trxcons{\trxcons{\trx}{\trxpushmark}}{\widetilde{\trx}}}{\tr_0}}{\stcons{\stcons{\kappa}{\mkstframe{\rho_f}{f}}}{\widetilde{\kappa}}}{\rho}{\rcmd}
    \iltsteprel^n
    \iltstate{\sigma'}{\trz{\trx'}{\tr_0'}}{\stcons{\kappa}{\widehat{\kappa}}}{\rho'}{\rcmd'}$,
    then:
    \begin{itemize}
    \item
      $\iltstate{\sigma}{\trz{\trxcons{\trxcons{\trx}{\trxpushmark}}{\widetilde{\trx}}}{\tr_0}}{\stcons{\stcons{\kappa}{\mkstframe{\rho_f}{f}}}{\widetilde{\kappa}}}{\rho}{\rcmd}
      \iltsteprel^{n_1}
      \iltstate{\sigma''}{\trz{\trxcons{\trxcons{\trx}{\trxpushmark}}{\widetilde{\trx}'}}{\tr_0''}}{\stcons{\kappa}{\mkstframe{\rho_f}{f}}}{\emptyenv}{\vec{\omega}}$
    \item
      $\rewindstepm{\trz{\trxcons{\trxcons{\trx}{\trxpushmark}}{\widetilde{\trx}'}}{\tr_0''}}{\trend}{\trz{\trxcons{\trx}{\trxpushmark}}{\tr_0'''}}{\widetilde{\tr}}$
    \item $\rho_f(f) = \ilfun{f}{\vec{x}}{e_f}$
    \item
      $\iltstate{\sigma''}{\trz{\trxcons{\trxcons{\trx}{\trxpushmark}}{\widetilde{\trx'}}}{\tr_0''}}{\stcons{\kappa}{\mkstframe{\rho_f}{f}}}{\emptyenv}{\vec{\omega}}
      \iltsteprel
      \iltstate{\sigma''}{\trz{\trxcons{\trx}{\trpush{\widetilde{\tr}}}}{\tr_0'''}}{\kappa}{\rho_f[\vec{x
          \mapsto \omega}]}{e_f}$
    \item
      $\iltstate{\sigma''}{\trz{\trxcons{\trx}{\trpush{\widetilde{\tr}}}}{\tr_0'''}}{\kappa}{\rho_f[\vec{x
          \mapsto \omega}]}{e_f} \iltsteprel^{n_2}
      \iltstate{\sigma'}{\trz{\trx'}{\tr_0'}}{\stcons{\kappa}{\widehat{\kappa}}}{\rho'}{\rcmd'}$
    \item $n = n_1 + 1 + n_2$
    \end{itemize}
  \item If
    $\iltstate{\sigma}{\trz{\trxcons{\trxcons{\trx}{\trxpushmark}}{\widetilde{\trx}}}{\tr_0}}{\stcons{\stcons{\kappa}{\mkstframe{\rho_f}{f}}}{\widetilde{\kappa}}}{\emptyenv}{\ilprop}
    \iltsteprel^n
    \iltstate{\sigma'}{\trz{\trx'}{\tr_0'}}{\stcons{\kappa}{\widehat{\kappa}}}{\rho'}{\rcmd'}$,
    then:
    \begin{itemize}
    \item
      $\iltstate{\sigma}{\trz{\trxcons{\trxcons{\trx}{\trxpushmark}}{\widetilde{\trx}}}{\tr_0}}{\stcons{\stcons{\kappa}{\mkstframe{\rho_f}{f}}}{\widetilde{\kappa}}}{\emptyenv}{\ilprop}
      \iltsteprel^{n_1}
      \iltstate{\sigma''}{\trz{\trxcons{\trxcons{\trx}{\trxpushmark}}{\widetilde{\trx}'}}{\tr_0''}}{\stcons{\kappa}{\mkstframe{\rho_f}{f}}}{\emptyenv}{\vec{\omega}}$
    \item
      $\rewindstepm{\trz{\trxcons{\trxcons{\trx}{\trxpushmark}}{\widetilde{\trx}'}}{\tr_0''}}{\trend}{\trz{\trxcons{\trx}{\trxpushmark}}{\tr_0'''}}{\widetilde{\tr}}$
    \item $\rho_f(f) = \ilfun{f}{\vec{x}}{e_f}$
    \item
      $\iltstate{\sigma''}{\trz{\trxcons{\trxcons{\trx}{\trxpushmark}}{\widetilde{\trx'}}}{\tr_0''}}{\stcons{\kappa}{\mkstframe{\rho_f}{f}}}{\emptyenv}{\vec{\omega}}
      \iltsteprel
      \iltstate{\sigma''}{\trz{\trxcons{\trx}{\trpush{\widetilde{\tr}}}}{\tr_0'''}}{\kappa}{\rho_f[\vec{x
          \mapsto \omega}]}{e_f}$
    \item
      $\iltstate{\sigma''}{\trz{\trxcons{\trx}{\trpush{\widetilde{\tr}}}}{\tr_0'''}}{\kappa}{\rho_f[\vec{x
          \mapsto \omega}]}{e_f} \iltsteprel^{n_2}
      \iltstate{\sigma'}{\trz{\trx'}{\tr_0'}}{\stcons{\kappa}{\widehat{\kappa}}}{\rho'}{\rcmd'}$
    \item $n = n_1 + 1 + n_2$
    \end{itemize}
  \end{enumerate}
\end{lem}
\begin{proof}
  By mutual induction on $n$.  If $n = 0$, then we obtain a
  contradiction to $\dropum{\trxcons{\trx}{\trxpushmark}} \notin
  \prefixes{\dropum{\trx'}}$.  So consider $n > 0$.  In each part we
  inspect the first step of the reduction.
  \begin{enumerate}
  \item
    \begin{itemize}
    \item Case {\bf E.0--7}: Straightforward, using the inductive
      hypothesis.
    \item Case {\bf E.8}:
      \begin{itemize}
      \item Subcase $\numof{\pushmark}{\widetilde{\trx}} = 0$:
        \begin{itemize}
        \item Then
          $\stcons{\stcons{\kappa}{\mkstframe{\rho_f}{f}}}{\widetilde{\kappa}} \statesep \rho \statesep \rcmd
          =
          \stcons{\kappa}{\mkstframe{\rho_f}{f}} \statesep \emptyenv \statesep \vec{\omega}$
          and
          $\rewindstepm{\trz{\trxcons{\trxcons{\trx}{\trxpushmark}}{\widetilde{\trx}}}{\tr_0}}{\trend}{\trz{\trxcons{\trx}{\trxpushmark}}{\tr_2}}{\tr_1}$.
        \item Hence
          $\iltstate{\sigma}{\trz{\trxcons{\trxcons{\trx}{\trxpushmark}}{\widetilde{\trx}}}{\tr_0}}{\stcons{\kappa}{\mkstframe{\rho_f}{f}}}{\emptyenv}{\vec{\omega}}
          \iltsteprel
          \iltstate{\sigma}{\trz{\trxcons{\trx}{\trpush{\tr_1}}}{\tr_2}}{\kappa}{\rho_f[\vec{x
              \mapsto \omega}]}{e_f}$ with \\
          $\iltstate{\sigma}{\trz{\trxcons{\trx}{\trpush{\tr_1}}}{\tr_2}}{\kappa}{\rho_f[\vec{x
              \mapsto \omega}]}{e_f} \iltsteprel^{n-1}
          \iltstate{\sigma'}{\trz{\trx'}{\tr_0'}}{\stcons{\kappa}{\widehat{\kappa}}}{\rho'}{\rcmd'}$.
        \item Thus the claim holds for $n_1 = 0$, $n_2 = n - 1$.
        \end{itemize}
      \item Subcase $\numof{\pushmark}{\widetilde{\trx}} > 0$:
        \begin{itemize}
        \item Then
          $\stcons{\stcons{\kappa}{\mkstframe{\rho_f}{f}}}{\widetilde{\kappa}}
          \statesep \rho \statesep \rcmd =
          \stcons{\stcons{\kappa}{\mkstframe{\rho_f}{f}}}{\stcons{\widetilde{\kappa}'}{\mkstframe{\widehat{\rho}}{\widehat{f}}}}
          \statesep \emptyenv \statesep \vec{\omega}$ and
          $\rewindstepm{\trz{\trxcons{\trxcons{\trx}{\trxpushmark}}{\widetilde{\trx}}}{\tr_0}}{\trend}{\trz{\trxcons{\trxcons{\trx}{\trxpushmark}}{\trxcons{\widetilde{\trx}'}{\trxpushmark}}}{\tr_0''}}{\widehat{\tr}}$.
        \item So
          $\iltstate{\sigma}{\trz{\trxcons{\trxcons{\trx}{\trxpushmark}}{\widetilde{\trx}}}{\tr_0}}{\stcons{\stcons{\kappa}{\mkstframe{\rho_f}{f}}}{\widetilde{\kappa}}}{\emptyenv}{\vec{\omega}}
          \iltsteprel
          \iltstate{\sigma}{\trz{\trxcons{\trxcons{\trx}{\trxpushmark}}{\trxcons{\widetilde{\trx}'}{\trpush{\widehat{\tr}}}}}{\tr_0''}}{\stcons{\stcons{\kappa}{\mkstframe{\rho_f}{f}}}{\widetilde{\kappa}'}}{\widehat{\rho}}{\widehat{e}}$.
        \item And
          $\iltstate{\sigma}{\trz{\trxcons{\trxcons{\trx}{\trxpushmark}}{\trxcons{\widetilde{\trx}'}{\trpush{\widehat{\tr}}}}}{\tr_0''}}{\stcons{\stcons{\kappa}{\mkstframe{\rho_f}{f}}}{\widetilde{\kappa}'}}{\widehat{\rho}}{\widehat{e}}
          \iltsteprel^{n-1}
          \iltstate{\sigma'}{\trz{\trx'}{\tr_0'}}{\stcons{\kappa}{\widehat{\kappa}}}{\rho'}{\rcmd'}$.
        \item Hence the claim holds by induction.
        \end{itemize}
      \end{itemize}
    \item Case {\bf E.P}: By induction (part 2).
    \item Case {\bf P.E,1--7}: Not possible.
    \item Case {\bf P.8}:
      \begin{itemize}
      \item Then $\rho,\rcmd = \emptyenv,\vec{\omega}$ and
        $\rewindstepm{\trz{\trxcons{\trxcons{\trx}{\trxpushmark}}{\widetilde{\trx}}}{\trend}}{\trend}{\trz{\trxcons{\trxcons{\trx}{\trxpushmark}}{\trxcons{\widetilde{\trx}'}{\trxpropmark{\tr_2}}}}{\trend}}{\tr_1}$.
      \item So
        $\iltstate{\sigma}{\trz{\trxcons{\trxcons{\trx}{\trxpushmark}}{\widetilde{\trx}}}{\tr_0}}{\stcons{\stcons{\kappa}{\mkstframe{\rho_f}{f}}}{\widetilde{\kappa}}}{\emptyenv}{\vec{\omega}}
        \iltsteprel
        \iltstate{\sigma}{\trz{\trxcons{\trxcons{\trx}{\trxpushmark}}{\trxcons{\widetilde{\trx}}{\trpush{\tr_1}}}}{\tr_2}}{\stcons{\stcons{\kappa}{\mkstframe{\rho_f}{f}}}{\widetilde{\kappa}}}{\emptyenv}{\ilprop}$.
      \item And
        $\iltstate{\sigma}{\trz{\trxcons{\trxcons{\trx}{\trxpushmark}}{\trxcons{\widetilde{\trx}}{\trpush{\tr_1}}}}{\tr_2}}{\stcons{\stcons{\kappa}{\mkstframe{\rho_f}{f}}}{\widetilde{\kappa}}}{\emptyenv}{\ilprop}
        \iltsteprel^{n-1}
        \iltstate{\sigma'}{\trz{\trx'}{\tr_0'}}{\stcons{\kappa}{\widehat{\kappa}}}{\rho'}{\rcmd'}$.
      \item Hence the claim holds by induction (part 2).
      \end{itemize}
    \item Case {\bf U.1--4}: Straightforward by induction.
    \end{itemize}
  \item
    \begin{itemize}
    \item Case {\bf E.0--8,P}: Not possible.
    \item Case {\bf P.E}: By induction (part 1).
    \item Case {\bf P.1--7}: Straightforward by induction.
    \item Case {\bf U.1--4}: Not possible.
    \item Case {\bf P.8}: Not possible.
    \end{itemize}
  \end{enumerate}
\end{proof}

\begin{lem}
  \label{lem:propmark-decomposition}
  Suppose $\dropum{\trx} \in \prefixes{\dropum{\trx'}}$,
  $\dropum{\trxcons{\trx}{\trxpropmark{\tr'}}} \notin
  \prefixes{\dropum{\trx'}}$ and $|\widetilde{\kappa}| =
  \numof{\pushmark}{\widetilde{\trx}}$.
  \begin{enumerate}
  \item If
    $\iltstate{\sigma}{\trz{\trxcons{\trxcons{\trx}{\trxpropmark{\tr'}}}{\widetilde{\trx}}}{\tr_0}}{\stcons{\kappa}{\widetilde{\kappa}}}{\rho}{\rcmd}
    \iltsteprel^n
    \iltstate{\sigma'}{\trz{\trx'}{\tr_0'}}{\stcons{\kappa}{\widehat{\kappa}}}{\rho'}{\rcmd'}$,
    then:
    \begin{itemize}
    \item
      $\iltstate{\sigma}{\trz{\trxcons{\trxcons{\trx}{\trxpropmark{\tr'}}}{\widetilde{\trx}}}{\tr_0}}{\kappa}{\rho}{\rcmd}
      \iltsteprel^{n_1}
      \iltstate{\sigma''}{\trz{\trxcons{\trxcons{\trx}{\trxpropmark{\tr'}}}{\widetilde{\trx}'}}{\trend}}{\kappa}{\emptyenv}{\vec{\omega}}$
    \item
      $\rewindstepm{\trz{\trxcons{\trxcons{\trx}{\trxpropmark{\tr'}}}{\widetilde{\trx}'}}{\trend}}{\trend}{\trz{\trxcons{\trx}{\trxpropmark{\tr'}}}{\trend}}{\widetilde{\tr}}$
    \item
      $\iltstate{\sigma''}{\trz{\trxcons{\trxcons{\trx}{\trxpropmark{\tr'}}}{\widetilde{\trx}'}}{\trend}}{\kappa}{\emptyenv}{\vec{\omega}}
      \iltsteprel
      \iltstate{\sigma''}{\trz{\trxcons{\trx}{\trpush{\widetilde{\tr}}}}{\tr'}}{\kappa}{\emptyenv}{\ilprop}$
    \item
      $\iltstate{\sigma''}{\trz{\trxcons{\trx}{\trpush{\widetilde{\tr}}}}{\tr'}}{\kappa}{\emptyenv}{\ilprop}
      \iltsteprel^{n_2}
      \iltstate{\sigma'}{\trz{\trx'}{\tr_0'}}{\stcons{\kappa}{\widehat{\kappa}}}{\rho'}{\rcmd'}$
    \item $n = n_1 + 1 + n_2$
    \end{itemize}
  \item If
    $\iltstate{\sigma}{\trz{\trxcons{\trxcons{\trx}{\trxpropmark{\tr'}}}{\widetilde{\trx}}}{\tr_0}}{\stcons{\kappa}{\widetilde{\kappa}}}{\emptyenv}{\ilprop}
    \iltsteprel^n
    \iltstate{\sigma'}{\trz{\trx'}{\tr_0'}}{\stcons{\kappa}{\widehat{\kappa}}}{\rho'}{\rcmd'}$,
    then:
    \begin{itemize}
    \item
      $\iltstate{\sigma}{\trz{\trxcons{\trxcons{\trx}{\trxpropmark{\tr'}}}{\widetilde{\trx}}}{\tr_0}}{\stcons{\kappa}{\widetilde{\kappa}}}{\emptyenv}{\ilprop}
      \iltsteprel^{n_1}
      \iltstate{\sigma''}{\trz{\trxcons{\trxcons{\trx}{\trxpropmark{\tr'}}}{\widetilde{\trx}'}}{\trend}}{\kappa}{\emptyenv}{\vec{\omega}}$
    \item
      $\rewindstepm{\trz{\trxcons{\trxcons{\trx}{\trxpropmark{\tr'}}}{\widetilde{\trx}'}}{\trend}}{\trend}{\trz{\trxcons{\trx}{\trxpropmark{\tr'}}}{\trend}}{\widetilde{\tr}}$
    \item
      $\iltstate{\sigma''}{\trz{\trxcons{\trxcons{\trx}{\trxpropmark{\tr'}}}{\widetilde{\trx}'}}{\trend}}{\kappa}{\emptyenv}{\vec{\omega}}
      \iltsteprel
      \iltstate{\sigma''}{\trz{\trxcons{\trx}{\trpush{\widetilde{\tr}}}}{\tr'}}{\kappa}{\emptyenv}{\ilprop}$
    \item
      $\iltstate{\sigma''}{\trz{\trxcons{\trx}{\trpush{\widetilde{\tr}}}}{\tr'}}{\kappa}{\emptyenv}{\ilprop}
      \iltsteprel^{n_2}
      \iltstate{\sigma'}{\trz{\trx'}{\tr_0'}}{\stcons{\kappa}{\widehat{\kappa}}}{\rho'}{\rcmd'}$
    \item $n = n_1 + 1 + n_2$
    \end{itemize}
  \end{enumerate}
\end{lem}
\begin{proof}
  By mutual induction on $n$.  If $n = 0$, then we obtain a
  contradiction to $\dropum{\trxcons{\trx}{\trxpropmark{\tr'}}} \notin
  \prefixes{\dropum{\trx'}}$.  So consider $n > 0$.  In each part we
  inspect the first step of the reduction.
  \begin{enumerate}
  \item
    \begin{itemize}
    \item Case {\bf E.0--7}: Straightforward, using the inductive
      hypothesis.
    \item Case {\bf E.8}:
      \begin{itemize}
      \item Then $\stcons{\kappa}{\widetilde{\kappa}} \statesep \rho
        \statesep \rcmd =
        \stcons{\kappa}{\stcons{\widetilde{\kappa}'}{\mkstframe{\widehat{\rho}}{\widehat{f}}}}
        \statesep \emptyenv \statesep \vec{\omega}$ and
        $\rewindstepm{\trz{\trxcons{\trxcons{\trx}{\trxpropmark{\tr'}}}{\widetilde{\trx}}}{\tr_0}}{\trend}{\trz{\trxcons{\trxcons{\trx}{\trxpropmark{\tr'}}}{\trxcons{\widetilde{\trx}'}{\trxpushmark}}}{\tr_0''}}{\widehat{\tr}}$.
      \item So
        $\iltstate{\sigma}{\trz{\trxcons{\trxcons{\trx}{\trxpropmark{\tr'}}}{\widetilde{\trx}}}{\tr_0}}{\stcons{\kappa}{\widetilde{\kappa}}}{\emptyenv}{\vec{\omega}}
        \iltsteprel
        \iltstate{\sigma}{\trz{\trxcons{\trxcons{\trx}{\trxpropmark{\tr'}}}{\trxcons{\widetilde{\trx}'}{\trpush{\widehat{\tr}}}}}{\tr_0''}}{\stcons{\kappa}{\widetilde{\kappa}'}}{\widehat{\rho}}{\widehat{f}}$.
      \item And
        $\iltstate{\sigma}{\trz{\trxcons{\trxcons{\trx}{\trxpropmark{\tr'}}}{\trxcons{\widetilde{\trx}'}{\trpush{\widehat{\tr}}}}}{\tr_0''}}{\stcons{\kappa}{\widetilde{\kappa}'}}{\widehat{\rho}}{\widehat{f}}
        \iltsteprel^{n-1}
        \iltstate{\sigma'}{\trz{\trx'}{\tr_0'}}{\stcons{\kappa}{\widehat{\kappa}}}{\rho'}{\rcmd'}$.
      \item Hence the claim holds by induction.
      \end{itemize}
    \item Case {\bf E.P}: By part (2).
    \item Case {\bf P.E,1--7}: Not possible.
    \item Case {\bf P.8}:
      \begin{itemize}
      \item Subcase $\numof{\propmark}{\widetilde{\trx}} = 0$:
        \begin{itemize}
        \item Then
          $\rewindstepm{\trz{\trxcons{\trxcons{\trx}{\trxpropmark{\tr'}}}{\widetilde{\trx}}}{\trend}}{\trend}{\trz{\trxcons{\trx}{\trxpropmark{\tr'}}}{\trend}}{\tr_1}$
          and $\numof{\pushmark}{\widetilde{\trx}} = 0$ and thus
          $\widetilde{\kappa} = \emptystack$.
        \item So
          $\iltstate{\sigma}{\trz{\trxcons{\trxcons{\trx}{\trxpropmark{\tr'}}}{\widetilde{\trx}}}{\tr_0}}{\kappa}{\rho}{\rcmd}
          \iltsteprel
          \iltstate{\sigma}{\trz{\trxcons{\trx}{\trpush{\tr_1}}}{\tr_2}}{\kappa}{\emptyenv}{\ilprop}$.
        \item And
          $\iltstate{\sigma}{\trz{\trxcons{\trx}{\trpush{\tr_1}}}{\tr_2}}{\kappa}{\emptyenv}{\ilprop}
          \iltsteprel^{n-1}
          \iltstate{\sigma'}{\trz{\trx'}{\tr_0'}}{\stcons{\kappa}{\widehat{\kappa}}}{\rho'}{\rcmd'}$.
        \item Thus the claim holds for $n_1 = 0$, $n_2 = n - 1$.
       \end{itemize}
      \item Subcase $\numof{\propmark}{\widetilde{\trx}} > 0$:
        \begin{itemize}
        \item Then
          $\rewindstepm{\trz{\trxcons{\trxcons{\trx}{\trxpropmark{\tr'}}}{\widetilde{\trx}}}{\trend}}{\trend}{\trz{\trxcons{\trxcons{\trx}{\trxpropmark{\tr'}}}{\trxcons{\widetilde{\trx}'}{\trxpropmark{\tr_2}}}}{\trend}}{\tr_1}$
          with $\numof{\pushmark}{\widetilde{\trx}'} =
          \numof{\pushmark}{\widetilde{\trx}}$.
        \item So
          $\iltstate{\sigma}{\trz{\trxcons{\trxcons{\trx}{\trxpropmark{\tr'}}}{\widetilde{\trx}}}{\tr_0}}{\stcons{\kappa}{\widetilde{\kappa}}}{\rho}{\rcmd}
          \iltsteprel
          \iltstate{\sigma}{\trz{\trxcons{\trxcons{\trx}{\trxpropmark{\tr'}}}{\trxcons{\widetilde{\trx}'}{\trpush{\tr_1}}}}{\tr_2}}{\stcons{\kappa}{\widetilde{\kappa}}}{\emptyenv}{\ilprop}$.
        \item And
          $\iltstate{\sigma}{\trz{\trxcons{\trxcons{\trx}{\trxpropmark{\tr'}}}{\trxcons{\widetilde{\trx}'}{\trpush{\tr_1}}}}{\tr_2}}{\stcons{\kappa}{\widetilde{\kappa}}}{\emptyenv}{\ilprop}
          \iltsteprel^{n-1}
          \iltstate{\sigma'}{\trz{\trx'}{\tr_0'}}{\stcons{\kappa}{\widehat{\kappa}}}{\rho'}{\rcmd'}$.
        \item Hence the claim holds by induction (part 2).
        \end{itemize}
      \end{itemize}
    \item Case {\bf U.1--4}: Straightforward, using the inductive
      hypothesis.
    \end{itemize}
  \item
    \begin{itemize}
    \item Case {\bf E.0--8,P}: Not possible.
    \item Case {\bf P.E}: By part (1).
    \item Case {\bf P.1--7}: Straightforward, using the inductive
      hypothesis.
    \item Case {\bf P.8}: Not possible.
    \item Case {\bf U.1--4}: Not possible.
    \end{itemize}
  \end{enumerate}
\end{proof}

\begin{lem}
  \label{lem:stack-prefix}
  Suppose $\dropum{\trxcons{\trx}{\trxpushmark}} \in
  \prefixes{\dropum{\trx'}}$ and $\numof{\pushmark}{\widetilde{\trx}}
  = |\widetilde{\kappa}|$.
  \begin{enumerate}
  \item If
    $\iltstate{\sigma}{\trz{\trxcons{\trx}{\trxcons{\trxpushmark}{\widetilde{\trx}}}}{\tr}}{\stcons{\kappa}{\widetilde{\kappa}}}{\rho}{\rcmd}
    \iltsteprel^n
    \iltstate{\sigma'}{\trz{\trx'}{\tr'}}{\kappa'}{\rho'}{\rcmd'}$,
    then $\kappa \in \prefixes{\kappa'}$.
  \item If
    $\iltstate{\sigma}{\trz{\trxcons{\trx}{\trxcons{\trxpushmark}{\widetilde{\trx}}}}{\tr}}{\stcons{\kappa}{\widetilde{\kappa}}}{\emptyenv}{\ilprop}
    \iltsteprel^n
    \iltstate{\sigma'}{\trz{\trx'}{\tr'}}{\kappa'}{\rho'}{\rcmd'}$,
    then $\kappa \in \prefixes{\kappa'}$.
  \end{enumerate}
\end{lem}
\begin{proof}
  By mutual induction on $n$.  If $n = 0$, then we obtain a
  contradiction to $\dropum{\trxcons{\trx}{\trxpushmark}} \notin
  \prefixes{\dropum{\trx'}}$.  So consider $n > 0$.  In each part we
  inspect the first step of the reduction.
  \begin{enumerate}
  \item
    \begin{itemize}
    \item Case {\bf E.0--7}: Straightforward, using the inductive
      hypothesis.
    \item Case {\bf E.8}:
      \begin{itemize}
      \item Subcase $\numof{\pushmark}{\widetilde{\trx}} = 0$:
        \begin{itemize}
        \item Then
          $\iltstate{\sigma}{\trz{\trxcons{\trx}{\trpush{\tr_1}}}{\tr_2}}{\kappa_1}{\rho_f[\vec{x
              \mapsto \omega}]}{e_f} \iltsteprel^{n-1}
          \iltstate{\sigma'}{\trz{\trx'}{\tr'}}{\kappa'}{\rho'}{\rcmd'}$.
        \item Lemma~\ref{lem:prefix-single-action} yields
          $\dropum{\trxcons{\trx}{\trpush{\tr_1}}} \in
          \prefixes{\dropum{\trx'}}$, which contradicts the first
          assumption.
        \end{itemize}
      \item Subcase $\numof{\pushmark}{\widetilde{\trx}} > 0$:
        \begin{itemize}
        \item Then
          $\iltstate{\sigma}{\trz{\trxcons{\trxcons{\trx}{\trxpushmark}}{\trxcons{\widetilde{\trx}'}{\trpush{\widehat{\tr}}}}}{\tr}}{\stcons{\kappa}{\widetilde{\kappa}'}}{\widehat{\rho}}{\widehat{e}}
          \iltsteprel^{n-1}
          \iltstate{\sigma'}{\trz{\trx'}{\tr'}}{\kappa'}{\rho'}{\rcmd'}$
          with $|\widetilde{\kappa}'| = |\widetilde{\kappa}| - 1 =
          \numof{\pushmark}{\widetilde{\trx}} - 1 =
          \numof{\pushmark}{\widetilde{\trx}'} =
          \numof{\pushmark}{\trxcons{\widetilde{\trx}'}{\trpush{\widehat{\tr}}}}$.
        \item Hence the claim holds by induction.
        \end{itemize}
      \end{itemize}
    \item Case {\bf E.P}: By induction (part 2).
    \item Case {\bf P.E,1--7}: Not possible.
    \item Case {\bf P.8}:
      \begin{itemize}
      \item Then
        $\iltstate{\sigma}{\trz{\trxcons{\trxcons{\trx}{\trxpushmark}}{\trxcons{\widetilde{\trx}'}{\trpush{\widehat{\tr}}}}}{\tr}}{\stcons{\kappa}{\widetilde{\kappa}}}{\emptyenv}{\ilprop}
        \iltsteprel^{n-1}
        \iltstate{\sigma'}{\trz{\trx'}{\tr'}}{\kappa'}{\rho'}{\rcmd'}$
        with $|\widetilde{\kappa}| =
        \numof{\pushmark}{\widetilde{\trx}} =
        \numof{\pushmark}{\widetilde{\trx}'} =
        \numof{\pushmark}{\trxcons{\widetilde{\trx}'}{\trpush{\widehat{\tr}}}}$.
      \item Hence the claim holds by induction (part 2).
      \end{itemize}
    \item Case {\bf U.1--4}: Straightforward by induction.
    \end{itemize}
  \item
    \begin{itemize}
    \item Case {\bf E.0--8,P}: Not possible.
    \item Case {\bf P.E}: By induction (part 1).
    \item Case {\bf P.1--7}: Straightforward by induction.
    \item Case {\bf U.1--4}: Not possible.
    \item Case {\bf P.8}: Not possible.
    \end{itemize}
  \end{enumerate}
\end{proof}

\begin{lem}
  \label{lem:stack-surgery}
  Suppose $\dropum{\trx} \in \prefixes{\dropum{\trx'}}$.
  \begin{enumerate}
  \item If $\iltstate{\sigma}{\trz{\trx}{\tr}}{\kappa_1}{\rho}{\rcmd}
    \iltsteprel^n
    \iltstate{\sigma'}{\trz{\trx'}{\tr'}}{\stcons{\kappa_1}{\kappa}}{\rho'}{\rcmd'}$,
    then $\iltstate{\sigma}{\trz{\trx}{\tr}}{\kappa_2}{\rho}{\rcmd}
    \iltsteprel^n
    \iltstate{\sigma'}{\trz{\trx'}{\tr'}}{\stcons{\kappa_2}{\kappa}}{\rho'}{\rcmd'}$
    for any $\kappa_2$.
  \item If $\iltstate{\sigma}{\trz{\trx}{\tr}}{\kappa_1}{\emptyenv}{\ilprop}
    \iltsteprel^n
    \iltstate{\sigma'}{\trz{\trx'}{\tr'}}{\stcons{\kappa_1}{\kappa}}{\rho'}{\rcmd'}$,
    then $\iltstate{\sigma}{\trz{\trx}{\tr}}{\kappa_2}{\emptyenv}{\ilprop}
    \iltsteprel^n
    \iltstate{\sigma'}{\trz{\trx'}{\tr'}}{\stcons{\kappa_2}{\kappa}}{\rho'}{\rcmd'}$
    for any $\kappa_2$.
  \end{enumerate}
\end{lem}
\begin{proof}
  Mutually, by induction on $n$.  If $n = 0$, both parts hold
  trivially.  Now suppose $n > 0$.  We inspect the first step of each
  reduction.
  \begin{enumerate}
  \item
    \begin{itemize}
    \item Case {\bf E.0--5,7}: By Lemma~\ref{lem:prefix-single-action}
      (except E.0), induction, and application of the corresponding
      rule.
    \item Case {\bf E.6}:
      \begin{itemize}
      \item Subcase $\dropum{\trxcons{\trx}{\trxpushmark}} \in
        \prefixes{\dropum{\trx'}}$:
        \begin{itemize}
        \item We know $\rcmd = \ilpush{f}{e}$ and
          $\iltstate{\sigma}{\trz{\trxcons{\trx}{\trxpushmark}}{\tr}}{\stcons{\kappa_1}{\mkstframe{\rho}{f}}}{\rho}{e}
          \iltsteprel^{n-1}
          \iltstate{\sigma'}{\trz{\trx'}{\tr'}}{\stcons{\kappa_1}{\kappa}}{\rho'}{\rcmd'}$.
        \item By Lemma~\ref{lem:stack-prefix} we know
          $\stcons{\kappa_1}{\mkstframe{\rho}{f}} \in
          \prefixes{\stcons{\kappa_1}{\kappa}}$.
        \item Hence $\stcons{\kappa_1}{\kappa} =
          \stcons{\kappa_1'}{\kappa'}$ for $\kappa_1' =
          \stcons{\kappa_1}{\mkstframe{\rho}{f}}$ and some $\kappa'$.
        \item The claim then follows by induction and application of
          rule {\bf E.6}.
        \end{itemize}
      \item Subcase $\dropum{\trxcons{\trx}{\trxpushmark}} \notin
        \prefixes{\dropum{\trx'}}$: By
        Lemma~\ref{lem:pushmark-decomposition},
        Lemma~\ref{lem:rewind-dropum},
        Lemma~\ref{lem:rewind-trace-context},
        Lemma~\ref{lem:prefix-single-action}, induction (twice), and
        rule {\bf E.6}.
      \end{itemize}
    \item Case {\bf E.8}:
      Lemmas~\ref{lem:rewind-trace-context},~\ref{lem:rewind-dropum}
      and~\ref{lem:prefix-single-action} yield both
      $\dropum{\trxcons{\trx''}{\trxpushmark}} \in
      \prefixes{\dropum{\trx'}}$ and
      $\dropum{\trxcons{\trx''}{\trpush{\_}}} \in
      \prefixes{\dropum{\trx'}}$, which is a contradiction.
    \item Case {\bf E.P}: By Lemma~\ref{lem:prefix-single-action},
      induction (part 2), and application of {\bf E.P}.
    \item Case {\bf P.8}:
      Lemmas~\ref{lem:rewind-trace-context},~\ref{lem:rewind-dropum}
      and~\ref{lem:prefix-single-action} yield both
      $\dropum{\trxcons{\trx''}{\trxpropmark{\_}}} \in
      \prefixes{\dropum{\trx'}}$ and
      $\dropum{\trxcons{\trx''}{\trpush{\_}}} \in
      \prefixes{\dropum{\trx'}}$, which is a contradiction.
    \item Case {\bf U.1--4}: By induction and application of the
      corresponding rule.
    \item Case {\bf P.E,1--7}: Not possible.
    \end{itemize}
  \item
    \begin{itemize}
    \item Case {\bf E.0--8,P}: Not possible.
    \item Case {\bf P.1--5}: By Lemma~\ref{lem:prefix-single-action},
      induction, and application of the corresponding rule.
    \item Case {\bf P.6}:
      \begin{itemize}
      \item Subcase $\dropum{\trxcons{\trx}{\trxpropmark{\widehat{\tr}}}} \in
        \prefixes{\dropum{\trx'}}$: By induction and application of
        rule {\bf P.6}.
      \item Subcase $\dropum{\trxcons{\trx}{\trxpropmark{\widehat{\tr}}}} \notin
        \prefixes{\dropum{\trx'}}$: By
        Lemma~\ref{lem:propmark-decomposition},
        Lemma~\ref{lem:rewind-dropum},
        Lemma~\ref{lem:rewind-trace-context},
        Lemma~\ref{lem:prefix-single-action}, induction (twice), and
        rule {\bf P.6}.
      \end{itemize}
    \item Case {\bf P.7,E}: By Lemma~\ref{lem:prefix-single-action},
      induction (part 1), and application of the corresponding rule.
    \item Case {\bf P.8}: Not possible.
    \item Case {\bf U.1--4}: Not possible.
    \end{itemize}
  \end{enumerate}
\end{proof}

\begin{lem}[CSA preservation (traced)]
  \label{lem:csa-preservation}
  If
  \begin{enumerate}
  \item $\iltstate{\sigma}{\trz{\trx}{\trend}}{\kappa}{\rho}{e}
    \iltsteprel^\ast
    \iltstate{\sigma'}{\trz{\trx'}{\_}}{\stcons{\kappa}{\kappa'}}{\rho'}{e'}$
  \item $\dropum{\trx} \in \prefixes{\dropum{\trx'}}$
  \item $\noreuse{\trz{\trx}{\trend}}$
  \item $\csa{\rho}{e}{\sigma}$
  \end{enumerate}
  then $\csa{\rho'}{e'}{\sigma'}$.
\end{lem}
\begin{proof}~
  \begin{itemize}
  \item By Lemma~\ref{lem:stack-surgery} we get
    $\iltstate{\sigma}{\trz{\trx}{\trend}}{\emptystack}{\rho}{e}
    \iltsteprel^\ast
    \iltstate{\sigma'}{\trz{\trx'}{\_}}{\kappa'}{\rho'}{e'}$.
  \item By Lemma~\ref{lem:noreuse-preservation} that reduction does
    not use rules other than {\bf E.*} and {\bf U.*}.    
  \item Hence by Lemma~\ref{lem:traced-to-untraced} we get
    $\ilustate{\sigma}{\emptystack}{\rho}{e} \ilusteprel^\ast
    \ilustate{\sigma'}{\kappa'}{\rho'}{e'}$.
  \item The claim then follows by
    Lemma~\ref{lem:csa-preservation-untraced}.
  \end{itemize}
\end{proof}

\begin{lem}[Decomposition]
  \label{lem:decomposition}
  Suppose $\fsc{\tr}$, from initial configuration~
  $\iltstate{\sigma}{\trz{\trx}{\trend}}{\kappa}{\rho}{\rcmd}$ and
  producing $\vec{\nu}$.
  \begin{enumerate}
  \item If $\tr = \trcons{\tralloc{\ell}{m}}{\tr'}$, then:
    \begin{enumerate}
    \item $\iltstate{\sigma}{\trz{\trx}{\trend}}{\kappa}{\rho}{\rcmd}
      \iltsteprel^*
      \iltstate{\sigma}{\trz{\trx}{\trend}}{\kappa}{\rho'}{\illet{\ilvar{x}{\ilalloc{y}}}{e}}$
      using {\bf E.0} only
    \item $\fsc{\tr'}$ from
      $\iltstate{\sigma'}{\trz{\trxcons{\trx}{\tralloc{\ell}{m}}}{\trend}}{\kappa}{\rho'[x
        \mapsto \ell]}{e}$, producing $\vec{\nu}$
    \item $\storestin{\sigma}{\rho'}{\ilalloc{y}} \storestep
      \storestout{\sigma'}{\ell}$
    \item $\rho'(y) = m$
    \end{enumerate}
  \item If $\tr = \trcons{\trread{\nu}{\ell}{m}}{\tr'}$, then:
    \begin{enumerate}
    \item $\iltstate{\sigma}{\trz{\trx}{\trend}}{\kappa}{\rho}{\rcmd}
      \iltsteprel^*
      \iltstate{\sigma}{\trz{\trx}{\trend}}{\kappa}{\rho'}{\illet{\ilvar{x}{\ilread{y}{z}}}{e}}$
      using {\bf E.0} only
    \item $\fsc{\tr'}$ from
      $\iltstate{\sigma}{\trz{\trxcons{\trx}{\trread{\nu}{\ell}{m}}}{\trend}}{\kappa}{\rho'[x
        \mapsto \nu]}{e}$, producing $\vec{\nu}$
    \item $\storestin{\sigma}{\rho'}{\ilread{y}{z}} \storestep
      \storestout{\sigma}{\nu}$
    \item $\rho'(y) = \ell$
    \item $\rho'(z) = m$
    \end{enumerate}
  \item If $\tr = \trcons{\trwrite{\nu}{\ell}{m}}{\tr'}$, then:
    \begin{enumerate}
    \item $\iltstate{\sigma}{\trz{\trx}{\trend}}{\kappa}{\rho}{\rcmd}
      \iltsteprel^*
      \iltstate{\sigma}{\trz{\trx}{\trend}}{\kappa}{\rho'}{\illet{\ilvar{\_}{\ilwrite{x}{y}{z}}}{e}}$
      using {\bf E.0} only
    \item $\fsc{\tr'}$ from
      $\iltstate{\sigma'}{\trz{\trxcons{\trx}{\trwrite{\nu}{\ell}{m}}}{\trend}}{\kappa}{\rho'}{e}$,
      producing $\vec{\nu}$
    \item $\storestin{\sigma}{\rho'}{\ilwrite{x}{y}{z}} \storestep
      \storestout{\sigma'}{0}$
    \item $\rho'(x) = \ell$
    \item $\rho'(y) = m$
    \item $\rho'(z) = \nu$
    \end{enumerate}
  \item If $\tr = \trcons{\trmemo{\rho'}{e}}{\tr'}$, then:
    \begin{enumerate}
    \item $\iltstate{\sigma}{\trz{\trx}{\trend}}{\kappa}{\rho}{\rcmd}
      \iltsteprel^*
      \iltstate{\sigma}{\trz{\trx}{\trend}}{\kappa}{\rho'}{\ilmemo{e}}$
      using {\bf E.0} only
    \item $\fsc{\tr'}$ from
      $\iltstate{\sigma}{\trz{\trxcons{\trx}{\trwake{\rho'}{e}}}{\trend}}{\kappa}{\rho'}{e}$,
      producing $\vec{\nu}$
    \end{enumerate}
  \item If $\tr = \trcons{\trwake{\rho'}{e}}{\tr'}$, then:
    \begin{enumerate}
    \item $\iltstate{\sigma}{\trz{\trx}{\trend}}{\kappa}{\rho}{\rcmd}
      \iltsteprel^*
      \iltstate{\sigma}{\trz{\trx}{\trend}}{\kappa}{\rho'}{\ilwake{e}}$
      using {\bf E.0} only
    \item $\fsc{\tr'}$ from
      $\iltstate{\sigma}{\trz{\trxcons{\trx}{\trwake{\rho'}{e}}}{\trend}}{\kappa}{\rho'}{e}$,
      producing $\vec{\nu}$
    \end{enumerate}
  \item If $\tr = \trcons{\trpush{\tr_1}}{\tr_2}$, then:
    \begin{enumerate}
    \item
      $\iltstate{\sigma}{\trz{\trx}{\trend}}{\kappa}{\rho}{\rcmd}
      \iltsteprel^*
      \iltstate{\sigma}{\trz{\trx}{\trend}}{\kappa}{\rho'}{\ilpush{f}{e}}$
      using {\bf E.0} only
    \item $\fsc{\tr_1}$ from
      $\iltstate{\sigma}{\trz{\trxcons{\trx}{\trxpushmark}}{\trend}}{\stcons{\kappa}{\mkstframe{\rho'}{f}}}{\rho'}{e}$,
      producing $\vec{\omega}$
    \item $\fsc{\tr_2}$ from
      $\iltstate{\_}{\trz{\trxcons{\trx}{\trpush{\tr_1}}}{\trend}}{\kappa}{\rho'[\vec{x
          \mapsto \omega}]}{e'}$, producing $\vec{\nu}$
    \item $\rho'(f) = \ilfun{f}{\vec{x}}{e'}$
    \end{enumerate}
  \item If $\tr = \trxcons{\trpop{\vec{\omega}}}{\tr'}$, then:
    \begin{enumerate}
    \item $\iltstate{\sigma}{\trz{\trx}{\trend}}{\kappa}{\rho}{\rcmd} \iltsteprel^* \iltstate{\sigma}{\trz{\trx}{\trend}}{\kappa}{\rho'}{\ilpop{\vec{x}}}$ using only {\bf E.0}
    \item $\rho'(\vec{x}) = \vec{\omega}$
    \item $\tr' = \trend$
    \item $\vec{\omega} = \vec{\nu}$
    \end{enumerate}
  \end{enumerate}
\end{lem}
\begin{proof}
  From the assumption we know that:
  \begin{enumerate}[(i)]
  \item $\csa{\rho}{\rcmd}{\sigma}$
  \item $\noreuse{\trz{\trx}{\trend}}$
  \item
    $\iltstate{\sigma}{\trz{\trx}{\trend}}{\kappa}{\rho}{\rcmd}
    \iltsteprel^n
    \iltstate{\sigma'}{\trz{\trx'}{\trend}}{\kappa}{\emptyenv}{\vec{\nu}}$
  \item
    $\rewindstepm{\trz{\trx'}{\trend}}{\trend}{\trz{\trx}{\trend}}{\tr}$
  \end{enumerate}
  The proof is by induction on $n$.  We are only interested in cases
  where $\tr$ is nonempty and thus $n > 0$.  In each part we inspect
  the first step of the reduction in (iii).
  \begin{enumerate}
  \item $\tr = \trcons{\tralloc{\ell}{m}}{\tr'}$
    \begin{itemize}
    \item Case {\bf E.0}: By Lemma~\ref{lem:csa-preservation} and induction.
    \item Case {\bf E.1}:
      \begin{itemize}
      \item Then:
        \begin{enumerate}
        \item $\rcmd = \illet{\ilvar{x}{\ilalloc{y}}}{e}$
        \item
          $\iltstate{\sigma''}{\trz{\trxcons{\trx}{\tralloc{\ell'}{m'}}}{\trend}}{\kappa}{\rho[x
            \mapsto \ell']}{e} \iltsteprel^{n-1}
          \iltstate{\sigma'}{\trz{\trx'}{\trend}}{\kappa}{\emptyenv}{\vec{\nu}}$
        \item $\storestin{\sigma}{\rho}{\ilalloc{y}} \storestep \storestout{\sigma''}{\ell'}$
        \item $\rho(y) = m'$
        \end{enumerate}
      \item By (iv), Lemma~\ref{lem:rewind-dropum} and
        Lemma~\ref{lem:rewind-single-action} we get
        $\rewindstepm{\trz{\dropum{\trx'}}{\trend}}{\trend}{\trz{\dropum{\trxcons{\trx}{\tralloc{\ell'}{m'}}}}{\trend}}{\tr'}
        \trrewind \trz{\dropum{\trx}}{\trend};{\tr}$ with $\tr =
        \trcons{\tralloc{\ell'}{m'}}{\tr'}$, hence $\ell' = \ell$ and
        $m' = m$.
      \item By Lemma~\ref{lem:noreuse-preservation} we know
        $\dropum{\trx'} = \trx'$ and
        $\dropum{\trxcons{\trx}{\tralloc{\ell}{m}}} =
        \trxcons{\trx}{\tralloc{\ell}{m}}$.
      \item Finally, Lemma~\ref{lem:csa-preservation} yields
        $\csa{\rho[x \mapsto \ell]}{e}{\sigma''}$ and therefore
        $\fsc{\tr'}$ from
        $\iltstate{\sigma''}{\trz{\trxcons{\trx}{\tralloc{\ell'}{m'}}}{\trend}}{\kappa}{\rho[x
          \mapsto \ell']}{e}$, producing $\vec{\nu}$.
      \end{itemize}
    \item Case {\bf E.2--8}:
      \begin{itemize}
      \item Then
        $\iltstate{\sigma''}{\trz{\trxcons{\trx}{t}}{\trend}}{\kappa}{\rho''}{\rcmd''}
        \iltsteprel^\ast
        \iltstate{\sigma'}{\trz{\trx'}{\trend}}{\kappa}{\emptyenv}{\vec{\nu}}$
        with $t \neq \tralloc{\ell}{m}$ (using
        Lemma~\ref{lem:pushmark-decomposition} in case {\bf E.6}).
      \item By (iv), Lemma~\ref{lem:rewind-dropum} and
        Lemma~\ref{lem:rewind-single-action} we get
        $\rewindstepm{\trz{\dropum{\trx'}}{\trend}}{\trend}{\trz{\dropum{\trxcons{\trx}{t}}}{\trend}}{\tr'}
        \trrewind \trz{\dropum{\trx}}{\trend};{\tr}$ with $\tr =
        \trcons{t}{\tr'}$.
      \item This is a contradiction.
      \end{itemize}
    \item Case {\bf E.P,P.E,P.1--8,U.1--4}: Impossible due to (ii).
    \end{itemize}
  \item $\tr = \trcons{\trread{\nu}{\ell}{m}}{\tr'}$
    \begin{itemize}
    \item Case {\bf E.0}: By Lemma~\ref{lem:csa-preservation} and induction.
    \item Case {\bf E.2}: base case
    \item Case {\bf E.1,3--8}: contradiction
    \item Case {\bf E.P,P.E,P.1--8,U.1--4}: Impossible due to (ii).
    \end{itemize}
  \item $\tr = \trcons{\trwrite{\nu}{\ell}{m}}{\tr'}$
    \begin{itemize}
    \item Case {\bf E.0}: By Lemma~\ref{lem:csa-preservation} and induction.
    \item Case {\bf E.3}: base case
    \item Case {\bf E.1,2,4--8}: contradiction
    \item Case {\bf E.P,P.E,P.1--8,U.1--4}: Impossible due to (ii).
    \end{itemize}
  \item $\tr = \trcons{\trmemo{\rho'}{e}}{\tr'}$
    \begin{itemize}
    \item Case {\bf E.0}: By Lemma~\ref{lem:csa-preservation} and induction.
    \item Case {\bf E.4}: base case
    \item Case {\bf E.1--3,5--8}: contradiction
    \item Case {\bf E.P,P.E,P.1--8,U.1--4}: Impossible due to (ii).
    \end{itemize}
  \item $\tr = \trcons{\trwake{\rho'}{e}}{\tr'}$
    \begin{itemize}
    \item Case {\bf E.0}: By Lemma~\ref{lem:csa-preservation} and induction.
    \item Case {\bf E.5}: base case
    \item Case {\bf E.1--4,6--8}: contradiction
    \item Case {\bf E.P,P.E,P.1--8,U.1--4}: Impossible due to (ii).
    \end{itemize}
  \item $\tr = \trcons{\trpush{\tr_1}}{\tr_2}$
    \begin{itemize}
    \item Case {\bf E.0}: By Lemma~\ref{lem:csa-preservation} and induction.
    \item Case {\bf E.6}:
      \begin{itemize}
      \item Then $\rcmd = \ilpush{f}{e}$ and
        $\iltstate{\sigma}{\trz{\trxcons{\trx}{\trxpushmark}}{\trend}}{\stcons{\kappa}{\mkstframe{\rho}{f}}}{\rho}{e}
        \iltsteprel^{n-1}
        \iltstate{\sigma'}{\trz{\trx'}{\trend}}{\kappa}{\emptyenv}{\vec{\nu}}$.
      \item By (iv), Lemma~\ref{lem:rewind-dropum},
        Lemma~\ref{lem:pushmark-decomposition} and
        Lemma~\ref{lem:noreuse-preservation} we get:
        \begin{itemize}
        \item
          $\iltstate{\sigma}{\trz{\trxcons{\trx}{\trxpushmark}}{\trend}}{\stcons{\kappa}{\mkstframe{\rho}{f}}}{\rho}{e}
          \iltsteprel^{n_1}
          \iltstate{\sigma''}{\trz{\trx''}{\trend}}{\stcons{\kappa}{\mkstframe{\rho}{f}}}{\emptyenv}{\vec{\omega}}$
        \item
          $\rewindstepm{\trz{\trx''}{\trend}}{\trend}{\trz{\trxcons{\trx}{\trxpushmark}}{\trend}}{\widetilde{\tr}}$
        \item $\rho(f) = \ilfun{f}{x}{e_f}$
        \item
          $\iltstate{\sigma''}{\trz{\trx''}{\trend}}{\stcons{\kappa}{\mkstframe{\rho}{f}}}{\emptyenv}{\vec{\omega}}
          \iltsteprel
          \iltstate{\sigma''}{\trz{\trxcons{\trx}{\trpush{\widetilde{\tr}}}}{\trend}}{\kappa}{\rho[\vec{x
              \mapsto \omega}]}{e_f}$
        \item
          $\iltstate{\sigma''}{\trz{\trxcons{\trx}{\trpush{\widetilde{\tr}}}}{\trend}}{\kappa}{\rho[\vec{x
              \mapsto \omega}]}{e_f} \iltsteprel^{n_2}
          \iltstate{\sigma'}{\trz{\trx'}{\trend}}{\kappa}{\emptyenv}{\vec{\nu}}$
        \item $n - 1 = n_1 + 1 + n_2$
        \end{itemize}
      \item By (iv), Lemma~\ref{lem:rewind-dropum} and
        Lemma~\ref{lem:rewind-single-action} we get
        $\rewindstepm{\trz{\dropum{\trx'}}{\trend}}{\trend}{\trz{\dropum{\trxcons{\trx}{\trpush{\widetilde{\tr}}}}}{\trend}}{\tr'}
        \trrewind \trz{\dropum{\trx}}{\trend};{\tr}$ with $\tr =
        \trcons{\trpush{\widetilde{\tr}}}{\tr'}$ and thus $\tr_1 =
        \widetilde{\tr}$ and $\tr_2 = \tr'$.
      \item Hence by Lemma~\ref{lem:noreuse-preservation} and Lemma~\ref{lem:csa-preservation} we know:
        \begin{itemize}
        \item $\fsc{\tr_1}$ from
          $\iltstate{\sigma}{\trz{\trxcons{\trx}{\trxpushmark}}{\trend}}{\stcons{\kappa}{\mkstframe{\rho}{f}}}{\rho}{e}$
        \item $\fsc{\tr_2}$ from
          $\iltstate{\sigma''}{\trz{\trxcons{\trx}{\trpush{\widetilde{\tr}}}}{\trend}}{\kappa}{\rho[\vec{x
              \mapsto \omega}]}{e_f}$
        \end{itemize}
      \end{itemize}
    \item Case {\bf E.1--5,7,8}: contradiction
    \item Case {\bf E.P,P.E,P.1--8,U.1--4}: Impossible due to (ii).
    \end{itemize}
  \item $\tr = \trcons{\trpop{\vec{\omega}}}{\tr'}$
    \begin{itemize}
    \item Case {\bf E.0}: By Lemma~\ref{lem:csa-preservation} and induction.
    \item Case {\bf E.7}:
      \begin{itemize}
      \item Then $\rcmd = \ilpop{\vec{x}}$ and
        $\iltstate{\sigma}{\trz{\trx}{\trend}}{\kappa}{\rho}{\ilpop{\vec{x}}}
        \iltsteprel
        \iltstate{\sigma}{\trz{\trxcons{\trx}{\vec{\omega}'}}{\trend}}{\kappa}{\emptyenv}{\vec{\omega}'}
        \iltsteprel^{n-1}
        \iltstate{\sigma'}{\trz{\trx'}{\trend}}{\kappa}{\emptyenv}{\vec{\nu}}$,
        where $\vec{\omega}' = \rho(\vec{x})$.
      \item We show that $n - 1 = 0$:
        \begin{itemize}
        \item For a contradiction, suppose that $n - 1 > 0$.
        \item Note that then the next reduction step must be
          either~{\bf P.8} or~{\bf E.8}.
        \item In either case, using
          Lemmas~\ref{lem:rewind-trace-context},~\ref{lem:prefix-dropum},
          and~\ref{lem:prefix-single-action}, we would get a
          contradiction to
          $\rewindstepm{\trz{\trx'}{\trend}}{\trend}{\trz{\trx}{\trend}}{\tr}$.
        \end{itemize}
      \item Hence $\vec{\omega}' = \vec{\nu}$.
      \item Furthermore, Lemmas~\ref{lem:rewind-dropum}
        and~\ref{lem:rewind-single-action} yield $\tr = \vec{\omega}'$
        and thus $\vec{\omega} = \vec{\nu}$ and $\tr' = \trend$.
      \end{itemize}
    \item Case {\bf E.1--6,8}: contradiction
    \item Case {\bf E.P,P.E,P.1--8,U.1--4}: Impossible due to (ii).
    \end{itemize}
  \end{enumerate}
\end{proof}

\begin{defn}[Last element of a trace]
  \[\begin{array}{lcll}
    \trlast{\trcons{t}{\tr}} &=& \trlast{\tr} & \tr \ne \trend
    \\
    \trlast{\trcons{t}{\trend}} &=& t
    \\
    \trlast{\trend} & \multicolumn{3}{l}{\emph{undefined}}
  \end{array}\]
\end{defn}

\begin{lem}[Evaluation values]
\label{lem:eval-vals}
If $\fsc{\tr}$ producing values~$\vec\nu$ then $\trlast{\tr} =
\vec\nu$.
\end{lem}
\begin{proof}
By induction over the structure of $\tr$.
\begin{itemize}
\item Case $\tr = \trend$: 
  \begin{itemize}
  \item Not possible.
  \end{itemize}
\item Case $\tr = \trcons{\trpop{\vec\omega}}{\tr'}$: By Lemma~\ref{lem:decomposition}.

\item Case $\tr = \trcons{t}{\tr'}$ with $t$ not a value: By
  Lemma~\ref{lem:decomposition} and induction.
\end{itemize}
\end{proof}

\begin{lem}[Propagation values]
\label{lem:prop-vals}
If 
\begin{enumerate}[(a)]
\item 
  $\iltstate{\sigma}{\trz{\trx}{\tr_1}}{\kappa}{\epsilon}{\ilprop}
  \iltsteprel^n
  \iltstate{\sigma'}{\trz{\trx'}{\_}}{\kappa}{\epsilon}{\vec{\nu}}$
\item $\rewindstepm{\trz{\dropum{\trx'}}{\_}}{\epsilon}{\trz{\dropum{\trx}}{\_}}{\_}$
\item reduction (a) does not contain a use of {\bf P.E}
\end{enumerate}
Then $\trlast{\tr_1} = \vec{\nu}$
\end{lem}
\begin{proof} By induction on the number of reduction steps~$n$.  Note
  necessarily that $n > 0$.  We inspect the first reduction step of (a).
\begin{itemize}

\item Case~{\bf E.0}-{\bf E.8},{\bf U.1}-{\bf U.4},{\bf P.8,E} not possible, due to (c).

\item Case~{\bf P.1}-{\bf P.5}
  \begin{itemize}
  \item Then 
    $\iltstate{\sigma}{\trz{\trx}{\trcons{t}{\widehat{\tr_1}}}}{\kappa}{\epsilon}{\ilprop}
    \iltsteprel
    \iltstate{\sigma'}{\trz{\trxcons{\trx}{t}}{\widehat{\tr_1}}}{\kappa}{\epsilon}{\ilprop}
    \iltsteprel^{n -1}
    \iltstate{\sigma'}{\trz{\trx'}{\_}}{\kappa}{\epsilon}{\vec{\nu}}$
    with $\tr_1 = \trcons{t}{\widehat{\tr_1}}$
    \item By Lemma~\ref{lem:rewind-single-action}, we have that 
      ${\trz{\dropum{\trx'}}{\_}};{\epsilon} \trrewind^\ast 
      \trz{\dropum{\trxcons{\trx}{t}}}{\widehat{\tr_1}} \trrewind 
      {\trz{\dropum{\trx}}{\_}};{\_}$
    \item The claim follows by induction.
  \end{itemize}

\item Case~{\bf P.6}
  \begin{itemize}
  \item Then 
    $\iltstate{\sigma}{\trz{\trx}{\trcons{\trpush{\tr_2}}{\tr_3}}}{\kappa}{\epsilon}{\ilprop}
    \iltsteprel
    \iltstate{\sigma}{\trz{\trxcons{\trx}{\trxpropmark{\tr_3}}}{\tr_2}}{\kappa}{\epsilon}{\ilprop}
    \iltsteprel^{n - 1}
    \iltstate{\sigma'}{\trz{\trx'}{\_}}{\kappa}{\epsilon}{\vec{\nu}}$ with $\tr_1 = {\trcons{\trpush{\tr_2}}{\tr_3}}$
  \item From Lemma~\ref{lem:propmark-decomposition} we have
    $\iltstate{\sigma}{\trz{\trxcons{\trx}{\trxpropmark{\tr_3}}}{\tr_2}}{\kappa}{\epsilon}{\ilprop}
    \iltsteprel^{n_1}
    \iltstate{\widehat{\sigma}}{\trz{\trxcons{\trx}{\trpush{\tr_2'}}}{\tr_3}}{\kappa}{\emptyenv}{\ilprop}
    \iltsteprel^{n_2}
    \iltstate{\sigma'}{\trz{\trx'}{\_}}{\kappa}{\emptyenv}{\vec{\nu}}$
  \item From Lemma~\ref{lem:rewind-single-action}, we have that
    ${\trz{\dropum{\trx'}}{\_}};{\_} 
    \trrewind^\ast 
    \trz{\dropum{\trxcons{\trx}{\trpush{\tr_2'}}}}{\_};\_
    \trrewind 
    \trz{\dropum{\trx}}{\_};\_$
  \item The claim follows by induction.
  \end{itemize}

\item Case~{\bf P.7}:
  \begin{itemize}
  \item Then 
    $\iltstate{\sigma}{\trz{\trx}{\trcons{\trpop{\vec{\omega}}}{\trend}}}{\kappa}{\epsilon}{\ilprop}
    \iltsteprel
    \iltstate{\sigma}{\trz{\trxcons{\trx}{\trpop{\vec{\omega}}}}{\trend}}{\kappa}{\epsilon}{\vec{\omega}}
    \iltsteprel^{n - 1}
    \iltstate{\sigma'}{\trz{\trx'}{\trend}}{\kappa}{\epsilon}{\vec{\nu}}$
    ~with $\tr_1 = \trcons{\trpop{\vec{\omega}}}{\trend}$
  \item We show that $n - 1 = 0$:
    \begin{itemize}
    \item For a contradiction, suppose that $n - 1 > 0$.
    \item We inspect the next step in $n - 1$, which must be either~{\bf P.8} or~{\bf E.8}.  We assume~{\bf E.8}; {\bf P.8} is analogous.
    
    \item Hence
      $\iltstate{\sigma}{\trz{\trcons{\trx}{\trpop{\vec{\omega}}}}{\trend}}{\kappa}{\emptyenv}{\vec{\omega}}
      \iltsteprel
      \iltstate{\sigma}{\trz{\trcons{\trx''}{\trpush{\tr}}}{\tr'}}{\kappa'}{\rho_f}{e_f}
      \iltsteprel^{n-2}
      \iltstate{\sigma'}{\trz{\trx'}{\_}}{\kappa}{\emptyenv}{\vec{\nu}}$
      where
      $\rewindstepm{\trz{\dropum{\trcons{\trx}{\trpop{\vec{\omega}}}}}{\trend}}{\trend}{\trz{\dropum{\trcons{\trx''}{\trxpushmark}}}{\tr'}}{\tr}$.
    \item By Lemma~\ref{lem:rewind-trace-context} we know
      $\trcons{\trx''}{\trxpushmark} \in
      \prefixes{\trcons{\trx}{\trpop{\vec{\omega}}}}$, i.e.,
      $\trcons{\trx''}{\trxpushmark} \in \prefixes{\trx}$.
    \item Hence $\dropum{\trx''},
      \dropum{\trxcons{\trx''}{\trxpushmark}} \in
      \prefixes{\dropum{\trx}} \subseteq
      \prefixes{\dropum{\trx'}}$ using
      Lemma~\ref{lem:rewind-trace-context}, (d), and
      Lemma~\ref{lem:prefix-dropum}.
    \item Using Lemma~\ref{lem:prefix-single-action} we get
      $\dropum{\trcons{\trx''}{\trpush{\tr}}} \in
      \prefixes{\dropum{\trx'}}$, contradicting
      $\dropum{\trcons{\trx''}{\trxpushmark}} \in
      \prefixes{\dropum{\trx'}}$.
    \end{itemize}      

  \item Hence, since $n = 1$ we have that
    \begin{itemize}
    \item $\trxcons{\trx}{\trpop{\vec{\nu}}} = \trx'$
    \item $\vec{\omega} = \vec{\nu}$
    \end{itemize}
  \item Moreover, $\trlast{\tr_1} = \trlast{\trcons{\trpop{\vec{\nu}}}{\trend}} = \vec{\nu}$
  \end{itemize}
\end{itemize}
\end{proof}

\begin{lem}[Case analysis: waking up before push action]
\label{lem:active-update-or-prop}
If
\begin{enumerate}[(a)]
\item 
  $
  \iltstate{\sigma}{\trz{\trx}{\tr}}{\kappa}{\emptyenv}{\ilprop}
  \iltsteprel^n
  \iltstate{\sigma'}{\trz{\trx'}{\trend}}{\kappa}{\emptyenv}{\vec\nu}
  $
\item $\rewindstepm{\trz{\dropum{\trx'}}{\_}}\_{\trz{\dropum{\trx}}{\_}}\_$
\item $\okay{\trx}$
\item $\fsc{\tr}$
\item $\tr = \trcons{\tr_1}{\trcons{\trpush{\tr_2}}{\tr_3}}$
\item $\tr_1$ contains no parenthesis (i.e., $\trpush{\_} \not\in \tr_1$)
\end{enumerate}
Then either:
\begin{enumerate}
\item 
  \begin{itemize}
  \item $
    \iltstate{\sigma}{\trz{\trx}{\tr}}{\kappa}{\emptyenv}{\ilprop}
    \iltsteprel^{n_1}
    \iltstate{\sigma''}{\trz{\trx''}{\trcons{\trwake{\rho}{e}}{\trcons{\tr_1'}{\trcons{\trpush{\tr_2}}{\tr_3}}}}}{\kappa}{\emptyenv}{\ilprop}
    \iltsteprel
    \iltstate{\sigma''}{\trz{\trxcons{\trx''}{\trwake{\rho}{e}}}{\trcons{\tr_1'}{\trcons{\trpush{\tr_2}}{\tr_3}}}}{\kappa}{\rho}{e}
    \iltsteprel^{n_2}
    \iltstate{\sigma'}{\trz{\trx'}{\trend}}{\kappa}{\emptyenv}{\vec\nu}
    $
  \item $
    \trz{\dropum{\trx'}}{\_};\_
    \trrewind^\ast
    \trz{\dropum{\trxcons{\trx''}{\trwake{\rho}{e}}}}{\_};\_
    \trrewind^\ast
    \trz{\dropum{\trx}}{\_};\_$
  \item $\okay{\trx''}$
  \item $\fsc{\trcons{\trwake{\rho}{e}}{\trcons{\tr_1'}{\trcons{\trpush{\tr_2}}{\tr_3}}}}$
  \item $\trlast{\trcons{\tr_1'}{\trcons{\trpush{\tr_2}}{\tr_3}}} = \trlast{\tr}$
  \item $n = n_1 + 1 + n_2$
  \end{itemize}
\item
  \begin{itemize}
  \item $
    \iltstate{\sigma}{\trz{\trx}{\tr}}{\kappa}{\emptyenv}{\ilprop}
    \iltsteprel^{n_1}
    \iltstate{\sigma''}{\trz{\trx''}{\tr_3}}{\kappa}{\emptyenv}{\ilprop}
    \iltsteprel^{n_2}
    \iltstate{\sigma'}{\trz{\trx'}{\trend}}{\kappa}{\emptyenv}{\vec\nu}
    $
  \item $
    \trz{\dropum{\trx'}}{\_};\_
    \trrewind^\ast
    \trz{\dropum{\trx''}}{\_};\_
    \trrewind^\ast
    \trz{\dropum{\trx}}{\_};\_$
  \item $\okay{\trx''}$
  \item $\fsc{\tr_3}$
  \item $\trlast{\tr_3} = \trlast{\tr}$
  \item $n = n_1 + n_2$, $n_1 > 0$
  \end{itemize}
\end{enumerate}
\end{lem}
\begin{proof}
By induction on the number of reduction steps~$n$.
Note necessarily that $n > 0$.
We inspect the first step taken.
\begin{itemize}
\item Cases {\bf E.0}-{\bf E.8}, {\bf E.P}, {\bf U.1}-{\bf U.4}: not possible.

\item Case {\bf P.1}-{\bf P.5}: 
\begin{itemize}
\item Then $\tr = \trcons{t}{\tr'}$ and
  $
  \iltstate{\sigma}{\trz{\trx}{\trcons{t}{\tr'}}}{\kappa}{\emptyenv}{\ilprop}
  \iltsteprel
  \iltstate{\widehat{\sigma}}{\trz{\trxcons{\trx}{t}}{\tr'}}{\kappa}{\emptyenv}{\ilprop}
  \iltsteprel^{n-1}
  \iltstate{\sigma'}{\trz{\trx'}{\trend}}{\kappa}{\emptyenv}{\vec\nu}
  $
\item Hence, ${\trz{\dropum{\trx'}}{\_}};\_
  \trrewind^{\ast}
  {\trz{\dropum{\trxcons{\trx}{t}}}{\_}};\_
  \trrewind
  {\trz{\dropum{\trx}}{\_}};\_$ by Lemma~\ref{lem:rewind-single-action}
\item From $\okay{\trx}$, we have $\okay{\trxcons{\trx}{t}}$
\item Note that $\tr' = \trcons{\tr_1'}{\trcons{\trpush{\tr_2}}{\tr_3}}$ where $\tr_1 = \trcons{t}{\tr_1'}$
\item Hence, from (f) we have that $\tr_1'$ contains no paranthesis
\item From $\fsc{\tr}$ we have $\fsc{\tr'}$ using Lemma~\ref{lem:decomposition}
\item The claim then follows by induction. 
\end{itemize}

\item Case {\bf P.E}: We show claim (1) as follows:
  \begin{itemize}
  \item Then $\tr = \trcons{\trupdate{\rho}{e}}{\trcons{\tr_1'}{\trpush{\tr_2}{\tr_3}}}$ and
    $
    \iltstate{\sigma}{\trz{\trx}{\trcons{\trupdate{\rho}{e}}{\tr'}}}{\kappa}{\emptyenv}{\ilprop}
    \iltsteprel
    \iltstate{\sigma}{\trz{\trxcons{\trx}{\trupdate{\rho}{e}}}{\tr'}}{\kappa}{\emptyenv}{\ilprop}
    \iltsteprel^{n-1}
    \iltstate{\sigma'}{\trz{\trx'}{\trend}}{\kappa}{\emptyenv}{\vec\nu}
    $
  \item Hence, ${\trz{\dropum{\trx'}}{\_}};\_
    \trrewind^{\ast}
    {\trz{\dropum{\trxcons{\trx}{\trupdate{\rho}{e}}}}{\_}};\_
    \trrewind
    {\trz{\dropum{\trx}}{\_}};\_$  by Lemma~\ref{lem:rewind-single-action}
  \item With $n_1 = 0$, claim (1) follows immediately by assumptions (c), (d) and (e).
\end{itemize}

\item Case {\bf P.6}: We show claim (2) as follows:
  \begin{itemize}
  \item Then $\tr = \trcons{\trpush{\tr_2}}{\tr_3}$, $\tr_1 = \trend$, and
    $
    \iltstate{\sigma}{\trz{\trx}{\trcons{\trpush{\tr_2}}{\tr_3}}}{\kappa}{\emptyenv}{\ilprop}
    \iltsteprel
    \iltstate{\sigma}{\trz{\trxcons{\trx}{\trxpropmark{\tr_3}}}{\tr_2}}{\kappa}{\emptyenv}{\ilprop}  
    \iltsteprel^{n-1}
    \iltstate{\sigma'}{\trz{\trx'}{\trend}}{\kappa}{\emptyenv}{\vec\nu}
    $
  \item From Lemma~{\ref{lem:propmark-decomposition}} we have:
  \begin{enumerate}[(i)]
    \item $
    \iltstate{\sigma}{\trz{\trx}{\trcons{\trpush{\tr_2}}{\tr_3}}}{\kappa}{\emptyenv}{\ilprop}
    \iltsteprel
    \iltstate{\sigma}{\trz{\trxcons{\trx}{\trxpropmark{\tr_3}}}{\tr_2}}{\kappa}{\emptyenv}{\ilprop}
    \iltsteprel^{m_1}
    \iltstate{\sigma''}{\trz{\trxcons{\trx}{\trpush{\tr_2'}}}{\tr_3}}{\kappa}{\emptyenv}{\ilprop}
    \iltsteprel^{m_2}
    \iltstate{\sigma'}{\trz{\trx'}{\trend}}{\kappa}{\emptyenv}{\vec\nu}
    $
  \item $n = 1 + m_1 + m_2$
  \end{enumerate}
  \item Hence $
    \trz{\dropum{\trx'}}{\trend};\_
    \trrewind^{\ast}
    \trz{\dropum{\trxcons{\trx}{\trpush{\tr_2'}}}}{\_};\_
    \trrewind
    \trz{\dropum{\trx}}{\_};\_
    $
    using (b) and Lemma~\ref{lem:rewind-single-action}
  \item From $\okay{\trx}$ we have $\okay{\trxcons{\trx}{\trpush{\tr_2'}}}$
  \item From Lemma~\ref{lem:decomposition} and (d), we have $\fsc{\tr_3}$
  \item Finally, by definition $\trlast{\tr_3} = \trlast{\trcons{\trpush{\tr_2}}{\tr_3}} = \trlast{\tr}$.
  \item This completes the case, showing claim (2) with $n_1 = 1 + m_1$ and $n_2 = m_2$.
  \end{itemize}
\item Case {\bf P.7}: not possible; it contradicts assumption (e).
\item Case {\bf P.8}: not possible; it contradicts assumption (a).
\end{itemize}
\end{proof}

\begin{lem}[Case analysis: final (non-nested) awakening]
\label{lem:general-active-update-or-prop}
If
\begin{enumerate}[(a)]
\item 
  $
  \iltstate{\sigma}{\trz{\trx}{\tr}}{\kappa}{\emptyenv}{\ilprop}
  \iltsteprel^n
  \iltstate{\sigma'}{\trz{\trx'}{\trend}}{\kappa}{\emptyenv}{\vec\nu}
  $
\item $\rewindstepm{\trz{\dropum{\trx'}}{\_}}\_{\trz{\dropum{\trx}}{\_}}\_$
\item $\okay{\trx}$
\item $\fsc{\tr}$
\end{enumerate}
Then either:
\begin{enumerate}
\item 
  \begin{itemize}
  \item $
    \iltstate{\sigma}{\trz{\trx}{\tr}}{\kappa}{\emptyenv}{\ilprop}
    \iltsteprel^{n_1}
    \iltstate{\sigma''}{\trz{\trx''}{\trcons{\trwake{\rho}{e}}{\tr'}}}{\kappa}{\emptyenv}{\ilprop}
    \iltsteprel
    \iltstate{\sigma''}{\trz{\trxcons{\trx''}{\trwake{\rho}{e}}}{\tr'}}{\kappa}{\rho}{e}
    \iltsteprel^{n_2}
    \iltstate{\sigma'}{\trz{\trx'}{\trend}}{\kappa}{\emptyenv}{\vec\nu}
    $
  \item $
    \trz{\dropum{\trx'}}{\_};\_
    \trrewind^\ast
    \trz{\dropum{\trxcons{\trx''}{\trwake{\rho}{e}}}}{\_};\_
    \trrewind^\ast
    \trz{\dropum{\trx}}{\_};\_$
  \item $\okay{\trx''}$
  \item $\fsc{\trcons{{\trwake{\rho}{e}}}{\tr'}}$
  \item $\trlast{\tr} = \trlast{\tr'}$
  \item $n = n_1 + 1 + n_2$
  \end{itemize}
\item
  \begin{itemize}
  \item $
    \iltstate{\sigma}{\trz{\trx}{\tr}}{\kappa}{\emptyenv}{\ilprop}
    \iltsteprel^{n_1}
    \iltstate{\sigma''}{\trz{\trx''}{\tr'}}{\kappa}{\emptyenv}{\ilprop}
    \iltsteprel^{n_2}
    \iltstate{\sigma'}{\trz{\trx'}{\trend}}{\kappa}{\emptyenv}{\vec\nu}
    $
  \item $
    \trz{\dropum{\trx'}}{\_};\_
    \trrewind^\ast
    \trz{\dropum{\trx''}}{\_};\_
    \trrewind^\ast
    \trz{\dropum{\trx}}{\_};\_$
  \item $\okay{\trx''}$
  \item $\fsc{\tr'}$
  \item $\trlast{\tr} = \trlast{\tr'}$
  \item $n = n_1 + n_2$
  \item Reduction $n_2$ contains no use of {\bf P.E}
  \end{itemize}
\end{enumerate}
\end{lem}
\begin{proof}
Case analysis on the shape of trace~$\tr$:

\begin{itemize}
\item Case: $\exists \tr_1, \tr_2, \tr_3$ such that $\tr = \trcons{\tr_1}{\trcons{\trpush{\tr_2}}{\tr_3}}$ and $\tr_1$ contains no parenthesis.
\begin{itemize}
\item Applying lemma~\ref{lem:active-update-or-prop}, we get subcases (i) and (ii):
\item[(i)]
  \begin{itemize}
  \item $
    \iltstate{\sigma}{\trz{\trx}{\tr}}{\kappa}{\emptyenv}{\ilprop}
    \iltsteprel^{n_1}
    \iltstate{\sigma''}{\trz{\trx''}{\trcons{\trwake{\rho}{e}}{\trcons{\tr_1'}{\trcons{\trpush{\tr_2}}{\tr_3}}}}}{\kappa}{\emptyenv}{\ilprop}
    \iltsteprel
    \iltstate{\sigma''}{\trz{\trxcons{\trx''}{\trwake{\rho}{e}}}{\trcons{\tr_1'}{\trcons{\trpush{\tr_2}}{\tr_3}}}}{\kappa}{\rho}{e}
    \iltsteprel^{n_2}
    \iltstate{\sigma'}{\trz{\trx'}{\trend}}{\kappa}{\emptyenv}{\vec\nu}
    $
  \item $
    \trz{\dropum{\trx'}}{\_};\_
    \trrewind^\ast
    \trz{\dropum{\trxcons{\trx''}{\trwake{\rho}{e}}}}{\_};\_
    \trrewind^\ast
    \trz{\dropum{\trx}}{\_};\_$
  \item $\okay{\trx''}$
  \item $\fsc{\trcons{\trwake{\rho}{e}}{\trcons{\tr_1'}{\trcons{\trpush{\tr_2}}{\tr_3}}}}$
  \item $\trlast{\trcons{\tr_1'}{\trcons{\trpush{\tr_2}}{\tr_3}}} = \trlast{\tr}$
  \item $n = n_1 + 1 + n_2$
  \end{itemize}
\item This immediately shows claim (1).

\item[(ii)]
  \begin{itemize}
  \item $
    \iltstate{\sigma}{\trz{\trx}{\tr}}{\kappa}{\emptyenv}{\ilprop}
    \iltsteprel^{n_1}
    \iltstate{\sigma''}{\trz{\trx''}{\tr_3}}{\kappa}{\emptyenv}{\ilprop}
    \iltsteprel^{n_2}
    \iltstate{\sigma'}{\trz{\trx'}{\trend}}{\kappa}{\emptyenv}{\vec\nu}
    $
  \item $
    \trz{\dropum{\trx'}}{\_};\_
    \trrewind^\ast
    \trz{\dropum{\trx''}}{\_};\_
    \trrewind^\ast
    \trz{\dropum{\trx}}{\_};\_$
  \item $\okay{\trx''}$
  \item $\fsc{\tr_3}$
  \item $\trlast{\tr_3} = \trlast{\tr}$
  \item $n = n_1 + n_2, n_1 > 0$
  \end{itemize}

\item Since we have that $n_2 < n$, we continue by induction on reduction~$n_2$, which shows the claim.
\end{itemize}

\item Case: \emph{Otherwise}: Note necessarily that $\tr = \trcons{t_1}{\trcons{\ldots}{t_m}}$ such that $\forall i.~t_i \ne \trpush{\_}$
\begin{itemize}
\item Subcase: reduction~(a) contains a use of {\bf P.E}:
\begin{itemize}

\item Hence, $\exists t_i = \trwake{\rho}{e}$ such that $\tr =
  \trcons{t_1}{\trcons{\ldots}{\trcons{t_i}{\trcons{\ldots}{t_m}}}}$
  and $\trx'' =
  \trxcons{\trx}{\trxcons{t_1}{\trxcons{\ldots}{t_{i-1}}}}$

\item Then, since $\okay{\trx}$ we have $\okay{\trx''}$

\item Moreover, $\trlast{\tr} 
  =
  \trlast{\trcons{t_1}{\trcons{\ldots}{\trcons{t_i}{\trcons{\ldots}{t_m}}}}}
  =
  \trlast{\trcons{t_2}{\trcons{\ldots}{\trcons{t_i}{\trcons{\ldots}{t_m}}}}}
  =
  \trlast{{\trcons{t_i}{\trcons{\ldots}{t_m}}}} 
  =
  \trlast{{\trcons{t_{i+1}}{\trcons{\ldots}{t_m}}}}$

\item Since $\tr' = \trcons{t_{i+1}}{\trcons{\ldots}{t_m}}$, we have
    that $\trlast{\tr} = \trlast{\tr'}$

\item We get the rest of claim (1) from repeated use of
  Lemmas~\ref{lem:rewind-single-action}~and~\ref{lem:decomposition}.
  (the number of required uses is $i - 1$).

\end{itemize}
\item Subcase: reduction~(a) contains no use of {\bf P.E}:
\begin{itemize}
\item Then we have claim (2) immediately, with $n_1 = 0$.
\end{itemize}
\end{itemize}
\end{itemize}
\end{proof}

\begin{thm}[Consistency]~
  \label{thm:consistency}
  \begin{enumerate}
  \item If
    \begin{enumerate}[(a)]
    \item $\iltstate {\sigma_2} {\trz{\trx_2}{\tr'_1}} {\kappa_2}
      {\emptyenv} {\ilprop} \iltsteprel^n \iltstate {\sigma'_2}
      {\trz{\trx'_2}{\tr_1''}} {\kappa_2} {\emptyenv} {\vec{\nu_2}}$
    \item $\okay{\trx_2}$
    \item $\fsc{\tr'_1}$, from initial configuration~ $\iltstate
      {\sigma_1} {\trz{\trx_1}{\trend}} {\kappa_1}{\rho_1}{\rcmd_1}$
    \item
      $\rewindstepm{\trz{\dropum{\trx'_2}}{\_}}{\trend}{\trz{\dropum{\trx_2}}{\_}}{\tr'_2}$
    \end{enumerate}
    then for any $\trx_3$ there is $\trx_3'$ such that
    \begin{enumerate}[(i)]
    \item $\okay{\trz{\trx'_2}{\tr_1''}}$
    \item
      $\iltstate{\nongarbof{\sigma_2}}{\trz{\trx_3}{\trend}}{\kappa_3}{\rho_1}{\rcmd_1}
      \iltsteprel^\ast
      \iltstate{\nongarbof{\sigma'_2}}{\trz{\trx'_3}{\trend}}{\kappa_3}{\emptyenv}{\vec{\nu_2}}$
    \item
      $\rewindstepm{\trz{\trx'_3}{\trend}}{\trend}{\trz{\trx_3}{\trend}}{\tr'_2}$
    \end{enumerate}
  \item If
    \begin{enumerate}[(a)]
    \item
      $\iltstate{\sigma_2}{\trz{\trx_2}{\tr'_1}}{\kappa_2}{\rho_2}{\rcmd_2}\iltsteprel^n
      \iltstate{\sigma'_2}{\trz{\trx'_2}{\tr_1''}}{\kappa_2}{\emptyenv}{\vec{\nu_2}}$
    \item $\okay{\trz{\trx_2}{\tr'_1}}$
    \item $\rewindstepm{\trz{\dropum{\trx'_2}}{\_}}{\trend}
      {\trz{\dropum{\trx_2}}{\_}}{\tr'_2}$
    \end{enumerate}
    then for any $\trx_3$ there is $\trx_3'$ such that
    \begin{enumerate}[(i)]
    \item $\okay{\trz{\trx'_2}{\tr_1''}}$
    \item $\iltstate {\nongarbof{\sigma_2}} {\trz{\trx_3}{\trend}}
      {\kappa_3} {\rho_2} {\rcmd_2} \iltsteprel^\ast \iltstate
      {\nongarbof{\sigma'_2}} {\trz{\trx'_3}{\trend}} {\kappa_3}
      {\emptyenv} {\vec{\nu_2}}$
    \item $\rewindstepm {\trz{\trx'_3}{\trend}} {\trend}
      {\trz{\trx_3}{\trend}} {\tr'_2}$
    \end{enumerate}
  \end{enumerate}
\end{thm}
\begin{proof}
  By simultaneous induction on $n$.
  \begin{enumerate}
  \item Note that necessarily $n > 0$.  We inspect the first reduction step of (a).
    \begin{itemize}

    \item Case {\bf P.1}:                %
      \begin{itemize}
      \item
        Then $\tr_1' = \trcons{\tralloc{\ell}{m}}{\widehat{\tr_1}}$ and
        $\iltstate{\sigma_2}{\trz{\trx_2}{\tr'_1}}{\kappa_2}{\emptyenv}{\ilprop}
        \iltsteprel
        \iltstate{\widehat{\sigma_2}}{\trz{\trxcons{\trx_2}{\tralloc{\ell}{m}}}{\widehat{\tr_1}}}{\kappa_2}{\emptyenv}{\ilprop}
        \iltsteprel^{n-1}
        \iltstate{\sigma'_2}{\trz{\trx'_2}{\tr_1''}}{\kappa_2}{\emptyenv}{\vec{\nu_2}}$,
        where $\storestin{\sigma_2}{\emptyenv}{\ilalloc{m}} \storestep
        \storestout{\widehat{\sigma_2}}{\ell}$.
      \item Hence
        $\rewindstepm{\trz{\dropum{\trx'_2}}{\_}}{\trend}{\trz{\dropum{\trxcons{\trx_2}{\tralloc{\ell}{m}}}}{\_}}{\widehat{\tr_2}}
        \trrewind
        \trz{\dropum{\trx_2}}{\_};{\trcons{\tralloc{\ell}{m}}{\widehat{\tr_2}}}$
        with $\tr_2' = \trcons{\tralloc{\ell}{m}}{\widehat{\tr_2}}$ by
        Lemma~\ref{lem:rewind-single-action} and (d).
      \item By Lemma~\ref{lem:decomposition} and (c) we get:
        \begin{itemize}
        \item $\iltstate{\sigma_1}{\trz{\trx_1}{\trend}}{\kappa_1}{\rho_1}{\rcmd_1}
          \iltsteprel^*
          \iltstate{\sigma_1}{\trz{\trx_1}{\trend}}{\kappa_1}{\rho_1'}{\illet{\ilvar{x}{\ilalloc{y}}}{e}}$
          using {\bf E.0} only
        \item $\fsc{\widehat{\tr_1}}$ from
          $\iltstate{\sigma_1'}{\trz{\trxcons{\trx_1}{\tralloc{\ell}{m}}}{\trend}}{\kappa_1}{\rho_1'[x \mapsto \ell]}{e}$
        \item $\storestin{\sigma_1}{\rho_1'}{\ilalloc{y}} \storestep
          \storestout{\sigma_1'}{\ell}$
        \item $\rho_1'(y) = m$
        \end{itemize}
      \item From (b) we get $\okay{\trxcons{\trx_2}{\tralloc{\ell}{m}}}$.
      \item By induction then:
        \begin{enumerate}[(i)]
        \item $\okay{\trz{\trx'_2}{\tr_1''}}$
        \item $\iltstate{\nongarbof{\widehat{\sigma_2}}}{\trz{\trxcons{\trx_3}{\tralloc{\ell}{m}}}{\trend}}{\kappa_3}{\rho_1'[x \mapsto \ell]}{e} \iltsteprel^\ast \iltstate{\nongarbof{\sigma'_2}}{\trz{\trx'_3}{\trend}}{\kappa_2}{\emptyenv}{\vec{\nu_2}}$
        \item
          $\rewindstepm{\trz{\trx'_3}{\trend}}{\trend}{\trz{\trxcons{\trx_3}{\tralloc{\ell}{m}}}{\trend}}{\widehat{\tr_2}}
          \trrewind \trz{\trx_3}{\trend};{\tr_2'}$
        \end{enumerate}
      \item From $\storestin{\sigma_2}{\emptyenv}{\ilalloc{m}}
        \storestep \storestout{\widehat{\sigma_2}}{\ell}$ and the
        knowledge about $\rho_1'$ follows
        $\storestin{\nongarbof{\sigma_2}}{\rho_1'}{\ilalloc{y}}
        \storestep \storestout{\nongarbof{\widehat{\sigma_2}}}{\ell}$.
      \item Hence, using Lemma~\ref{lem:purity},
        $\iltstate{\nongarbof{\sigma_2}}{\trz{\trx_3}{\trend}}{\kappa_3}{\rho_1}{\rcmd_1}
        \iltsteprel^\ast
        \iltstate{\nongarbof{\widehat{\sigma_2}}}{\trz{\trxcons{\trx_3}{\tralloc{\ell}{m}}}{\trend}}{\kappa_3}{\rho_1'[x
            \mapsto \ell]}{e} \iltsteprel^\ast
        \iltstate{\nongarbof{\sigma'_2}}{\trz{\trx'_3}{\trend}}{\kappa_2}{\emptyenv}{\vec{\nu_2}}$
      \end{itemize}

    \item Case {\bf P.2}:
      \begin{itemize}
      \item Then $\tr_1' =
        \trcons{\trread{\nu}{\ell}{m}}{\widehat{\tr_1}}$ and
        $\iltstate{\sigma_2}{\trz{\trx_2}{\tr'_1}}{\kappa_2}{\emptyenv}{\ilprop}
        \iltsteprel
        \iltstate{\sigma_2}{\trz{\trxcons{\trx_2}{\trread{\nu}{\ell}{m}}}{\widehat{\tr_1}}}{\kappa_2}{\emptyenv}{\ilprop}
        \iltsteprel^{n-1}
        \iltstate{\sigma'_2}{\trz{\trx'_2}{\tr_1''}}{\kappa_2}{\emptyenv}{\vec{\nu_2}}$,
        where $\storestin{\sigma_2}{\emptyenv}{\ilread{\ell}{m}}
        \storestep \storestout{\sigma_2}{\nu}$.
      \item Hence
        $\rewindstepm{\trz{\dropum{\trx'_2}}{\_}}{\trend}{\trz{\dropum{\trxcons{\trx_2}{\trread{\nu}{\ell}{m}}}}{\_}}{\widehat{\tr_2}}
        \trrewind
        \trz{\dropum{\trx_2}}{\_};{\trcons{\trread{\nu}{\ell}{m}}{\widehat{\tr_2}}}$
        with $\tr_2' =
        \trcons{\trread{\nu}{\ell}{m}}{\widehat{\tr_2}}$ by
        Lemma~\ref{lem:rewind-single-action} and (d).
      \item By Lemma~\ref{lem:decomposition} and (c) we get:
        \begin{itemize}
        \item $\iltstate{\sigma_1}{\trz{\trx_1}{\trend}}{\kappa_1}{\rho_1}{\rcmd_1}
          \iltsteprel^*
          \iltstate{\sigma_1}{\trz{\trx_1}{\trend}}{\kappa_1}{\rho_1'}{\illet{\ilvar{x}{\ilread{y}{z}}}{e}}$
          using {\bf E.0} only
        \item $\fsc{\widehat{\tr_1}}$ from
          $\iltstate{\sigma_1}{\trz{\trxcons{\trx_1}{\trread{\nu}{\ell}{m}}}{\trend}}{\kappa_1}{\rho_1'[x
            \mapsto \nu]}{e}$
        \item $\storestin{\sigma_1}{\rho_1'}{\ilread{y}{z}} \storestep
          \storestout{\sigma_1}{\nu}$
        \item $\rho_1'(y) = \ell$
        \item $\rho_1'(z) = m$
        \end{itemize}
      \item From (b) we get
        $\okay{\trxcons{\trx_2}{\trread{\nu}{\ell}{m}}}$.
      \item By induction then:
        \begin{enumerate}[(i)]
        \item $\okay{\trz{\trx'_2}{\tr_1''}}$
        \item
          $\iltstate{\nongarbof{\sigma_2}}{\trz{\trxcons{\trx_3}{\trread{\nu}{\ell}{m}}}{\trend}}{\kappa_3}{\rho_1'[x
            \mapsto \nu]}{e} \iltsteprel^\ast
          \iltstate{\nongarbof{\sigma'_2}}{\trz{\trx'_3}{\trend}}{\kappa_2}{\emptyenv}{\vec{\nu_2}}$
        \item
          $\rewindstepm{\trz{\trx'_3}{\trend}}{\trend}{\trz{\trxcons{\trx_3}{\trread{\nu}{\ell}{m}}}{\trend}}{\widehat{\tr_2}} \trrewind \trz{\trx_3}{\trend};{\tr_2'}$
        \end{enumerate}
      \item From $\storestin{\sigma_2}{\emptyenv}{\ilread{\ell}{m}}
        \storestep \storestout{\sigma_2}{\nu}$ and the knowledge about
        $\rho_1'$ follows
        $\storestin{\nongarbof{\sigma_2}}{\rho_1'}{\ilread{y}{z}}
        \storestep \storestout{\nongarbof{\sigma_2}}{\nu}$.
      \item Hence, by Lemma~\ref{lem:purity},
        $\iltstate{\nongarbof{\sigma_2}}{\trz{\trx_3}{\trend}}{\kappa_3}{\rho_1}{\rcmd_1}
        \iltsteprel^\ast
        \iltstate{\nongarbof{\sigma_2}}{\trz{\trxcons{\trx_3}{\trread{\nu}{\ell}{m}}}{\trend}}{\kappa_3}{\rho_1'[x
            \mapsto \nu]}{e} \iltsteprel^\ast
        \iltstate{\nongarbof{\sigma'_2}}{\trz{\trx'_3}{\trend}}{\kappa_2}{\emptyenv}{\vec{\nu_2}}$
      \end{itemize}
      
    \item Case {\bf P.3}:
      \begin{itemize}
      \item
        Then $\tr_1' =
        \trcons{\trwrite{\nu}{\ell}{m}}{\widehat{\tr_1}}$ and
        $\iltstate{\sigma_2}{\trz{\trx_2}{\tr'_1}}{\kappa_2}{\emptyenv}{\ilprop}
        \iltsteprel
        \iltstate{\widehat{\sigma_2}}{\trz{\trxcons{\trx_2}{\trwrite{\nu}{\ell}{m}}}{\widehat{\tr_1}}}{\kappa_2}{\emptyenv}{\ilprop}
        \iltsteprel^{n-1}
        \iltstate{\sigma'_2}{\trz{\trx'_2}{\tr_1''}}{\kappa_2}{\emptyenv}{\vec{\nu_2}}$,
        where $\storestin{\sigma_2}{\emptyenv}{\ilwrite{\ell}{m}{\nu}}
        \storestep \storestout{\widehat{\sigma_2}}{0}$.
      \item Hence
        $\rewindstepm{\trz{\dropum{\trx'_2}}{\_}}{\trend}{\trz{\dropum{\trxcons{\trx_2}{\trwrite{\nu}{\ell}{m}}}}{\_}}{\widehat{\tr_2}}
        \trrewind
        \trz{\dropum{\trx_2}}{\_};{\trcons{\trwrite{\nu}{\ell}{m}}{\widehat{\tr_2}}}$
        with $\tr_2' =
        \trcons{\trwrite{\nu}{\ell}{m}}{\widehat{\tr_2}}$ by
        Lemma~\ref{lem:rewind-single-action} and (d).
      \item By Lemma~\ref{lem:decomposition} and (c) we get:
        \begin{itemize}
        \item
          $\iltstate{\sigma_1}{\trz{\trx_1}{\trend}}{\kappa_1}{\rho_1}{\rcmd_1}
          \iltsteprel^*
          \iltstate{\sigma_1}{\trz{\trx_1}{\trend}}{\kappa_1}{\rho_1'}{\illet{\ilvar{\_}{\ilwrite{x}{y}{z}}}{e}}$
          using {\bf E.0} only
        \item $\fsc{\widehat{\tr_1}}$ from
          $\iltstate{\sigma_1'}{\trz{\trxcons{\trx_1}{\trwrite{\nu}{\ell}{m}}}{\trend}}{\kappa_1}{\rho_1'}{e}$
        \item $\storestin{\sigma_1}{\rho_1'}{\ilwrite{x}{y}{z}}
          \storestep \storestout{\sigma_1'}{0}$
        \item $\rho_1'(x) = \ell$
        \item $\rho_1'(y) = m$
        \item $\rho_1'(z) = \nu$
        \end{itemize}
      \item From (b) we get
        $\okay{\trxcons{\trx_2}{\trwrite{\nu}{\ell}{m}}}$.
      \item By induction then:
        \begin{enumerate}[(i)]
        \item $\okay{\trz{\trx'_2}{\tr_1''}}$
        \item
          $\iltstate{\nongarbof{\sigma_2}}{\trz{\trxcons{\trx_3}{\trwrite{\nu}{\ell}{m}}}{\trend}}{\kappa_3}{\rho_1'}{e}
          \iltsteprel^\ast
          \iltstate{\nongarbof{\sigma'_2}}{\trz{\trx'_3}{\trend}}{\kappa_2}{\emptyenv}{\vec{\nu_2}}$.
        \item
          $\rewindstepm{\trz{\trx'_3}{\trend}}{\trend}{\trz{\trxcons{\trx_3}{\trwrite{\nu}{\ell}{m}}}{\trend}}{\widehat{\tr_2}} \trrewind \trz{\trx_3}{\trend};{\tr_2'}$
        \end{enumerate}
      \item From $\storestin{\sigma_2}{\emptyenv}{\ilwrite{\nu}{\ell}{m}}
        \storestep \storestout{\sigma_2'}{0}$ and the knowledge about
        $\rho_1'$ follows
        $\storestin{\nongarbof{\sigma_2}}{\rho_1'}{\ilwrite{x}{y}{z}}
        \storestep \storestout{\nongarbof{\sigma_2'}}{0}$.
      \item Hence, using Lemma~\ref{lem:purity},
        $\iltstate{\nongarbof{\sigma_2}}{\trz{\trx_3}{\trend}}{\kappa_3}{\rho_1}{\rcmd_1}
        \iltsteprel^\ast
        \iltstate{\nongarbof{\sigma_2}}{\trz{\trxcons{\trx_3}{\trwrite{\nu}{\ell}{m}}}{\trend}}{\kappa_3}{\rho_1'}{e}
        \iltsteprel^\ast
        \iltstate{\nongarbof{\sigma'_2}}{\trz{\trx'_3}{\trend}}{\kappa_2}{\emptyenv}{\vec{\nu_2}}$
      \end{itemize}

    \item Case {\bf P.4}:
      \begin{itemize}
      \item Then $\tr_1' = \trcons{\trmemo{\rho}{e}}{\widehat{\tr_1}}$
        and
        $\iltstate{\sigma_2}{\trz{\trx_2}{\tr'_1}}{\kappa_2}{\emptyenv}{\ilprop}
        \iltsteprel
        \iltstate{\sigma_2}{\trz{\trxcons{\trx_2}{\trmemo{\rho}{e}}}{\widehat{\tr_1}}}{\kappa_2}{\emptyenv}{\ilprop}
        \iltsteprel^{n-1}
        \iltstate{\sigma'_2}{\trz{\trx'_2}{\tr_1''}}{\kappa_2}{\emptyenv}{\vec{\nu_2}}$.
      \item Hence
        $\rewindstepm{\trz{\dropum{\trx'_2}}{\_}}{\trend}{\trz{\dropum{\trxcons{\trx_2}{\trmemo{\rho}{e}}}}{\_}}{\widehat{\tr_2}}
        \trrewind
        \trz{\dropum{\trx_2}}{\_};{\trcons{\trmemo{\rho}{e}}{\widehat{\tr_2}}}$
        with $\tr_2' = \trcons{\trmemo{\rho}{e}}{\widehat{\tr_2}}$ by
        Lemma~\ref{lem:rewind-single-action} and (d).
      \item By Lemma~\ref{lem:decomposition} and (c) we get:
        \begin{itemize}
        \item
          $\iltstate{\sigma_1}{\trz{\trx_1}{\trend}}{\kappa_1}{\rho_1}{\rcmd_1}
          \iltsteprel^*
          \iltstate{\sigma_1}{\trz{\trx_1}{\trend}}{\kappa_1}{\rho}{\ilmemo{e}}$
          using {\bf E.0} only
        \item $\fsc{\widehat{\tr_1}}$ from
          $\iltstate{\sigma_1}{\trz{\trxcons{\trx_1}{\trmemo{\rho}{e}}}{\trend}}{\kappa_1}{\rho}{e}$
        \end{itemize}
      \item From (b) we get
        $\okay{\trxcons{\trx_2}{\trmemo{\rho}{e}}}$.
      \item By induction then:
        \begin{enumerate}[(i)]
        \item $\okay{\trz{\trx'_2}{\tr_1''}}$
        \item
          $\iltstate{\nongarbof{\sigma_2}}{\trz{\trxcons{\trx_3}{\trmemo{\rho}{e}}}{\trend}}{\kappa_3}{\rho}{e}
          \iltsteprel^\ast
          \iltstate{\nongarbof{\sigma'_2}}{\trz{\trx'_3}{\trend}}{\kappa_2}{\emptyenv}{\vec{\nu_2}}$
        \item
          $\rewindstepm{\trz{\trx'_3}{\trend}}{\trend}{\trz{\trxcons{\trx_3}{\trmemo{\rho}{e}}}{\trend}}{\widehat{\tr_2}}
          \trrewind \trz{\trx_3}{\trend};{\tr_2'}$
        \end{enumerate}
      \item Finally, using Lemma~\ref{lem:purity},
        $\iltstate{\nongarbof{\sigma_2}}{\trz{\trx_3}{\trend}}{\kappa_3}{\rho_1}{\rcmd_1}
        \iltsteprel^\ast
        \iltstate{\nongarbof{\sigma_2}}{\trz{\trxcons{\trx_3}{\trmemo{\rho}{e}}}{\trend}}{\kappa_3}{\rho}{e}
        \iltsteprel^\ast
        \iltstate{\nongarbof{\sigma'_2}}{\trz{\trx'_3}{\trend}}{\kappa_2}{\emptyenv}{\vec{\nu_2}}$
      \end{itemize}

    \item Case {\bf P.5}:
      \begin{itemize}
      \item Then $\tr_1' = \trcons{\trwake{\rho}{e}}{\widehat{\tr_1}}$
        and
        $\iltstate{\sigma_2}{\trz{\trx_2}{\tr'_1}}{\kappa_2}{\emptyenv}{\ilprop}
        \iltsteprel
        \iltstate{\sigma_2}{\trz{\trxcons{\trx_2}{\trwake{\rho}{e}}}{\widehat{\tr_1}}}{\kappa_2}{\emptyenv}{\ilprop}
        \iltsteprel^{n-1}
        \iltstate{\sigma'_2}{\trz{\trx'_2}{\tr_1''}}{\kappa_2}{\emptyenv}{\vec{\nu_2}}$.
      \item Hence
        $\rewindstepm{\trz{\dropum{\trx'_2}}{\_}}{\trend}{\trz{\dropum{\trxcons{\trx_2}{\trwake{\rho}{e}}}}{\_}}{\widehat{\tr_2}}
        \trrewind
        \trz{\dropum{\trx_2}}{\_};{\trcons{\trwake{\rho}{e}}{\widehat{\tr_2}}}$
        with $\tr_2' = \trcons{\trwake{\rho}{e}}{\widehat{\tr_2}}$ by
        Lemma~\ref{lem:rewind-single-action} and (d).
      \item By Lemma~\ref{lem:decomposition} and (c) we get:
        \begin{itemize}
        \item
          $\iltstate{\sigma_1}{\trz{\trx_1}{\trend}}{\kappa_1}{\rho_1}{\rcmd_1}
          \iltsteprel^*
          \iltstate{\sigma_1}{\trz{\trx_1}{\trend}}{\kappa_1}{\rho}{\ilwake{e}}$
          using {\bf E.0} only
        \item $\fsc{\widehat{\tr_1}}$ from
          $\iltstate{\sigma_1}{\trz{\trxcons{\trx_1}{\trwake{\rho}{e}}}{\trend}}{\kappa_1}{\rho}{e}$
        \end{itemize}
      \item From (b) we get
        $\okay{\trxcons{\trx_2}{\trwake{\rho}{e}}}$.
      \item By induction then:
        \begin{enumerate}[(i)]
        \item $\okay{\trz{\trx'_2}{\tr_1''}}$
        \item
          $\iltstate{\nongarbof{\sigma_2}}{\trz{\trxcons{\trx_3}{\trwake{\rho}{e}}}{\trend}}{\kappa_3}{\rho}{e}
          \iltsteprel^\ast
          \iltstate{\nongarbof{\sigma'_2}}{\trz{\trx'_3}{\trend}}{\kappa_2}{\emptyenv}{\vec{\nu_2}}$
        \item
          $\rewindstepm{\trz{\trx'_3}{\trend}}{\trend}{\trz{\trxcons{\trx_3}{\trwake{\rho}{e}}}{\trend}}{\widehat{\tr_2}}
          \trrewind \trz{\trx_3}{\trend};{\tr_2'}$
        \end{enumerate}
      \item Finally, using Lemma~\ref{lem:purity},
        $\iltstate{\nongarbof{\sigma_2}}{\trz{\trx_3}{\trend}}{\kappa_3}{\rho_1}{\rcmd_1}
        \iltsteprel^\ast
        \iltstate{\nongarbof{\sigma_2}}{\trz{\trxcons{\trx_3}{\trwake{\rho}{e}}}{\trend}}{\kappa_3}{\rho}{e}
        \iltsteprel^\ast
        \iltstate{\nongarbof{\sigma'_2}}{\trz{\trx'_3}{\trend}}{\kappa_2}{\emptyenv}{\vec{\nu_2}}$
      \end{itemize}

    \item Case {\bf P.6}:
      \begin{itemize}
      \item Then $\tr_1' =
        \trcons{\trpush{\widetilde{\tr_1}}}{\widehat{\tr_1}}$ and
        $\iltstate{\sigma_2}{\trz{\trx_2}{\tr'_1}}{\kappa_2}{\emptyenv}{\ilprop}
        \iltsteprel
        \iltstate{\sigma_2}{\trz{\trxcons{\trx_2}{\trxpropmark{\widehat{\tr_1}}}}{\widetilde{\tr_1}}}{\kappa_2}{\emptyenv}{\ilprop}
        \iltsteprel^{n-1}
        \iltstate{\sigma'_2}{\trz{\trx'_2}{\tr_1''}}{\kappa_2}{\emptyenv}{\vec{\nu_2}}$.
      \item By Lemmas~\ref{lem:rewind-trace-context}
        and~\ref{lem:propmark-decomposition} we get:
        \begin{itemize}
        \item
          $\iltstate{\sigma_2}{\trz{\trxcons{\trx_2}{\trxpropmark{\widehat{\tr_1}}}}{\widetilde{\tr_1}}}{\kappa_2}{\emptyenv}{\ilprop}
          \iltsteprel^{n_1}
          \iltstate{\widehat{\sigma_2}}{\trz{\widehat{\trx_2}}{\trend}}{\kappa_2}{\emptyenv}{\vec{\nu}}$
        \item
          $\rewindstepm{\trz{\widehat{\trx_2}}{\trend}}{\trend}{\trz{\trxcons{\trx_2}{\trxpropmark{\widehat{\tr_1}}}}{\trend}}{\tr}$
        \item
          $\iltstate{\widehat{\sigma_2}}{\trz{\widehat{\trx_2}}{\trend}}{\kappa_2}{\emptyenv}{\vec{\nu}}
          \iltsteprel
          \iltstate{\widehat{\sigma_2}}{\trz{\trxcons{\trx_2}{\trpush{\tr}}}{\widehat{\tr_1}}}{\kappa_2}{\emptyenv}{\ilprop}$
        \item
          $\iltstate{\widehat{\sigma_2}}{\trz{\trxcons{\trx_2}{\trpush{\tr}}}{\widehat{\tr_1}}}{\kappa_2}{\emptyenv}{\ilprop}
          \iltsteprel^{n_2}
          \iltstate{\sigma'_2}{\trz{\trx'_2}{\tr_1''}}{\kappa_2}{\emptyenv}{\vec{\nu_2}}$
        \item $n - 1 = n_1 + 1 + n_2$
        \end{itemize}
      \item By Lemma~\ref{lem:decomposition} and (c) we get:
        \begin{itemize}
        \item
          $\iltstate{\sigma_1}{\trz{\trx_1}{\trend}}{\kappa_1}{\rho_1}{\rcmd_1}
          \iltsteprel^*
          \iltstate{\sigma_1}{\trz{\trx_1}{\trend}}{\kappa_1}{\rho}{\ilpush{f}{e}}$
          using {\bf E.0} only
        \item $\fsc{\widetilde{\tr_1}}$ from
          $\iltstate{\sigma_1}{\trz{\trxcons{\trx_1}{\trxpushmark}}{\trend}}{\stcons{\kappa_1}{\mkstframe{\rho}{f}}}{\rho}{e}$, producing value $\vec{\omega}$
        \item $\fsc{\widehat{\tr_1}}$ from
          $\iltstate{\sigma_1'}{\trz{\trxcons{\trx_1}{\trpush{\widetilde{\tr_1}}}}{\trend}}{\kappa_1}{\rho'}{e'}$
        \item $\rho(f) = \ilfun{f}{\vec{x}}{e'}$
        \item $\rho' = \rho[\vec{x\mapsto\omega}]$
        \end{itemize}
      \item Since $\okay{\trx_2}$ and $\fsc{\widehat{\tr_1}}$, we know
        $\okay{\trxcons{\trx_2}{\trxpropmark{\widehat{\tr_1}}}}$.
      \item Furthermore,
        $\rewindstepm{\trz{\widehat{\trx_2}}{\trend}}{\trend}{\trz{\trxcons{\trx_2}{\trxpropmark{\widehat{\tr_1}}}}{\trend}}{\tr}$
        implies
        $\rewindstepm{\trz{\dropum{\widehat{\trx_2}}}{\trend}}{\trend}{\trz{\dropum{\trxcons{\trx_2}{\trxpropmark{\widehat{\tr_1}}}}}{\trend}}{\tr}$
        by Lemma~\ref{lem:rewind-dropum}.
      \item Induction with $n_1$ then yields:
        \begin{itemize}
        \item $\okay{\trz{\widehat{\trx_2}}{\trend}}$
        \item
          $\iltstate{\nongarbof{\sigma_2}}{\trz{\trxcons{\trx_3}{\trxpushmark}}{\trend}}{\stcons{\kappa_3}{\mkstframe{\rho}{f}}}{\rho}{e}
          \iltsteprel^\ast
          \iltstate{\nongarbof{\widehat{\sigma_2}}}{\trz{\widehat{\trx_3}}{\trend}}{\stcons{\kappa_3}{\mkstframe{\rho}{f}}}{\emptyenv}{\vec{\nu}}$
        \item
          $\rewindstepm{\trz{\widehat{\trx_3}}{\trend}}{\trend}{\trz{\trxcons{\trx_3}{\trxpushmark}}{\trend}}{\tr}$
        \end{itemize}
      \item Since $\okay{\trx_2}$ we get
        $\okay{\trxcons{\trx_2}{\trpush{\tr}}}$.
      \item We get
        $\rewindstepm{\trz{\dropum{\trx'_2}}{\_}}{\trend}{\trz{\dropum{\trxcons{\trx_2}{\trpush{\tr}}}}{\_}}{\widehat{\tr_2}}
        \trrewind
        \trz{\dropum{\trx_2}}{\_};{\trcons{\trpush{\tr}}{\widehat{\tr_2}}}$
        with $\tr_2' = \trcons{\trpush{\tr}}{\widehat{\tr_2}}$ by
        Lemma~\ref{lem:rewind-single-action} and (d).
      \item Induction with $n_2$ then yields:
        \begin{itemize}
        \item $\okay{\trz{\trx_2'}{\tr_1''}}$
        \item
          $\iltstate{\nongarbof{\widehat{\sigma_2}}}{\trz{\trxcons{\trx_3}{\trpush{\tr}}}{\trend}}{\kappa_3}{\rho'}{e'}
          \iltsteprel^\ast
          \iltstate{\nongarbof{\sigma_2'}}{\trz{\trx_3'}{\trend}}{\kappa_3}{\emptyenv}{\vec{\nu_2}}$
        \item
          $\rewindstepm{\trz{\trx_3'}{\trend}}{\trend}{\trz{\trxcons{\trx_3}{\trpush{\tr}}}{\trend}}{\widehat{\tr_2}}
            \trrewind \trz{\trx_3}{\trend};{\tr_2'}$
        \end{itemize}
      \item Finally, using Lemma~\ref{lem:purity},
        $\iltstate{\nongarbof{\sigma_2}}{\trz{\trx_3}{\trend}}{\kappa_3}{\rho_1}{\rcmd_1}
        \iltsteprel^\ast
        \iltstate{\nongarbof{\sigma_2}}{\trz{\trxcons{\trx_3}{\trxpushmark}}{\trend}}{\stcons{\kappa_3}{\mkstframe{\rho}{f}}}{\rho}{e}$
      \item
        $\iltstate{\nongarbof{\sigma_2}}{\trz{\trxcons{\trx_3}{\trxpushmark}}{\trend}}{\stcons{\kappa_3}{\mkstframe{\rho}{f}}}{\rho}{e}
        \iltsteprel^\ast
        \iltstate{\nongarbof{\widehat{\sigma_2}}}{\trz{\widehat{\trx_3}}{\trend}}{\stcons{\kappa_3}{\mkstframe{\rho}{f}}}{\emptyenv}{\vec{\nu}}$
      \item
        $\iltstate{\nongarbof{\widehat{\sigma_2}}}{\trz{\widehat{\trx_3}}{\trend}}{\stcons{\kappa_3}{\mkstframe{\rho}{f}}}{\emptyenv}{\vec{\nu}}
        \iltsteprel
        \iltstate{\nongarbof{\widehat{\sigma_2}}}{\trz{\trxcons{\trx_3}{\trpush{\tr}}}{\trend}}{\kappa_3}{\rho''}{e'}$,
        where $\rho'' = \rho[\vec{x \mapsto \nu}]$
        
        \vspace*{5mm}

      \item It remains to show that $\vec{\omega} = \vec{\nu}$ and thus $\rho' = \rho''$.
        
      \item Recall that we have:
        \begin{itemize}
        \item
          $\iltstate{\sigma_2}{\trz{\trxcons{\trx_2}{\trxpropmark{\widehat{\tr_1}}}}{\widetilde{\tr_1}}}{\kappa_2}{\emptyenv}{\ilprop}
          \iltsteprel^{n_1}
          \iltstate{\widehat{\sigma_2}}{\trz{\widehat{\trx_2}}{\trend}}{\kappa_2}{\emptyenv}{\vec{\nu}}$
        \item
          $\rewindstepm{\trz{\widehat{\trx_2}}{\trend}}{\trend}{\trz{\trxcons{\trx_2}{\trxpropmark{\widehat{\tr_1}}}}{\trend}}{\tr}$
        \item $\okay{\trxcons{\trx_2}{\trxpropmark{\widehat{\tr_1}}}}$
        \item $\fsc{\widetilde{\tr_1}}$
        \end{itemize}
      \item By Lemmas~\ref{lem:rewind-dropum} and~\ref{lem:general-active-update-or-prop}, we get two subcases:
        \begin{enumerate}
        \item First subcase (of two)
          \begin{itemize}
          \item In this subcase, we have that:
            \begin{itemize}
            \item $
              \iltstate{\sigma_2}{\trz{\trxcons{\trx_2}{\trxpropmark{\widehat{\tr_1}}}}{\widetilde{\tr_1}}}{\kappa_2}{\emptyenv}{\ilprop}
              \iltsteprel^{m_1}
              \iltstate{\widehat{\sigma_2'}}{\trz{\trx''}{\trcons{\trwake{\widehat{\rho}}{\widehat{e}}}{\tr'}}}{\kappa_2}{\emptyenv}{\ilprop}
              $     
            \item
              $
              \iltstate{\widehat{\sigma_2'}}{\trz{\trx''}{\trcons{\trwake{\widehat{\rho}}{\widehat{e}}}{\tr'}}}{\kappa_2}{\emptyenv}{\ilprop}
              \iltsteprel
              \iltstate{\widehat{\sigma_2'}}{\trz{\trxcons{\trx''}{\trwake{\widehat{\rho}}{\widehat{e}}}}{\tr'}}{\kappa_2}{\widehat{\rho}}{\widehat{e}}
              $               
            \item
              $
              \iltstate{\widehat{\sigma_2'}}{\trz{\trxcons{\trx''}{\trwake{\widehat{\rho}}{\widehat{e}}}}{\tr'}}{\kappa_2}{\widehat{\rho}}{\widehat{e}}
              \iltsteprel^{m_2}
              \iltstate{\widehat{\sigma_2}}{\trz{\widehat{\trx_2}}{\trend}}{\kappa_2}{\emptyenv}{\vec\nu}
              $
            \item $
              \trz{\dropum{\widehat{\trx_2}}}{\_};\_
              \trrewind^\ast
              \trz{\dropum{\trxcons{\trx''}{\trwake{\rho}{e}}}}{\_};\_
              \trrewind^\ast
              \trz{\dropum{\trxcons{\trx_2}{\trxpropmark{\widehat{\tr_1}}}}}{\_};\_$
            \item $\okay{{\trx''}}$
            \item $\fsc{\trcons{\trwake{\widehat{\rho}}{\widehat{e}}}{\tr'}}$
            \item $\trlast{\widetilde{\tr_1}} = \trlast{\tr'}$
            \item $n_1 = m_1 + 1 + m_2$
            \end{itemize}
          \item By Lemma~\ref{lem:decomposition} we get
            $\okay{\trz{\trxcons{\trx''}{\trwake{\widehat{\rho}}{\widehat{e}}}}{\tr'}}$.
          \item From induction on reduction $m_2$ using part (2), we get:
            \begin{itemize}      
            \item $
              \iltstate{\nongarbof{\widehat{\sigma_2'}}}{\trz{\trend}{\trend}}{\kappa_2}{\widehat{\rho}}{\widehat{e}}
              \iltsteprel^\ast
              \iltstate{\nongarbof{\widehat{\sigma_2}}}{\trz{\trx_4}{\trend}}{\kappa_2}{\emptyenv}{\vec\nu}
              $
            \item $
              {\trz{\trx_4}{\trend}};\trend
              \trrewind^\ast
              {\trz{\trend}{\trend}};\_
              $ 
            \end{itemize}
          \item From Lemma~\ref{lem:stack-surgery}, we have that
            $
            \iltstate{\nongarbof{\widehat{\sigma_2'}}}{\trz{\trend}{\trend}}{\emptystack}{\widehat{\rho}}{\widehat{e}}
            \iltsteprel^\ast
            \iltstate{\nongarbof{\widehat{\sigma_2}}}{\trz{\trx_4}{\trend}}{\emptystack}{\emptyenv}{\vec\nu}
            $        
          \item From Lemma~\ref{lem:noreuse-preservation} and
            Lemma~\ref{lem:traced-to-untraced} we have that: $
            \ilustate{\nongarbof{\widehat{\sigma_2'}}}{\emptystack}{\widehat{\rho}}{\widehat{e}}
            \ilusteprel^\ast
            \ilustate{\nongarbof{\widehat{\sigma_2}}}{\emptystack}{\emptyenv}{\vec{\nu}}
            $
          \item Next, since
            $\fsc{\trcons{\trwake{\widehat{\rho}}{\widehat{e}}}{\tr'}}$,
            with the help of Lemma~\ref{lem:decomposition} there
            exists $\trx_5$, $\sigma_5$, $\kappa_5$, $\rho_5$, $e_5$,
            $\trx_5'$, $\sigma_5'$ and $\vec{\omega'}$ such that
            \begin{itemize}
            \item $\csa{\rho_5}{e_5}{\sigma_5}$
            \item $\noreuse{\trz{\trx_5}{\trend}}$
            \item
              $\iltstate{\sigma_5}{\trz{\trx_5}{\trend}}{\kappa_5}{\rho_5}{e_5}
              \iltsteprel^\ast
              \iltstate{\sigma_5}{\trz{\trx_5}{\trend}}{\kappa_5}{\widehat{\rho}}{\ilwake{\widehat{e}}}$
              using {\bf E.0} only
            \item $\fsc{\tr'}$ from
              $\iltstate{\sigma_5}{\trz{\trxcons{\trx_5}{\trwake{\widehat{\rho}}{\widehat{e}}}}{\trend}}{\kappa_5}{\widehat{\rho}}{\widehat{e}}$
              producing $\vec{\omega'}$
            \item
              $\iltstate{\sigma_5}{\trz{\trxcons{\trx_5}{\trwake{\widehat{\rho}}{\widehat{e}}}}{\trend}}{\kappa_5}{\widehat{\rho}}{\widehat{e}}
              \iltsteprel^\ast
              \iltstate{\sigma_5'}{\trz{\trx_5'}{\trend}}{\kappa_5}{\emptyenv}{\vec{\omega'}}
              $
            \item ${\trz{\trx_5'}{\trend}};\trend \trrewind^\ast
              \trz{\trxcons{\trx_5}{\trwake{\widehat{\rho}}{\widehat{e}}}}{\trend};\tr'$
            \end{itemize}
          \item From $\fsc{\tr'}$, $\fsc{\widetilde{\tr_1}}$ and
            $\trlast{\tr'} = \trlast{\widetilde{\tr_1}}$ we have
            $\vec{\omega}' = \trlast{\tr'} =
            \trlast{\widetilde{\tr_1}} = \vec\omega$ by
            Lemma~\ref{lem:eval-vals}.
          \item By Lemma~\ref{lem:csa-preservation} we get
            $\csa{\widehat{\rho}}{\ilwake{\widehat{e}}}{\sigma_5}$ and
            thus
            $\sa{\widehat{\rho}}{\ilwake{\widehat{e}}}{\sigma_5}$.
          \item From
            Lemmas~\ref{lem:rewind-trace-context},~\ref{lem:prefix-dropum},~\ref{lem:noreuse-preservation},~\ref{lem:stack-surgery},
            and ~\ref{lem:traced-to-untraced} we have
            $\ilustate{\sigma_5}{\emptystack}{\widehat{\rho}}{\widehat{e}}
            \ilusteprel^\ast
            \ilustate{\sigma_5'}{\emptystack}{\emptyenv}{\vec\omega}$.
          \item Since also $
            \ilustate{\nongarbof{\widehat{\sigma_2'}}}{\emptystack}{\widehat{\rho}}{\widehat{e}}
            \ilusteprel^\ast
            \ilustate{\nongarbof{\widehat{\sigma_2}}}{\emptystack}{\emptyenv}{\vec{\nu}}
            $ we have that $\vec\nu = \vec\omega$ by definition of
            \SA.
          \end{itemize}            
          
        \item Second (and last) subcase:
          \begin{itemize}
          \item In this subcase, we have that:
            \begin{itemize}
            \item $
              \iltstate{\sigma_2}{\trz{\trxcons{\trx_2}{\trxpropmark{\widehat{\tr_1}}}}{\widetilde{\tr_1}}}{\kappa_2}{\emptyenv}{\ilprop}
              \iltsteprel^{m_1}
              \iltstate{\widehat{\sigma_2'}}{\trz{\trx''}{\tr'}}{\kappa_2}{\emptyenv}{\ilprop}
              \iltsteprel^{m_2}
              \iltstate{\widehat{\sigma_2}}{\trz{\widehat{\trx_2}}{\trend}}{\kappa_2}{\emptyenv}{\vec\nu}
              $
            \item $
              \trz{\dropum{\widehat{\trx_2}}}{\_};\_
              \trrewind^\ast
              \trz{\dropum{\trx''}{\_}};\_
              \trrewind^\ast
              \trz{\dropum{\trxcons{\trx_2}{\trxpropmark{\widehat{\tr_1}}}}}{\_};\_$
            \item $\trlast{\widetilde{\tr_1}} = \trlast{\tr'}$
            \item Reduction $m_2$ contains no use of {\bf P.E}
            \end{itemize}
            
          \item Applying Lemma~\ref{lem:prop-vals} to the reduction~$m_2$ we have that $\trlast{\tr'} = \vec\nu$.
          \item Putting this together, we have that $\trlast{\widetilde{\tr_1}} = \trlast{\tr'} = \vec\nu$.
          \item Finally, by applying Lemma~\ref{lem:eval-vals} to
            ``$\fsc{\widetilde{\tr_1}}$ from $\ldots$ producing $\vec{\omega}$'', 
            we have that $\trlast{\widetilde{\tr_1}} = \vec{\omega}$.
          \item Hence, $\vec{\omega} = \vec{\nu}$.        
          \end{itemize}
        \end{enumerate}

      \end{itemize}

    \item Case {\bf P.7}:
      \begin{itemize}
      \item Then $\tr_1' = \trpop{\vec{\nu_1}}$ and
        $\iltstate{\sigma_2}{\trz{\trx_2}{\tr_1'}}{\kappa_2}{\emptyenv}{\ilprop}
        \iltsteprel
        \iltstate{\sigma_2}{\trz{\trcons{\trx_2}{\trpop{\vec{\nu_1}}}}{\trend}}{\kappa_2}{\emptyenv}{\vec{\nu_1}}
        \iltsteprel^{n-1}
        \iltstate{\sigma'_2}{\trz{\trx'_2}{\tr_1''}}{\kappa_2}{\emptyenv}{\vec{\nu_2}}$.
      \item We show that $n-1 = 0$:
        \begin{itemize}
        \item Assume the contrary.  The only reduction rules that
          apply to
          $\iltstate{\sigma_2}{\trz{\trcons{\trx_2}{\trpop{\vec{\nu_1}}}}{\trend}}{\kappa_2}{\emptyenv}{\vec{\nu_1}}$
          are {\bf E.8} and {\bf P.8}.  We consider only the former
          case; the latter is analogous.
        \item Hence
          $\iltstate{\sigma_2}{\trz{\trcons{\trx_2}{\trpop{\vec{\nu_1}}}}{\trend}}{\kappa_2}{\emptyenv}{\vec{\nu_1}}
          \iltsteprel
          \iltstate{\sigma_2}{\trz{\trcons{\trx_2''}{\trpush{\tr}}}{\tr'}}{\kappa_2'}{\rho_f}{e_f}
          \iltsteprel^{n-2}
          \iltstate{\sigma'_2}{\trz{\trx'_2}{\tr_1''}}{\kappa_2}{\emptyenv}{\vec{\nu_2}}$
          where
          $\rewindstepm{\trz{\trcons{\trx_2}{\trpop{\vec{\nu_1}}}}{\trend}}{\trend}{\trz{\trcons{\trx_2''}{\trxpushmark}}{\tr'}}{\tr}$.
        \item By Lemma~\ref{lem:rewind-trace-context} we know
          $\trcons{\trx_2''}{\trxpushmark} \in
          \prefixes{\trcons{\trx_2}{\trpop{\vec{\nu_1}}}}$, i.e.,
          $\trcons{\trx_2''}{\trxpushmark} \in \prefixes{\trx_2}$.
        \item Hence $\dropum{\trx_2''},
          \dropum{\trxcons{\trx_2''}{\trxpushmark}} \in
          \prefixes{\dropum{\trx_2}} \subseteq
          \prefixes{\dropum{\trx_2'}}$ using
          Lemma~\ref{lem:rewind-trace-context}, (d), and
          Lemma~\ref{lem:prefix-dropum}.
        \item Using Lemma~\ref{lem:prefix-single-action} we get
          $\dropum{\trcons{\trx_2''}{\trpush{\tr}}} \in
          \prefixes{\dropum{\trx_2'}}$, contradicting
          $\dropum{\trcons{\trx_2''}{\trxpushmark}} \in
          \prefixes{\dropum{\trx_2'}}$.
        \end{itemize}
      \item Hence $\vec{\nu_1} = \vec{\nu_2}$ and $\sigma_2' =
        \sigma_2$ and $\trx_2' = \trcons{\trx_2}{\vec{\nu_2}}$.
      \item By inversion on (d) we get $\tr_2' = \vec{\nu_2}$.
      \item $\okay{\trz{\trxcons{\trx_2}{\vec{\nu_2}}}{\trend}}$
        follows from (b).
      \item By Lemma~\ref{lem:decomposition} and (c) we get
        $\iltstate{\sigma_1}{\trz{\trx_1}{\trend}}{\kappa_1}{\rho_1}{\rcmd_1}
        \iltsteprel^*
        \iltstate{\sigma_1}{\trz{\trx_1}{\trend}}{\kappa_1}{\rho_1'}{\ilpop{\vec{x}}}$
        using only {\bf E.0}, where $\rho_1'(\vec{x}) = \vec{\nu_1}$.
      \item Hence, using Lemma~\ref{lem:purity},
        $\iltstate{\nongarbof{\sigma_2}}{\trz{\trx_3}{\trend}}{\kappa_3}{\rho_1}{\rcmd_1}
        \iltsteprel^*
        \iltstate{\nongarbof{\sigma_2}}{\trz{\trx_3}{\trend}}{\kappa_3}{\rho_1'}{\ilpop{\vec{x}}}
        \iltsteprel
        \iltstate{\nongarbof{\sigma_2'}}{\trz{\trxcons{\trx_3}{\vec{\nu_2}}}{\trend}}{\kappa_3}{\emptyenv}{\nu_2}$.
      \item Finally,
        $\rewindstepm{\trz{\trcons{\trx_3}{\trpop{\vec{\nu_2}}}}{\trend}}{\trend}{\trz{\trx_3}{\trend}}{\vec{\nu_2}}$.
      \end{itemize}

    \item Case {\bf P.E}:
      \begin{itemize}
      \item Then $\tr_1' = \trcons{\trwake{\rho}{e}}{\widehat{\tr_1}}$
        and
        $\iltstate{\sigma_2}{\trz{\trx_2}{\tr_1'}}{\kappa_2}{\emptyenv}{\ilprop}
        \iltsteprel
        \iltstate{\sigma_2}{\trz{\trcons{\trx_2}{\trwake{\rho}{e}}}{\widehat{\tr_1}}}{\kappa_2}{\rho}{e}
        \iltsteprel^{n-1}
        \iltstate{\sigma'_2}{\trz{\trx'_2}{\tr_1''}}{\kappa_2}{\emptyenv}{\vec{\nu_2}}$.
      \item Hence
        $\rewindstepm{\trz{\dropum{\trx'_2}}{\_}}{\trend}{\trz{\dropum{\trxcons{\trx_2}{\trwake{\rho}{e}}}}{\_}}{\widehat{\tr_2}}
        \trrewind
        \trz{\dropum{\trx_2}}{\_};{\trcons{\trwake{\rho}{e}}{\widehat{\tr_2}}}$
        with $\tr_2' = \trcons{\trwake{\rho}{e}}{\widehat{\tr_2}}$ by
        Lemma~\ref{lem:rewind-single-action} and (d).
      \item By Lemma~\ref{lem:decomposition} and (a) we get:
        \begin{itemize}
        \item
          $\iltstate{\sigma_1}{\trz{\trx_1}{\trend}}{\kappa_1}{\rho_1}{\rcmd_1}
          \iltsteprel^*
          \iltstate{\sigma_1}{\trz{\trx_1}{\trend}}{\kappa_1}{\rho}{\ilwake{e}}$
          using {\bf E.0} only
        \item $\fsc{\widehat{\tr_1}}$ from
          $\iltstate{\sigma_1}{\trz{\trxcons{\trx_1}{\trwake{\rho}{e}}}{\trend}}{\kappa_1}{\rho}{e}$
          and thus $\okay{\widehat{\tr_1}}$
        \end{itemize}
      \item $\okay{\trxcons{\trx_2}{\trwake{\rho}{e}}}$ follows from
        $\okay{\trx_2}$.
      \item Induction and part (2) then yield:
        \begin{enumerate}[(i)]
        \item $\okay{\trz{\trx'_2}{\tr_1''}}$
        \item
          $\iltstate{\nongarbof{\sigma_2}}{\trz{\trxcons{\trx_3}{\trwake{\rho}{e}}}{\trend}}{\kappa_3}{\rho}{e}
          \iltsteprel^\ast
          \iltstate{\nongarbof{\sigma'_2}}{\trz{\trx'_3}{\trend}}{\kappa_3}{\emptyenv}{\vec{\nu_2}}$
        \item
          $\rewindstepm{\trz{\trx'_3}{\trend}}{\trend}{\trz{\trxcons{\trx_3}{\trwake{\rho}{e}}}{\trend}}{\widehat{\tr_2}}
          \trrewind \trz{\trx_3}{\trend};{\tr_2'}$
        \end{enumerate}
      \item Finally, using Lemma~\ref{lem:purity},
        $\iltstate{\nongarbof{\sigma_2}}{\trz{\trx_3}{\trend}}{\kappa_3}{\rho_1}{\rcmd_1}
        \iltsteprel^\ast
        \iltstate{\nongarbof{\sigma_2}}{\trz{\trxcons{\trx_3}{\trwake{\rho}{e}}}{\trend}}{\kappa_3}{\rho}{e}$.

      \end{itemize}

    \item Cases {\bf E.0--8}, {\bf E.P}, {\bf P.8}, {\bf U.1--4}: not possible
    \end{itemize}
    
  \item Case analysis on $n$.  First, we handle the simple case when $n = 0$:
    \begin{itemize}
    \item Since $n = 0$, we have that:
      \begin{itemize}
      \item $\trz{\trx_2}{\tr_1'} = \trz{\trx_2'}{\tr_1''}$
      \item $\sigma_2 = \sigma_2'$
      \item $\rho_2 = \epsilon$
      \item $\rcmd_2 = \vec \nu_2$
      \item $\tr_2' = \trend$, by inversion on (c)
      \end{itemize}
    \item $\okay{\trz{\trx_2'}{\tr_1''}}$ is given.
    \item Pick $\trx_3' = \trx_3$.
    \item Then reflexively we have that
      $\iltstate {\nongarbof{\sigma_2}} {\trz{\trx_3}{\trend}}
      {\kappa_3} {\rho_2} {\rcmd_2} \iltsteprel^\ast \iltstate
      {\nongarbof{\sigma'_2}} {\trz{\trx'_3}{\trend}} {\kappa_3}
      {\emptyenv} {\vec{\nu_2}}$.
    \item Similarly, reflexively we have that
      $\rewindstepm {\trz{\trx'_3}{\trend}} {\trend}
      {\trz{\trx_3}{\trend}} {\tr'_2}$.
    \end{itemize}
    
    For $n > 0$ we inspect the first reduction step of (a):
    \begin{itemize} 
      
    \item Case {\bf E.0}.
      \begin{itemize} 
      \item Then 
        $\iltstate 
        {\sigma_2} {\trz{\trx_2}{\tr'_1}} {\kappa_2}
        {\rho_2} {\iluexp}
        \iltsteprel
        \iltstate {\sigma_2} {\trz{\trx_2}{\tr'_1}} {\kappa_2}
        {\rho_2'} {\rcmd_2'}
        \iltsteprel^{n - 1} 
        \iltstate 
        {\sigma'_2} {\trz{\trx'_2}{\tr_1''}} {\kappa_2} 
        {\emptyenv} {\vec{\nu_2}}$
      \item Induction yields:
        \begin{enumerate}[(i)]
        \item $\okay{\trz{\trx'_2}{\tr_1''}}$
        \item $\iltstate {\nongarbof{\sigma_2}} {\trz{\trx_3}{\trend}}
          {\kappa_3} {\rho'_2} {\rcmd'_2} \iltsteprel^\ast \iltstate
          {\nongarbof{\sigma'_2}} {\trz{\trx'_3}{\trend}} {\kappa_3}
          {\emptyenv} {\vec{\nu_2}}$
        \item $\rewindstepm {\trz{\trx'_3}{\trend}} {\trend}
          {\trz{\trx_3}{\trend}} {\tr'_2}$
        \end{enumerate}
      \item Finally, using Lemma~\ref{lem:purity} and (ii) we have that
        $\iltstate 
        {\nongarbof{\sigma_2}} {\trz{\trx_3}{\trend}} {\kappa_3}
        {\rho_2} {\iluexp}
        \iltsteprel^\ast
        \iltstate {\nongarbof{\sigma_2}} {\trz{\trx_3}{\trend}}
        {\kappa_3} {\rho'_2} {\rcmd'_2} \iltsteprel^\ast \iltstate
        {\nongarbof{\sigma'_2}} {\trz{\trx'_3}{\trend}} {\kappa_3}
        {\emptyenv} {\vec{\nu_2}}$
      \end{itemize}

    \item Case {\bf E.1}:
      \begin{itemize}
      \item Then 
        $\iltstate 
        {\sigma_2} {\trz{\trx_2}{\tr'_1}} {\kappa_2}
        {\rho_2} {\illet{\ilvar{x}{\ilalloc{y}}}{e}}
        \iltsteprel
        \iltstate {\widehat{\sigma_2}}
        {\trz{\trxcons{\trx_2}{\tralloc{\ell}{m}}}{\tr'_1}} {\kappa_2}
        {\rho_2'} {e}
        \iltsteprel^{n - 1} 
        \iltstate 
        {\sigma'_2} {\trz{\trx'_2}{\tr_1''}} {\kappa_2} 
        {\emptyenv} {\vec{\nu_2}}$
        \\
        Where:
        \begin{itemize}
        \item $\storestin{\sigma_2}{\rho_2}{\ilalloc{y}} \storestep \storestout{\widehat{\sigma_2}}{\ell}$
        \item $\rho_2(y) = m$
        \item $\rho_2' = \rho_2[x \mapsto \ell]$
        \end{itemize}
      \item From $\okay{\trx_2}$ we have that $\okay{\trxcons{\trx_2}{\tralloc{\ell}{m}}}$
      \item By Lemma~\ref{lem:rewind-single-action} and (c) we have
        $\rewindstepm{\trz{\dropum{\trx'_2}}{\_}}{\trend}{\trz{\dropum{\trxcons{\trx_2}{\tralloc{\ell}{m}}}}{\_}}{\widehat{\tr_2}}
        \trrewind
        \trz{\dropum{\trx_2}}{\_};{\tr'_2}$
        with $\tr_2' = \trcons{\tralloc{\ell}{m}}{\widehat{\tr_2}}$ 
      \item By induction then:
        \begin{enumerate}[(i)]
        \item $\okay{\trz{\trx'_2}{\tr_1''}}$
        \item
          $\iltstate{\nongarbof{\widehat{\sigma_2}}}{\trz{\trxcons{\trx_3}{\tralloc{\ell}{m}}}{\trend}}{\kappa_3}{\rho_2'}{e}
          \iltsteprel^\ast
          \iltstate{\nongarbof{\sigma'_2}}{\trz{\trx'_3}{\trend}}{\kappa_3}{\emptyenv}{\vec{\nu_2}}$
        \item
          $\rewindstepm{\trz{\trx'_3}{\trend}}{\trend}{\trz{\trxcons{\trx_3}{\tralloc{\ell}{m}}}{\trend}}{\widehat{\tr_2}}
          \trrewind \trz{\trx_3}{\trend};{\tr_2'}$
        \end{enumerate}
      \item Hence,
        $\iltstate{\nongarbof{\sigma_2}}{\trz{\trx_3}{\trend}}{\kappa_3}{\rho_2}{\illet{\ilvar{x}{\ilalloc{y}}}{e}}
        \iltsteprel
        \iltstate{\nongarbof{\widehat{\sigma_2}}}{\trz{\trxcons{\trx_3}{\tralloc{\ell}{m}}}{\trend}}{\kappa_3}{\rho_2'}{e}
        \iltsteprel^\ast
        \iltstate{\nongarbof{\sigma'_2}}{\trz{\trx'_3}{\trend}}{\kappa_3}{\emptyenv}{\vec{\nu_2}}$
      \end{itemize}
      
    \item Case {\bf E.2}:
      \begin{itemize}
      \item Then 
        $\iltstate 
        {\sigma_2} {\trz{\trx_2}{\tr'_1}} {\kappa_2}
        {\rho_2} {\illet{\ilvar{x}{\ilread{y}{z}}}{e}}
        \iltsteprel
        \iltstate {\sigma_2}
        {\trz{\trxcons{\trx_2}{\trread{\nu}{\ell}{m}}}{\tr'_1}} {\kappa_2}
        {\rho_2'} {e}
        \iltsteprel^{n - 1} 
        \iltstate 
        {\sigma'_2} {\trz{\trx'_2}{\tr_1''}} {\kappa_2} 
        {\emptyenv} {\vec{\nu_2}}$
      \\Where:
        \begin{itemize}
        \item $\storestin{\sigma_2}{\rho_2}{\ilread{y}{z}} \storestep \storestout{\sigma_2}{\nu}$
        \item $\rho_2(y) = \ell$
        \item $\rho_2(z) = m$
        \item $\rho_2' = \rho_2[x \mapsto \nu]$
        \end{itemize}
      \item From $\okay{\trx_2}$ we have that $\okay{\trxcons{\trx_2}{\trread{\nu}{\ell}{m}}}$
      \item By Lemma~\ref{lem:rewind-single-action} and (c) we have
        $\rewindstepm{\trz{\dropum{\trx'_2}}{\_}}{\trend}{\trz{\dropum{\trxcons{\trx_2}{\trread{\nu}{\ell}{m}}}}{\_}}{\widehat{\tr_2}}
        \trrewind
        \trz{\dropum{\trx_2}}{\_};{\tr'_2}$
        with $\tr_2' = \trcons{\trread{\nu}{\ell}{m}}{\widehat{\tr_2}}$ 
      \item By induction then:
        \begin{enumerate}[(i)]
        \item $\okay{\trz{\trx'_2}{\tr_1''}}$
        \item
          $\iltstate{\nongarbof{\sigma_2}}{\trz{\trxcons{\trx_3}{\trread{\nu}{\ell}{m}}}{\trend}}{\kappa_3}{\rho_2'}{e}
          \iltsteprel^\ast
          \iltstate{\nongarbof{\sigma'_2}}{\trz{\trx'_3}{\trend}}{\kappa_3}{\emptyenv}{\vec{\nu_2}}$
        \item
          $\rewindstepm{\trz{\trx'_3}{\trend}}{\trend}{\trz{\trxcons{\trx_3}{\trread{\nu}{\ell}{m}}}{\trend}}{\widehat{\tr_2}}
          \trrewind \trz{\trx_3}{\trend};{\tr_2'}$
        \end{enumerate}
      \item Hence,
        $\iltstate{\nongarbof{\sigma_2}}{\trz{\trx_3}{\trend}}{\kappa_3}{\rho_2}{\illet{\ilvar{x}{\ilread{y}{z}}}{e}}
        \iltsteprel
        \iltstate{\nongarbof{\sigma_2}}{\trz{\trxcons{\trx_3}{\trread{\nu}{\ell}{m}}}{\trend}}{\kappa_3}{\rho_2'}{e}
        \iltsteprel^\ast
        \iltstate{\nongarbof{\sigma'_2}}{\trz{\trx'_3}{\trend}}{\kappa_3}{\emptyenv}{\vec{\nu_2}}$
      \end{itemize}

    \item Case {\bf E.3}:
      \begin{itemize}
      \item Then 
        $\iltstate 
        {\sigma_2} {\trz{\trx_2}{\tr'_1}} {\kappa_2}
        {\rho_2} {\illet{\ilvar{\_}{\ilwrite{x}{y}{z}}}{e}}
        \iltsteprel
        \iltstate {\widehat{\sigma_2}}
        {\trz{\trxcons{\trx_2}{\trwrite{\nu}{\ell}{m}}}{\tr'_1}} {\kappa_2}
        {\rho_2} {e}
        \iltsteprel^{n - 1} 
        \iltstate 
        {\sigma'_2} {\trz{\trx'_2}{\tr_1''}} {\kappa_2} 
        {\emptyenv} {\vec{\nu_2}}$
      \\ Where:
        \begin{itemize}
        \item $\storestin{\sigma_2}{\rho_2}{\ilwrite{x}{y}{z}} \storestep \storestout{\widehat{\sigma_2}}{0}$
        \item $\rho_2(x) = \ell$
        \item $\rho_2(y) = m$
        \item $\rho_2(z) = \nu$
        \end{itemize}
      \item From $\okay{\trx_2}$ we have that $\okay{\trxcons{\trx_2}{\trwrite{\nu}{\ell}{m}}}$
      \item By Lemma~\ref{lem:rewind-single-action} and (c) we have
        $\rewindstepm{\trz{\dropum{\trx'_2}}{\_}}{\trend}{\trz{\dropum{\trxcons{\trx_2}{\trwrite{\nu}{\ell}{m}}}}{\_}}{\widehat{\tr_2}}
        \trrewind
        \trz{\dropum{\trx_2}}{\_};{\tr'_2}$
        with $\tr_2' = \trcons{\trwrite{\nu}{\ell}{m}}{\widehat{\tr_2}}$ 
      \item By induction then:
        \begin{enumerate}[(i)]
        \item $\okay{\trz{\trx'_2}{\tr_1''}}$
        \item
          $\iltstate{\nongarbof{\widehat{\sigma_2}}}{\trz{\trxcons{\trx_3}{\trwrite{\nu}{\ell}{m}}}{\trend}}{\kappa_3}{\rho_2}{e}
          \iltsteprel^\ast
          \iltstate{\nongarbof{\sigma'_2}}{\trz{\trx'_3}{\trend}}{\kappa_3}{\emptyenv}{\vec{\nu_2}}$
        \item
          $\rewindstepm{\trz{\trx'_3}{\trend}}{\trend}{\trz{\trxcons{\trx_3}{\trwrite{\nu}{\ell}{m}}}{\trend}}{\widehat{\tr_2}}
          \trrewind \trz{\trx_3}{\trend};{\tr_2'}$
        \end{enumerate}
      \item Hence,
        $\iltstate{\nongarbof{\sigma_2}}{\trz{\trx_3}{\trend}}{\kappa_3}{\rho_2}{\illet{\ilvar{\_}{\ilwrite{x}{y}{z}}}{e}}
        \iltsteprel
        \iltstate{\nongarbof{\widehat{\sigma_2}}}{\trz{\trxcons{\trx_3}{\trwrite{\nu}{\ell}{m}}}{\trend}}{\kappa_3}{\rho_2}{e}
        \iltsteprel^\ast
        \iltstate{\nongarbof{\sigma'_2}}{\trz{\trx'_3}{\trend}}{\kappa_3}{\emptyenv}{\vec{\nu_2}}$
      \end{itemize}      
      
    \item Case {\bf E.4}:
      \begin{itemize}
      \item Then 
        $\iltstate 
        {\sigma_2} {\trz{\trx_2}{\tr'_1}} {\kappa_2}
        {\rho_2} {\ilmemo{e}}
        \iltsteprel
        \iltstate {\sigma_2}
        {\trz{\trxcons{\trx_2}{\trmemo{\rho_2}{e}}}{\tr'_1}} {\kappa_2}
        {\rho_2} {e}
        \iltsteprel^{n - 1} 
        \iltstate 
        {\sigma'_2} {\trz{\trx'_2}{\tr_1''}} {\kappa_2} 
        {\emptyenv} {\vec{\nu_2}}$
      \item From $\okay{\trx_2}$ we have that $\okay{\trxcons{\trx_2}{\trmemo{\rho_2}{e}}}$
      \item By Lemma~\ref{lem:rewind-single-action} and (c) we have
        $\rewindstepm{\trz{\dropum{\trx'_2}}{\_}}{\trend}{\trz{\dropum{\trxcons{\trx_2}{\trmemo{\rho_2}{e}}}}{\_}}{\widehat{\tr_2}}
        \trrewind
        \trz{\dropum{\trx_2}}{\_};{\tr'_2}$
        with $\tr_2' = \trcons{\trmemo{\rho_2}{e}}{\widehat{\tr_2}}$ 
      \item By induction then:
        \begin{enumerate}[(i)]
        \item $\okay{\trz{\trx'_2}{\tr_1''}}$
        \item
          $\iltstate{\nongarbof{\sigma_2}}{\trz{\trxcons{\trx_3}{\trmemo{\rho_2}{e}}}{\trend}}{\kappa_3}{\rho_2}{e}
          \iltsteprel^\ast
          \iltstate{\nongarbof{\sigma'_2}}{\trz{\trx'_3}{\trend}}{\kappa_3}{\emptyenv}{\vec{\nu_2}}$
        \item
          $\rewindstepm{\trz{\trx'_3}{\trend}}{\trend}{\trz{\trxcons{\trx_3}{\trmemo{\rho_2}{e}}}{\trend}}{\widehat{\tr_2}}
          \trrewind \trz{\trx_3}{\trend};{\tr_2'}$
        \end{enumerate}
      \item Hence,
        $\iltstate{\nongarbof{\sigma_2}}{\trz{\trx_3}{\trend}}{\kappa_3}{\rho_2}{\ilmemo{e}}
        \iltsteprel
        \iltstate{\nongarbof{\sigma_2}}
        {\trz{\trxcons{\trx_3}{\trmemo{\rho_2}{e}}}{\trend}}
        {\kappa_3}{\rho_2}{e}
        \iltsteprel^\ast
        \iltstate{\nongarbof{\sigma'_2}}{\trz{\trx'_3}{\trend}}{\kappa_3}{\emptyenv}{\vec{\nu_2}}$
      \end{itemize}      

    \item Case {\bf E.5}:
      \begin{itemize}
      \item Then 
        $\iltstate 
        {\sigma_2} {\trz{\trx_2}{\tr'_1}} {\kappa_2}
        {\rho_2} {\ilupdate{e}}
        \iltsteprel
        \iltstate {\sigma_2}
        {\trz{\trxcons{\trx_2}{\trupdate{\rho_2}{e}}}{\tr'_1}} {\kappa_2}
        {\rho_2} {e}
        \iltsteprel^{n - 1} 
        \iltstate 
        {\sigma'_2} {\trz{\trx'_2}{\tr_1''}} {\kappa_2} 
        {\emptyenv} {\vec{\nu_2}}$
      \item From $\okay{\trx_2}$ we have that $\okay{\trxcons{\trx_2}{\trupdate{\rho_2}{e}}}$
      \item By Lemma~\ref{lem:rewind-single-action} and (c) we have
        $\rewindstepm{\trz{\dropum{\trx'_2}}{\_}}{\trend}{\trz{\dropum{\trxcons{\trx_2}{\trupdate{\rho_2}{e}}}}{\_}}{\widehat{\tr_2}}
        \trrewind
        \trz{\dropum{\trx_2}}{\_};{\tr'_2}$
        with $\tr_2' = \trcons{\trupdate{\rho_2}{e}}{\widehat{\tr_2}}$ 
      \item By induction then:
        \begin{enumerate}[(i)]
        \item $\okay{\trz{\trx'_2}{\tr_1''}}$
        \item
          $\iltstate{\nongarbof{\sigma_2}}{\trz{\trxcons{\trx_3}{\trupdate{\rho_2}{e}}}{\trend}}{\kappa_3}{\rho_2}{e}
          \iltsteprel^\ast
          \iltstate{\nongarbof{\sigma'_2}}{\trz{\trx'_3}{\trend}}{\kappa_3}{\emptyenv}{\vec{\nu_2}}$
        \item
          $\rewindstepm{\trz{\trx'_3}{\trend}}{\trend}{\trz{\trxcons{\trx_3}{\trupdate{\rho_2}{e}}}{\trend}}{\widehat{\tr_2}}
          \trrewind \trz{\trx_3}{\trend};{\tr_2'}$
        \end{enumerate}
      \item Hence,
        $\iltstate{\nongarbof{\sigma_2}}{\trz{\trx_3}{\trend}}{\kappa_3}{\rho_2}{\ilupdate{e}}
        \iltsteprel
        \iltstate{\nongarbof{\sigma_2}}
        {\trz{\trxcons{\trx_3}{\trupdate{\rho_2}{e}}}{\trend}}
        {\kappa_3}{\rho_2}{e}
        \iltsteprel^\ast
        \iltstate{\nongarbof{\sigma'_2}}{\trz{\trx'_3}{\trend}}{\kappa_3}{\emptyenv}{\vec{\nu_2}}$
      \end{itemize}      

    \item Case {\bf E.6}:
      \begin{itemize}
      \item Then $\iltstate {\sigma_2} {\trz{\trx_2}{\tr'_1}}
        {\kappa_2} {\rho_2} {\ilpush{f}{e}} \iltsteprel \iltstate
        {\sigma_2} {\trz{\trxcons{\trx_2}{\trxpushmark}}{\tr'_1}}
        {\stcons{\kappa_2}{\mkstframe{\rho_2}{f}}} {\rho_2} {e} \iltsteprel^{n - 1} \iltstate
        {\sigma'_2} {\trz{\trx'_2}{\tr_1''}} {\kappa_2} {\emptyenv}
        {\vec{\nu_2}}$
      \item Note that also
        $\iltstate{\nongarbof{\sigma_2}}{\trz{\trx_3}{\trend}}{\kappa_3}{\rho_2}{\ilpush{f}{e}}
        \iltsteprel
        \iltstate{\nongarbof{\sigma_2}}{\trz{\trxcons{\trx_3}{\trxpushmark}}{\trend}}{\stcons{\kappa_3}{\mkstframe{\rho_2}{f}}}{\rho_2}{e}$.
      \item By Lemma~\ref{lem:pushmark-decomposition} the $n-1$
        reduction above decomposes as follows:
        \begin{itemize}
        \item $\iltstate {\sigma_2}
          {\trz{\trxcons{\trx_2}{\trxpushmark}}{\tr'_1}}
          {\stcons{\kappa_2}{\mkstframe{\rho_2}{f}}} {\rho_2} {e}
          \iltsteprel^{n_1}
          \iltstate{\widehat{\sigma_2}}{\trz{\widehat{\trx_2}}{\widehat{\tr'_1}}}{\stcons{\kappa_2}{\mkstframe{\rho_2}{f}}}{\emptyenv}{\vec{\nu}}$
        \item
          $\rewindstepm{\trz{\widehat{\trx_2}}{\widehat{\tr'_1}}}{\trend}{\trz{\trxcons{\trx_2}{\trxpushmark}}{\widetilde{\tr'_1}}}{\tr}$
        \item
          $\iltstate{\widehat{\sigma_2}}{\trz{\widehat{\trx_2}}{\widehat{\tr'_1}}}{\stcons{\kappa_2}{\mkstframe{\rho_2}{f}}}{\emptyenv}{\vec{\nu}}
          \iltsteprel
          \iltstate{\widehat{\sigma_2}}{\trz{\trxcons{\trx_2}{\trpush{\tr}}}{\widetilde{\tr'_1}}}{\kappa_2}{\widehat{\rho_2}}{e_f}$
        \item
          $\iltstate{\widehat{\sigma_2}}{\trz{\trxcons{\trx_2}{\trpush{\tr}}}{\widetilde{\tr'_1}}}{\kappa_2}{\widehat{\rho_2}}{e_f}
          \iltsteprel^{n_2} \iltstate {\sigma'_2}
          {\trz{\trx'_2}{\tr_1''}} {\kappa_2} {\emptyenv}
          {\vec{\nu_2}}$
        \item $n = n_1 + 1 + n_2$
        \end{itemize}
      \item From $\okay{\trz{\trx_2}{\tr'_1}}$ we get
        $\okay{\trz{\trxcons{\trx_2}{\trxpushmark}}{\tr'_1}}$.
      \item From
        $\rewindstepm{\trz{\widehat{\trx_2}}{\widehat{\tr'_1}}}{\trend}{\trz{\trxcons{\trx_2}{\trxpushmark}}{\widetilde{\tr'_1}}}{\tr}$
        follows
        $\rewindstepm{\trz{\dropum{\widehat{\trx_2}}}{\_}}{\trend}{\trz{\dropum{\trxcons{\trx_2}{\trxpushmark}}}{\_}}{\tr}$
        by Lemma~\ref{lem:rewind-dropum}.
      \item Hence induction with $n_1$ yields:
        \begin{itemize}
        \item $\okay{\trz{\widehat{\trx_2}}{\widehat{\tr'_1}}}$
        \item
          $\iltstate{\nongarbof{\sigma_2}}{\trz{\trxcons{\trx_3}{\trxpushmark}}{\trend}}{\stcons{\kappa_3}{\mkstframe{\rho_2}{f}}}{\rho_2}{e}
          \iltsteprel^\ast
          \iltstate{\nongarbof{\widehat{\sigma_2}}}{\trz{\trx_3''}{\trend}}{\stcons{\kappa_3}{\mkstframe{\rho_2}{f}}}{\emptyenv}{\vec{\nu}}$
        \item
          $\rewindstepm{\trz{\trx_3''}{\trend}}{\trend}{\trz{\trxcons{\trx_3}{\trxpushmark}}{\trend}}{\tr}$
        \end{itemize}
      \item Note that
        $\iltstate{\nongarbof{\widehat{\sigma_2}}}{\trz{\trx_3''}{\trend}}{\stcons{\kappa_3}{\mkstframe{\rho_2}{f}}}{\emptyenv}{\vec{\nu}}
        \iltsteprel
        \iltstate{\nongarbof{\widehat{\sigma_2}}}{\trz{\trxcons{\trx_3}{\trpush{\tr}}}{\trend}}{\kappa_3}{\widehat{\rho_2}}{e_f}$.
      \item
        $\okay{\trz{\trxcons{\trx_2}{\trpush{\tr}}}{\widetilde{\tr'_1}}}$
        follows from $\okay{\trz{\widehat{\trx_2}}{\widehat{\tr'_1}}}$
        by Lemma~\ref{lem:rewind-okay}.
      \item By Lemma~\ref{lem:rewind-single-action} and (c) we have
        $\rewindstepm{\trz{\dropum{\trx'_2}}{\_}}{\trend}{\trz{\dropum{\trxcons{\trx_2}{\trpush{\tr}}}}{\_}}{\widehat{\tr_2}}
        \trrewind \trz{\dropum{\trx_2}}{\_};{\tr'_2}$ with $\tr_2' =
        \trcons{\trpush{\tr}}{\widehat{\tr_2}}$
      \item So induction with $n_2$ yields:
        \begin{itemize}
        \item $\okay{\trz{\widehat{\trx_2'}}{\tr''_1}}$
        \item
          $\iltstate{\nongarbof{\widehat{\sigma_2}}}{\trz{\trxcons{\trx_3}{\trpush{\tr}}}{\trend}}{\kappa_3}{\widehat{\rho_2}}{e_f}
          \iltsteprel^\ast
          \iltstate{\nongarbof{\sigma_2'}}{\trz{\trx_3'}{\trend}}{\kappa_3}{\emptyenv}{\vec{\nu_2}}$
        \item
          $\rewindstepm{\trz{\trx_3'}{\trend}}{\trend}{\trz{\trxcons{\trx_3}{\trpush{\tr}}}{\trend}}{\widehat{\tr_2}}
          \trrewind \trz{\trx_3}{\trend};{\tr_2'}$
        \end{itemize}
      \item Finally,
        $\iltstate{\nongarbof{\sigma_2}}{\trz{\trx_3}{\trend}}{\kappa_3}{\rho_2}{\ilpush{f}{e}}
        \iltsteprel^\ast
        \iltstate{\nongarbof{\sigma_2'}}{\trz{\trx_3'}{\trend}}{\kappa_3}{\emptyenv}{\vec{\nu_2}}$
        by putting the pieces together.
      \end{itemize}
    \item Case {\bf E.7} 
      \begin{itemize}
      \item Then 
        $\iltstate 
        {\sigma_2} {\trz{\trx_2}{\tr'_1}} {\kappa_2}
        {\rho_2} {\ilpop{\vec{x}}}
        \iltsteprel
        \iltstate {\sigma_2}
        {\trz{\trxcons{\trx_2}{\trpop{\vec{\nu}}}}{\tr'_1}} {\kappa_2}
        {\emptyenv} {\vec{\nu}}
        \iltsteprel^{n - 1} 
        \iltstate 
        {\sigma'_2} {\trz{\trx'_2}{\tr_1''}} {\kappa_2} 
        {\emptyenv} {\vec{\nu_2}}$
      \\ Where: $\fromto{\rho_2(x_i)}{1}{|\vec{x}|} = \vec{\nu}$
      \item From $\okay{\trx_2}$ we have that $\okay{\trxcons{\trx_2}{\trpop{\vec{\nu}}}}$
      \item By Lemma~\ref{lem:rewind-single-action} and (c) we have
        $\rewindstepm{\trz{\dropum{\trx'_2}}{\_}}{\trend}{\trz{\dropum{\trxcons{\trx_2}{\trpop{\vec{\nu}}}}}{\_}}{\widehat{\tr_2}}
        \trrewind
        \trz{\dropum{\trx_2}}{\_};{\tr'_2}$
        with $\tr_2' = \trcons{\trpop{\vec{\nu}}}{\widehat{\tr_2}}$ 
      \item By induction then:
        \begin{enumerate}[(i)]
        \item $\okay{\trz{\trx'_2}{\tr_1''}}$
        \item
          $\iltstate{\nongarbof{\sigma_2}}{\trz{\trxcons{\trx_3}{\trpop{\vec{\nu}}}}{\trend}}{\kappa_3}{\emptyenv}{\vec{\nu}}
          \iltsteprel^\ast
          \iltstate{\nongarbof{\sigma'_2}}{\trz{\trx'_3}{\trend}}{\kappa_3}{\emptyenv}{\vec{\nu_2}}$
        \item
          $\rewindstepm{\trz{\trx'_3}{\trend}}{\trend}{\trz{\trxcons{\trx_3}{\trpop{\vec{\nu}}}}{\trend}}{\widehat{\tr_2}}
          \trrewind \trz{\trx_3}{\trend};{\tr_2'}$
        \end{enumerate}
      \item Hence,
        $\iltstate{\nongarbof{\sigma_2}}{\trz{\trx_3}{\trend}}{\kappa_3}{\rho_2}{\ilpop{\vec{x}}}
        \iltsteprel
        \iltstate{\nongarbof{\sigma_2}}
        {\trz{\trxcons{\trx_3}{\trpop{\vec{\nu}}}}{\trend}}
        {\kappa_3}{\emptyenv} {\vec{\nu}}
        \iltsteprel^\ast
        \iltstate{\nongarbof{\sigma'_2}}{\trz{\trx'_3}{\trend}}{\kappa_3}{\emptyenv}{\vec{\nu_2}}$
      \end{itemize}      

    \item Case {\bf E.8}: We show that this case does not arise.
      \begin{itemize}
      \item Then
        \begin{itemize}
        \item
          $\iltstate{\sigma_2}{\trz{\trx_2}{\tr_1'}}
          {\stcons{\widehat{\kappa_2}}{\mkstframe{\widehat{\rho}}{f}}}
          {\epsilon}{\vec{\nu}}
          \iltsteprel
          \iltstate{\sigma_2}{\trz{\trcons{\widehat{\trx_2}}{\trpush{\tr_3}}}{\widehat{\tr_1'}}}{\widehat{\kappa_2}}{\_}{\_}
          \iltsteprel^{n-1}
          \iltstate{\sigma'_2}{\trz{\trx'_2}{\tr_1''}}
          {\stcons{\widehat{\kappa_2}}{\mkstframe{\widehat{\rho}}{f}}}
          {\emptyenv}{\vec{\nu_2}}$        
        \item $\rewindstepm{\trz{\dropum{\trx_2}}{\tr_1'}}{\trend}{\trz{\dropum{\trxcons{\widehat{\trx_2}}{\trxpushmark}}}{\widehat{\tr_1'}}}{\tr_3}$
      \end{itemize}
      \item By Lemmas~\ref{lem:prefix-single-action}
        and~\ref{lem:rewind-trace-context} we get both
        \begin{itemize}
        \item $\dropum{\trxcons{\widehat{\trx_2}}{\trxpushmark}} \in \prefixes{\dropum{\trx_2'}}$ 
        \item $\dropum{\trxcons{\widehat{\trx_2}}{\trpush{\tr_3}}} \in \prefixes{\dropum{\trx_2'}}$.
        \end{itemize}
      \item This is a contradiction and thus rules out this case.
      \end{itemize}

    \item Case {\bf P.8}: We show that this case does not arise.
      \begin{itemize}
      \item Then
        \begin{itemize}
        \item
          $\iltstate{\sigma_2}{\trz{\trx_2}{\tr_1'}}{\kappa_2}{\rho_2}{\rcmd_2}
          \iltsteprel
          \iltstate{\sigma_2}{\trz{\trcons{\widehat{\trx_2}}{\trpush{\tr_4}}}{\tr_3}}{\kappa_2}{\emptyenv}{\ilprop}
          \iltsteprel^{n-1}
          \iltstate{\sigma'_2}{\trz{\trx'_2}{\tr_1''}}{\kappa_2}{\emptyenv}{\vec{\nu_2}}$
        \item $\rewindstepm{\trz{\dropum{\trx_2}}{\trend}}{\trend}{\trz{\dropum{\trxcons{\widehat{\trx_2}}{\trxpropmark{\tr_3}}}}{\trend}}{\tr_4}$
      \end{itemize}
      \item By Lemmas~\ref{lem:prefix-single-action}
        and~\ref{lem:rewind-trace-context} we get both
        \begin{itemize}
        \item $\dropum{\trxcons{\widehat{\trx_2}}{\trxpropmark{\tr_3}}} \in \prefixes{\dropum{\trx_2'}}$ 
        \item $\dropum{\trxcons{\widehat{\trx_2}}{\trpush{\tr_4}}} \in \prefixes{\dropum{\trx_2'}}$.
        \end{itemize}
      \item This is a contradiction and thus rules out this case.
      \end{itemize}

    \item Case {\bf E.P} 
      \begin{itemize}
      \item Then $\tr_1' = \trcons{\trmemo{\rho_2}{e}}{\widehat{\tr_1}}$
        and
        $\iltstate{\sigma_2}{\trz{\trx_2}{\tr_1'}}{\kappa_2}{\rho_2}{\ilmemo e}
        \iltsteprel
        \iltstate 
        {\sigma_2}{\trz{\trxcons{\trx_2}{\trmemo{\rho_2}{e}}}{\widehat{\tr_1}}}{\kappa_2}{\rho_2}{e}
        \iltsteprel^{n-1}
        \iltstate{\sigma'_2}{\trz{\trx'_2}{\tr_1''}}{\kappa_2}{\emptyenv}{\vec{\nu_2}}$
      
      \item Hence
        $\rewindstepm{\trz{\dropum{\trx'_2}}{\_}}{\trend}{\trz{\trxcons{\dropum{\trx_2}}{\trmemo{\rho_2}{e}}}{\_}}{\widehat{\tr_2}}
        \trrewind
        \trz{\dropum{\trx_2}}{\_};{\tr_2'}$
        with $\tr_2' = \trcons{\trmemo{\rho}{e}}{\widehat{\tr_2}}$ by
        Lemma~\ref{lem:rewind-single-action} and (c).
      \item From (b) we have that $\okay{\trcons{\trmemo{\rho_2}{e}}{\widehat{\tr_1}}}$,
        and by inversion we have that $\fsc{\trcons{\trmemo{\rho_2}{e}}{\widehat{\tr_1}}}$.
      \item Hence, by Lemma~\ref{lem:decomposition} we know there exists some components~$\trx_1, \sigma_1, \kappa_1, \rho_1, \rcmd_1$ such that
        \begin{itemize}
        \item
          $\iltstate{\sigma_1}{\trz{\trx_1}{\trend}}{\kappa_1}{\rho_1}{\rcmd_1}
          \iltsteprel^*
          \iltstate{\sigma_1}{\trz{\trx_1}{\trend}}{\kappa_1}{\rho_2}{\ilmemo{e}}$
          using {\bf E.0} only
        \item $\fsc{\widehat{\tr_1}}$ from
          $\iltstate{\sigma_1}{\trz{\trxcons{\trx_1}{\trmemo{\rho_2}{e}}}{\trend}}{\kappa_1}{\rho_2}{e}$
        \end{itemize}
      \item $\okay{\trxcons{\trx_2}{\trmemo{\rho_2}{e}}}$ follows from $\okay{\trx_2}$.
      \item Induction and part (1) then yield:
        \begin{enumerate}[(i)]
        \item $\okay{\trz{\trx'_2}{\tr_1''}}$
        \item
          $\iltstate{\nongarbof{\sigma_2}}{\trz{\trxcons{\trx_3}{\trmemo{\rho_2}{e}}}{\trend}}{\kappa_3}{\rho_2}{e}
          \iltsteprel^\ast
          \iltstate{\nongarbof{\sigma'_2}}{\trz{\trx'_3}{\trend}}{\kappa_3}{\emptyenv}{\vec{\nu_2}}$
        \item
          $\rewindstepm{\trz{\trx'_3}{\trend}}{\trend}{\trz{\trxcons{\trx_3}{\trmemo{\rho_2}{e}}}{\trend}}{\widehat{\tr_2}}
          \trrewind \trz{\trx_3}{\trend};{\tr_2'}$
        \end{enumerate}
      \item Finally, using Lemma~\ref{lem:purity},
        $\iltstate{\nongarbof{\sigma_2}}{\trz{\trx_3}{\trend}}{\kappa_3}{\rho_1}{\rcmd_1}
        \iltsteprel^\ast
        \iltstate{\nongarbof{\sigma_2}}{\trz{\trxcons{\trx_3}{\trmemo{\rho_2}{e}}}{\trend}}{\kappa_3}{\rho_2}{e}$.
      \end{itemize}

    \item Case {\bf U.1}
      \begin{itemize}
      \item Then
        $\iltstate 
        {\sigma_2} {\trz{\trx_2}{\trcons{\tralloc{\ell}{m}}{\widehat{\tr'_1}}}} {\kappa_2}
        {\rho_2} {\rcmd_2}
        \- \iltsteprel
        \iltstate {\sigma_2[\ell \mapsto \storegarb]} {\trz{\trx_2}{\widehat{\tr'_1}}} {\kappa_2}
        {\rho_2} {\rcmd_2}
        \iltsteprel^{n - 1} 
        \iltstate 
        {\sigma'_2} {\trz{\trx'_2}{\tr_1''}} {\kappa_2} 
        {\emptyenv} {\vec{\nu_2}}$ 
      \item By inversion on (b) we have that $\fsc{\trcons{\tralloc{\ell}{m}}{\widehat{\tr'_1}}}$
      \item By Lemma~\ref{lem:decomposition}, we have that~$\fsc{\widehat{\tr'_1}}$
      \item Hence, $\okay{\widehat{\tr'_1}}$
      \item Induction yields:
        \begin{enumerate}[(i)]
        \item $\okay{\trz{\trx'_2}{\tr_1''}}$
        \item $\iltstate {\nongarbof{\sigma_2[\storegarb \mapsto \ell]}} {\trz{\trx_3}{\trend}}
          {\kappa_3} {\rho_2} {\rcmd_2} \iltsteprel^\ast \iltstate
          {\nongarbof{\sigma'_2}} {\trz{\trx'_3}{\trend}} {\kappa_3}
          {\emptyenv} {\vec{\nu_2}}$
        \item $\rewindstepm {\trz{\trx'_3}{\trend}} {\trend}
          {\trz{\trx_3}{\trend}} {\tr'_2}$
        \end{enumerate}
      \item By definition, $\nongarbof{\sigma_2[\ell \mapsto \storegarb]} = \nongarbof{\sigma_2}$
      \item Hence, 
        $\iltstate {\nongarbof{\sigma_2}} {\trz{\trx_3}{\trend}}
        {\kappa_3} {\rho_2} {\rcmd_2} \iltsteprel^\ast \iltstate
        {\nongarbof{\sigma'_2}} {\trz{\trx'_3}{\trend}} {\kappa_3}
        {\emptyenv} {\vec{\nu_2}}$
      \end{itemize}

    \item Case {\bf U.2}
      \begin{itemize}
      \item Then $\iltstate {\sigma_2}
        {\trz{\trx_2}{\trcons{t}{\widehat{\tr'_1}}}} {\kappa_2}
        {\rho_2} {\rcmd_2} \- \iltsteprel \iltstate {\sigma_2}
        {\trz{\trx_2}{\widehat{\tr'_1}}} {\kappa_2} {\rho_2}
        {\rcmd_2} \iltsteprel^{n - 1} \iltstate {\sigma'_2}
        {\trz{\trx'_2}{\tr_1''}} {\kappa_2} {\emptyenv} {\vec{\nu_2}}$
      \item By inversion on (b), with the knowledge that $t \ne \trpush{\_}$, we have that $\fsc{\trcons{t}{\widehat{\tr'_1}}}$
      \item By Lemma~\ref{lem:decomposition}, we have that~$\fsc{\widehat{\tr'_1}}$
      \item Hence, $\okay{\widehat{\tr'_1}}$
      \item The claim then follows by induction.
      \end{itemize}

    \item Case {\bf U.3}
      \begin{itemize}
      \item Then $\iltstate {\sigma_2}
        {\trz{\trx_2}{\trcons{\trpush{\widehat{\tr'_1}}}{\widehat{\tr'_2}}}}
        {\kappa_2} {\rho_2} {\rcmd_2} \- \iltsteprel \iltstate
        {\sigma_2}
        {\trz{\trxcons{\trx_2}{\trxundomark{\widehat{\tr'_2}}}}{\widehat{\tr'_1}}}
        {\kappa_2} {\rho_2} {\rcmd_2} \iltsteprel^{n - 1} \iltstate
        {\sigma'_2} {\trz{\trx'_2}{\tr_1''}} {\kappa_2} {\emptyenv}
        {\vec{\nu_2}}$
      \item By inversion of (b) we show that $\okay{\widehat{\tr'_1}}$ and $\okay{\widehat{\tr'_2}}$:
        \begin{itemize}
        \item Subcase: $\okay{\widehat{\tr'_1}}$ and $\okay{\widehat{\tr'_2}}$.
          \begin{itemize}
          \item Immediate.
          \end{itemize}
        \item Subcase:
          $\fsc{\trcons{\trpush{\widehat{\tr'_1}}}{\widehat{\tr'_2}}}$.
          \begin{itemize}
          \item From Lemma~\ref{lem:decomposition} we have
            $\fsc{\widehat{\tr'_1}}$ and $\fsc{\widehat{\tr'_2}}$.
          \item The claim then follows immediately.
          \end{itemize}
        \end{itemize}
      \item Hence from (b), we have $\okay{\trxcons{\trx_2}{\trxundomark{\widehat{\tr'_2}}}}$
      \item Note that $\dropum{\trxcons{\trx_2}{\trxundomark{\widehat{\tr'_2}}}} = \dropum{\trx_2}$
      \item Hence, from (c) we have
        $\rewindstepm{\trz{\dropum{\trx'_2}}{\_}}{\trend} 
        {\trz{\dropum{\trxcons{\trx_2}{\trxundomark{\widehat{\tr_2'}}}}}{\_}}{\tr'_2}$
      \item The claim then follows by induction.
      \end{itemize}
 
    \item Case {\bf U.4} 
      \begin{itemize}
      \item Then
        $\iltstate 
        {\sigma_2}
        {\trz{\trxcons{\widehat{\trx_2}}{\trxundomark{\widehat{\tr}}}}{\trend}}
        {\kappa_2} {\rho_2} {\rcmd_2}
        \- \iltsteprel
        \iltstate {\sigma_2}         
        {\trz{\widehat{\trx_2}}{\widehat{\tr}}}
        {\kappa_2} {\rho_2} {\rcmd_2}
        \iltsteprel^{n - 1} 
        \iltstate 
        {\sigma'_2} {\trz{\trx'_2}{\tr_1''}} {\kappa_2} 
        {\emptyenv} {\vec{\nu_2}}$ 
      \item From (b) we have both $\okay{\widehat{\trx_2}}$ and $\okay{\widehat{\tr}}$
      \item Note that $\dropum{\trxcons{\widehat{\trx_2}}{\trxundomark{\widehat{\tr}}}} = \dropum{\widehat{\trx_2}}$
      \item Hence, from (c) we have $\rewindstepm{\trz{\dropum{\trx'_2}}{\_}}{\trend} {\trz{\dropum{\widehat{\trx_2}}}{\_}}{\tr'_2}$
      \item The claim then follows by induction.
      \end{itemize}      
    \end{itemize}
  \end{enumerate}
\end{proof}

\begin{defn}[Big-step Sugar]
  \begin{mathpar}
    \inferrule*[right=BigEval]{
      \iltstepm
      {\trz{\trstart}{\tr}}{\sigma}{\epsilon}{\rho\statesep\rcmd}
      {\trz{\trx}{\trend}}{\sigma'}{\epsilon}{\epsilon\statesep\vec\nu}
      \\\\
      \rewindstepm{\trz{\trx}{\trend}}{\trend}{\trz{\trstart}{\trend}}{\tr'}
    }
    {
      \iltbigstep{\tr}{\sigma}{\rho\statesep\rcmd}{\tr'}{\sigma'}{\vec\nu}
    }
    \and
    \inferrule*[right=BigProp]{
      \iltstepm
      {\trz{\trstart}{\tr}}{\sigma}{\epsilon}{\epsilon\statesep\ilprop}
      {\trz{\trx}{\trend}}{\sigma'}{\epsilon}{\epsilon\statesep\vec\nu}
      \\\\
      \rewindstepm{\trz{\trx}{\trend}}{\trend}{\trz{\trstart}{\trend}}{\tr'}
    }
    {
      \iltbigprop{\tr}{\sigma}{\tr'}{\sigma'}{\vec\nu}
    }
  \end{mathpar}
\end{defn}

\begin{cor}[Big-step Consistency]
  Suppose~$\iltbigstep{\trend}{\sigma_1}{\rho_1\statesep\rcmd_1}{\tr_1}{\sigma_1'}{\vec\nu_1}$
  and $\csa{\rho_1}{\rcmd_1}{\sigma_1}$.
  
  \begin{enumerate}
  
  \item
    If~$\iltbigstep{\tr_1}{\sigma_2}{\rho_2\statesep\rcmd_2}{\tr_2}{\sigma_2'}{\vec\nu_2}$
    then~$\iltbigstep{\trend}{\nongarbof{\sigma_2}}{\rho_2\statesep\rcmd_2}{\tr_2}{\nongarbof{\sigma_2'}}{\vec\nu_2}$

  \item
    If~$\iltbigprop{\tr_1}{\sigma_2}{\tr_2}{\sigma_2'}{\vec\nu_2}$
    then~$\iltbigstep{\trend}{\nongarbof{\sigma_2}}{\rho_1\statesep\rcmd_1}{\tr_2}{\nongarbof{\sigma_2'}}{\vec\nu_2}$
  \end{enumerate}
\end{cor}
\begin{proof}
  Immediate corollary of Theorem~\ref{thm:consistency}
\end{proof}

  \clearpage
  \section{Proofs for DPS Conversion}

In this section, let $\DPSfun{f}{x}$ denote the auxiliary function
that is used in the DPS translation of a \kw{push} command (where $n =
\arityof{f}$):
$$
\DPSfun{f}{x} = \begin{aligned}[t]
  &(\kw{fun}~f'(y). \ilwake{}
  \\[-1mm]
  &~~~~\illet{\ilvar{y_1}{\ilread{y}{1}}} {~\cdots}
  \\[-1mm]
  &~~~~\illet{\ilvar{y_n}{\ilread{y}{n}}} {}
  \\[-1mm]
  &~~~~\ilapp{f}{y_1, \ldots, y_n, x})
\end{aligned}
$$

Furthermore, we write $\FreeLabels{X}$ to denote the set of function
names free in the syntactic object $X$.  %

\subsection{DPS Conversion Preserves Extensional Semantics}
\begin{defn}
  $$
  \frac{}{\staterel{\emptyenv}{\epsilon \mapsto \epsilon}{\emptyenv}}
  $$
  $$
  \frac{
    \staterel{\kappa_1}{\vec x \mapsto \vec \ell}{\kappa_2}
    \qquad
    \DPSenv{\rho_f} \subseteq \rho_f' \land \rho_f'(x_f) = \ell_f \land \rho_f'(f') = \DPSfun{f}{x_f}
  }{
    \staterel{\stcons{\kappa_1}{\mkstframe{\rho_f}{f}}}{\vec x \append x_f \mapsto \vec \ell \append \ell_f}{\stcons{\kappa_2}{\mkstframe{\rho_f'}{f'}}}
  }
  $$
\end{defn}

\begin{thm}\label{thm:dps-extensional}
  If
  \begin{itemize}
  \item $\ilustepm{\sigma_1}{\kappa_1}{\rho_1 \statesep e}{\sigma_1'}{\emptyenv}{\emptyenv \statesep \vec \nu}$
  \item $\domof{\sigma_2} = \{\vec \ell, \ell\}$
  \item $\vec \ell, \ell \notin \domof{\sigma_1'}$
  \item $\staterel{\kappa_1}{\vec x \mapsto \vec \ell}{\kappa_2}$
  \item $\DPSenv{\rho_1} \subseteq \rho_2$
  \item $\rho_2(x) = \ell$
  \end{itemize}
  then $\ilustepm{\sigma_1 \uplus \sigma_2}{\kappa_2}{\rho_2 \statesep
    \DPSexp{e}{x}}{\sigma_1' \uplus \sigma_2'}{\emptyenv}{\emptyenv
    \statesep \ell'}$ where $\ell' = \mathsf{head}(\vec \ell \append \ell)$
  and $\sigma_2'(\ell', i) = \nu_i$ for all $i$.
\end{thm}
\begin{proof}
  By induction on the length of the reduction chain.
  \begin{itemize}
  \item Case \fbox{$e = \illet{\ilfun{f}{\vec z}{e_1}}{e_2}$}:
    \begin{itemize}
    \item Then $\ilustep{\sigma_1}{\kappa_1}{\rho_1 \statesep
        e}{\sigma_1}{\kappa_1}{\rho_1' \statesep e_2}
      \ilustepmr{\sigma_1'}{\emptyenv}{\emptyenv \statesep \vec \nu}$,
      where $\rho_1' = \rho_1[f \mapsto \ilfun{f}{\vec z}{e_1}]$.
    \item We know $\ilustep{\sigma_1 \uplus \sigma_2}{\kappa_2}{\rho_2
        \statesep \DPSexp{e}{x}}{\sigma_1 \uplus
        \sigma_2}{\kappa_2}{\rho_2' \statesep \DPSexp{e_2}{x}}$, where
      $\rho_2' = \rho_2[f \mapsto \ilfun{f}{\vec z \append
        y}{\DPSexp{e_1}{y}}]$.
    \item It is easy to see that $\DPSenv{\rho_1'} \subseteq \rho_2'$
      follows from $\DPSenv{\rho_1} \subseteq \rho_2$.
    \item The claim then follows by induction.
    \end{itemize}
  \item Case \fbox{$e = \ilif{x}{e_1}{e_2}$}:
    \begin{itemize}
    \item Suppose $\rho_1(x) = 0$ (the other case is analogous).
    \item Then $\ilustep{\sigma_1}{\kappa_1}{\rho_1 \statesep
        e}{\sigma_1}{\kappa_1}{\rho_1 \statesep e_1}
      \ilustepmr{\sigma_1'}{\emptyenv}{\emptyenv \statesep \vec \nu}$.
    \item $\DPSenv{\rho_1} \subseteq \rho_2$ implies $\rho_2(x) = 0$.
    \item Hence we know $\ilustep{\sigma_1 \uplus
        \sigma_2}{\kappa_2}{\rho_2 \statesep \DPSexp{e}{x}}{\sigma_1
        \uplus \sigma_2}{\kappa_2}{\rho_2 \statesep \DPSexp{e_1}{x}}$.
    \item The claim then follows by induction.
    \end{itemize}
  \item Case \fbox{$e = \ilapp{f}{\vec z}$}:
    \begin{itemize}
    \item Then $\ilustep{\sigma_1}{\kappa_1}{\rho_1 \statesep
        e}{\sigma_1}{\kappa_1}{\rho_1' \statesep e_f}
      \ilustepmr{\sigma_1'}{\emptyenv}{\emptyenv \statesep \vec \nu}$,
      where $\rho_1' = \rho_1\fromto{[y_i \mapsto
        \rho_1(z_i)]}{i=1}{\mathsf{length}(\vec z)}$ and $\rho_1(f) =
      \ilfun{f}{\vec y}{e_f}$.
    \item From $\DPSenv{\rho_1} \subseteq \rho_2$ we know $\rho_2(f) =
      \ilfun{f}{\vec y \append x}{\DPSexp{e_f}{x}}$.
    \item Hence $\ilustep{\sigma_1 \uplus \sigma_2}{\kappa_2}{\rho_2
        \statesep \DPSexp{e}{x}}{\sigma_1 \uplus
        \sigma_2}{\kappa_2}{\rho_2' \statesep{\DPSexp{e_f}{x}}}$,
      where $\rho_2' = \rho_2\fromto{[y_i \mapsto
        \rho_2(z_i)]}{i=1}{\mathsf{length}(\vec z)}$.
    \item It is easy to see that $\DPSenv{\rho_1'} \subseteq \rho_2'$
      follows from $\DPSenv{\rho_1} \subseteq \rho_2$.
    \item The claim then follows by induction.
    \end{itemize}
  \item Case \fbox{$e = \illet{\ilvar{y}{\iota}}{e'}$}:
    \begin{itemize}
    \item Then $\ilustep{\sigma_1}{\kappa_1}{\rho_1 \statesep
        e}{\sigma_1''}{\kappa_1}{\rho_1' \statesep e'}
      \ilustepmr{\sigma_1'}{\emptyenv}{\emptyenv \statesep \vec \nu}$,
      where $\rho_1' = \rho_1[y \mapsto \nu']$ and
      $\storestin{\sigma_1}{\rho_1}{\iota}\storestep\storestout{\sigma_1''}{\nu'}$.
    \item Since $\domof{\sigma_2} = \{\vec{\ell},\ell\}$ and
      $\vec{\ell},\ell \notin \domof{\sigma_1'} \supseteq
      \domof{\sigma_1''}$ and $\DPSenv{\rho_1} \subseteq \rho_2$, we
      get $\storestin{\sigma_1 \uplus
        \sigma_2}{\rho_2}{\iota}\storestep\storestout{\sigma_1''
        \uplus \sigma_2}{\nu'}$.
    \item Hence we know $\ilustep{\sigma_1 \uplus
        \sigma_2}{\kappa_2}{\rho_2 \statesep \DPSexp{e}{x}}{\sigma_1''
        \uplus \sigma_2}{\kappa_2}{\rho_2' \statesep \DPSexp{e'}{x}}$,
      where $\rho_2' = \rho_2[y \mapsto \nu']$.
    \item It is easy to see that $\DPSenv{\rho_1'} \subseteq \rho_2'$
      follows from $\DPSenv{\rho_1} \subseteq \rho_2$.
    \item The claim then follows by induction.
    \end{itemize}
  \item Case \fbox{$e = \ilpush{f}{e'}$}:
    \begin{itemize}
    \item Then $\ilustep{\sigma_1}{\kappa_1}{\rho_1 \statesep
        e}{\sigma_1}{\kappa_1'}{\rho_1 \statesep e'}
      \ilustepmr{\sigma_1'}{\emptyenv}{\emptyenv \statesep \vec \nu}$,
      where $\kappa_1' = \stcons{\kappa_1}{\mkstframe{\rho_1}{f}}$.
    \item We know $\ilustepm{\sigma_2}{\kappa_2}{\rho_2 \statesep
        \DPSexp{e}{x}}{\sigma_2'}{\kappa_2'}{\rho_2'
        \statesep{\DPSexp{e'}{x'}}}$, where
      \begin{itemize}
      \item $\sigma_2' = \sigma_2\fromto{[(\ell',i) \mapsto
          \bot]}{i=1}{n}$, so $\domof{\sigma_2'} = \{\vec \ell, \ell,
        \ell'\}$
      \item $\ell' \notin \domof{\sigma_2} \cup \domof{\sigma_1'}$
      \item $\kappa_2' = \stcons{\kappa_2}{\mkstframe{\rho_f'}{f'}}$
      \item $\rho_f' = \rho_2[f' \mapsto \DPSfun{f}{x}]$
      \item $\rho_2' = \rho_f'[x' \mapsto \ell']$
      \end{itemize}
    \item We show $\staterel{\kappa_1'}{\vec x \append x \mapsto \vec
        \ell \append \ell}{\kappa_2'}$:
      \begin{itemize}
      \item $\staterel{\kappa_1}{\vec x \mapsto \vec \ell}{\kappa_2}$
        is given.
      \item $\DPSenv{\rho_1} \subseteq \rho_f'$ follows from
        $\DPSenv{\rho_1} \subseteq \rho_2$.
      \item $\rho_f'(x) = \ell$ follows from $\rho_2(x) = \ell$.
      \item $\rho_f'(f') = \DPSfun{f}{x}$ is obvious.
      \end{itemize}
    \item Also, $\DPSenv{\rho_1} \subseteq \rho_2'$ follows from
      $\DPSenv{\rho_1} \subseteq \rho_2$
    \item Finally, $\rho_2'(x') = \ell'$.
    \item The claim then follows by induction (note that
      $\mathsf{head}(\vec \ell \append \ell) = \mathsf{head}(\vec \ell
      \append \ell \append \ell')$).
    \end{itemize}
  \item Case \fbox{$e = \ilpop{\vec z}$} and $\kappa_1 = \emptyenv$:
    \begin{itemize}
    \item Then $\ilustep{\sigma_1}{\kappa_1}{\rho_1 \statesep
        e}{\sigma_1}{\emptyenv}{\emptyenv \statesep \vec \nu}$ and
      $\sigma_1' = \sigma_1$ and $\nu_i = \rho_1(z_i)$ for all $i$.
    \item From $\staterel{\kappa_1}{\vec x \mapsto \vec
        \ell}{\kappa_2}$ we get $\kappa_2 = \emptyenv$ and $\vec \ell
      = \emptyenv$.
    \item Thus we know $\ilustepm{\sigma_1 \uplus
        \sigma_2}{\kappa_2}{\rho_2 \statesep \DPSexp{e}{x}}{\sigma_1
        \uplus \sigma_2'}{\emptyenv}{\emptyenv \statesep \ell}$, where
      $\sigma_2' = \sigma_2\fromto{[(\ell,i) \mapsto
        \rho_2(z_i)]}{i=1}{\mathsf{length}(\vec z)}$.
    \item Note that $\sigma_2'(\ell,i) = \rho_2(z_i) = \rho_1(z_i) =
      \nu_i$, for any $i$.
    \item Finally, note that $\ell = \mathsf{head}(\ell) =
      \mathsf{head}(\vec \ell \append \ell)$.
    \end{itemize}
  \item Case \fbox{$e = \ilpop{\vec z}$} and $\kappa_1 =
    \stcons{\kappa_1'}{\mkstframe{\rho_f}{f}}$:
    \begin{itemize}
    \item Then $\ilustepm{\sigma_1}{\kappa_1}{\rho_1 \statesep
        e}{\sigma_1}{\kappa_1'}{\rho_1' \statesep e_f}
      \ilustepmr{\sigma_1'}{\emptyenv}{\emptyenv \statesep \vec \nu}$,
      where $\rho_1' = \rho_f\fromto{[y_i \mapsto
        \rho_1(z_i)]}{i=1}{\mathsf{length}(\vec z)}$ and $\rho_f(f) =
      \ilfun{f}{\vec y}{e_f}$.
    \item From $\staterel{\kappa_1}{\vec x \mapsto \vec
        \ell}{\kappa_2}$ we know
      \begin{itemize}
      \item $\kappa_2 = \stcons{\kappa_2'}{\mkstframe{\rho_f'}{f'}}$
      \item $\vec x = \vec x' \append x_f$ and $\vec \ell = \vec \ell'
        \append \ell_f$
      \item $\DPSenv{\rho_f} \subseteq \rho_f' \land \rho_f'(x_f) = \ell_f$
      \item $\rho_f'(f') = \DPSfun{f}{x_f}$
      \end{itemize}
    \item Therefore $\ilustepm{\sigma_1 \uplus
        \sigma_2}{\kappa_2}{\rho_2 \statesep \DPSexp{e}{x}}{\sigma_1
        \uplus \sigma_2'}{\kappa_2}{\emptyenv \statesep \ell}$, where
      $\sigma_2' = \sigma_2\fromto{[(\ell,i) \mapsto
        \rho_2(z_i)]}{i=1}{\mathsf{length}(\vec z)}$.
    \item And $\ilustepm{\sigma_1 \uplus
        \sigma_2'}{\kappa_2}{\emptyenv \statesep \ell}{\sigma_1 \uplus
        \sigma_2'}{\kappa_2'}{\rho_2' \statesep
        \ilapp{f}{y_1,\dots,y_n,x_f}}$, where $\rho_2' = \rho_f'[y
      \mapsto \ell]\fromto{[y_i \mapsto
        \rho_2(z_i)]}{i=1}{\mathsf{length}(\vec z)}$.
    \item And $\ilustepm{\sigma_1 \uplus \sigma_2'}{\kappa_2'}{\rho_2'
        \statesep \ilapp{f}{y_1,\dots,y_n,x_f}}{\sigma_1 \uplus
        \sigma_2'}{\kappa_2'}{\rho_2' \statesep \DPSexp{e_f}{x_f}}$.
    \item Note that $\DPSenv{\rho_1'} \subseteq \rho_2'$ follows from
      $\DPSenv{\rho_f} \subseteq \rho_f'$ and $\DPSenv{\rho_1}
      \subseteq \rho_2$.
    \item The claim then follows by induction.
    \end{itemize}
  \end{itemize}
\end{proof}

\begin{cor}
  If $\ilustepm{\sigma_1}{\emptyenv}{\rho \statesep
    e}{\sigma_1'}{\emptyenv}{\emptyenv \statesep \vec \nu}$, then
  $\ilustepm{\sigma_1}{\emptyenv}{\DPSenv{\rho} \statesep
    \illet{\ilvar{x}{\ilalloc{n}}}{\DPSexp{e}{x}}}{\sigma_1' \uplus
    \sigma_2'}{\emptyenv}{\emptyenv \statesep \ell}$ with
  $\sigma_2'(\ell, i) = \nu_i$ for all $i$.
\end{cor}

\subsection{DPS Conversion Produces CSA Programs}

\begin{defn}
  $$
  \isdpsenv{\rho}{\rho'} \iff \DPSenv{\rho'} \subseteq \rho \land \forall
  f \in \domof{\rho}.\ \FreeLabels{\rho(f)} \subseteq \domof{\rho'}
  $$
\end{defn}

\begin{defn}
  $$
  \frac{
  }{
    \emptyenv \heartsuit
  }
  $$
  $$
  \frac{
    \kappa \heartsuit \qquad
    \rho_f'(x_f) = \ell_f \land \rho_f'(f') = \DPSfun{f}{x_f} \land \exists \rho_1.\
    \isdpsenv{\rho_f'}{\rho_1}
  }{
    \stcons{\kappa}{\mkstframe{\rho_f'}{f'}} \heartsuit
  }
  $$
\end{defn}

\begin{lem}
  \label{lem:dps-to-update}
  If
  \begin{enumerate}
  \item $\ilustepm{\sigma}{\kappa}{\rho \statesep
    \DPSexp{e}{x}}{\sigma'}{\kappa'}{\rho' \statesep \ilupdate{e'}}$
  \item $\exists \rho_1.\ \isdpsenv{\rho}{\rho_1} \land \FreeLabels{\DPSexp{e}{x}} \subseteq \domof{\rho_1}$
  \item $\kappa \heartsuit$
  \item $\rho(x) = \ell$
  \end{enumerate}
  then:
  \begin{itemize}
  \item $e' = \DPSexp{e''}{y}$
  \item $\rho'(y) = \ell'$
  \item $\exists \rho_2.\ \isdpsenv{\rho'}{\rho_2} \land \FreeLabels{\DPSexp{e''}{y}} \subseteq \domof{\rho_2}$
  \end{itemize}
\end{lem}
\begin{proof}
  By induction on the length of the reduction chain in (1).
  \begin{itemize}
  \item Case \fbox{$e = \illet{\ilfun{f}{\vec z}{e_1}}{e_2}$}:
    \begin{itemize}
    \item Then $\ilustep{\sigma}{\kappa}{\rho \statesep
        \DPSexp{e}{x}}{\sigma}{\kappa}{\hat{\rho} \statesep
        \DPSexp{e_2}{x}} \ilustepmr{\sigma'}{\kappa'}{\rho' \statesep
        \ilupdate{e'}}$, where $\hat{\rho} = \rho[f \mapsto
      \ilfun{f}{\vec z \append y}{\DPSexp{e_1}{y}}]$.
    \item Note that (2) has been preserved.
    \item The claim then follows by induction.
    \end{itemize}
  \item Case \fbox{$e = \ilif{x}{e_1}{e_2}$}:
    \begin{itemize}
    \item Suppose $\rho(x) = 0$ (the other case is analogous).
    \item Then $\ilustep{\sigma}{\kappa}{\rho \statesep
        \DPSexp{e}{x}}{\sigma}{\kappa}{\rho \statesep \DPSexp{e_1}{x}}
      \ilustepmr{\sigma'}{\kappa'}{\rho' \statesep \ilupdate{e'}}$.
    \item Note that (2) has been preserved.
    \item The claim then follows by induction.
    \end{itemize}
  \item Case \fbox{$e = \ilapp{f}{\vec z}$}:
    \begin{itemize}
    \item From (2) we know $\rho(f) = \ilfun{f}{\vec y \append
        x}{\DPSexp{e_f}{x}}$.
    \item Hence $\ilustep{\sigma}{\kappa}{\rho \statesep
        \DPSexp{e}{x}}{\sigma}{\kappa}{\hat{\rho} \statesep
        \DPSexp{e_f}{x}} \ilustepmr{\sigma'}{\kappa'}{\rho' \statesep
        e'}$, where $\hat{\rho} = \rho\fromto{[y_i \mapsto
        \rho(z_i)]}{i=1}{\mathsf{length}(\vec z)}$. %
    \item Note that (2) has been preserved.
    \item The claim then follows by induction.
    \end{itemize}
  \item Case \fbox{$e = \illet{\ilvar{z}{\iota}}{\hat{e}}$}:
    \begin{itemize}
    \item Then $\ilustep{\sigma}{\kappa}{\rho \statesep
        \DPSexp{e}{x}}{\hat{\sigma}}{\kappa}{\hat{\rho} \statesep
        \DPSexp{\hat{e}}{x}} \ilustepmr{\sigma'}{\kappa'}{\rho'
        \statesep \ilupdate{e'}}$, where $\hat{\rho} = \rho[z \mapsto
      \nu]$.
    \item Note that (2) has been preserved.
    \item The claim then follows by induction.
    \end{itemize}
  \item Case \fbox{$e = \ilmemo{\hat{e}}$}:
    \begin{itemize}
    \item Then $\ilustep{\sigma}{\kappa}{\rho \statesep
        \DPSexp{e}{x}}{\sigma}{\kappa}{\rho \statesep
        \DPSexp{\hat{e}}{x}} \ilustepmr{\sigma'}{\kappa'}{\rho'
        \statesep \ilupdate{e'}}$.
    \item Note that (2) has been preserved.
    \item The claim then follows by induction.
    \end{itemize}
  \item Case \fbox{$e = \ilwake{\hat{e}}$}:
    \begin{itemize}
    \item If the length of the reduction is $0$, then $e' =
      \DPSexp{\hat{e}}{x}$ and we are done.
    \item Otherwise we know $\ilustep{\sigma}{\kappa}{\rho \statesep
        \DPSexp{e}{x}}{\sigma}{\kappa}{\rho \statesep
        \DPSexp{\hat{e}}{x}} \ilustepmr{\sigma'}{\kappa'}{\rho'
        \statesep \ilupdate{e'}}$.
    \item Note that (2) has been preserved.
    \item The claim then follows by induction.
    \end{itemize}
  \item Case \fbox{$e = \ilpush{f}{\hat{e}}$}:
    \begin{itemize}
    \item Then $\ilustepm{\sigma}{\kappa}{\rho \statesep
        \DPSexp{e}{x}}{\sigma}{\kappa}{\rho_f' \statesep
        \ilpush{f'}{\ilmemo{\illet{\ilvar{z}{\ilalloc{n}}}{\DPSexp{\hat{e}}{z}}}}}$,
      where $\rho_f' = \rho[f' \mapsto \DPSfun{f}{x}]$.
    \item Now $\ilustepm{\sigma}{\kappa}{\rho_f' \statesep
        \ilpush{f'}{\ilmemo{\illet{\ilvar{z}{\ilalloc{n}}}{\DPSexp{\hat{e}}{z}}}}}{\tilde{\sigma}}{\tilde{\kappa}}{\tilde{\rho}
        \statesep \DPSexp{\hat{e}}{z}}$, where:
      \begin{itemize}
      \item $\tilde{\kappa} = \stcons{\kappa}{\mkstframe{\rho_f'}{f'}}$
      \item $\tilde{\rho} = \rho_f'[z \mapsto \ell']$
      \end{itemize}
    \item Note that $\DPSenv{\rho_1} \subseteq \tilde{\rho} \land
      \FreeLabels{\DPSexp{\hat{e}}{z}} \subseteq \domof{\rho_1} \land
      \forall f \in \domof{\tilde{\rho}}.\
      \FreeLabels{\tilde{\rho}(f)} \subseteq \domof{\rho_1}$ follows
      from (2).
    \item Furthermore we know
      $\ilustepm{\tilde{\sigma}}{\tilde{\kappa}}{\tilde{\rho}
        \statesep \DPSexp{\hat{e}}{z}}{\sigma'}{\kappa'}{\rho'
        \statesep \ilupdate{e'}}$.
    \item The claim thus follows by induction if we can show
      $\tilde{\kappa} \heartsuit$.
    \item And yes, we can!
    \end{itemize}
  \item Case \fbox{$e = \ilpop{\vec z}$} and $\kappa = \emptyenv$:
    impossible due to (1)
  \item Case \fbox{$e = \ilpop{\vec z}$} and $\kappa =
    \stcons{\tilde{\kappa}}{\mkstframe{\rho_f'}{f'}}$:
    \begin{itemize}
    \item From $\kappa \heartsuit$ we know:
      \begin{enumerate}
      \item $\DPSenv{\rho_2} \subseteq \rho_f' \land
        \forall g \in \domof{\rho_f'}.\ \FreeLabels{\rho_f'(g)}
        \subseteq \domof{\rho_2}$
      \item $\rho_f'(f') = \DPSfun{f}{x_f}$
      \item $\tilde{\kappa} \heartsuit$
      \end{enumerate}
    \item Hence we know $\rho_f'(f) = \ilfun{f}{\vec y \append
        x_f}{\DPSexp{e_f}{x_f}}$.
    \item So $\ilustepm{\sigma}{\kappa}{\rho \statesep
        \DPSexp{e}{x}}{\tilde{\sigma}}{\tilde{\kappa}}{\tilde{\rho}
        \statesep \ilapp{f}{y_1,\dots,y_n,x_f}}
      \ilustepmr{\tilde{\sigma}}{\tilde{\kappa}}{\tilde{\rho}
        \statesep \DPSexp{e_f}{x_f}}$, where:
      \begin{itemize}
      \item $\tilde{\rho} = \rho_f'[y \mapsto \ell]\fromto{[y_i \mapsto
          \rho(z_i)]}{i=1}{\mathsf{length}(\vec z)}$
      \end{itemize}
    \item Note that $\DPSenv{\rho_2} \subseteq \rho_f' \land \forall g
      \in \domof{\rho_f'}.\ \FreeLabels{\rho_f'(g)} \subseteq
      \domof{\rho_2}$ implies $\DPSenv{\rho_2} \subseteq \tilde{\rho}
      \land \FreeLabels{\DPSexp{e_f}{x_f}} \subseteq \domof{\rho_2}
      \land \forall f \in \domof{\tilde{\rho}}.\
      \FreeLabels{\tilde{\rho}(f)} \subseteq \domof{\rho_2}$.
    \item Furthermore we know
      $\ilustepm{\tilde{\sigma}}{\tilde{\kappa}}{\tilde{\rho}
        \statesep \DPSexp{e_f}{x_f}}{\sigma'}{\kappa'}{\rho' \statesep
        \ilupdate{e'}}$ and the claim thus follows by induction.
    \end{itemize}
  \end{itemize}
\end{proof}

\begin{defn}
  $$
  \frac{
  }{
    \emptyenv \triangleright^\ell \ell
  }
  $$
  $$
  \frac{
    \kappa \triangleright^{\ell_f} \ell'
    \qquad
    \rho_f'(x_f) = \ell_f \land \rho_f'(f') = \DPSfun{f}{x_f} \land
 \exists \rho_1.\ \isdpsenv{\rho_f'}{\rho_1}
  }{
    \stcons{\kappa}{\mkstframe{\rho_f'}{f'}} \triangleright^\ell \ell'
  }
  $$
\end{defn}

\begin{lem}\label{lem:dps-to-value}
  If 
  \begin{enumerate}
  \item $\kappa \triangleright^\ell \ell'$
  \item $\rho(x) = \ell$
  \item $\exists \rho_1.\ \isdpsenv{\rho}{\rho_1} \land
    \FreeLabels{\DPSexp{e}{x}} \subseteq \domof{\rho_1}$
  \item $\ilustepm{\sigma}{\kappa}{\rho \statesep
      \DPSexp{e}{x}}{\sigma'}{\emptyenv}{\emptyenv \statesep \vec \nu}$
  \end{enumerate}
  then $\vec \nu = \ell'$.
\end{lem}
\begin{proof}
  By induction on the length of the reduction chain.
  \begin{itemize}
  \item Case \fbox{$e = \illet{\ilfun{f}{\vec z}{e_1}}{e_2}$}:
    \begin{itemize}
    \item Then $\ilustep{\sigma}{\kappa}{\rho \statesep
        \DPSexp{e}{x}}{\sigma}{\kappa}{\rho' \statesep
        \DPSexp{e_2}{x}} \ilustepmr{\_}{\emptyenv}{\emptyenv \statesep
        \vec \nu}$, where $\rho' = \rho[f \mapsto \ilfun{f}{\vec z
        \append y}{\DPSexp{e_1}{y}}]$.
    \item Note that (3) has been preserved.
    \item The claim then follows by induction.
    \end{itemize}
  \item Case \fbox{$e = \ilif{x}{e_1}{e_2}$}:
    \begin{itemize}
    \item Suppose $\rho(x) = 0$ (the other case is analogous).
    \item Then $\ilustep{\sigma}{\kappa}{\rho \statesep
      \DPSexp{e}{x}}{\sigma}{\kappa}{\rho \statesep \DPSexp{e_1}{x}}
      \ilustepmr{\sigma'}{\emptyenv}{\emptyenv \statesep \vec \nu}$.
    \item The claim then follows by induction.
    \end{itemize}
  \item Case \fbox{$e = \ilapp{f}{\vec z}$}:
    \begin{itemize}
    \item From (3) we know $\rho(f) = \ilfun{f}{\vec y \append
        x'}{\DPSexp{e_f}{x'}}$.
    \item Hence $\ilustep{\sigma}{\kappa}{\rho \statesep
      \DPSexp{e}{x}}{\sigma}{\kappa}{\rho' \statesep \DPSexp{e_f}{x}}
      \ilustepmr{\sigma'}{\emptyenv}{\emptyenv \statesep \vec \nu}$,
      where $\rho' = \rho\fromto{[y_i \mapsto
          \rho(z_i)]}{i=1}{\mathsf{length}(\vec z)}$. %
    \item Note that (3) has been preserved.
    \item The claim then follows by induction.
    \end{itemize}
  \item Case \fbox{$e = \illet{\ilvar{y}{\iota}}{e'}$}:
    \begin{itemize}
    \item Then $\ilustep{\sigma}{\kappa}{\rho \statesep
      \DPSexp{e}{x}}{\sigma'}{\kappa}{\rho' \statesep \DPSexp{e'}{x}}
      \ilustepmr{\sigma'}{\emptyenv}{\emptyenv \statesep \vec \nu}$,
      where $\rho' = \rho[y \mapsto \nu']$.
    \item Note that (3) has been preserved.
    \item The claim then follows by induction.
    \end{itemize}
  \item Case \fbox{$e = \ilpush{f}{e'}$}:
    \begin{itemize}
    \item Then $\ilustepm{\sigma}{\kappa}{\rho \statesep
      \DPSexp{e}{x}}{\sigma''}{\kappa'}{\rho' \statesep
      \DPSexp{e'}{x_f}} \ilustepmr{\sigma'}{\emptyenv}{\emptyenv
      \statesep \vec \nu}$, where
      \begin{itemize}
      \item $\kappa' = \stcons{\kappa}{\mkstframe{\rho_f'}{f'}}$
      \item $\rho_f' = \rho[f' \mapsto \DPSfun{f}{x}]$
      \item $\rho' = \rho_f'[x_f \mapsto \ell_f]$
      \end{itemize}
    \item We show $\kappa' \triangleright^{\ell_f} \ell'$:
      \begin{itemize}
      \item $\exists \rho_1.\ \DPSenv{\rho_1} \subseteq \rho_f' \land
        \forall g \in \domof{\rho_f'}.\ \FreeLabels{\rho_f'(g)}
        \subseteq \domof{\rho_1}$ follows from (3).
      \item $\rho_f'(f') = \DPSfun{f}{x}$ is obvious.
      \item $\rho_f'(x) = \ell$ follows from $\rho(x) = \ell$. %
      \item $\kappa \triangleright^{\ell} \ell'$ is given.
      \end{itemize}
    \item Note that (3) has been preserved.
    \item The claim then follows by induction.
    \end{itemize}
  \item Case \fbox{$e = \ilpop{\vec z}$} and $\kappa = \emptyenv$:
    \begin{itemize}
    \item Then $\ilustepm{\sigma}{\kappa}{\rho \statesep
      \DPSexp{e}{x}}{\sigma'}{\emptyenv}{\emptyenv \statesep \ell}$
      and thus $\vec \nu = \ell$.
    \item From $\kappa \triangleright^\ell \ell'$ we know $\ell =
      \ell'$.
    \end{itemize}
  \item Case \fbox{$e = \ilpop{\vec z}$} and $\kappa =
    \stcons{\kappa'}{\mkstframe{\rho_f'}{f'}}$:
    \begin{itemize}
    \item From $\kappa \triangleright^\ell \ell'$ we know
      \begin{enumerate}
      \item $\exists \rho_1.\ \DPSenv{\rho_1} \subseteq \rho_f' \land \forall g \in \domof{\rho_f'}.\ \FreeLabels{\rho_f'(g)} \subseteq \domof{\rho_1}$
      \item $\rho_f'(f') = \DPSfun{f}{x_f}$
      \item $\rho_f'(x_f) = \ell_f$
      \item $\kappa' \triangleright^{\ell_f} \ell'$
      \end{enumerate}
    \item So $\ilustepm{\sigma}{\kappa}{\rho \statesep
      \DPSexp{e}{x}}{\sigma''}{\kappa}{\emptyenv \statesep \ell}
      \ilustepmr{\sigma''}{\kappa'}{\rho' \statesep
        \ilapp{f}{y_1,\dots,y_n,x_f}}$, where $\rho' = \rho_f'[y
        \mapsto \ell]\fromto{[y_i \mapsto
          \rho(z_i)]}{i=1}{\mathsf{length}(\vec z)}$.
    \item From (2) and (1) we know $\rho'(f) = \ilfun{f}{\vec y
      \append x_f}{\DPSexp{e_f}{x_f}}$.
    \item Thus $\ilustep{\sigma''}{\kappa'}{\rho' \statesep
      \ilapp{f}{y_1,\dots,y_n,x_f}}{\sigma''}{\kappa'}{\rho' \statesep
      \DPSexp{e_f}{x_f}} \ilustepmr{\sigma'}{\emptyenv}{\emptyenv
      \statesep \vec \nu}$.
    \item The claim then follows by induction.
    \end{itemize}
  \end{itemize}
\end{proof}

\begin{thm}\label{thm:dps-csa}
  $\csa{\DPSenv{\rho}}{\illet{\ilvar{x}{\ilalloc{n}}}{\DPSexp{e}{x}}}{\sigma}$
\end{thm}
\begin{proof}~ Suppose
  $\ilustate{\sigma}{\emptystack}{\DPSenv{\rho}}{\illet{\ilvar{x}{\ilalloc{n}}}{\DPSexp{e}{x}}}
  \ilusteprel^m \ilustate{\sigma'}{\kappa}{\rho'}{e'}$.  We must show
  $\sa{\rho'}{e'}{\sigma'}$.  We distinguish two cases:
  \begin{itemize}
  \item Case $m = 0$:
    \begin{itemize}
    \item So suppose
      $\ilustate{\sigma'}{\emptystack}{\DPSenv{\rho}}{\illet{\ilvar{x}{\ilalloc{n}}}{\DPSexp{e}{x}}}
      \ilusteprel^\ast
      \ilustate{\sigma''}{\kappa'}{\rho''}{\ilupdate{e''}}$.
    \item Hence $\ilustepm{\sigma'\fromto{[(\ell,i) \mapsto
          \bot]}{i=1}{n}}{\emptystack}{\DPSenv{\rho}[x \mapsto \ell]
        \statesep \DPSexp{e}{x}}{\sigma''}{\kappa'}{\rho'' \statesep
        \ilupdate{e''}}$.
    \item Since $\emptystack \heartsuit$,
      Lemma~\ref{lem:dps-to-update} yields:
      \begin{itemize}
      \item $e'' = \DPSexp{\widehat{e}}{y}$
      \item $\rho''(y) = \ell'$
      \item $\DPSenv{\rho_2} \subseteq \rho'' \land
        \FreeLabels{\DPSexp{\widehat{e}}{y}} \subseteq \domof{\rho_2}
        \land \forall f \in \domof{\rho''}.\ \FreeLabels{\rho''(f)}
        \subseteq \domof{\rho_2}$
      \end{itemize}
    \item Now suppose
      $\ilustate{\_}{\emptystack}{\rho''}{\DPSexp{\widehat{e}}{y}}
      \ilusteprel^\ast
      \ilustate{\_}{\emptystack}{\emptyenv}{\vec{\nu}}$.
    \item Since $\emptystack \triangleright^{\ell'} \ell'$,
      Lemma~\ref{lem:dps-to-value} yields $\vec{\nu} = \ell'$.
    \end{itemize}
  \item Case $m > 0$:
    \begin{itemize}
    \item Then $\ilustate{\sigma\fromto{[(\ell,i) \mapsto
          \bot]}{i=1}{n}}{\emptystack}{\DPSenv{\rho}[x \mapsto
        \ell]}{\DPSexp{e}{x}} \ilusteprel^{m-1}
      \ilustate{\sigma'}{\kappa}{\rho'}{e'}$.
    \item So suppose $\ilustate{\sigma'}{\emptystack}{\rho'}{e'}
      \ilusteprel^\ast
      \ilustate{\sigma''}{\kappa'}{\rho''}{\ilupdate{e''}}$.
    \item By Lemma~\ref{lem:extended-stack},
      $\ilustate{\sigma'}{\kappa}{\rho'}{e'} \ilusteprel^\ast
      \ilustate{\sigma''}{\kappa @ \kappa'}{\rho''}{\ilupdate{e''}}$.
    \item Hence $\ilustate{\sigma\fromto{[(\ell,i) \mapsto
          \bot]}{i=1}{n}}{\emptystack}{\DPSenv{\rho}[x \mapsto
        \ell]}{\DPSexp{e}{x}} \ilusteprel^\ast
      \ilustate{\sigma''}{\kappa @ \kappa'}{\rho''}{\ilupdate{e''}}$.
    \item The rest goes as in the first case.
    \end{itemize}
  \end{itemize}
\end{proof}

  \clearpage
  \section{Cost Semantics Proofs}

\begin{lem}\label{lem:rewind-undomark}
  If $\trz{\Pi}{\trstart} ; {\trstart} \trrewind^\ast
  \trz{\trcons{\trx'}{\trxpushmark}}{T_2}; T_1$ and $\undomark \notin
  \Pi$, then $T_2 = \trstart$.
\end{lem}

\begin{thm}\label{thm:ilu-ilt-cost}
  If
  \begin{itemize}
  \item $\ilustepm{\sigma}{\kappa}{\rho \statesep
      e}{\_}{\emptyenv}{\emptyenv \statesep \seqof \nu_1}$, described
    by $\seqof{s_1}$
  \item $\iltstepm{\trzip{\Pi}{\trend}}{\sigma}{\kappa}{\rho \statesep
      e}{\_}{\_}{\emptyenv}{\emptyenv \statesep \seqof \nu_2}$,
    described by $\seqof{s_2}$
  \item $\propmark, \undomark \notin \Pi$
  \end{itemize}
  then:
  \begin{itemize}
  \item $\gamma_s~\seqof{s_1} = \gamma_s~\seqof{s_2}$
  \item $\gamma_\sigma~\seqof{s_1} = \gamma_\sigma~\seqof{s_2}$
  \item $\gamma_\kappa~\seqof{s_1} = \gamma_\kappa~\seqof{s_2}$
  \end{itemize}
\end{thm}
\begin{proof}
  By induction on the length of $\seqof{s_1}$.  Note that $\seqof{s_1}
  = \empseq$ is not possible, so $\seqof{s_1} = \usup{s} \cons
  \seqof{s_3}$.  We analyze $\usup{s}$ and in each case observe that
  $\seqof{s_2}$ must start with the step $\tsup{s}$ that is associated
  with the corresponding {\bf E} rule.  Note that:
  \begin{itemize}
  \item Rules {\bf E.P} and {\bf P.E} never apply because the reuse
    trace is empty.
  \item Rules {\bf P.1--7} never apply because $\ilprop$ is
    not an expression.
  \item Rule {\bf P.8} never applies because its premise would imply
    $\propmark \in \Pi$.
  \item Rules {\bf U.1--3} never apply because the reuse
    trace is empty.
  \item Rule {\bf U.4} never applies because $\undomark \notin \Pi$.
  \end{itemize}

  In each case, we find that $\tsup{s}$ has the same cost as
  $\usup{s}$ in the three models, i.e., $\gamma~\usup{s} =
  \gamma~\tsup{s}$ for $\gamma \in \{\gamma_s, \gamma_\sigma,
  \gamma_\kappa\}$.  Also, each step preserves the assumptions.  In
  particular, in rule {\bf E.8}, $T_2 = \trstart$ implies $T_2' =
  \trstart$ by Lemma~\ref{lem:rewind-undomark}.

\end{proof}

\begin{cor}
  If
  \begin{itemize}
  \item $\ilustepm{\sigma}{\emptyenv}{\rho \statesep
      e}{\_}{\emptyenv}{\emptyenv \statesep \seqof \nu_1}$, described
    by $\seqof{s_1}$
  \item $\iltstepm{\trzip{\trstart}{\trend}}{\sigma}{\emptyenv}{\rho
      \statesep e}{\_}{\_}{\emptyenv}{\emptyenv \statesep \seqof
      \nu_2}$, described by $\seqof{s_2}$
  \end{itemize}
  then:
  \begin{itemize}
  \item $\gamma_s~\seqof{s_1}~\zero_s = \gamma_s~\seqof{s_2}~\zero_s$
  \item $\gamma_\sigma~\seqof{s_1}~\zero_\sigma =
    \gamma_\sigma~\seqof{s_2}~\zero_\sigma$
  \item $\gamma_\kappa~\seqof{s_1}~\zero_\kappa =
    \gamma_\kappa~\seqof{s_2}~\zero_\kappa$
  \end{itemize}
\end{cor}

\begin{thm}
  Suppose the following:
  \begin{itemize}
  \item $\ilustepm{\sigma_1}{\emptyenv}{\rho \statesep
      e}{\sigma_1'}{\emptyenv}{\emptyenv \statesep \vec \nu}$,
    described by $\seqof{s_1}$
  \item $\ilustepm{\sigma_1}{\emptyenv}{\DPSenv{\rho} \statesep
      \illet{\ilvar{x}{\ilalloc{n}}}{\DPSexp{e}{x}}}{\sigma_1' \uplus
      \sigma_2'}{\emptyenv}{\emptyenv \statesep \ell}$, described by
    $s_{\text{alloc}} \cons \seqof{s_2}$
  \item $\tuple{u, d, \_, \_} =
    \gamma_\kappa~\seqof{s_1}~\zero_\kappa$
  \item $\tuple{a_1, r_1, w_1} =
    \gamma_\sigma~\seqof{s_1}~\zero_\sigma$
  \item $\tuple{a_2, r_2, w_2} =
    \gamma_\sigma~\seqof{s_2}~\zero_\sigma$
  \item $N$ is the maximum arity of any \kw{pop} taken in
    $\seqof{s_1}$
  \end{itemize}
  Then:
  \begin{enumerate}
  \item $\gamma_\kappa~\seqof{s_1}~\zero_\kappa =
    \gamma_\kappa~\seqof{s_2}~\zero_\kappa$
  \item $a_2 - a_1 = u$
  \item $r_2 - r_1 \leq N * d$
  \item $w_2 - w_1 \leq N * (d + 1)$
  \item $\gamma_s~\seqof{s_2}~\zero_s - \gamma_s~\seqof{s_1}~\zero_s
    \leq (2N+5) * u + N$
  \end{enumerate}
\end{thm}
\begin{proof}
  Informally, this is easy to see from the definition of DPS
  conversion as explained below.  The only interesting cases are
  \kw{push}s and \kw{pop}s.  Note that since both computations start
  and end with an empty stack, we know $u = d$.
  \begin{enumerate}
  \item Observe that the conversion preserves the number and order of
    \kw{push}s and \kw{pop}s and that both computations end in an
    empty stack.
  \item Observe that the translation of a \kw{push}
    introduces a single additional \kw{alloc}.
  \item Observe that the translation of a \kw{push} introduces a
    function containing at most $N$ additional \kw{read}s.  Each time
    the stack is popped, such a function is executed.  This happens
    $d$ times.
  \item Observe that the translation of a \kw{pop} introduces at most
    $N$ additional \kw{write}s.  We know that $d+1$ \kw{pop}s are
    executed (the last one when the stack is already empty, thereby
    terminating the program).
  \item Observe that: executing the translation of a \kw{push} takes 3
    additional steps (function definition, \kw{memo}, \kw{alloc}) to
    reach its body; executing the translation of a \kw{pop} (of which
    $d+1$ are executed) takes at most $N$ steps before actually doing
    the \kw{pop}; in the $d$ cases where the stack is popped, the
    function generated by the corresponding \kw{push} is executed,
    which takes at most $1+N+1$ steps.  In total, this adds up to $3 *
    u + N * (d+1) + (1+N+1) * d = (2N+5) * u + N$ additional steps.
  \end{enumerate}
  Formally, this can be proven by a very tedious induction, similar
  to---but much more space consuming than---the proof of
  Theorem~\ref{thm:dps-extensional}.
\end{proof}

}

\end{document}